# Collision Free Navigation of a Multi-Robot Team for Intruder Interception

by

*Ali Marzoughi*

*August 2018*

# Abstract

The ability of mobile robots to work as a team in hard and hazardous environments and consequently their widespread use in various industries is a strong incentive for researchers to develop practical algorithm and methods for increasing the performance of mobile robots. The ability of autonomous decision-making for navigation and path planning is the important problem, which has been investigated by researchers to improve the performance of a team of mobile robots in a certain mission.

The contribution of this study is classified as follows; In the first stage, we propose a decentralised motion control algorithm for the mobile robots to intercept an intruder entering (*k-intercepting*) or escaping (*e-intercepting*) a protected region. In continue, we propose a decentralized navigation strategy (*dynamic-intercepting*) for a multi-robot team known as predators to intercept the intruders or in the other words, preys, from escaping a siege ring which is created by the predators. A necessary and sufficient condition for the existence of a solution of this problem is obtained. At the second stage, we propose an intelligent game-based decision-making algorithm (IGD) for a fleet of mobile robots to maximize the probability of detection in a bounded region. We prove that the proposed decentralised cooperative and non-cooperative game-based decision-making algorithm enables each robot to make the best decision to choose the shortest path with minimum local information. Third, we propose a leader-follower based collision-free navigation control method for a fleet of mobile robots to traverse an unknown cluttered environment. Fourth, we propose a decentralised navigation algorithm for a team of multi-robot to traverse an area where occupied by multiple obstacles to trap a target. We prove that each individual team member is able to traverse safely in the region, which is cluttered by many obstacles with any shapes to trap the target while using the sensors in some indefinite switching points and not continuously, which leads to saving energy consumption and increasing the battery life of the robots consequently. And finally, we propose a novel navigation strategy for a unicycle mobile robot in a cluttered area with moving obstacles based on virtual field force algorithm. The mathematical proof of the navigation laws and the computer simulations are provided to confirm the validity, robustness, and reliability of the proposed methods.

# Contents

# Chapter 1

# Introduction

## 1.1 A brief history of mobile robots' development

Invention of the primary generation of mobile robots could be traced back to the years 1948 –1950, when the William Grey Walter (1910–1977) and his wife Vivian Dovey Walter unveiled their first tortoises Elmer and Elise, which were built in the backroom laboratory of their house [1], [2], [3], [4].

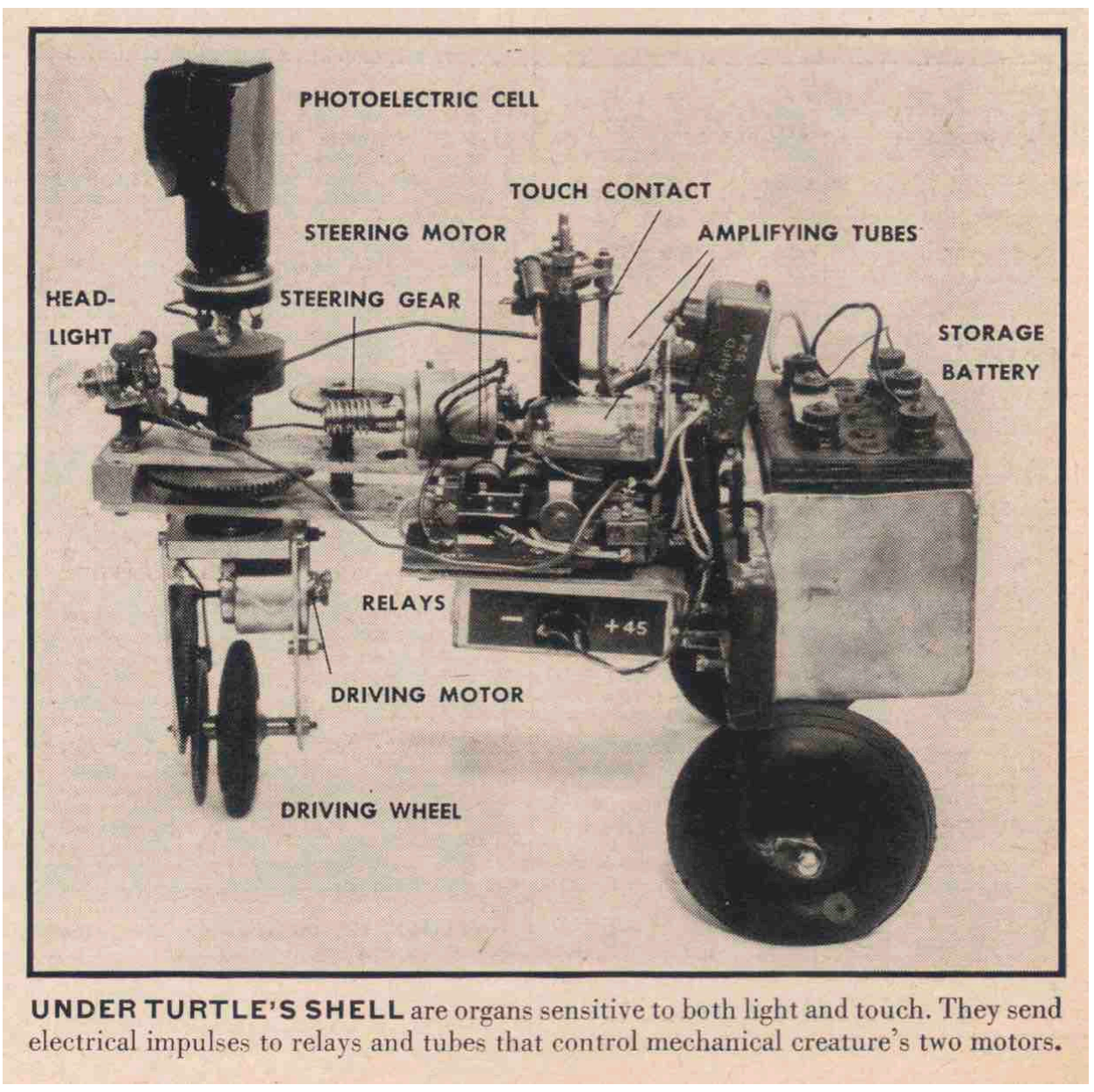

Figure 1.1: 1948 – ELSIE (Electro-mechanical robot, Light Sensitive with Internal and External stability) – W. Grey Walter (*Source: cyberneticzoo.com*)

After a decade, in early 1960s, the Johns Hopkins Beast was created at the applied physics laboratory at John Hopkins University by a group of brain researchers. The cybernetic Beast was able to wander the area to feed itself by finding a distinctive black power outlet on the wall by relying on its photocell eye and sonar sensor [5]. The robot was much more complex than its older siblings Elmer and Elise.

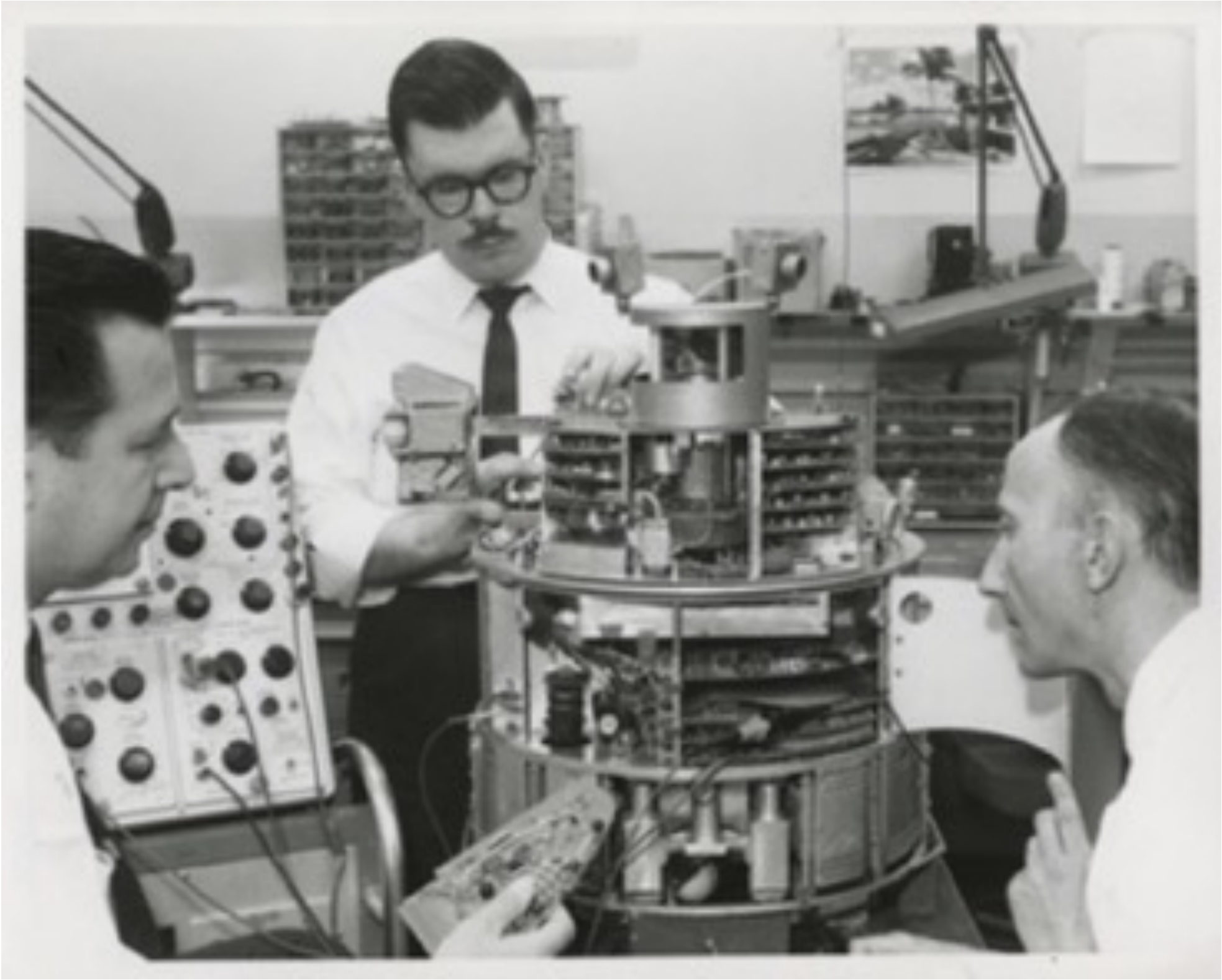

Figure 1.2: "Mar 31 1965" Photo No. 73738; UNDER THE SKIN…. Leonard Scheer, left, William Whitmore with automaton, and Dennis Walters have removed the bumper and cover from automaton to examine the complex electronics.". Photo of three men examining and working on the electronics of a robot in a laboratory/shop environment. (*Source: cyberneticzoo.com*)

The 1970s could be considered as the starting point of the significant development in design and production of complex autonomous systems. Advent of the digital control, a massive drop in the price of sensors and processors and the deeper perception of the artificial intelligence has led to an increase in the interest of research, development, and deployment of the autonomous systems in various modalities such as air, ground, sea, and space[6]. There are many examples out of which we refer to some for our assistance. Shakey is one of the most famous of them, which was built at the Stanford Research Institute (SRI). Shakey who was named after its jerky motion, known as the first robot with the ability to do more complex tasks required to be planned, navigation, and object rearrangement. It was a cutting-edge technology due to its Artificial Intelligent capabilities and its robustness in action execution [7], [8].

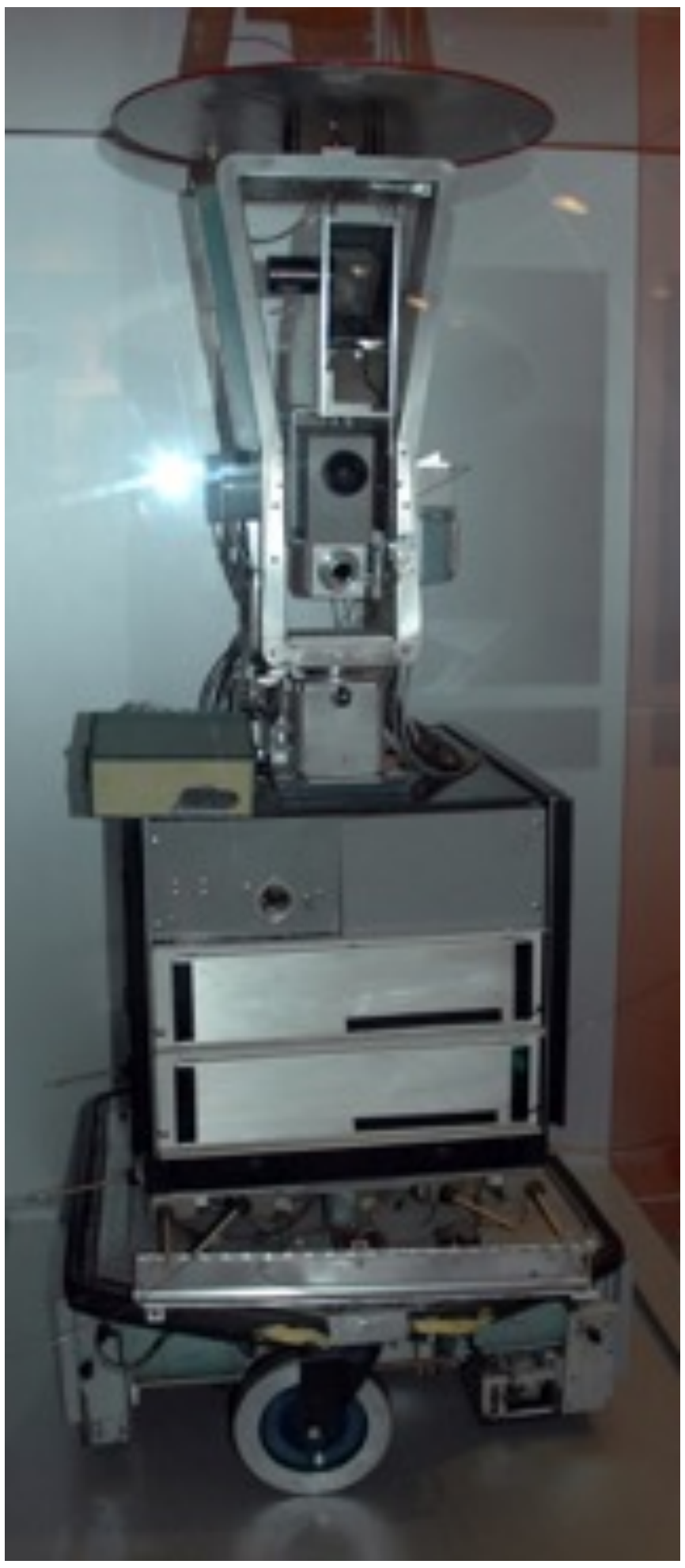

Figure 1.3: A photo of Shakey the Robot in its case at the Computer History Museum. (*Source: www.wikipedia.org*)

The advent of the Lunokhod-1, launched by the Soviet Union on November 10, 1970, as part of the Lunokhod program, surprised the world. The lunar automatic vehicle weighing 756 kg, 2.2 m long, and about 2.2 m wide was controlled by radio commands from the earth, successfully traversed 47 km on the surface of the moon during its 14 months operation [9], [10], [11].

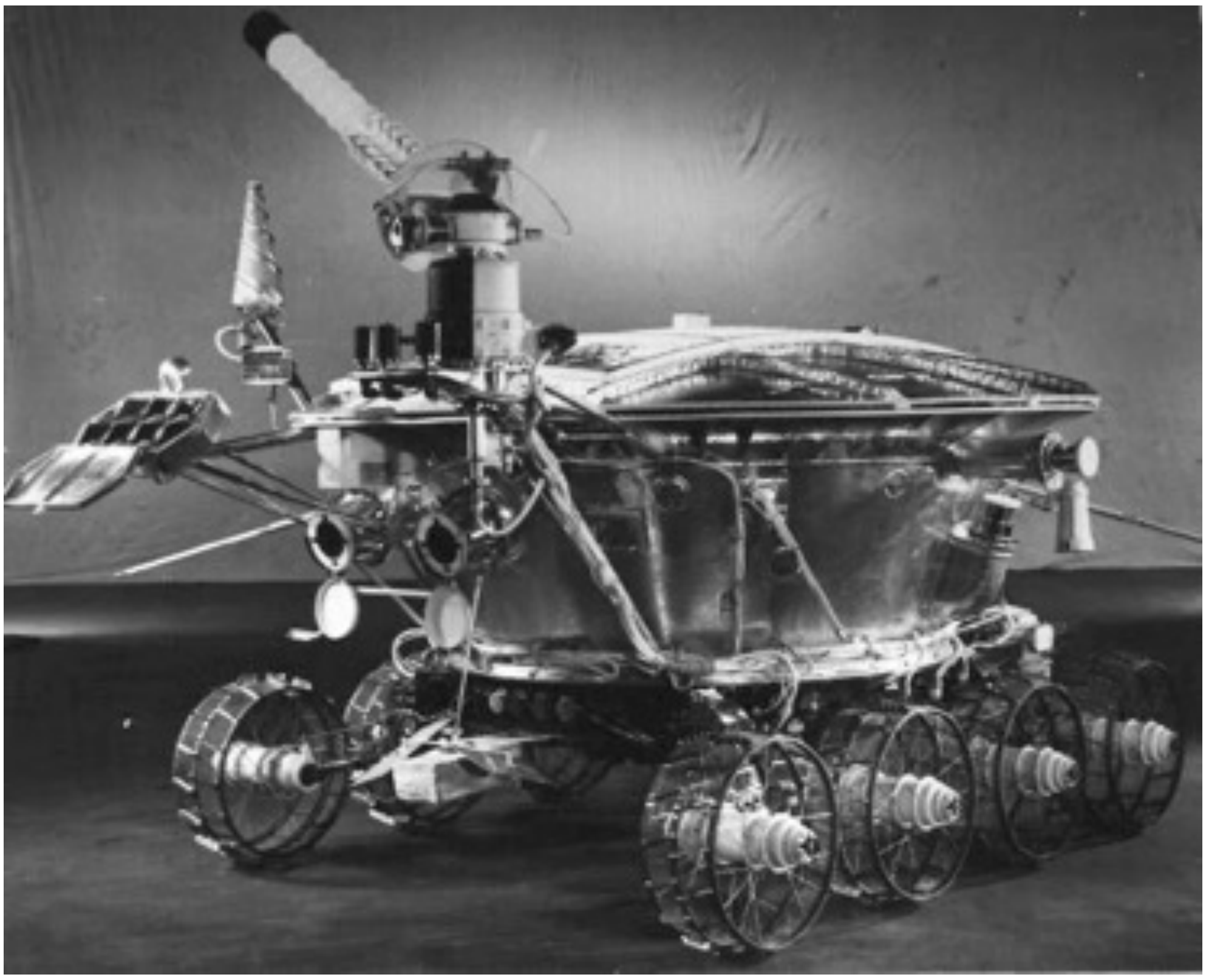

Figure 1.4: Soviet Union Lunar Rover. (*Source: www.nasa.gov*)

After 1980s, the number of autonomous vehicles with the ability of handling the complex tasks increased. For example, we can refer to the Road-Following robot, which was built in the robotic institute of Carnegie-Melon University. The first result confirmed that the robot successfully ran over a curving 20 m path and 10 m segments of straight side-walks. The robot found the path by tracking the edges of the road [12], [13].

Over the past three decades, the application range of mobile robots has increased drastically. They operate in homes as vacuum cleaner robots as well as in hazardous environments for complex operations.

## 1.2 Overview

Generally, mobile robots are categorised as non-autonomous, semi-autonomous, and autonomous. Non-autonomous mobile robots need to be fully controlled by operators. Remote-control camera drones and Intelligent Pig robot are some of the examples of non-autonomous mobile robots. An intelligent Pig is run inside the pipeline to find any corrosion or cracking inside the oil and gas pipelines. The robot is located in the pipeline by the operators and the propulsion force of the robot is the fluid flow. On the other hand, Semi-autonomous mobile robots are able to do some subtasks without the operators' interferences. Autopilot equipped drones and some types of Intelligent Pigs such as PigWave, could be noted as prominent examples of a semi-autonomous robots. Finally, the autonomous mobile robots can perform tasks independently. Many of them have been created and implemented in different areas such as factories, warehouses, healthcare, and agriculture such as the robot created by Ecorobotix for weeding a land [14], [15].

All types of mobile robots have a common problem known as navigation control. The navigation control is more complex in autonomous mobile robots as the robots are expected to plan without the operators' interferences. For example, an autonomous car has to travel between two points in a street. First, the self-driving car needs to find the path between two points, which is known as high-level task. The high-level task could be done before the trip or in a real-time fashion by receiving information from the GPS system. Second, the self-driving car should be able to act instantaneously for any unpredictable events such as avoiding collision to the dynamic or static objects in the environment. These types of actions, which need to have real-time information frequently in the region are known as low-level task and normally contains the decision-making strategies to have a safe and reliable travel in the environment[16], [17], [18].

The control method of such a complex system can be categorised as centralised, decentralised, and hybrid. In fact, information distribution and decision-making fashion can be considered as two key distinguishing factors in each of the mentioned control method.

In a fully centralised control method, all the information either collected by the robots or any other data collection devices should be sent to a central control station for further process and to make the final decision. Once the decision is made by the central control station, it will be distributed between every agent for further action. Opposed to centralised control method, in a decentralised control method, every decision is made by each agent individually based on the available information. And finally, the hybrid control method is a combination of centralised and decentralised methods where the robots are able to make some decision individually, whereas, they can negotiate with other team members to make a cooperative decision or use the central controller information in some cases[19], [20].

In this study, we propose novel decentralised and hybrid control methods to improve the performance of autonomous mobile robots in terms of low-level tasks e.g. low-level motion planning[21],[22],[23] while cooperating as a team or individually in a certain mission. Many controllers in this report belongs to the class of sliding mode controllers [24], [25], [26].

## 1.3 Chapter outline

The problem statement of each chapter is outlined as follows:

Chapter 2 presents a review of the literature regarding the problem of mobile robots' intrusion detection and target tracking in a bounded region, decision making, static and dynamic obstacles collision avoidance and energy saving.

Chapter 3 is concerned with the problem of intruder interception in an unbounded smooth region by a multi robot team based on [27]. The contribution of this chapter is presenting a novel decentralised intrusion detection algorithm called *e-intercepting* intrusion detection model. The model proposes with all necessary and sufficient condition which results in always *e-intercepting* the intruder in every single point that the intruder tries to cross the boundary. The proposed decentralized navigation law is easy to implement in real time boundary protection applications, result from its non-demanding computational quiddity.

Chapter 4 is concerned with the problem of intrusion detection by a multi-robot team in a boundary region based on [28]. The contribution of this chapter is presenting a novel decentralised intrusion detection algorithm called *k-intercepting* intrusion detection model. The model proposes with all necessary and sufficient condition which results in always *k-intercepting* the intruder in the boundary region. The proposed decentralized navigation law is easy to implement in real time boundary protection applications, result from its non-demanding computational quiddity.

Chapter 5 is concerned with the problem of hunters and a prey which is trapped in a siege ring which is created by a team of mobile robots. The contribution of this chapter is presenting a novel decentralized navigation method which guarantees to maintain the intruder inside the siege ring for all time. On the other hands, the intruder is intercepted by at least one robot in its every attempt to escape the region.

Chapter 6 is concerned with the problem of hunters and preys. The contribution of this chapter is developing a novel and robust algorithm to intercept the multiple intruders in a region which can arbitrary move and tend to escape the region without being trapped by the guardian robots.

Chapter 7 is based on [29] and is concerned with the problem of intrusion detection by a multi-robot team with a communication limitation. The contribution of this chapter is developing an intelligent game-based decision-making strategy (IGD) for a group of mobile robots' results in maximizing the probability of intrusion detection with either minimum or no communication between the team members.

Chapter 8 is concerned with a multi-robot team navigation in an unknown area occupied by a static obstacle based on [30]. The contribution of this chapter is to develop a semi-decentralized leader-follower based navigation strategy called position estimation switching algorithm (PSEA) which allows the mobile robots safely maneuver in the environment by estimating the next switching position. All the necessary measurement and computation for the planning the safest path is done by the leader in each switching steps called sojourn time.

Chapter 9 is based on [31] which is concerned with a multi-robot team navigation in an unknown area occupied by multiple static obstacles. The contribution of this chapter is modification of the PSEA to allows a multi-robot team moves among multiple obstacles in an unknown region while avoiding collision with the obstacles.

Chapter 10 is based on [32], that is concerned with a decentralized multi-robot navigation strategy in a cluttered area with the purpose of target trapping. The contribution of this chapter is developing a fully decentralized navigation method in which the robots sagely maneuver in the region to trap the target, autonomously with a significant improvement in energy consumption by the robots.

Chapter 11 is based on [33], that is concerned with the problem of collision avoidance with the dynamic obstacles. The contribution of this chapter is developing an artificial potential field-based navigation method which allows a mobile robot avoid collision with the obstacles while moving in an unknown area occupied by multiple dynamic obstacles with the capability of merging and rotation in any direction. In this method the robot is able to find the safest path between the obstacles regardless of the direction of the motion of the obstacles.

And finally, Chapter 12, presents a conclusion of this report and gives a recommendation of potential future works.

# Chapter 2

# Literature review

In this chapter, different control methods presented by the researchers in terms of navigation control and decision-making have been reviewed. In chapter 1, we briefly gave an overview of the autonomous vehicles' development and their general applications. As we explained, mobile robots are autonomous machines, which are equipped by the sensors for data gathering and communication and the capability of working individually or as a member of a team. In other words, it could be considered that mobile sensors are able to work in a network and cooperate with the other static or dynamic sensors within the network to complete missions such as intruder detection and target tracking in an environment where occupied by the dynamic or static obstacles. Therefore, some fundamentals such as sensor networks, environment, sensors, vehicles, and obstacles have been discussed in this chapter.

## 2.1 Wireless sensor networks

A group of sensor nodes that cooperatively work in a network, either static, dynamic, or in a combination of both is known as a wireless sensor network (WSN) [34],[35]. A WSN could consists of various types of sensors such as seismic sensors, which measure the seismic vibrations, thermal sensors, lasers, sonars, infrareds, visual sensors, and radars, which are used to monitor the environment [36],[29].

WSNs have been operated in a wide variety of applications such as Military applications for example in hazard exploration and in nuclear, biological, or chemical attacks, target tracking, intrusion detection, and surveillance [37], Environmental applications for prediction of a natural disaster, bush fire detection and pollution studies [38]–[48], and Healthcare applications such as robotic beds, neurosurgery, tele monitoring for data gathering, diagnostics, patient monitoring, and drug administration [49]–[54].

In hazardous and unreachable environments, using mobile sensors could be a sufficient alternative rather than distributing static sensors by dropping them in the environment as they are able to collect and transfer data while moving within the region. A team of mobile robots which is navigated by a certain navigation law can be viewed as an example of network control systems [55]–[64], [65]. In fact, their ability to move in addition to monitoring, data gathering, and cooperation, enables them to act like a human, but in dangerous and inaccessible environments by the human [66]. Mobile wireless sensor networks can be categorised as ground-based robots, aerial robots, and aquatic robots.

In this study, the problem of navigation and path planning for ground based mobile robot is presented.

## 2.2 Path planning and navigation strategies

Safe manoeuvring in an unknown cluttered environment is an essential requirement to mobile robots to complete a mission successfully. Extensive research in this area has improved the performance of the mobile robots; however, there are still many challenges and shortcomings in many cases resulting from uncertainties and ambiguities of available knowledge [67]. The path planning algorithm could be categorised as Global Path Planning and Local Path Planning.

### 2.2.1 Global path planner (GPP)

In GPP, the exact location and orientation of the target priori is calculated and provided to the robots by the detectors. Many path planning algorithms have been developed based on this approach such as roadmaps, voronoi, Dijkstra algorithm, non-holonomic planner, $A^*$ or best first algorithm, ,velocity-obstacles, cell decomposition, random trees, neural network based algorithm, particle swarm optimisation hierarchical algorithm, state time-space, and heterogeneous-ants [68]–[78] and papers therein all the developed algorithm with the GPP approach guarantees successfully meat the goal including target tracking and trapping, intrusion detection while avoiding collision with obstacles in the region. However, real-time implementation is hardly achievable. Furthermore, the GPP's are computationally complex specially in an uncertain and unpredictable environment [79]–[87].

### 2.2.2 Local path planner

In Local path planning approach, the robots estimate the feasible trajectory to the target based on the real-time information gathered by their on-board sensors [88]–[90]. In contrary to the GPP approach, the robots plan a short portion of the path iteratively. Therefore, the real time performance in an unknown environment and simple computation is achieved by using this approach. Similar to the GPP approach, there are many successful techniques using LPP approach such as dynamic window [91]–[93], collision cones [94], [95], and inevitable collision states [96]. However, as the LPPs are the steepest descent optimisation method, their drawback comes from their excessive caution to prevent any collision with the obstacles, which leads the robots to be susceptible to the local minima result in increasing the possibility of getting stuck in cluttered environments [97]–[99]. Many researchers developed novel methods to overcome the shortcoming of the path planning algorithm which are based on LPP's approach. For example, *Li et al.* proposed an improved dynamic window approached that consider the relation of the size of each agent and the space between the obstacles [100]. An image-based position control algorithm proposed in [101] which is completely with no dependency to the camera’s parameter. Another robust and reliable incremental simultaneous localisation and mapping problem (SLAM) algorithm, which has developed by *F.Bai et al.* successfully solved the problem of local minima and the outliers [102]. On the other hand, implementing LPP at reactive controllers while using GPP approach for a priori information about the region could compensate for the drawback of the pure LPP, however, the robots can’t work in a completely unknown environment. Furthermore, in a purely reactive LPP based method we can point to the biologically inspired methods presented in [103]–[107].

## 2.3 Centralised and decentralised control structures

In a multi-agent system, the control architecture could be categorised as centralised or decentralised based on the relationship and interaction among the robots and the strategy of task allocation to the team members [108].

### 2.3.1 Centralised control structure

Based on this control structure, every individual team mate maintains its connection with a central commander that is responsible for allocating and distributing the tasks to the agents during the mission. In this case, it is the duty of each agent to communicate with the central commander and transfer the data collected from the environment at certain time intervals for reprocessing and reallocating the task.

Because there is one decision maker in the system, designing control methods based on centralised structure could prevent any duplication of efforts saving time and cost. Therefore, for a multi-robot team with a limited number of agents who work in a known environment where the global information could be easily accessible, a control strategy based centralised structure could be a very well-suited choice [109], [110]. Lots of literatures proposed the control strategies for a multi-robot team navigation control based on centralised structure. For example, Bicchi et al. presented a centralised control method to solve the problem of mobile robot task allocation problems to increase the life time of the network [111]. In case of task allocation problem for a group of inspector mobile robots working in an industrial plant, we can point to the literature [112]. In the method presented in [113] a single global task is allocated to a group of heterogeneous mobile robots. The purely centralised control method are exemplified by [114]–[124]. Apart from all the advantages of the centralised control structure, the most important disadvantage of this method is that the robots are highly dependent on the central commander, which could be a member of the team or a control centre out of the team. It's obvious that the team works fine as long as they have been received the task from the administrative centre. Therefore, any malfunction in the control centre affects to the performance of the team directly, which in turn disrupts the entire team from completing the mission [125], [108].

### 2.3.2 Decentralised control structure

In this category, every individual agent in the group is responsible for decision making based on the available local sensory information or any prior information of the environment. In this method, a central commander is allocated to each robot in the group, therefore, any malfunction in the control centre or any other team member operation doesn't affect the performance of the team [126],[32]. In this control structure, the team members also are able to work as a distributed system to communicate with each other when required or exchange the information to help the other team mates for making the right decision in some critical situations, without any dependency of the other team members' information or task allocation. There are many literatures that have proposed decentralised control methods to improve the performance of a multi-agent team. For example, path planning problems in search and rescue, intruder detection and boundary protection, and static and dynamic obstacles are exemplified by [30], [127]–[133], [29], [134]–[143] and references therein. Different consensus based methods are presented in [144], [139], [145]–[147] to solve the problem of multi robot task allocation and path planning. As some other advantages of the decentralised control structure, we can denote the flexibility, robustness and working in an environment with minimum communication requirements.

## 2.4 Decision-making of a multi-robot system

Decision-making is a key factor in controlling a group of mobile robots in an environment to achieve a reliable and robust performance in their mission. As a matter of fact, in a multi agent system, the decision-making of each agent should meet the goal of the whole team and not the individual member. A decision could be made either by a central commander as a centralised decision-making strategy and distributed between the members of the team [148], or by each team member individually based on the local information and limited communication and negotiation by the other team members [149]. In a centralised decision-making procedure, a solid connection between the robots and the central commander is essential, which makes it impractical for the situation with communication difficulties while it's not required in a distributed decision-making procedure. In this case, game theory is a sufficient tool, which provides a bag of analytical solutions for a rational and strategic decision-making by the team members [150].

A game-based decision could be made either cooperatively or non-cooperatively based on the condition and constrictions of the players and the environment. In a game theoretic approach, a player could be considered as an individual or as a group of individuals and a group could be considered as a group of individual players or a group of subgroups of players. Therefore, if the action of a player is primitive the model referred as non-cooperative and in case any joint actions of a group is primitive it referred as cooperative game. As examples of the solving the problem of navigation, target following and intruder detection by the mobile robots, we can address the works that have been presented in [151]–[153]. The non-cooperative game strategies, which have been proposed in [151],[152] for a fleet of planner to track an unauthorised target in the environment result in a fast and a robust communication between the agents. However, they need to maintain their connection with a centralised sensory subsystem during the whole mission.

Furthermore, a collection of non-deterministic, distributed approaches for the purpose of security matters of critical facilities, which have been presented by Hernandez et al., [153] demonstrate a significant performance improvement of the team. However, the number of the nodes and edges, which are connected to chosen by the robots were not considered.

## 2.5 Intrusion detection

Intrusion detection is a fundamental problem of multi-robot navigation control in various security mission such as border security[104],[154],[133]. Before diving into the studies that have been done in case of intrusion detection, a brief explanation of intrusion detection system is required merely for readers' knowledge. A system, which is able to analyse and identify any abnormal behaviour in an environment where equipped with a multi agent network is defined as an intrusion detection system [155]. Any intrusion detection system consists of data collection, data analysis, and a proper action by the team members that could be purely centralised, purely decentred, or a combination of both. Various methods have been proposed for IDS; however, discussing them is out of interest of this research. In this study, the intruder assumes to be predefined as a heterogonous agent in the network in contrast with the team members who are homogenous. Therefore, analysing the intruder is not required. However, the team members detect the intruder and take the best action in a decentralised fashion.

A key component of area protection against any intrusion is coverage control problem. In this case, the barrier coverage problem and swipe coverage problem are exemplified by [147], [136]–[138]. In barrier coverage problem, robots are deployed as a static barrier in the region boundary to detect any unwanted intruder. On the other hand, in the swiping coverage problem, the mobile robots swiping the region to protect every point in the environment of any unwanted intrusion. A decentralised randomised navigation control method, which is proposed in [140], shows a robust coverage of the region with probability one. There is neither a requirement of predefined leader nor a requirement of initialising the position of the robots; however, a solid communication is required nonetheless.

A non-cooperative game-based method is presented in [151]. In the proposed method a fleet of mobile robots using a non-cooperative game-based strategy to track a dynamic target in the region based on centralised control approach. An anthropomorphic behavioural based planning for target tracking and intrusion detection has been presented in [156], [157]. However, errors, repetitive motions, and confusion in decision-making sometimes takes place by the robots. A method known as territorial wok division, which is based on behavioural based method presented in [158]. In this model, a game-based strategy has been used to solve the problem of the decision-making's conflict between the team mates. The agents can make the best decision independently, with no communication or minimum communication in some special cases, however, it suffers from a high intrusion cost in the environment.

## 2.6 Safe manoeuvring in a cluttered region

As a matter of fact, mostly, the real environments are occupied by dynamic and static obstacles. Therefore, a safe manoeuvring for a fleet of mobile robots in cluttered environments has encouraged researchers to develop navigation strategies that enable the autonomous vehicles to complete their excepted mission while avoiding any type of obstacle in the area of interest [159], [160]. Virtual structure navigation methods are proposed in [161] and [162] for a safe manoeuvring fleet of mobile robots in a cluttered region. The robustness of the works has been confirmed in the result, however, the robots need to maintain their communication rigorously based on a continues measurement. In the literatures [163]–[167], various behavioural navigation methods are presented. Based on the proposed methods, every individual member has the capability of maintaining the heading align with the field's vector orientation. A desired behaviour should be prescribed to each team member to avoid obstacles in the environment. Various types of artificial potential field methods to avoid dynamic and static obstacles have been proposed in the literatures [168]–[180], [33] and any references therein. In the case of proposed methods, obstacles are considered as repulsive force sources and target(s) are considered to be sources of attractive force. The simplicity of the proposed model makes them to be practical in real time applications. Earl and Andrea [181], presented a mixed-integer linear programming-based algorithm, which leads to avoid obstacles successfully, but, nondeterministic polynomial time problem result in computational complexity of the method. Literatures [182]–[185], presented leader-follower based algorithm to avoid obstacles by mobile robots. As the leader is the only one that measures the distances and calculates the best heading and the safe path, the error would be minimised, however, a solid connection between the robots and the leader is essential. Therefore, a faulty member causes problem to the entire team. A mathematically rigorous navigation strategy is presented in [104] and [103] for mobile robots in a bounded region. The results confirm the feasibility of the model to avoid the obstacles during the mission however the problem of a multi-robot team has not been considered in the model.

# Chapter 3

# The problem of *e-intercepting* an intruder on a region boundary by a multi-robot team

In this chapter we present a problem of intruder interception on the boundary of a planar region through the use of a network of mobile robots. A necessary and sufficient condition for the existence of a solution of this problem is obtained. We propose a decentralized motion control algorithm for the mobile robots to intercept an intruder leaving the region. The algorithm is developed based on some simple rules that are computationally efficient and easily implementable in real time. The important recent technological developments in robotics greatly increase the number of real-world applications that are suitable for multi-robot teams. Therefore, in recent years, the use of teams of autonomous unmanned vehicles in patrolling, monitoring and surveillance tasks has been increased significantly. A fundamental problem of robotics research is navigation of mobile robots for patrolling a boundary of a region in various border-security missions; see e.g. [104], [133], [154], [186]–[188].The most common approach to protection a region from intruders is coverage control in which the barrier coverage problem and the sweep coverage problem are studied. The barrier coverage problem is to deploy a group of mobile robots with sensing capabilities to form a static sensor barrier that detects any object trying to enter a protected region [147],[137] On the other hand, the sweep coverage problem is to steer a group of mobile robots along the boundary of the protected region so that every point in some neighborhood of the boundary is detected by some robot [136],[138].

In this paper, we consider a team of mobile robots moving along the boundary of a planar region. the multi robot team with the proposed navigation law belongs to the class of hybrid dynamical systems [189]–[192].

The robots move in a decentralized fashion, i.e. each robot navigates independently and has information about current coordinates of just several closest other robots [27]. Unlike coverage control problems of [147],[138],[136],[29] we assume that the intruder becomes visible to the robots at some time , i.e. all the robots know the planar coordinates of the intruder after a certain time moment. The objective of the multi-robot team is to intercept the intruder which means that when the interceptor crosses the boundary of the planar region, there should be at least one robot close to the interception point. The proposed problem statement is relevant to various problems of asset guarding in which a team of autonomous unmanned surface vehicles (USVs) patrols and guards an asset in an environment with hostile boats. Such problems require the team of USVs to cooperatively patrol the area around the asset, identify intruders, and actively block them [193], [194]. An important example is safeguarding civilian harbors from terrorist attacks coming from the blue border (i.e. the sea-side) [195]–[197].

The reminder of this chapter will be organized as follows; In section 3.1, we present the problem of e-intercepting for an intruder which is trapped in a bounded region and tries to escape the region by crossing the boundary of the region while avoiding intercepting by the guardians' robots. Section 3.2 presents a proposed decentralized navigation method to guarantee the interception of the intruder in all the time by at least on robot in the boundary of the region. In section 3.3, the simulations results show the successful performance of the presented method and finally, we give a summary of the chapter in section 3.4.

## 3.1 Problem statement

Let R be a closed convex planar region with a piecewise smooth boundary. Notice that $R$ may be unbounded. Furthermore, let $S$ be a segment of the boundary of the region $R$ between some points $P_1$ and $P_2$; see Fig. 3.1.

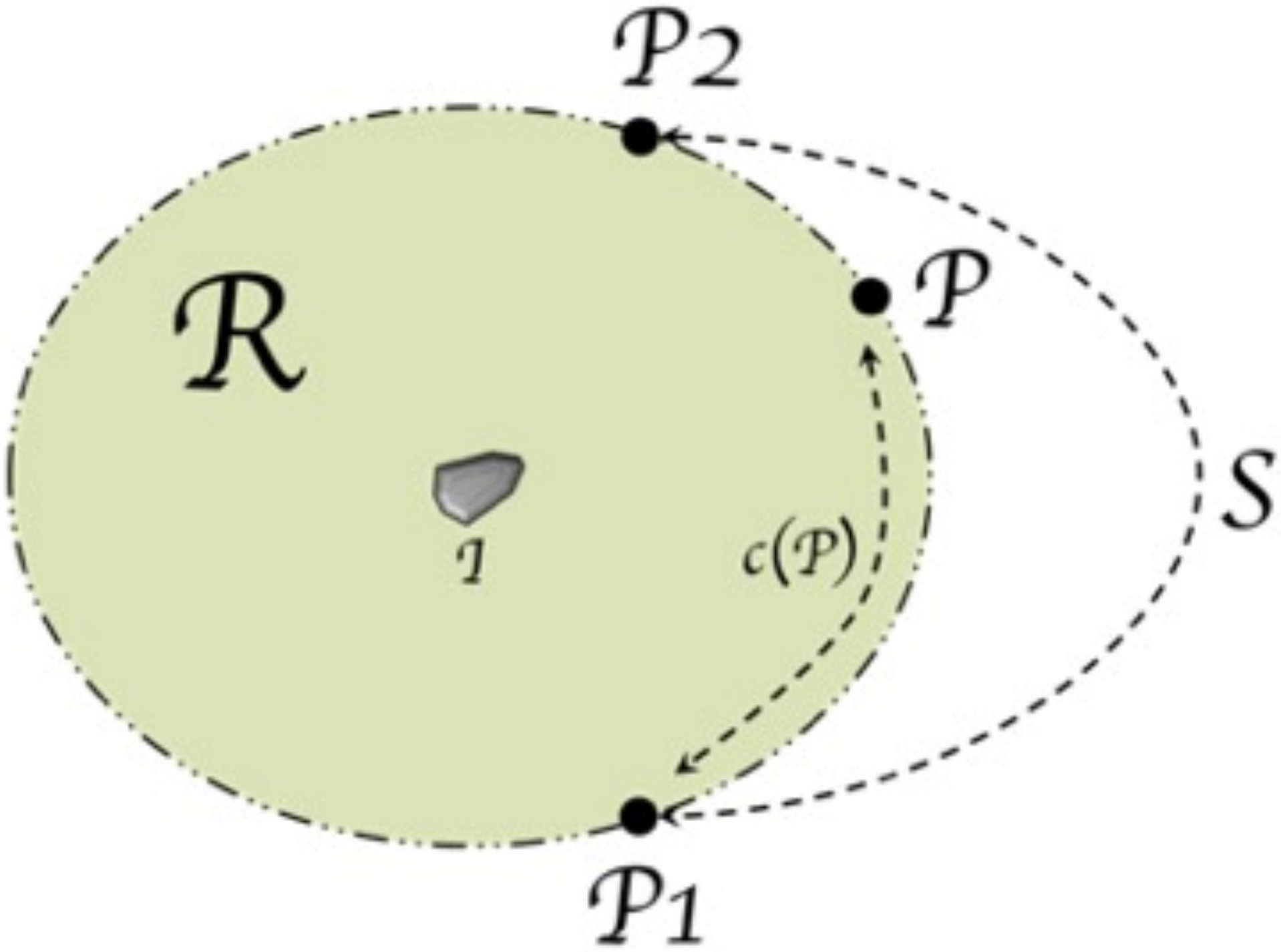

Figure 3.1: Bounded region *R*

We consider a moving in the region R point-wise intruder $I$ that aims to exit the region $R$ through the segment $S$; see Fig.3.1. We assume that the intruder cannot cross the boundary of R outside the segment $S$. Let $xI(t)$ be planar coordinates of the intruder. The intruder is moving with an arbitrary time-varying vector velocity $v_I(t) = \dot{x}_I(t)$ satisfying the constraint

$$\|v_I(t)\| \leq V_I^{max} \;\; \forall t \geq 0 \tag{3.1}$$

where $V_I^{max} > 0$ is a given constant, $\|\cdot\|$ denotes the standard Euclidean vector norm. Also, $x_I(t)$ denotes the planar coordinates of the intruder. Moreover, let $n > 1$ be a given positive integer. We consider n mobile point-wise robots labelled $1,2,\ldots,n$ that prevent the intruder from leaving the region $R$ through the segment $S$. Unlike the intruder that can move in any direction in the plane, the robots can move only along the segment $S$ in the both directions. Furthermore, $x_1(t), x_2(t), \ldots, x_n(t)$ denote the planar coordinates of the robots $1, 2, \ldots, n$.

We introduce the curvilinear coordinate $c(P)$ for any point $P \in S$ such that $c(P)$ is the length of the portion of the segment $S$ between the points $P1$ and $P$; see Fig.4.1. This implies that $c(P_1) = 0$ and $c(P_2) = L$ where $L$ is the length of the segment $S$. We assume that $c_1(t) := c(x_1(t)), c_2(t) := c(x_2(t)), \ldots, c_n(t) := c(x_n(t))$ are the curvilinear coordinates of the robots $1,2,\ldots,n$ at time $t \geq 0$.

Furthermore, we suppose that the robots labelled according to their curvilinear coordinates so that $0 \leq c_1(t) \leq c_2(t) \leq \ldots \leq c_n(t) \leq L \; \forall t \geq 0$, which means that the robots never change their order on the segment $S$.

We assume that the motion of the robots along $S$ is described by the equation

$$\dot{c}_i(t) = u_i(t) \qquad \forall i = 1, \ldots, n \tag{3.2}$$

where $ui(t)$ is the control input of the robot $i$. We assume that the control inputs $ui(t)$ satisfy the constraint

$$|u_i(t)| \leq V_R^{max} \quad \forall t \geq 0 \tag{3.3}$$

where $V\ max\ >\ 0$ is a given constant.

## *Available measurements:*

At any time $t$, each robot $i, 2\ \leq\ i\ \leq\ n\ -\ 1$ knows the curvilinear coordinates $c_{i\ -\ 1}(t), c_{i\ +\ 1}(t)$ of the robots $i\ -\ 1$ and $i\ +\ 1$, respectively. The robots 1 and $n$ know the curvilinear coordinates $c_2(t), c_{n\ -\ 1}(t)$ of the robots 2 and $n-1$, respectively. Moreover, each robot i knows its own coordinate $c_i(t)$. Furthermore, the intruder becomes visible to the robots at some time $t_0\ \geq\ 0$, i.e. all the robots know the planar coordinates $x_I\,(t)$ of the intruder for all $t\ \geq\ t_0$.

## *Definition 3.1:*

Let $\varepsilon\ >\ 0$ be a given constant. Suppose that the intruder crosses the segment $S$ at time $t^*$i.e. $x_I(t^*) \in S$. We say that the multi-robot team ε−intercepts the intruder at time $t^*$ if there exists some index $i\ =\ 1,2,\ldots,n$ such that $|c(x_I(t_\star)) - c_i(t_\star)|\ \leq\ \varepsilon$. Furthermore, a multi- robot team navigation strategy that is based on the available information is called *ε-intercepting* if for any movement of the intruder, the multi-robot team *ε-intercepts* it when the intruder crosses the segment $S$.

In other words, *ε-intercept* means that when the interceptor crosses the segment $S$, there should be at least one robot close enough to the interception point.

The intruder's objective is to exit the region $R$ through the segment $S$ while avoiding *ε-intercept* by the multi-robot team. The objective of the multi-robot team is to *ε-intercept* the intruder. The problem under consideration in this paper is to derive a necessary and sufficient condition under which *ε-intercept* is possible for any motion of the intruder and design a decentralized navigation strategy for the multi-robot team that is based on the available information and will always result in *ε-intercept* of the intruder.

## 3.2 *e-Intercepting* navigation algorithm

Let $P \in S$ and $x$ be an interior point of the region $R$. Then $L(x, P)$ denotes the straight-line segment connecting $x$ and $P$. Since $R$ is convex, $L(x, P)$ is in $R$ and the intersection of $L(x, P)$ and the boundary of $R$ contains only the point $P$. Furthermore, let $\alpha(x, P)$ denote the length of $L(x, P)$. Furthermore, let i be an index such that $|c_i(t) - c(P)| \leq |c_j(t) - c(P)|$ for all $j = 1,,n$. Then, introduce the variable $\beta(t, P) := |c_i(t) - c(P)|$. In other words, $i$ is the robot closest to the point $P$ at time $t, \beta(t, P)$ is the length of the sub-segment of the segment $S$ between the closest robot's current location and the point $P$.

Introduce the function $F(s)$ from the interval $[0, L]$ to the segment $S$ such that for any number $s \in [0, L]$, $F(s)$ is the point $P \in S$ such that $c(P) = s$. Furthermore, let $[A_1, A_2]$ denote the closed sub-segment of the segment $S$ between the points $A_1$ and $A_2$. For $i = 1, \dots, n$, introduce sub-segments $Si(t)^-, Si(t)^+$of the segment $S$ as follows:

$$S_i(t)^- := \left[F\left(\frac{c_{i-1}(t) + c_i(t)}{2}\right), F(c_i(t))\right]$$

$$if\ i = 2, \dots, n;$$

$$S_i(t)^- := [P_1, F(c_i(t))]$$

$$if\ i = 1;$$

$$S_i(t)^+ := \left[F(c_i(t)), F\left(\frac{c_i(t) + c_{i+1}(t)}{2}\right)\right]$$

$$if\ i = 1, \dots, n-1;$$

$$S_i(t)^+ := [F(c_i(t)), P_2]$$

$$if\ i = n. \tag{3.4}$$

Moreover, for $i = 1, 2, \ldots, n$, introduce the numbers $M_i^-(t)$ and $M_i^+(t)$ as

$$M_i^-(t) := \sup_{P \in S_i(t)^-} \left(\beta(t,P) - \frac{\alpha(x_I(t),P)V_R^{max}}{V_I^{max}}\right);$$

$$M_i^+(t) := \sup_{P \in S_i(t)^+} \left(\beta(t,P) - \frac{\alpha(x_I(t),P)V_R^{max}}{V_I^{max}}\right) \tag{3.5}$$

Now we can introduce the following decentralized navigation law:

$$u_i(t) := V_R^{max} \quad if \quad M_i^-(t) < M_i^+(t)$$

$$u_i(t) := -V_R^{max} \quad if\ M_i^-(t) > M_i^+(t)$$

$$u_i(t) := 0 \ \ if\ M_i^-(t) = M_i^+(t) \tag{3.6}$$

for all $i = 1, \ldots, n$.

## *Remark 3.1:*

The intuition behind the decentralized navigation law (3.6) can be explained as follows.

The sub-segments $S_i^-(t), S_i^+(t)$ are sets of points of the curve $S$ for which the robot i is the closest robot at time $t$. The robot moves with the maximum allowed speed towards the one of these segments that is more" dangerous" at the current time, i.e. it has the biggest possible distance between the intruder and the closest robot at the moment of crossing $S$ by the intruder. This biggest possible distance is described by (6).

## *Theorem 3.1:*

Consider the multi-robot team satisfying (3.3) and the intruder satisfying (3.1). Then there exists an ε-intercepting multi-robot team navigation strategy if and only if

$$\sup_{P\in S}\left(\beta(t_0,P)-\frac{\alpha(x_I(t_0),P)V_R^{max}}{V_I^{max}}\right)\leq\varepsilon \tag{3.7}$$

where $t0 \geq 0$ is the time at which the intruder becomes visible to the robots.

Moreover, if the inequality (3.7) holds, then the navigation law (3.6) is an ε-intercepting navigation strategy.

## *Remark 3.2:*

Notice that since the region $R$ is convex, and the segment $S$ is compact, the supremum in (3.7) is achieved for some point $P$.

## *Proof:*

First, we prove that if the inequality (3.7) does not hold, then the intruder can always cross the segment $S$ without ε−intercepting by the multi-robot team. Indeed, if (3.7) does not hold, then there exists a point $P \in S$ such that

$$\left(\beta(t_0,P)-\frac{|L(x_I(t_0),P)|V_R^{max}}{V_I^{max}}\right)>\varepsilon \tag{3.8}$$

Now let the intruder move along the straight-line segment $|L(x_I(t_0),P)|$ connecting the points $x_I(t_0)$ and $P$ with its maximum speed $V_I^{max}$ . In this case, the intruder reaches the point $P$ at the time $t^* = t_0 + \frac{|L(x_I(t_0),P)|}{V_I^{max}}$ . It obviously follows from (3.8) that, the closest robot to the point $P$ cannot be closer to $P$ at the time $t^*$ than $\varepsilon$. Therefore, the *ε-neighborhood* of the point $P = x_I(t^*)$ at the segment $S$ cannot contain any robot at time$t^*$. This implies that the multi-robot team does not ε–intercept the intruder.

We now prove that if the inequality (3.7) holds, the multi-robot team navigated by the law (3.6) always *ε-intercepts* the intruder when it crosses the segment $S$. First, we prove the following claim.

Indeed, for any trajectory $[x_I(t), c_1(t), \ldots, c_n(t)]$ of the intruder-multi-robot introduce the Lyapunov function

$$W[x_I(t), c_1(t), \ldots, c_n(t)] := sup_{P \in S} \left( \beta(t,P) - \frac{\alpha(x_I(t),P) V_R^{max}}{V_I^{max}} \right) \quad (3.9)$$

Notice that since the region $R$ is convex, and the segment $S$ is compact, the supremum in (3.9) is achieved for some point $P$. Furthermore, by definition, $\alpha(x_I(t), P)$ is the length of the straight segment $L(x_I(t), P)$ connecting $x_I(t)$ and $P$. Hence, it is obvious that

$$\alpha(x_I(t), P) = \inf_{M(x_I(t),P) \in \mathcal{M}(x_I(t),P)} |M(x_I(t), P| \quad (3.10)$$

where $\mathcal{M}(x_I(t), P)$ is the set of all smooth paths $M(x_I(t), P$ inside $R$ connecting $xI(t)$ and $P$, and $|M(x_I(t), P|$ denotes the length of the path $M(x_I(t), P$. In other words, $\mathcal{M}(x_I(t), P)$ is the set of all possible paths of the intruder between $x_I(t)$ and $P$ . Furthermore, it immediately follows from (3.9), (3.10) and (3.6) that

$$W[x_I(t_1), c_1(t_1), \dots, c_n(t_1)] \le W[x_I(t_2), c_2(t_2), \dots, c_n(t_2)]\,, \forall t_2 \ge t_1 \ge t_0 \quad (3.11)$$

Now (3.11) and (3.8) imply that if the intruder reaches a point $P \in S$ at some time $t^* \ge t_0$ , the robot closest to the point $P$ at time $t^*$ cannot be further from $P$ than $\varepsilon$. Therefore, the *ε-neighborhood* of the point $P = x_I(t^*)$ at the segment $S$ contains at least one robot at time $t^*$. This implies that the multi-robot team ε–intercepts the intruder. This completes the proof of Theorem 3.1.

The inequality (3.7) and the navigation law (3.6) can be made computationally simpler under the following assumption.

## *Assumption 3.1:*

The following inequality holds:

$$V_I^{max} \ge V_R^{max} \quad (3.12)$$

For $i = k, k+1, \dots, n-k+1$, introduce points $D_i(t)^-, D_i(t)^+$ of the segment $S$ as follows:

$$D_i(t)^- := F\left(\frac{c_{i-1}(t)+c_i(t)}{2}\right) \qquad if\ i = 2, \dots, n;$$

$$D_i(t)^- := P_1 \qquad if\ i = 1;$$

$$D_i(t)^+ := F\left(\frac{c_i(t)+c_{i+1}(t)}{2}\right) \qquad if\ i = k, \dots, n-1;$$

$$D_i(t)^+ := P_2 \qquad if\ i = n. \quad (3.13)$$

Moreover, for $i = k, k+1, \dots, n-k+1$, introduce the numbers $H_i(t)^-$ and $H_i(t)^+$ as follows:

$$H_i^-(t) := \frac{c_i(t)-c_{i-1}(t)}{2} - \frac{\alpha(x_I(t),D_i(t)^-)V_R^{max}}{V_I^{max}}$$

$$if\ i = 2, \dots, n;$$

$$H_i^-(t) := c_i(t) - \frac{\alpha(x_I(t),P_1)V_R^{max}}{V_I^{max}}$$

$$if\ i = 1;$$

$$H_i^+(t) := \frac{c_{i+k}(t)-c_i(t)}{2} - \frac{\alpha(x_I(t),D_i(t)^+)V_R^{max}}{V_I^{max}}$$

$$if\ i = 1, \dots, n-1;$$

$$H_i^+(t) := L - c_i(t) - \frac{\alpha(x_I(t),P_2)V_R^{max}}{V_I^{max}}$$

$$if\ i = n \tag{3.14}$$

For $i = 1,\dots,n$, the simplified navigation law (3.6) becomes:

$$u_i(t) := V_R^{max} \quad if \quad H_i^-(t) < H_i^+(t)$$

$$u_i(t) := -V_R^{max} \quad if \quad H_i^-(t) > H_i^+(t)$$

$$u_i(t) := 0 \quad if \quad H_i^-(t) = H_i^+(t) \tag{3.15}$$

## *Theorem 3.2:*

Consider the multi-robot team (3.3) and the intruder satisfying (3.1) and Assumption 3.1. Furthermore, let H be the set of numbers $H_i(t)^-$ and $H_i(t)^+$where $i = k, k + 1,\dots, n - k + 1$ and $t0 \geq 0$ is the time at which the intruder becomes visible to the robots. Then there exists an ε–intercepting multi-robot team navigation strategy if and only if

$$max\, \mathcal{H} \leq \varepsilon. \tag{3.16}$$

Moreover, if the inequality (3.16) holds, then the navigation law (3.15) is an ε-intercepting navigation strategy.

*Proof:*

We prove that if Assumption 3.1 holds, then

$$M_i^-(t) = H_i^-(t)$$

$$M_i^+(t) = H_i^+(t) \tag{3.17}$$

where $M_i^-(t), H_i^-(t)$, $M_i^+(t), H_i^+(t)$ are defined by (3.5), (3.14). Indeed, let $P_3, P_4 \in S_i(t)^-$ and $c(P_3) < c(P_4$ where $S_i(t)^-$ is defined by (3.4). Then, for any $x$, we obviously have that $\alpha(x, P_3) \leq \alpha(x, P_4) + c(P_4) - c(P_3)$. This and Assumption 3.1 imply that

$$\beta(t, P_3) - \frac{\alpha(x,P_3)V_R^{max}}{V_I^{max}} \geq (t, P_4) - \frac{\alpha(x,P_4)V_R^{max}}{V_I^{max}} \tag{3.18}$$

For any $x$. This implies that

$$\sup_{P \in S_i(t)^-} \left(\beta(t, P) - \frac{\alpha(x(t),P)V_R^{max}}{V_I^{max}}\right) \tag{3.19}$$

is achieved at the left end of the interval $S_i(t)^-$. Therefore, $M_i^-(t) = H_i^-(t)$. Analogously, $M_i^+(t) = H_i^+(t)$. Hence, (3.17) holds and the statement of Theorem 3.2 follows from Theorem 3.1.

## 3.3 Simulations and discussion

In this section, we present an example that illustrates the main results of the paper. We consider a team of five mobile robots deployed on the boundary of a planar region to intercept the intruder that aims to exit this region.

We assume that $V_I^{max} = 4.2$ and $V_R^{max} = 3.0$, hence, Assumption 3.1 holds. Therefore, we apply Theorem 3.2 and the navigation law (3.15).

Fig.3.2 shows the positions and the motion directions of the robots when the intruder tends to exit the region R and is at the points a, b, c, and d. The robots are indexed in counter-clockwise direction from point $P1$ to point $P2$.

Fig.3.3, shows the evolution of the y-coordinates of the intruder and the robots when the intruder is moving along the trajectory shown in Fig.3.2.

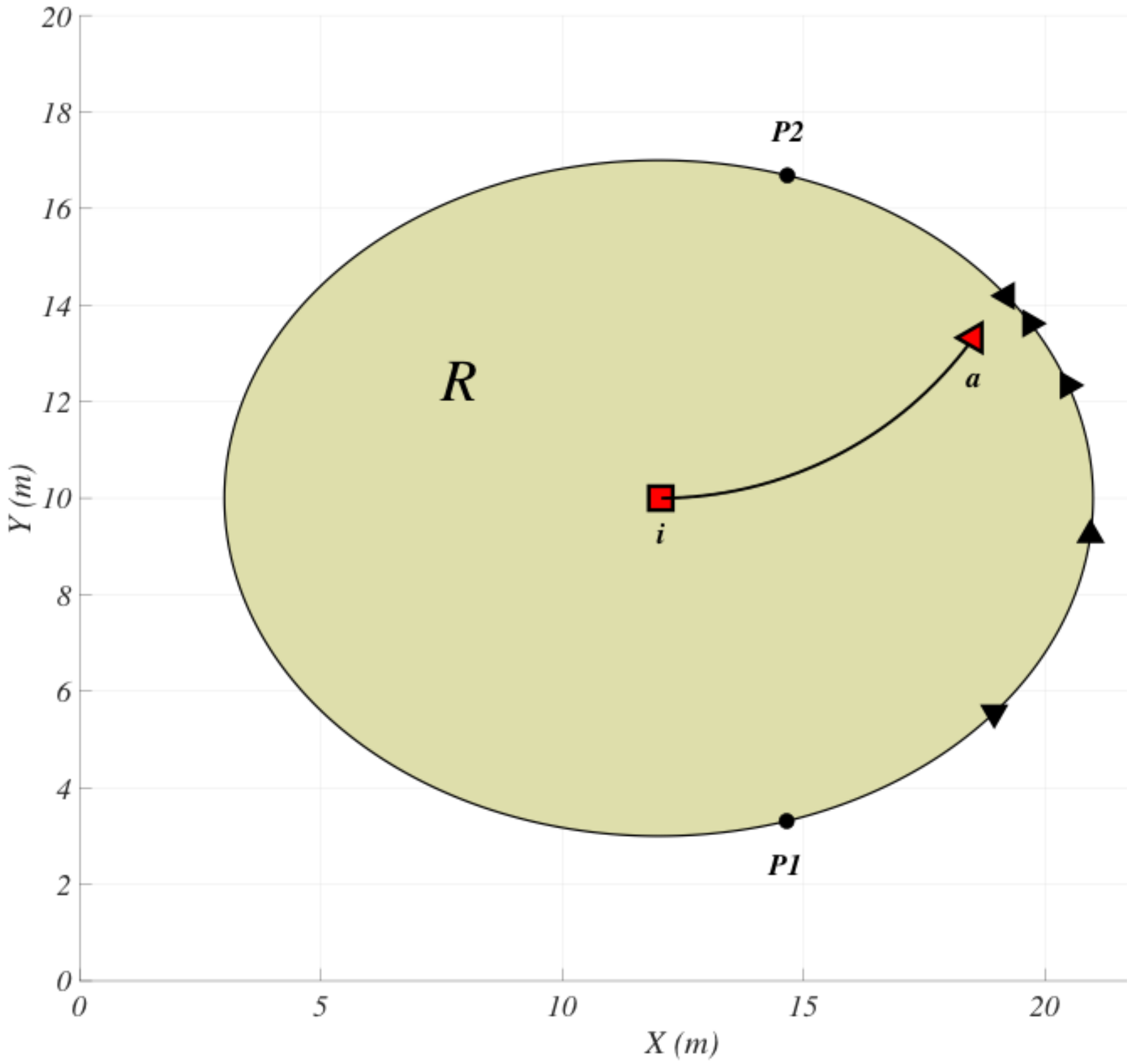

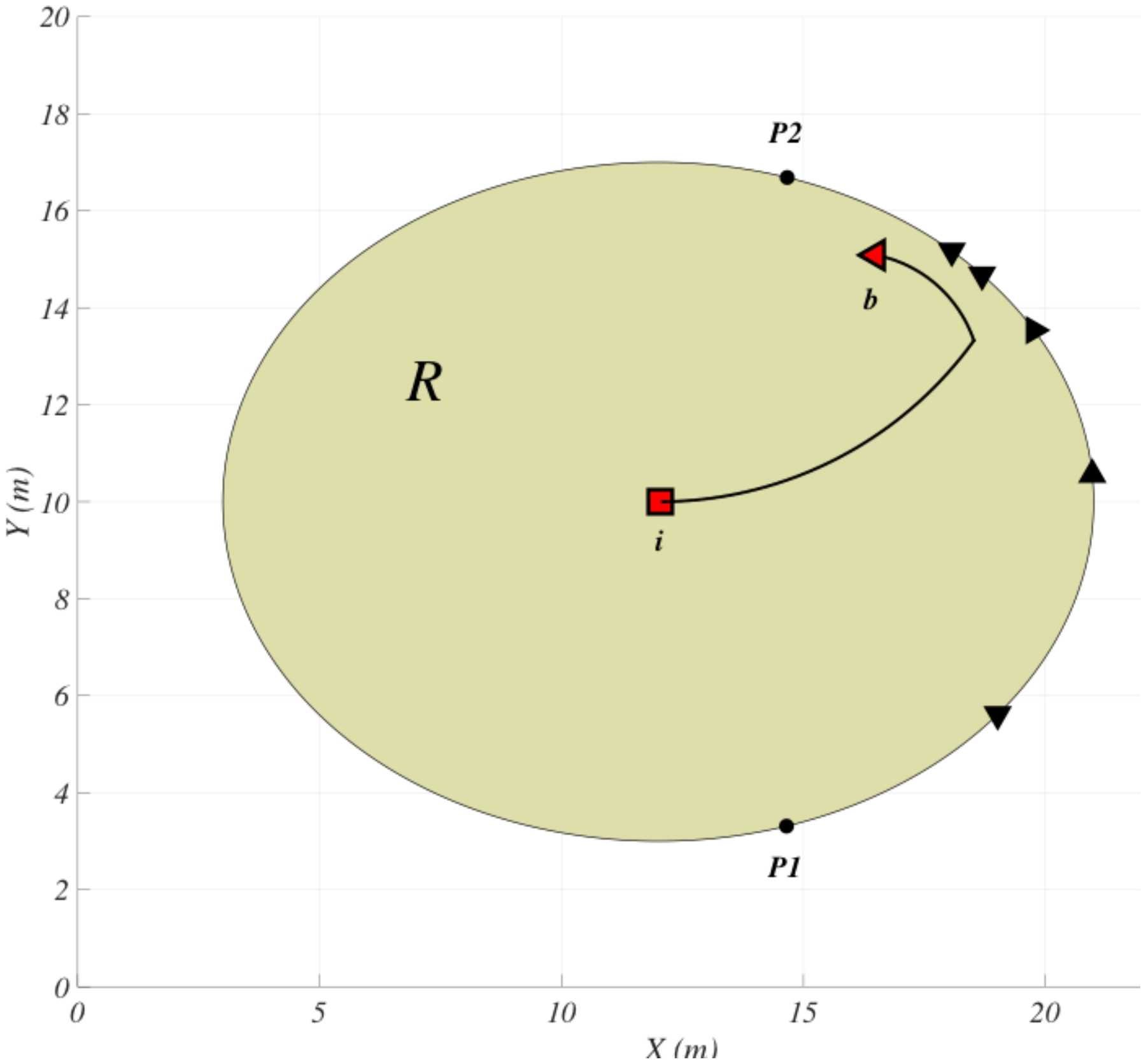

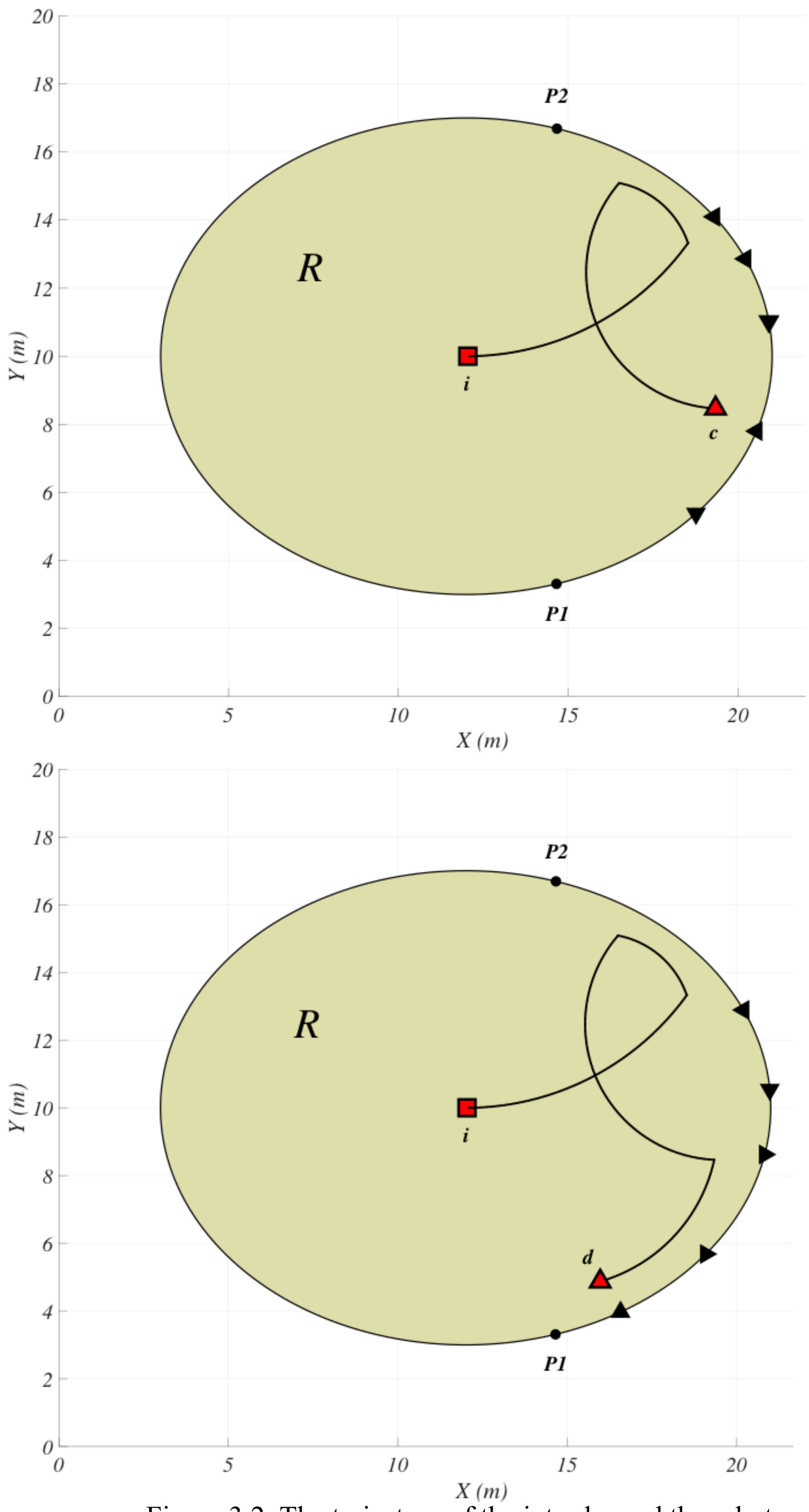

Figure 3.2: The trajectory of the intruder and the robots

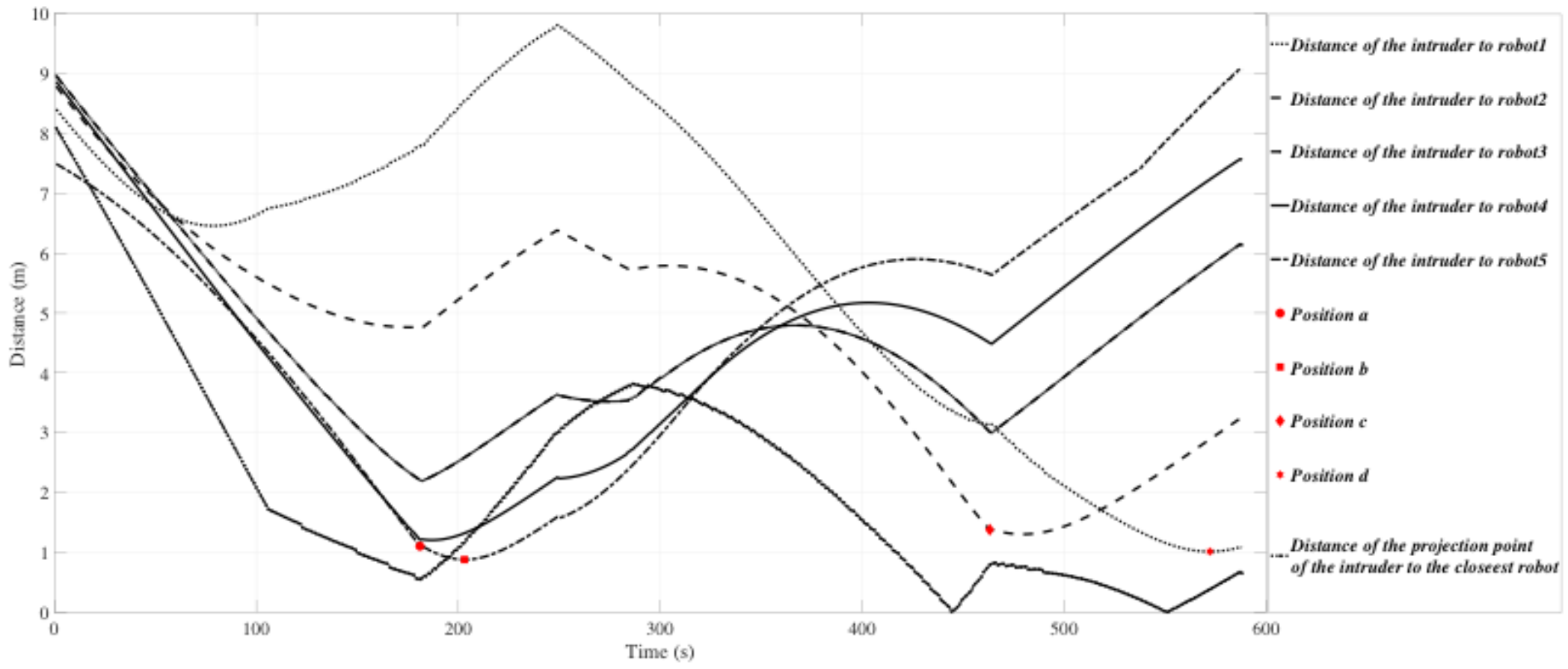

Figure 3.3: y-coordinates of the intruder and the robots

## 3.4 Summary

We proposed a decentralized motion control algorithm for a network of mobile robots to intercept an intruder on the boundary of a planar region. A necessary and sufficient condition for the existence of such an algorithm was derived. The proposed algorithm is based on some simple rules that only require information about the intruder and the closest neighbors of each robot. Computer simulations confirmed the efficiency of the developed navigation algorithm.

# Chapter 4

# *k-intercepting* navigation strategy

The problem of intruder interception in a boundary region by a multi-robot team is presented in this chapter. In this case, we propose a decentralized navigation algorithm for a fleet of multi robot with necessary and sufficient condition which led to always intercepting an unwanted intruder in the boundary of the region. The proposed decentralized navigation law is easy to implement in real time boundary protection applications, result from its non-demanding computational quiddity. The remainder of this section is organized as follows. We investigate the problem statements in section 4.1. The main results and proof of the theorems are presented in Section 4.2. The simulations that confirms the validation of the proposed navigation strategy are given in Section 4.3; and finally, Section 4.4 gives a summary of the chapter.

## 4.1 Problem statement

We consider a closed linearly connected region $R$ which is bounded with a piecewise smooth boundary. The segment $S$ of the boundary is supposed to be a portion which is located between two arbitrary points denoted as $P_1$ and $P_2$ as it shows in Fig.4.1.

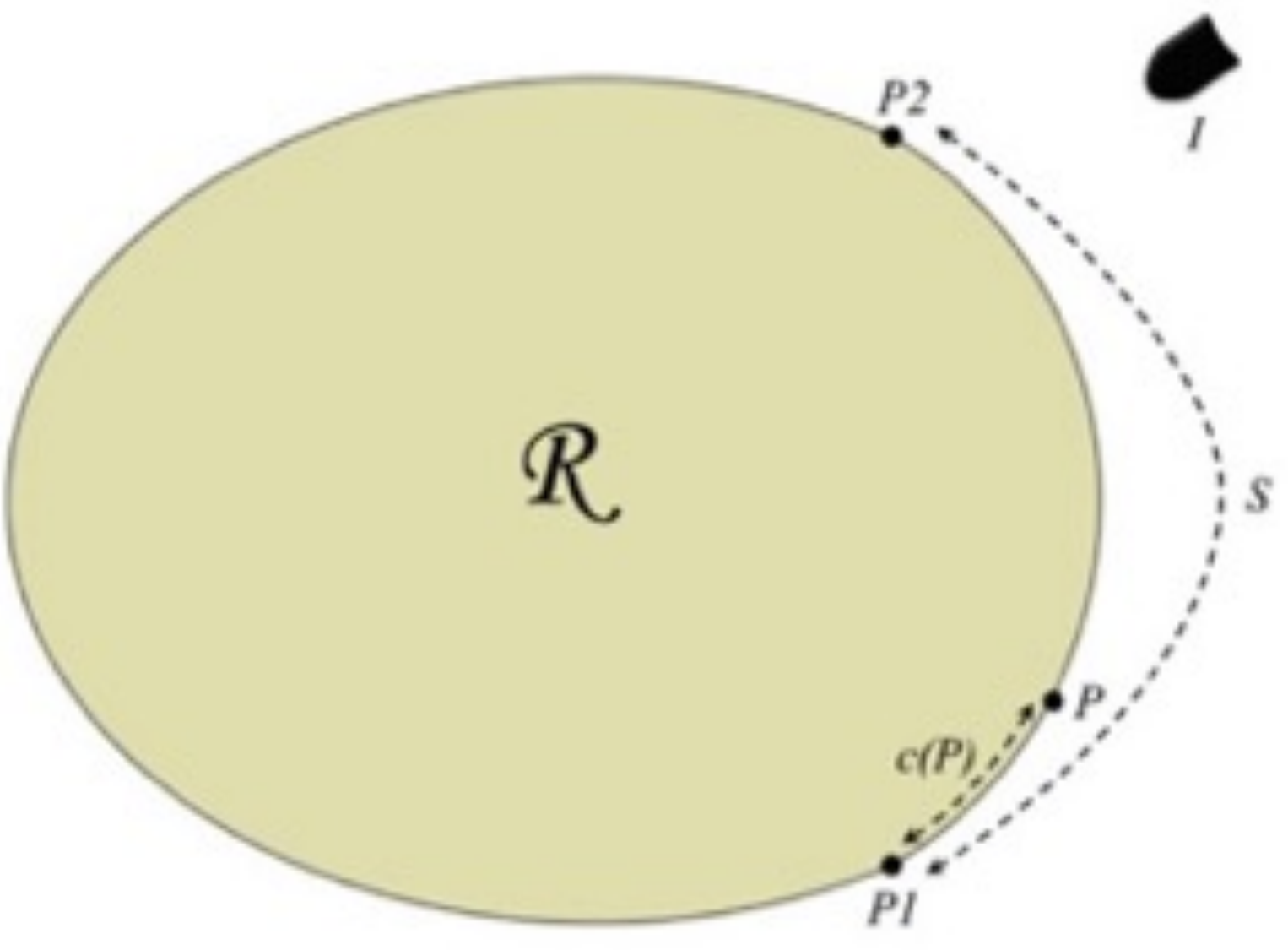

Figure 4.1: Bounded region *R*, Intruder *I* and boundary segment *S*

An intruder $I$, considered to be a point-wise object moves towards the segment $S$ of bounded region $R$ to intrude the region of interest from the outside for a subversive mission.

If we suppose $x_I(t)$ be planar coordinates of the intruder $I$ that is moving in the plane with an arbitrary vector velocity $v_I(t) = \dot{x}_I(t)$, then the Euclidian norm of the vector velocity $v_I(t)$ should not exceed a given constant velocity at any time as follows:

$$\|v_I(t)\| \leq V^{max} \qquad \forall\, t \geq 0 \tag{4.1}$$

Where $V^{max} > 0$ is a given constant and $\|\cdot\|$ represents the standard Euclidean vector norm.

Furthermore, a given positive integer $n$ denotes the number of robots in the team which are responsible for protecting the boundary of region $R$ against any intruder's attack. In this scenario the robots can move on the segment $S$ in both orientation towards the points $P_1$ and $P_2$ while the intruder can move in any direction in the plane. The planar coordinates of the robots *1,2, ..., n,* are denoted by $x_1(t), x_2(t), \dots, x_n(t)$. In continue, $c(P)$, denotes the curvilinear coordinate of any point $P \in S$ . If we assume $L$ is the full length of the segment $S$, then $c(P)$ is the length of any portion of $S$ which lies between point $P_1$ and $P$, as shown in Fig 4.1.

It connotes that $c(P_1) = 0$, and $c(P_2) = L$. The curvilinear coordinates of robots 1,2,...,n are shown by $c_1(t) := c(x_1(t)), c_2(t) := c(x_2(t)), \dots, c_n(t) := c(x_n(t))$ , at any time $t \geq 0$. Meanwhile, the robots are labeled according to their curvilinear coordinates as in:

$$0 \leq c_1(t) \leq c_2(t) \leq \cdots \leq c_n(t) \leq L \qquad \forall t \geq 0. \tag{4.2}$$

The condition (4.2) implies that the order of the robots have never changed on the segment $S$. Equation (4.3) describes the motion of the robots among the segment $S$ as follows:

$$\dot{c}_i(t) = u_i(t) \qquad \forall i = 1, \dots, n \tag{4.3}$$

where $u_i$ , is the control input of the robot $i$, that satisfies the following constraint:

$$|u_i(t)| \leq V_R^{max} \quad \forall t \geq 0 \tag{4.4}$$

In (4.4), $V_R^{max} \geq 0$ is a given constant.

## *Available measurements:*

We suppose a given positive integer $1 \leq k \leq \frac{n}{2}$ . Then, for all $i - k \leq j \leq i + k$ at any time $t$, each robot $i$ knows the curvilinear coordinates $c_j(t)$ of any robot $j$ as wel as its own curvilinear coordinate $c_i(t)$. Furthermore, all the robots know the planar coordinate $x_I(t)$ of the intruder for all the time $t \geq t_0$, where $t_0 \geq 0$ indicates the time that the intruder becomes visible to the multi robot team.

*Definition 4.1:*

Let $\varepsilon > 0$ be a given constant. Suppose that the intruder $I$ cross the segment $S$ at time $t^*$i.e. $x_I(t^*) \in S$. The multi-robot team *k-intercepts* the intruder at time $t^*$if there exist some index $i = 1,2, \dots, n$ such that $\left|c\big(x_I(t^*)\big) - c_j(t^*)\right| \le \varepsilon \;\; \forall\, i \le j \le i + k - 1$.

Therefore, a navigation strategy of a multi-robot team is called *k-intercepting* strategy if the multi-robot team *k-intercepts* the intruder at the time the intruder crosses the segment $S$.

Now we state the problem. In this scenario, the objective of the intruder is crossing the segment $S$ to enter region $R$ while avoiding being *k-intercepted* by the multi-robot boundary protection team. On the other hand, the aim of the robots is to protect the region $R$, by *k-intercepting* the intruder. In the next section, we show that how the proposed decentralized navigation strategy always results in *k-intercepting* for any movement of the intruder in the plane.

## 4.2 *k-intercepting* algorithm and the main results

consider $P$ be a point on segment $S$ and $x$ be a point outside the region $R$. Then $\mathcal{P}(x, P)$ denotes a set of all continues piecwise smooth paths $L(x, P)$ that connects point $x$ to the $P$, where the point $P$ is te only intersection point of the path $L(x, P)$ and the region $R$. Furthermore, let $\alpha(x, P) = inf\ (L(x, P)) \in \mathcal{P}(x, P)$.

Moreover, $i(1), i(2), \dots, i(n)$, supposed to be the permutation of indices *1, 2,...,n,* which satisfy $\left|c_{i(1)}(t) - c(P)\right| \le \left|c_{i(2)}(t) - c(P)\right| \le \dots \le \left|c_{i(n)}(t) - c(P)\right|$.Then we introduce a variable $\beta(t, P)$, such that, $\beta(t, P) \coloneqq \left|c_{i(k)}(t) - c(P)\right|$. In the other word, $\beta(t, P)$ is the *k*-th among all the robots according to the length of sub-segments of the segment $S$ from the current position of each robot to the point $P$. In continue,

we introduce the function $F(s)$ from the interval $[0, L]$ to the segment $S$ such that for any number $s \in [0, L], F(s)$, there is a point $P \in S$ where $c(P) = s$. Furthermore, suppose a closed sub-segment $|c(A_2) - c(A_1)|$ for $c(A_2) > c(A_1)$, of the segment $S$, which lies between points $A_1$and $A_2$. For $i = k, k+1, \dots, n-k+1$, introduce sub-segments $S_i(t)^-, S_i(t)^+$ of the segment $S$ as follows:

$$S_i(t)^- := \left[F\left(\frac{c_i(t) + c_{i-k}(t)}{2}\right), F\left(\frac{c_i(t) + c_{i-k+1}(t)}{2}\right)\right]$$

$$if\ i = k+1, \dots, n-k+1;$$

$$S_i(t)^- := \left[P_1, F\left(\frac{c_i(t) + c_{i-k+1}(t)}{2}\right)\right]$$

$$if\ i = k;$$

$$S_i(t)^+ := \left[F\left(\frac{c_i(t) + c_{i+k-1}(t)}{2}\right), F\left(\frac{c_i(t) + c_{i+k}(t)}{2}\right)\right]$$

$$if\ i = k, \dots, n-k;$$

$$S_i(t)^+ := \left[F\left(\frac{c_i(t) + c_{i+k-1}(t)}{2}\right), P_2\right]$$

$$if\ i = n-k+1. \tag{4.5}$$

Furthermore, we introduce the numbers $M_i^-(t)$ and $M_i^+(t)$ for $i = k, k+1, \dots, n-k+1$ as in:

$$M_i^-(t) := \sup_{P \in S_i(t)^-} \left(\beta(t, P) - \frac{\alpha(x_I(t), P) V_R^{max}}{V_I^{max}}\right);$$

$$M_i^+(t) := \sup_{P \in S_i(t)^+} \left(\beta(t, P) - \frac{\alpha(x_I(t), P) V_R^{max}}{V_I^{max}}\right). \tag{4.6}$$

Now we introduce the following decentralized navigation law:

$$\begin{cases} \begin{cases} u_i(t) := V_R^{max} & if \quad 0 < i < k \\ u_i(t) := -V_R^{max} & if\ \ n-k+1 < i \le n \end{cases} & | \quad M_i^-(t) \neq M_i^+(t) \\ \{u_i(t) := 0 & |M_i^-(t) = M_i^+(t) \end{cases} \tag{4.7}$$

For $i = k, k+1, \dots, n-k+1$,

$$u_i(t) := V_R^{max} \quad if \quad M_i^-(t) < M_i^+(t)$$

$$u_i(t) := -V_R^{max} \quad if \quad M_i^-(t) > M_i^+(t)$$

$$u_i(t) := 0 \quad if \;\; M_i^-(t) = M_i^+(t) \tag{4.8}$$

## *Remark 4.1*:

The intuition behind the decentralized navigation law can be explained as follows. The sub-segments $S_i(t)^-, S_i(t)^+$ are sets of point of the curve $S$ for which the robot $i$ is the $k-th$ furthest robot at time *t*. The robot moves with the maximum allowed speed towards the one of these segments that is more" dangerous" at the current time, i.e. it has the biggest possible distance between the intruder and the $k-th$ robot at the moment of crossing $S$ by the intruder. This biggest possible distance is described by (4.6). Moreover, the following assumption is required.

## *Assumption 4.1:*

For any trajectory $[x_I(t), c_1(t), \dots, c_n(t)]$ of the proposed system, there is no more than finite number of time instant $\tau$ where $M_i^-(\tau) = M_i^+(\tau)$

## *Theorem 4.1:*

Consider the multi-robot team (4.1) and the intruder are satisfying (4.1), (4.4) and Assumption 4.1. Let $1 \leq k < \frac{n}{2}$ be a given positive integer. Then there exists a k−intercepting multi-robot team navigation strategy if and only if:

$$\sup_{P \in S} \left( \beta(t_0, P) - \frac{\alpha(x_I(t_0), P) V_R^{max}}{V_I^{max}} \right) \leq \varepsilon \tag{4.9}$$

In which, the intruder becomes visible to the robots at time $t_0 \geq 0$. Moreover, the navigation law (4.7) and (4.8) is a *k-intercepting* navigation strategy if the inequality (4.9) holds.

## *Proof:*

At the first stage, we suppose that inequality (4.9) dos not hold. Then we prove that the intruder always can cross the segment $S$ without being *k-intercepted* by the multi-robot team.

Indeed, if (4.9) does not hold, then there exists a point $P \in S$ and a continuous piecewise smooth paths $L(x_I(t_0), P)$ connecting $x_I(t_0)$ and $P$ such that the intersection of $L(x_I(t_0), P)$ and $R$ contains only the point $P$ , and

$$\left(\beta(t_0, P) - \frac{|L(x_I(t_0), P)| V_R^{max}}{V_I^{max}}\right) > \varepsilon \tag{4.10}$$

Where $|L(x_I(t_0), P)|$ indicate the length of piecewise smooth path $L(x_I(t_0), P)$. Now, if the intruder moves with the linear velocity $V_I^{max}$ along the path $L(x_I(t_0), P)$, it reaches the intersection point $P$ at time $t_* = t_0 + \frac{|L(x_I(t_0), P)|}{V_I^{max}}$. It obviously follows from (4.10) that the $k$-th robot, cannot be closer than $\varepsilon$ to the point $P$ at time $t^*$ , and the intruder is not *k-intercepted* by the multi-robot consequently.

At the second stage, we prove that the multi-robot team which is navigated by the navigation law (4.7), (4.8) always *k-intercepts* the intruder at the time it crosses the segment $S$, if inequality (4.9) holds. For any trajectory $[x_I(t), c_1(t), \dots, c_n(t)]$ of the intruder-multi-robot system, introduce the Lyapunov function as in:

$$W[x_I(t), c_1(t), \dots, c_n(t)] := \sup_{P \in S} \left(\beta(t, P) - \frac{|L(x_I(t), P)| V_R^{max}}{V_I^{max}}\right) \tag{4.11}$$

Therefore, the inequality (4.12) follows from navigation law (4.7), (4,8) and Assumption (4.1) as follows:

$$W[x_I(t_1), c_1(t_1), \dots, c_n(t_1)] \leq W[x_I(t_2), c_2(t_2), \dots, c_n(t_2)]$$

$$\forall t_2 \geq t_1 \geq t_0 \tag{4.12}$$

From (4.10) and (4.12) imply that if the intruder reaches a point $P \in S$ at some time $t^* \geq t_0$, the $k$-th robot cannot be further than $\varepsilon$ to the point $P$ at the time $t^*$. Thus, the $\varepsilon$-neighbourhood of the intersection point $P = x_I(t^*)$ on the segment $S$ is occupied by at least $k$ robot at time $t^*$, which implies that the intruder is $k$-intercepted by the multi robot team. Therefore, the Theorem 4.1 is proved.

The inequality (4.9) and the navigation law (4.7), (4.8) could be simplifies if the following assumption holds which results in simplicity in computation.

## *Assumption 4.2:*

We assume that the maximum linear velocity of each agent in the multi robot team never exceeds the maximum linear velocity of the intruder such that $V_I^{max} \geq V_R^{max}$ holds.

Then we introduce points $D_i(t)^-, D_i(t)^+$ of the segment $S$ for $i = k, k+1, \dots, n-k+1$ as follows:

$$D_i(t)^- := F\left(\frac{c_i(t)+c_{i-k}(t)}{2}\right) \qquad if\ i = k+1, \dots, n-k+1;$$

$$D_i(t)^- := P_1 \qquad if\ i = k;$$

$$D_i(t)^+ := F\left(\frac{c_i(t)+c_{i+k}(t)}{2}\right) \qquad if\ i = k, \dots, n-k;$$

$$D_i(t)^+ := P_2\ \ if\ i = n-k+1. \tag{4.13}$$

Moreover, we introduce some numbers $H_i^-(t), H_i^+(t)$ for $i = k, k+1, \dots, n-k+1$ as follows:

$$H_i^-(t) := \frac{c_i(t)-c_{i-k}(t)}{2} - \frac{\alpha(x_I(t), D_i(t)^-)V_R^{max}}{V_I^{max}}$$

$$if\ i = k+1, \dots, n-k+1;$$

$$H_i^-(t) := c_i(t) - \frac{\alpha(x_I(t), P_1)V_R^{max}}{V_I^{max}}$$

$$if\ i = k;$$

$$H_i^+(t) := \frac{c_{i+k}(t)-c_i(t)}{2} - \frac{\alpha\left(x_I(t), D_i(t)^+\right)V_R^{max}}{V_I^{max}}$$

$$if\ i = k, \dots, n-k;$$

$$H_i^+(t) := L - c_i(t) - \frac{\alpha(x_I(t), P_2)V_R^{max}}{V_I^{max}}$$

$$if\ i = n-k+1. \tag{4.14}$$

The navigation law (4.7) stays the same for $i < k$ or $i > n-k$, otherwise, for $i = k, k+1, \dots, n-k+1$, the navigation law (3.8) is simplified as follows:

$$u_i(t) \coloneqq V_R^{max} \quad if \quad H_i^-(t) < H_i^+(t)$$

$$u_i(t) \coloneqq -V_R^{max} \quad if \quad H_i^-(t) > H_i^+(t)$$

$$u_i(t) \coloneqq 0 \quad if \quad H_i^-(t) = H_i^+(t). \tag{4.15}$$

Furthermore, the Assumption 4.1 is modified as follows:

## *Assumption 4.3:*

For any trajectory $[x_I(t), c_1(t), \dots, c_n(t)]$ of the proposed system, there is no more than finite number of time instant $\tau$ where $H_i^-(\tau) = H_i^+(\tau)$.

## *Theorem 4.2:*

Consider the multi-robot team (4.1) and the intruder are satisfying (4.1), (4.4) and Assumptions 4.2, 4.3. Let $1 \leq k < \frac{n}{2}$ be a given positive integer. Furthermore suppose $\mathcal{H}$ be the set of numbers $H_i^-(t_0), H_i^+(t_0)$ for $i = k, k+1, \dots, n-k+1$. Let $t_0 \geq 0$ be the time that the intruder becomes visible to the robots. Then there exists a k−intercepting multi-robot team navigation strategy if and only if

$$max\, \mathcal{H} \leq \varepsilon. \tag{4.16}$$

Furthermore, the navigation law (4.7), (4.15) is a *k-intercepting* navigation strategy if the inequality (4.16) holds.

## *Proof of theorem 4.2:*

We prove that:

$$M_i^-(t) = H_i^-(t) \text{ and } M_i^+(t) = H_i^+(t) \tag{4.17}$$

If Assumption 4.2 holds, where $M_i^-(t), M_i^+(t)$, $H_i^-(t)$, and $H_i^+(t)$ are defined by (4.6) and (4.14). indeed, there exist two points $P_3$ , $P_4$ are elements of $S_i(t)^-$ such that $c(P_3) < c(P_4)$, where $S_i(t)^-$ is defined by (4.5).

then it's obvious that for any $x$:

$$\alpha(x, P_3) \leq \alpha(x, P_4) + c(P_4) - c(P_3) \tag{4.18}$$

Therefore, the assumption 4.2, and inequality (4.18) imply that (4.19) is achieved at the left end of the interval $S_i(t)^-$.

$$\sup_{P \in S_i(t)^-} \left( \beta(t, P) - \frac{\alpha(x_I(t), P) V_R^{max}}{V_I^{max}} \right) \tag{4.19}$$

This prove that $M_i^-(t) = H_i^-(t)$, and similarly $M_i^+(t) = H_i^+(t)$. Therefore, inequality (4.17) holds and the statement of Theorem 4.2 follows from theorem 4.1.

## 4.3 Simulations

In this scenario, the area supposed to be bounded except in one side where is monitored by a team of pointwise mobile robots.

At the first stage, we assumed that the gateway of the area is located between two possible points P1 and P2, and the team of interceptors moves along a straight line or a curve between these two points. The simulations result shows the validation of the navigation law (4.7) and (4.15) while inequality (4.16) holds.

### 4.3.1 Protecting a straight boundary region

At the first stage, we consider the conditions that $k = 1$ and the multi-robot motion path is a straight line.

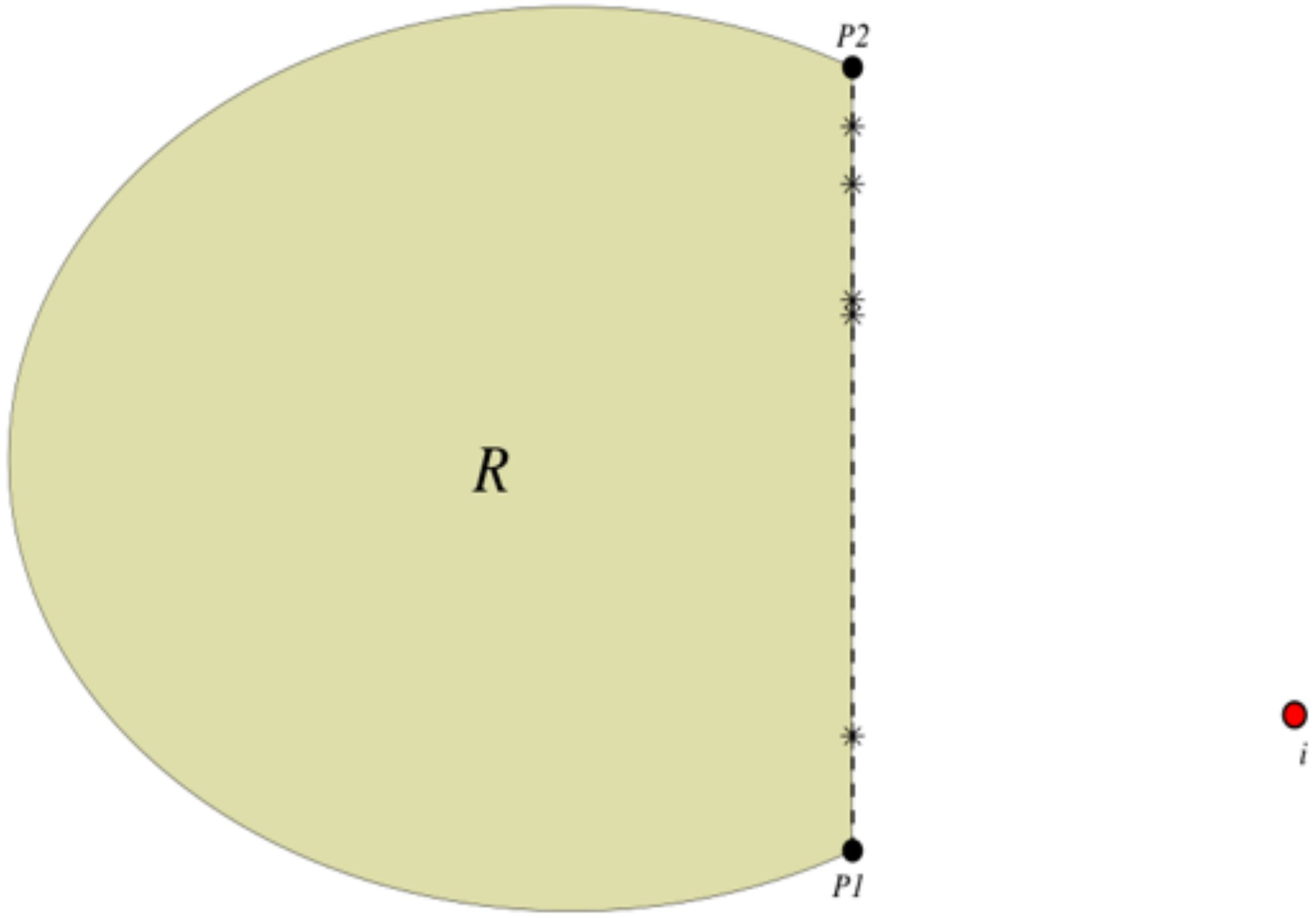
P2
R
i
P1

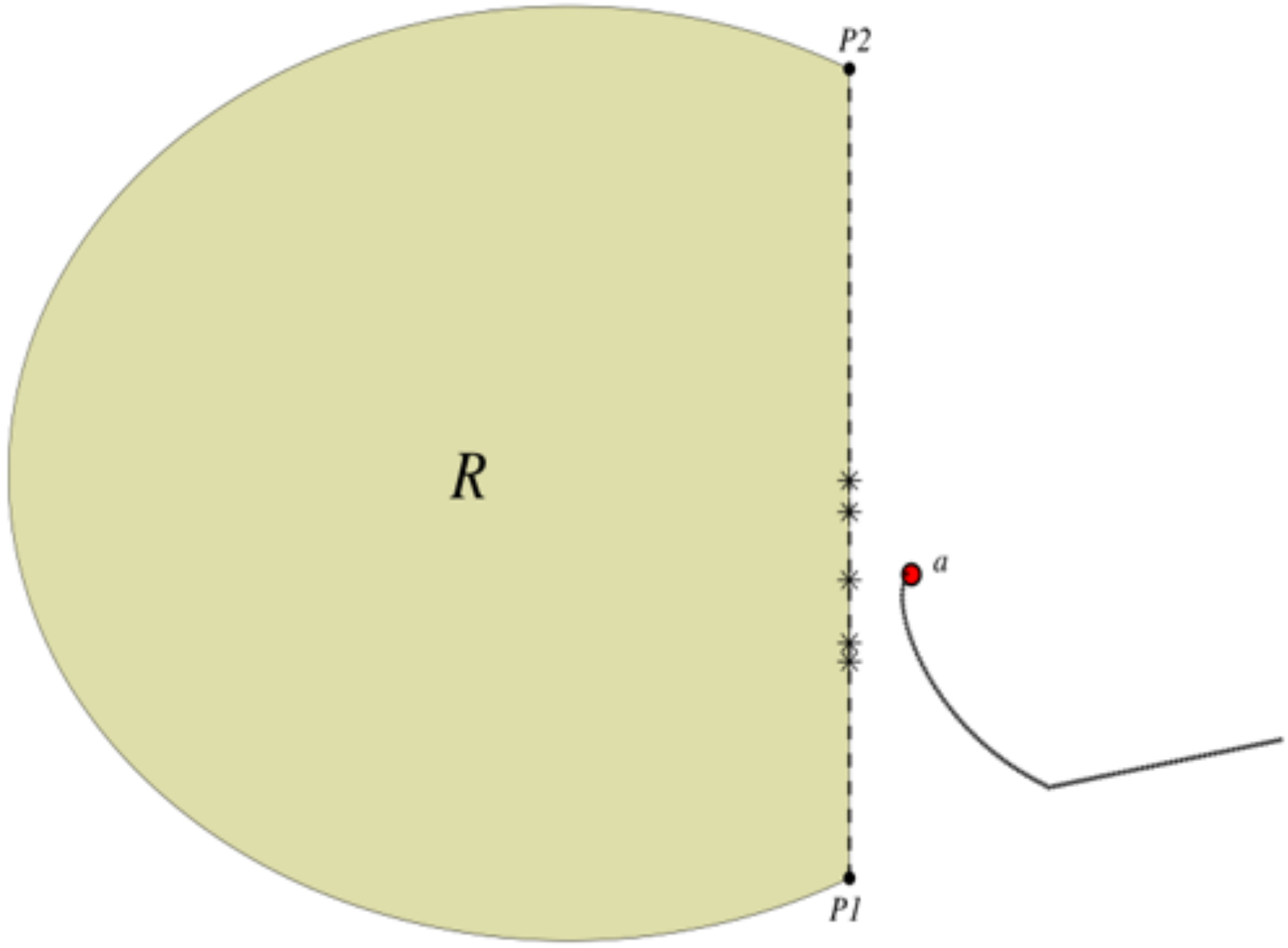
P2
R
a
P1

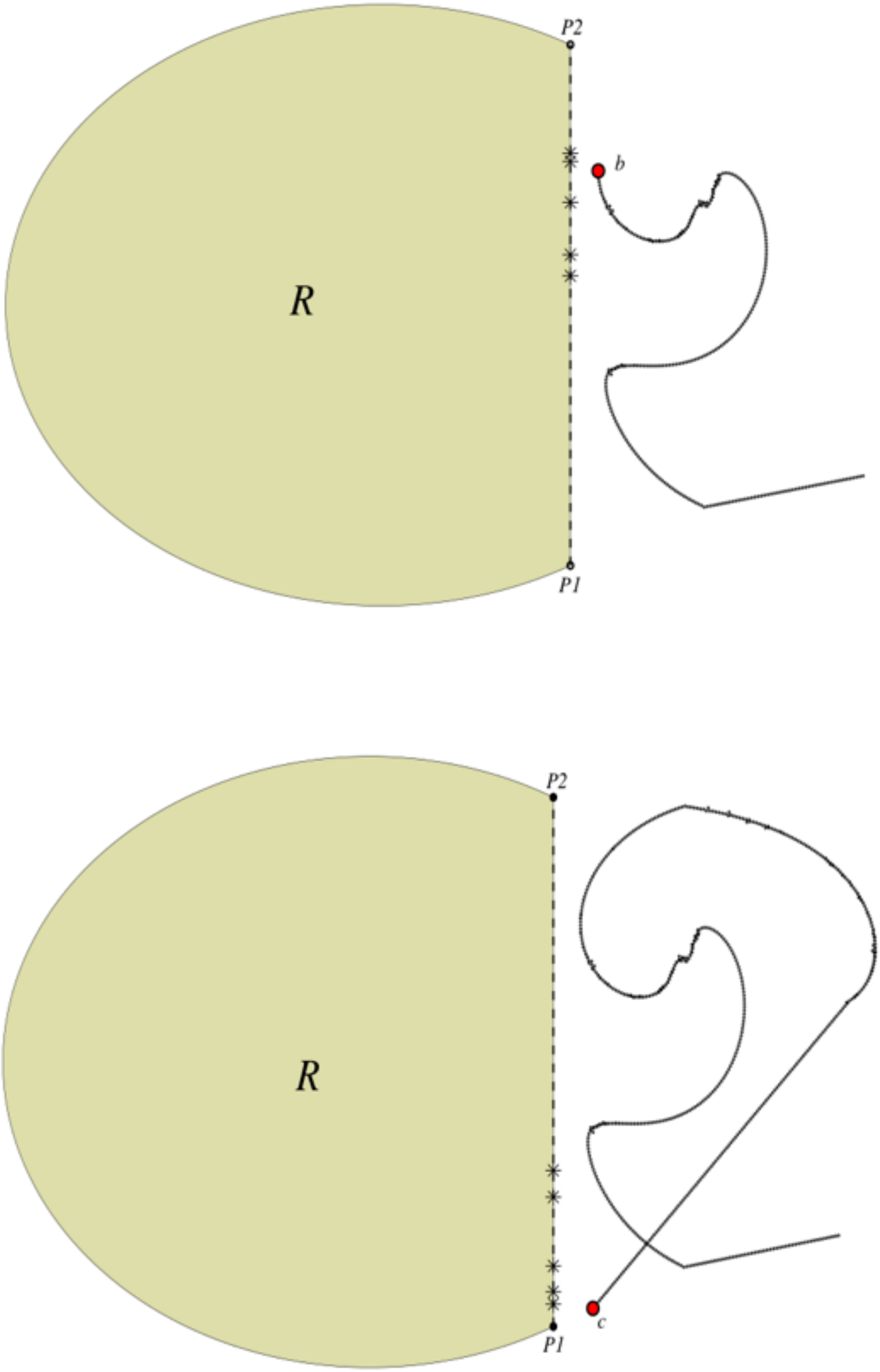

Figure 4.2 (a,b,c,d): *k-intercepting* the intruder on a straight boundary region

Fig.4.2 shows the area of interest and the multi-robot team versus the intruder. As it is illustrated in this figure, there are three points, where the intruder has its minimum distance with the gateway and the team of interceptors consequently. Furthermore, based on the value of k=1, it is expected at least one robot intercepts the intruder while it is getting closer to the gateway. Analysing Fig4.3 (a, b, c) explains how the mobile robot team acts against the intruder in critical points.

Fig.4.3 (a, b, c) illustrates three sections where the intruder has its minimum distance with the entry of the environment respectively.

Furthermore, each bar represents the distance of each robot from point P1. Moreover, the green bar which belongs to the intruder represents the distance between the intersection point of the perpendicular line from the intruder to the entrance and the position P1 either. It is explicitly obvious that the robot four is intercepting the intruder when the intruder is getting close to the area for the first time. Fig.4.3(b) shows the robot 4 is the closest interceptor to the intruder while the intruder is getting close to the environment for the second time either and finally robot 1 is the closest interceptor to the intruder when the intruder is getting close to the field for the third time.

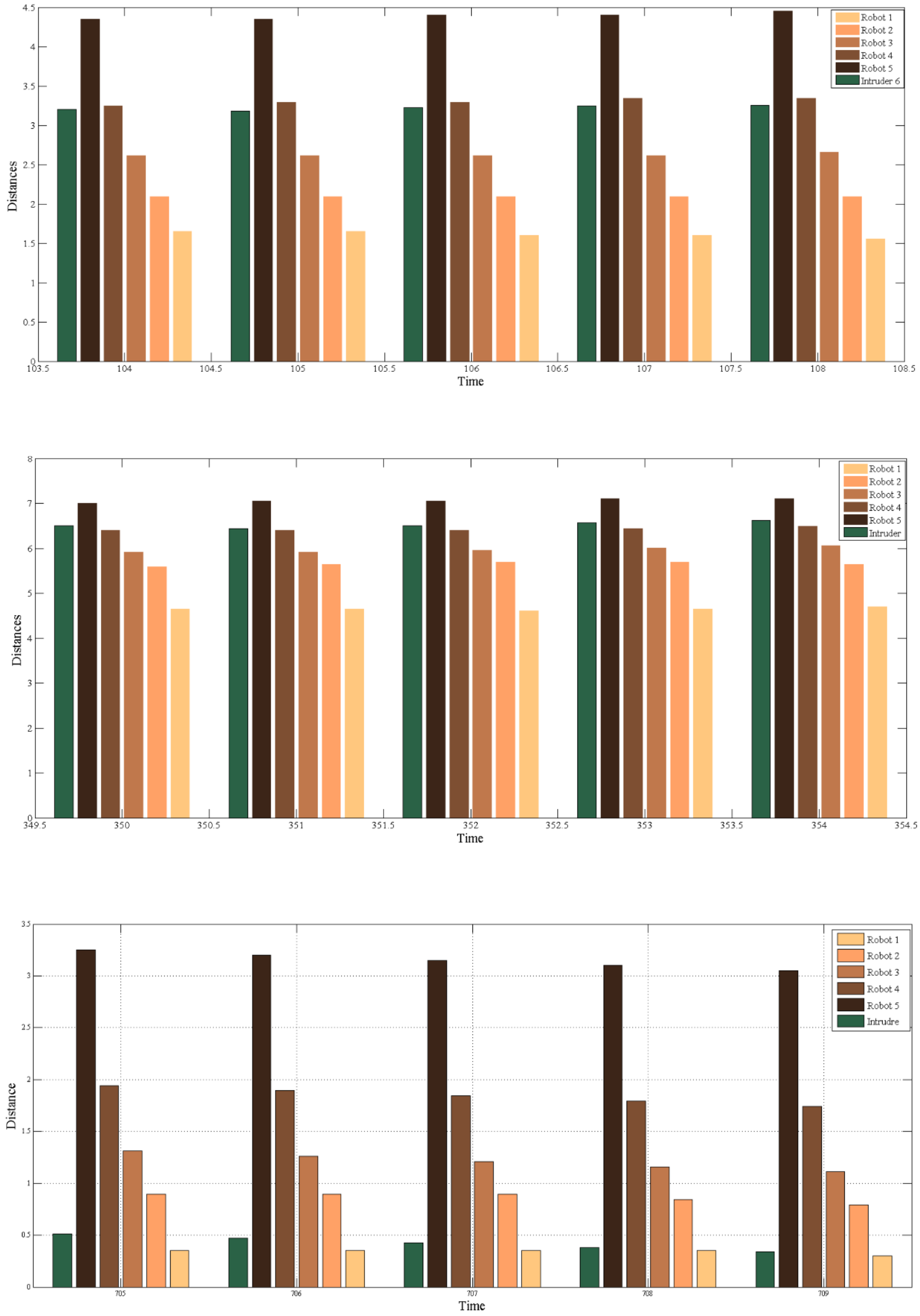

Figure 4.3 (a, b, c): The position of the intruder and the interceptors while the intruder is in its closest position to the straight region boundary for *k=1*.

As it shows in Fig.4.4, robot 4 which is plotted on the light blue graph is intercepting the intruder when the intruder has the minimum distance with the entry of the field between time intervals [103,…,108] and [349,…,355]. On the other hand, the robot 1 intercepts the intruder in time interval [704,…,709].

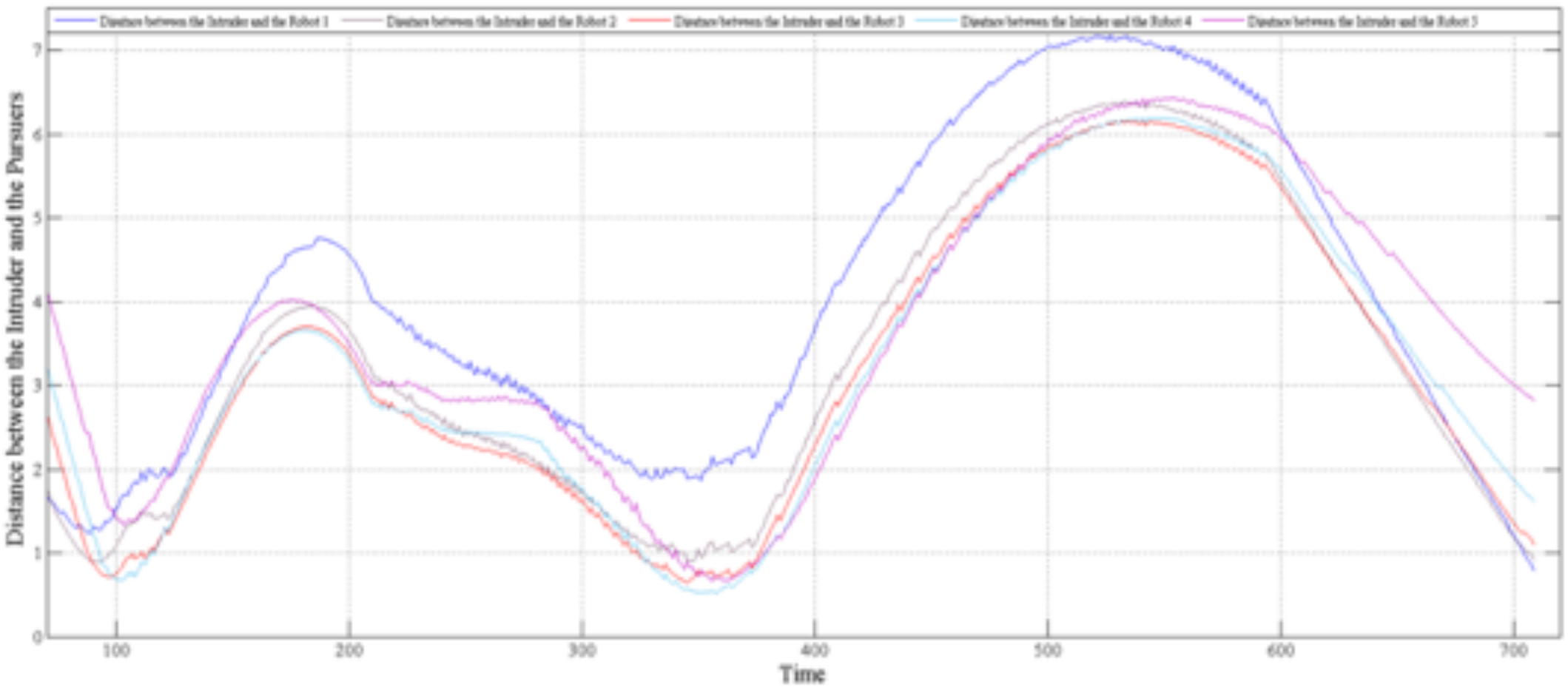

Figure 4.4 Distance between the intruder and each member of the multi-robot team in a straight region boundary when for *k=1*.

In Fig.4.5, we considered a virtual point which is supposed to be the correspondence of the intruder coordinates in its path to the gateway which is located between point P1 and P2 at time $t = (0,1,2,\dots)$. As it shows in the Fig.3.5, the corresponding coordinates of the intruder is intersecting with the robot 4 and 1 in critical time intervals [103,…,108],[349,…,355] and [704,…,709] respectively, therefore, there is at least one robot intercepts the intruder when it is trying to intrude to the environment.

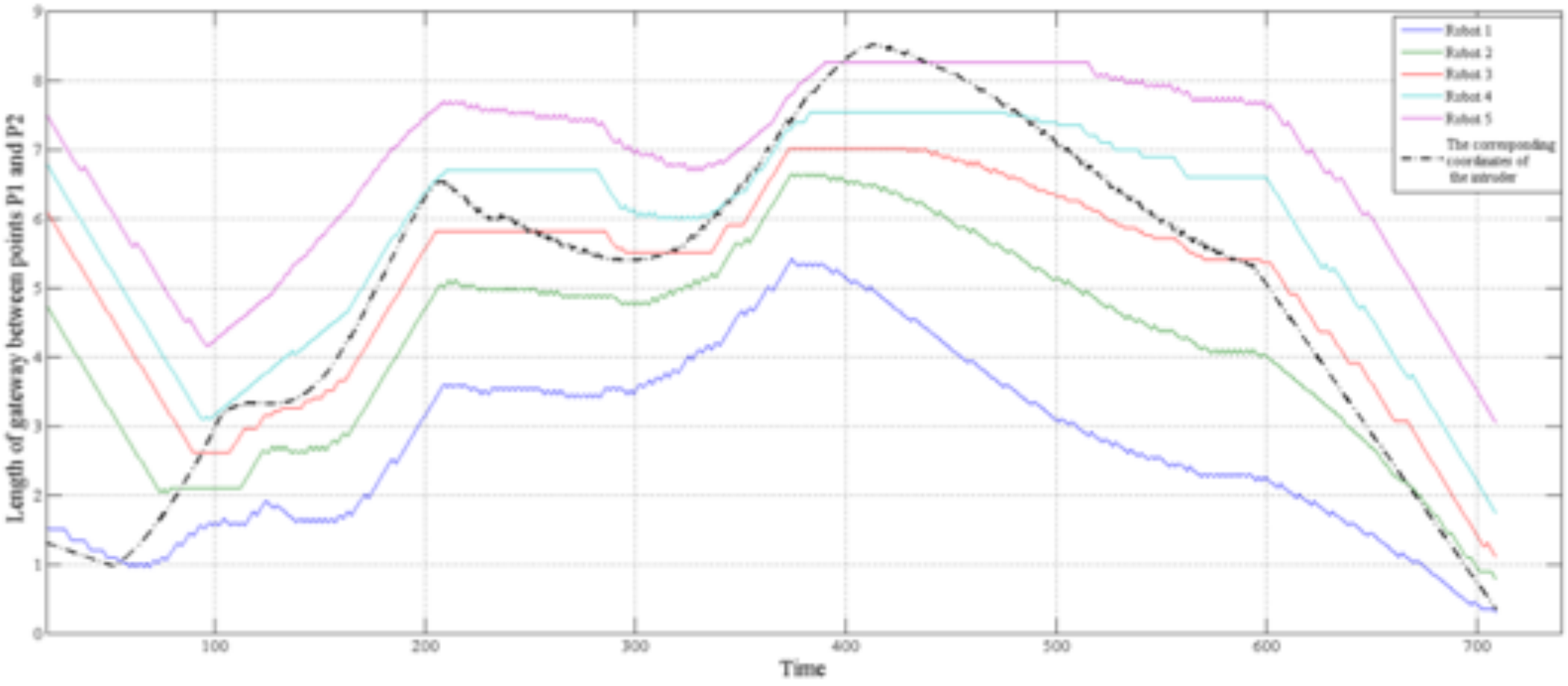

Figure 4.5: *k-intercepting* the intruder in a straight region boundary for *k=1*.

In continue, we consider the condition that $k = 2$ but the enetry is a straight line located between points P1 and P2.

In this case, Fig.4.6 (a, b, c) confirms the validation of the navigation laws (4.7) and (4.15). Comparing Fig.4.6 and Fig.4.3 illustrates how the team of robots acts differently when the value of k has changed. According to the Fig.4.6 (a, b, c), two robots are intercepting the intruder at the critical sections. On the other hand, Fig.4.7, confirms that in time interval [103,…,109] the intruder is intercepted by the robots 3 and 4. In the next critical segment in time interval [340,…,346], the intruder is intercepted by the robots 4 and 5, and finally at the time interval [704,…,710], robots 1 and 2 intercepting the intruder. As a definite confirmation, we refer to Fig.3.8. In time interval [103,…,109],the Robot three which is plotted on red covers the point with the high possibility of intrusion in cooperative with robot four which tends to move towards the critical segment.

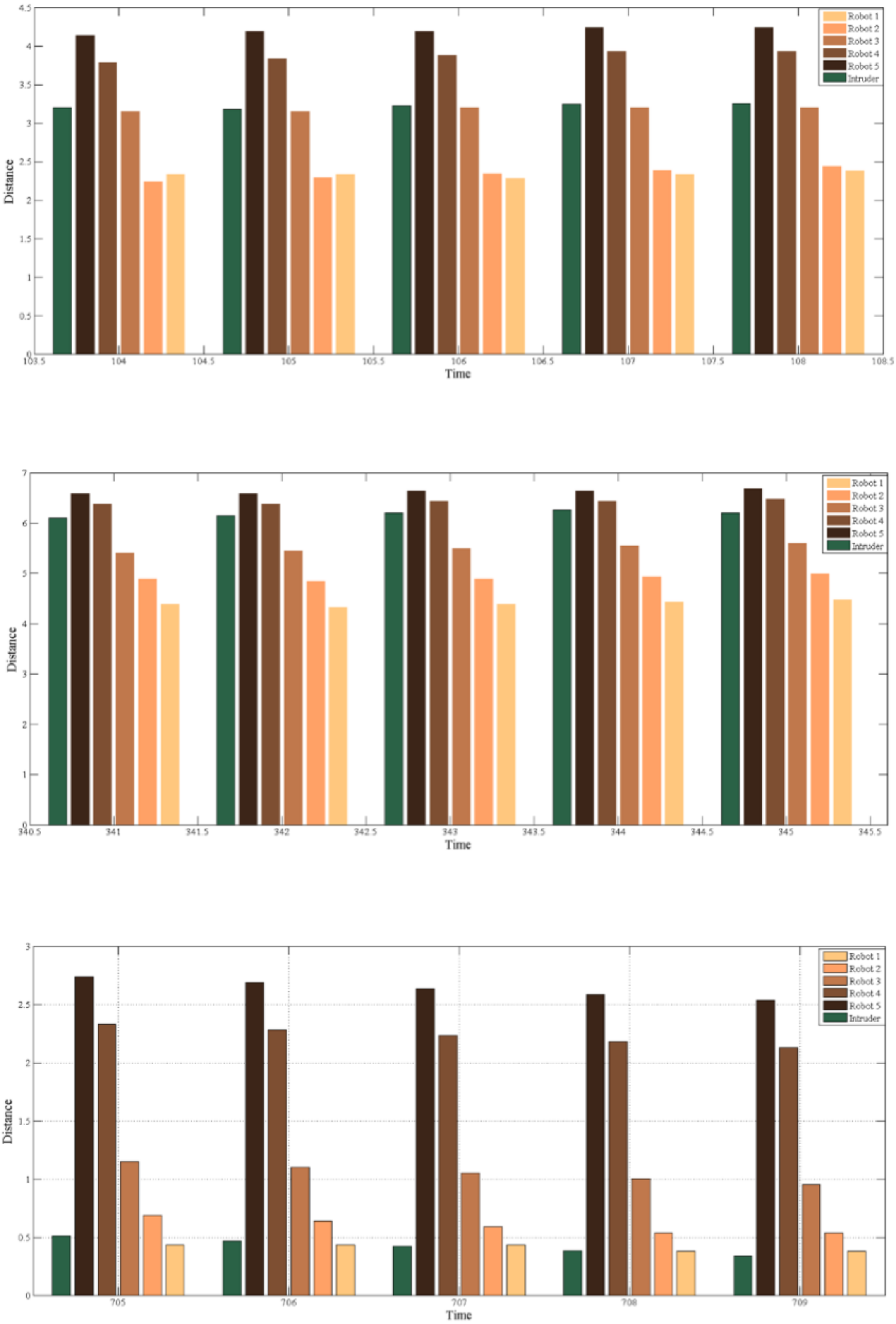

Figure 4.6 (a, b, c): The position of the intruder and the interceptors while the intruder is in its closest position to the straight region boundary for *k=2*.

In the second step, when the intruder is getting close to the entry, between time intervals [340,…,346], the robots 4, and 5 acts against the intruder properly. In final step in time interval [704,…,710], robot 1 and 2 intercept the intruder while it has the minimum distance with the entry.

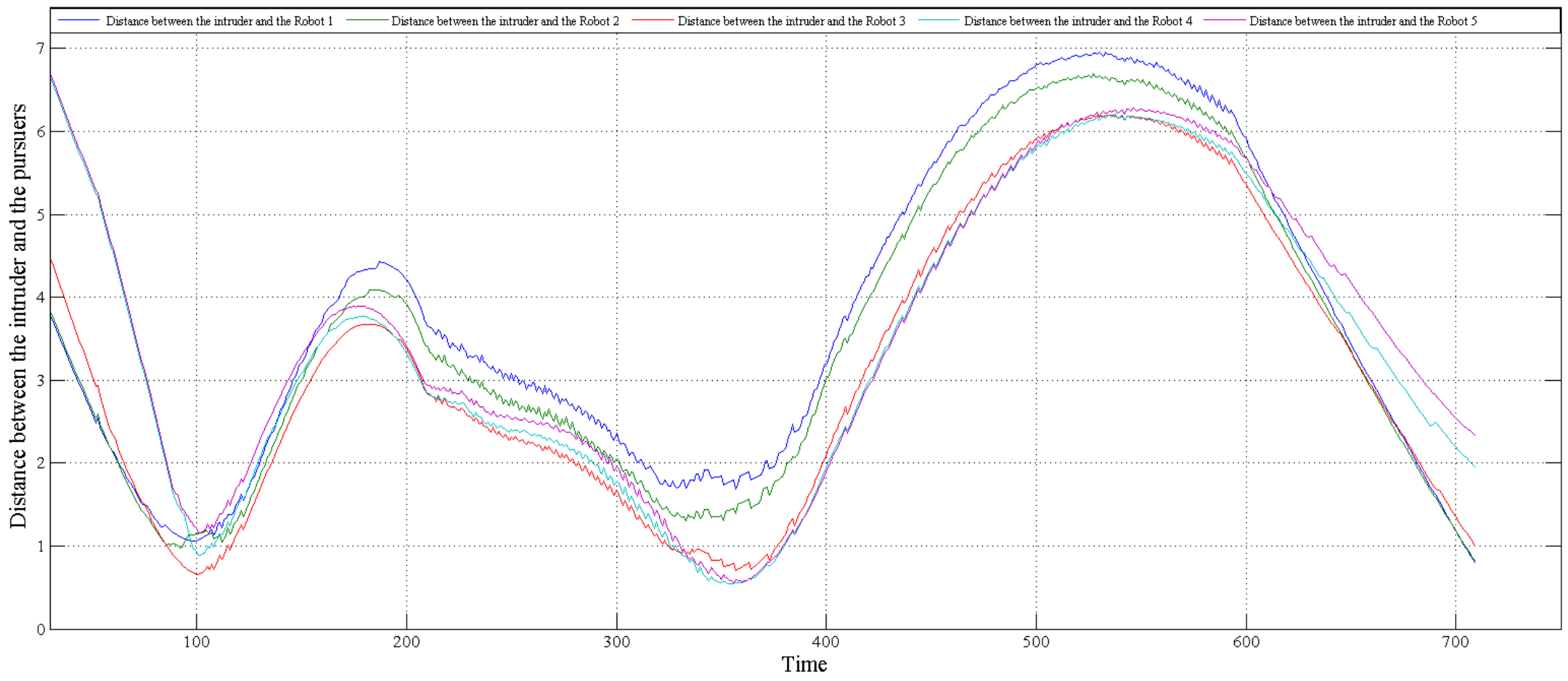

Figure 4.7: Distance between the intruder and each member of the multi-robot team in a straight region boundary when for *k=2*.

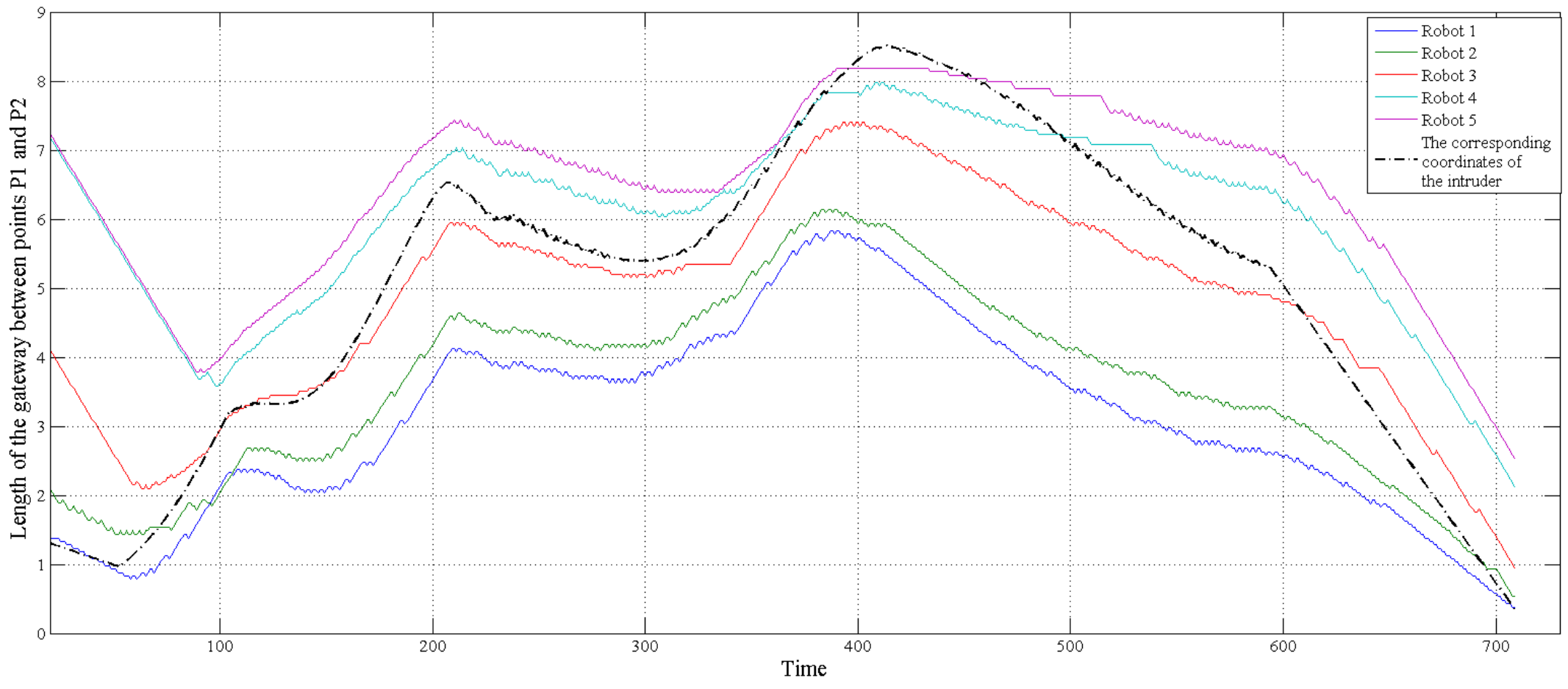

Figure 4.8: *k-intercepting* the intruder in a straight region boundary for *k=2*.

### 4.3.2 Protecting a curved boundary region

At the next stage, we considered the area with a curved entry which located between points P1 and P2. We compare the multi-robot team actions regarding intercepting the intruder. Fig.4.9 (a,b,c,d) shows the bounded region $R$, the multi-robot team and the intruder path towards the boundary region.

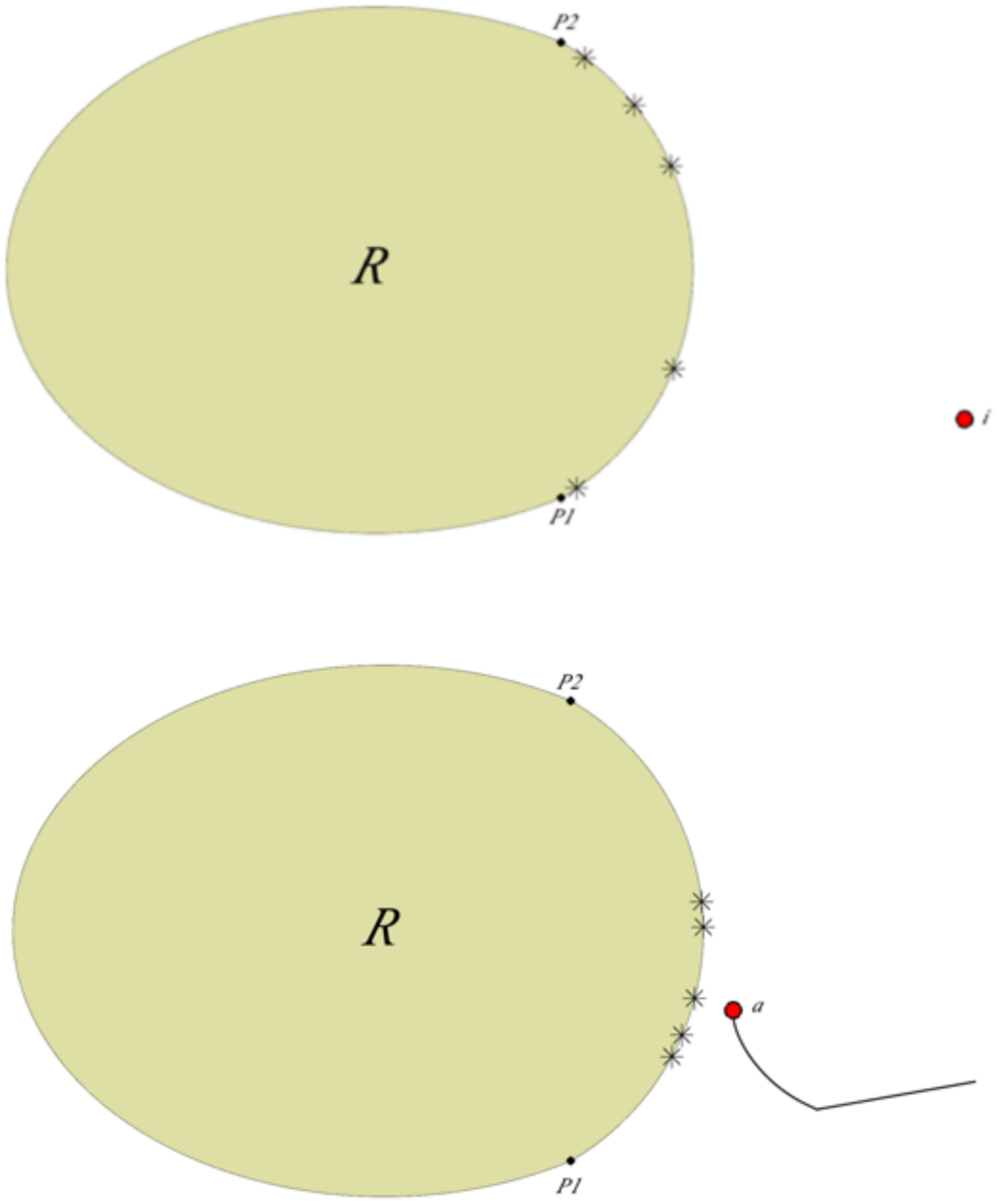

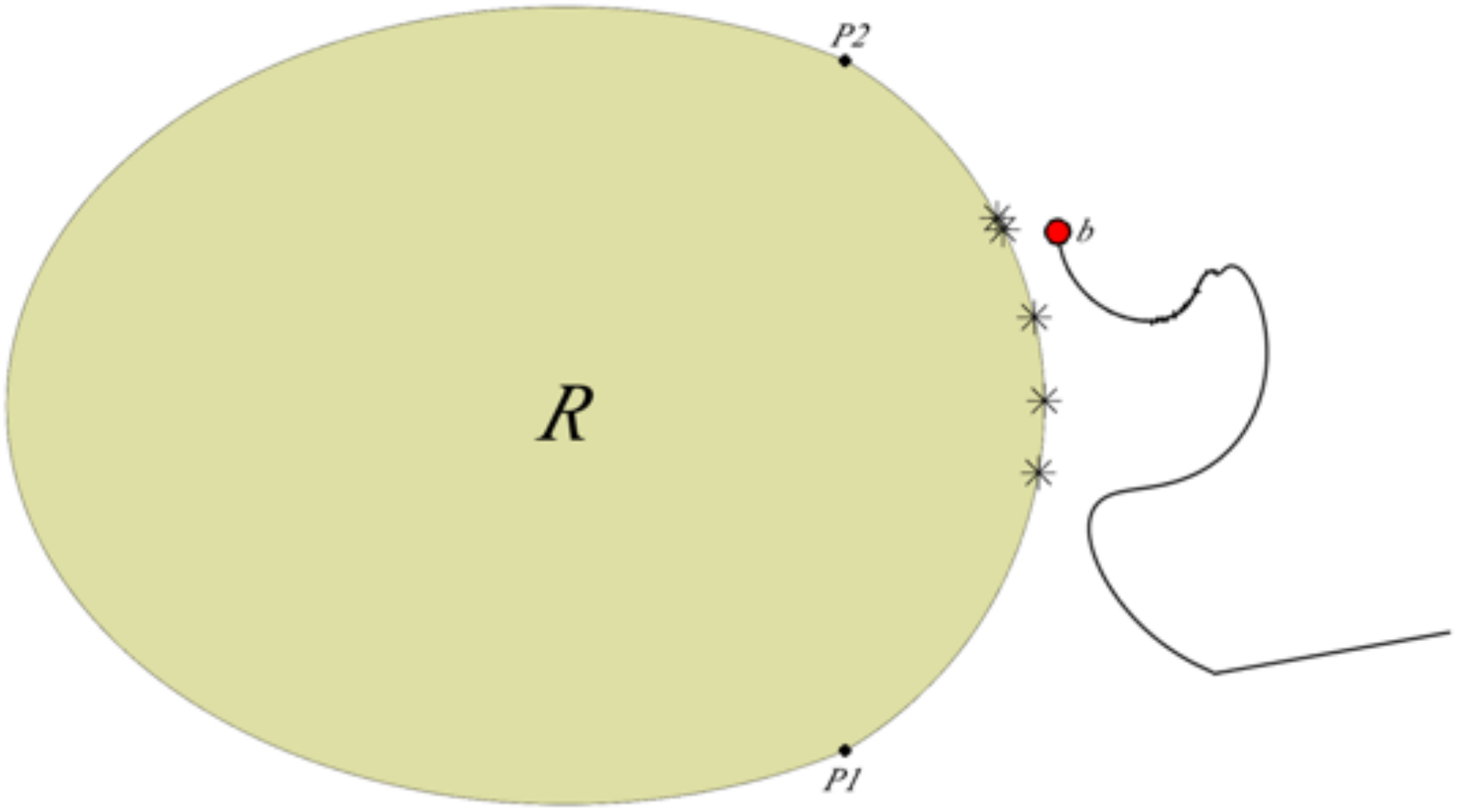

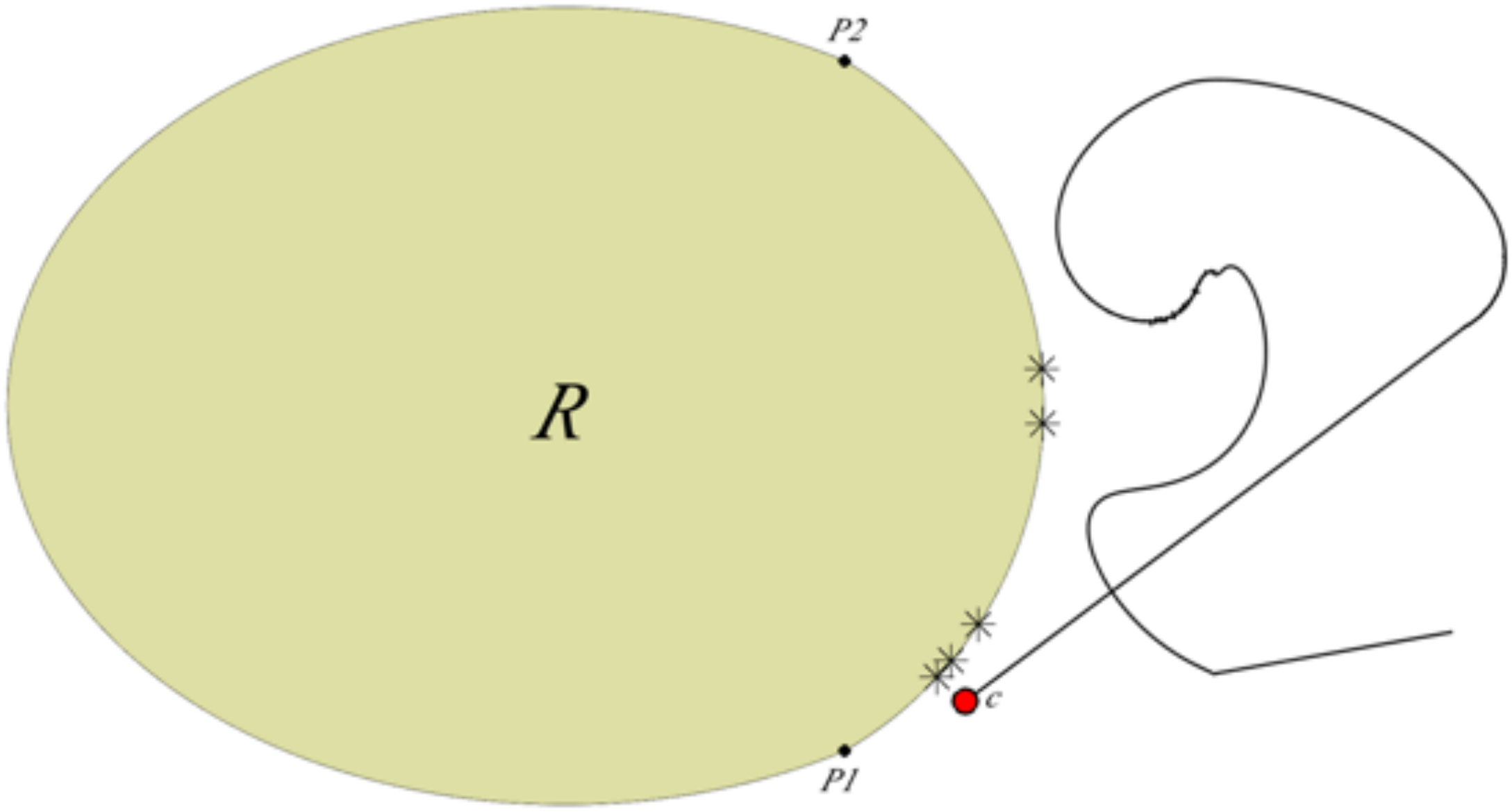

Figure 4.9: *k-intercepting* the intruder on a curved boundary region

Fig.4.10 (a, b, c) and 4.13 (a, b, c) show, how the robots intercept the intruder while the intruder is getting close to the entry. Once more, we considered two different values for k. Fig.4.10 (a, b, c) represents the behavior of the multi-robot team while k=1 and Fig.4.13 (a, b, c) represents the team action while k=2. As it shows in the Fig.4.10 (a, b, c), robots 1 ,3 and 4 are the closest interceptors to the intruder in the critical points respectively.

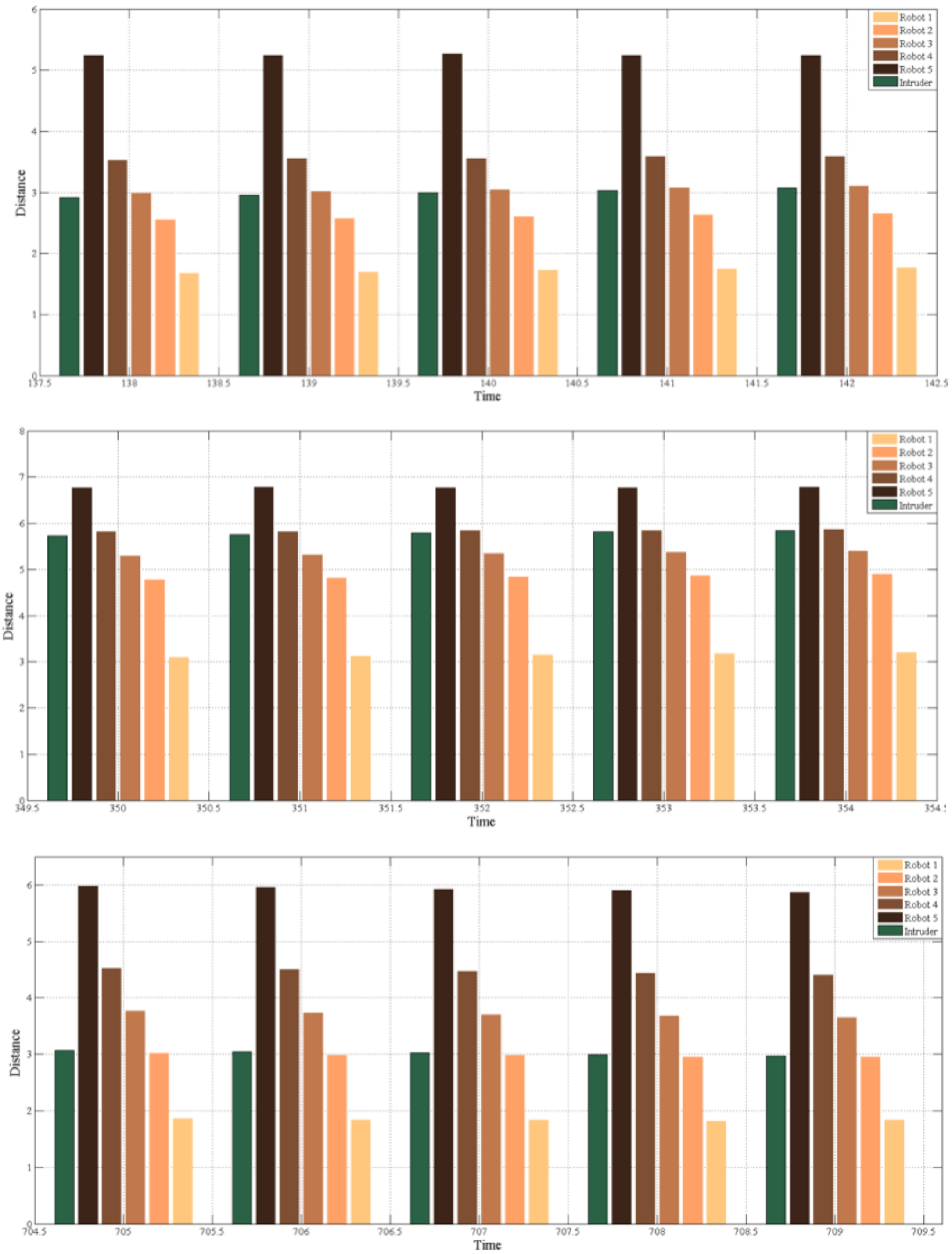

Figure 4.10 (a, b, c): The position of the intruder and the interceptors while the intruder is in its closest position to the curved region boundary for *k=1*.

Fig.4.11 and Fig.4.12, confirm the validation of the navigation law (4.7) and (4.15) while the robots move on a curved path and the value of k=1.

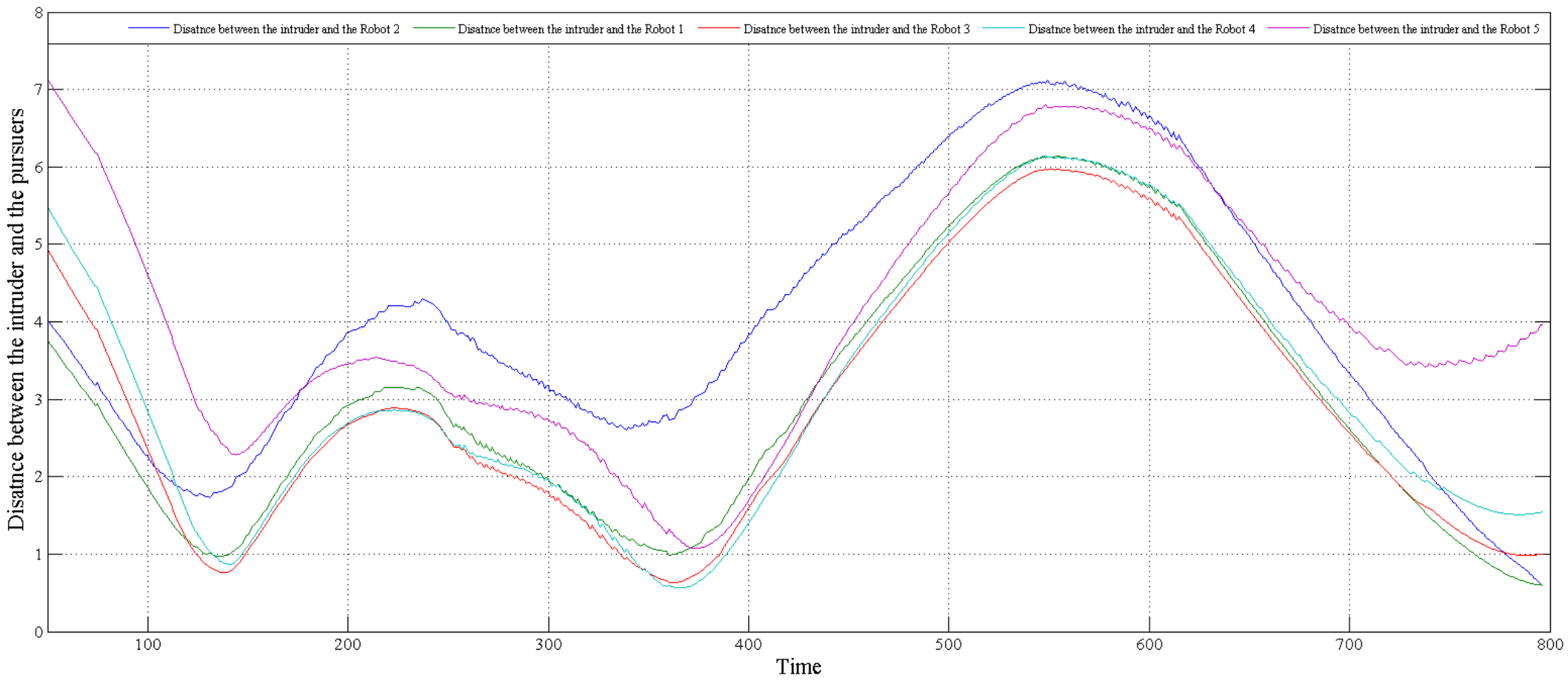

Figure 4.11: Distance between the intruder and each member of the multi-robot team in a curved region boundary when for *k=1*.

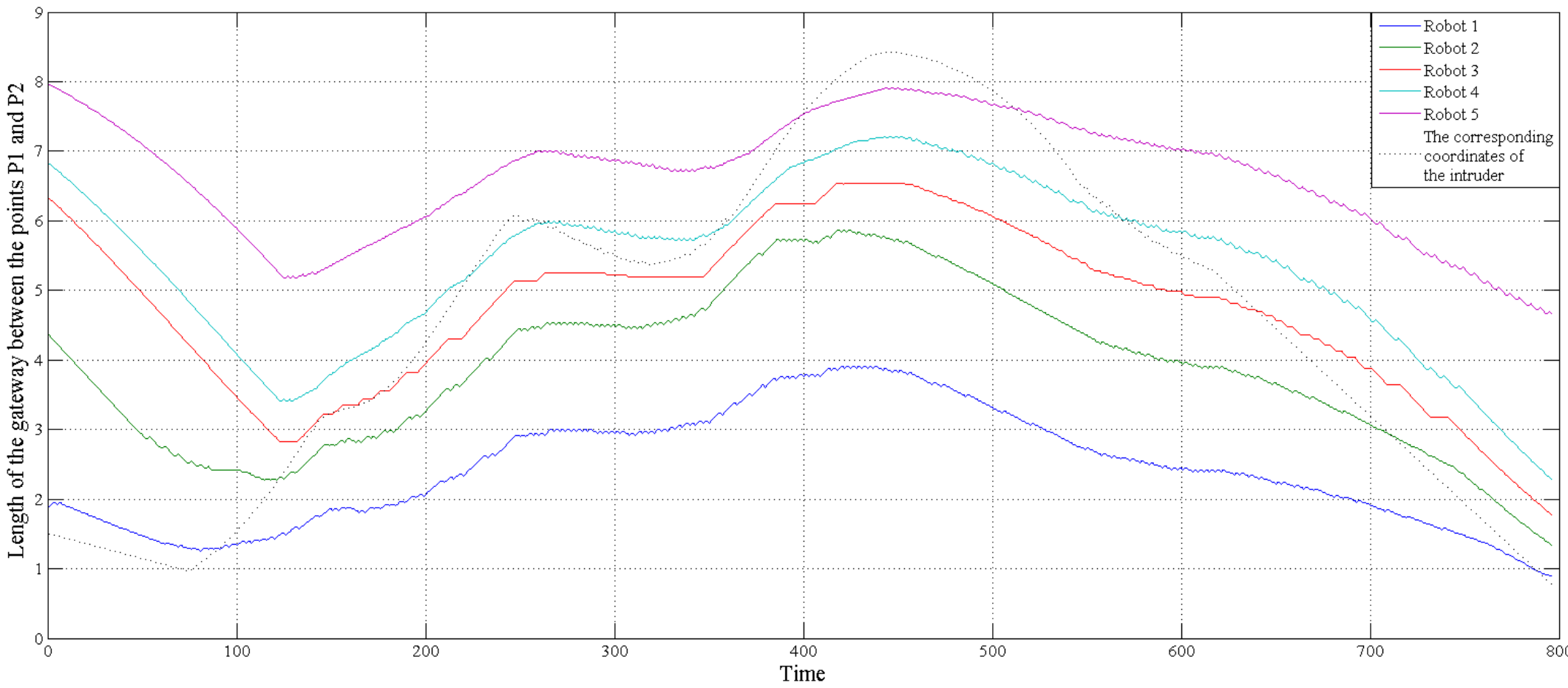

Figure 4.12: *k-intercepting* the intruder in a curved region boundary for *k=1*.

As shown in both the Fig.4.11 and Fig4.12, the intruder is intercepted by at least one robot when it is getting close to the gateway of the region.

Finally, we investigate the validity of the proposed navigation law considering k=2.

According to Fig.4.13.(a), the robot 1 and the robot 2 are moving close to each other and act as a team to intercept the intruder in time interval [126,…,132], when the intruder has its minimum distance with the entry of the region for the first time. As shown in Fig.4.13 (b), the mission is implemented by the robot 4 and the robot 5, and in the last critical situation which is shown in Fig.4.13 (c), both the robots one and two intercepting the intruder while it is located in its minimum distance with the entry of the field.

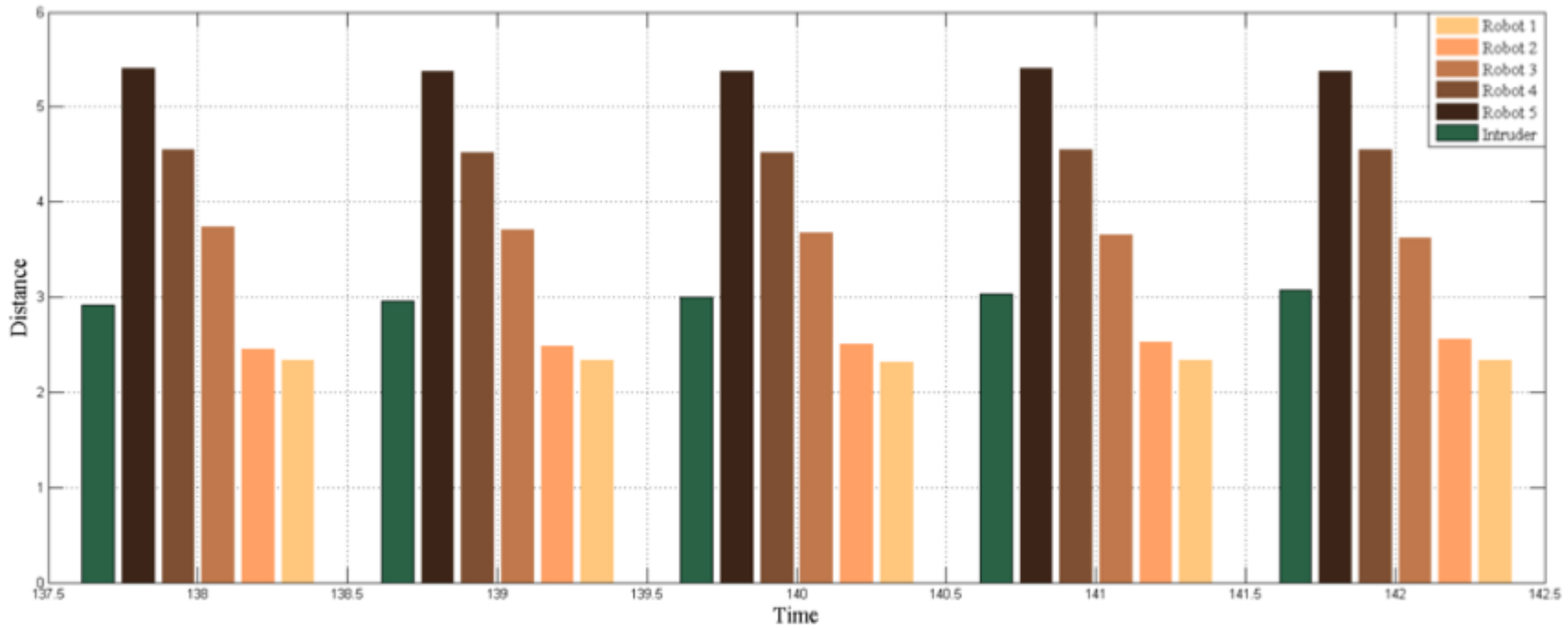

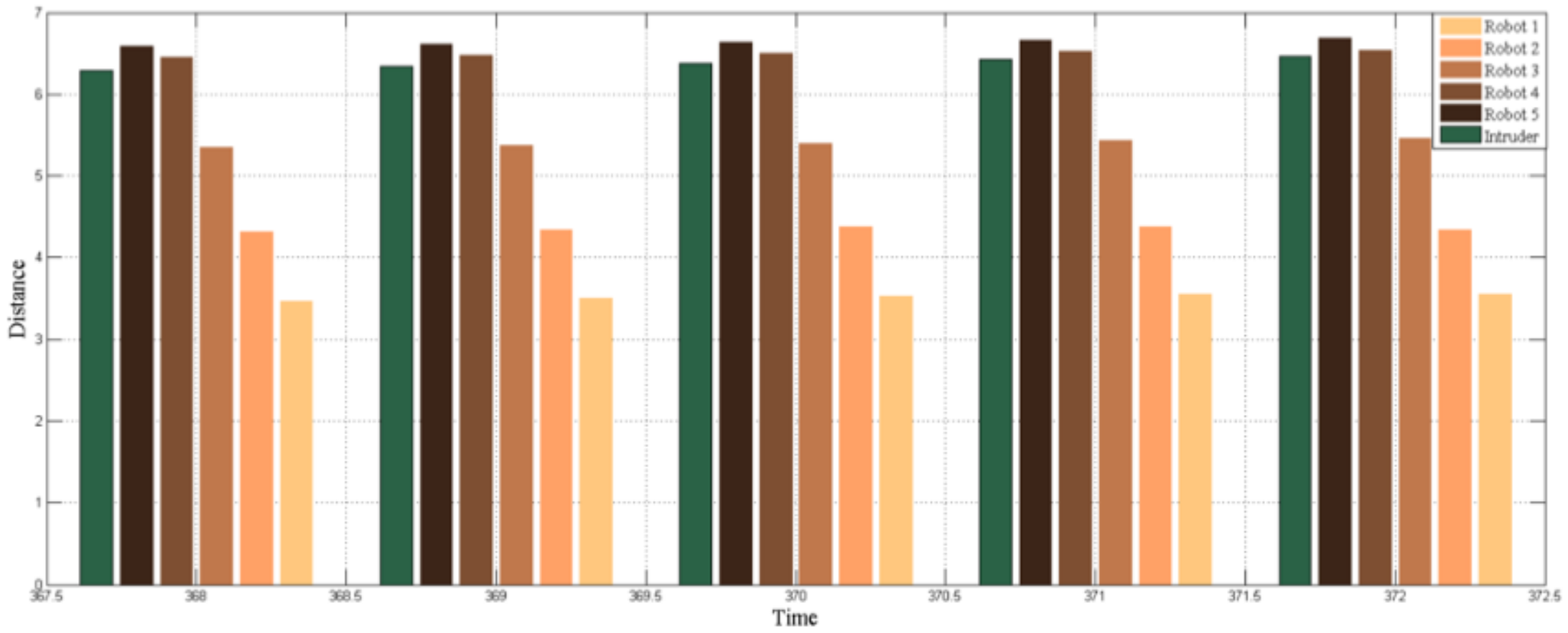

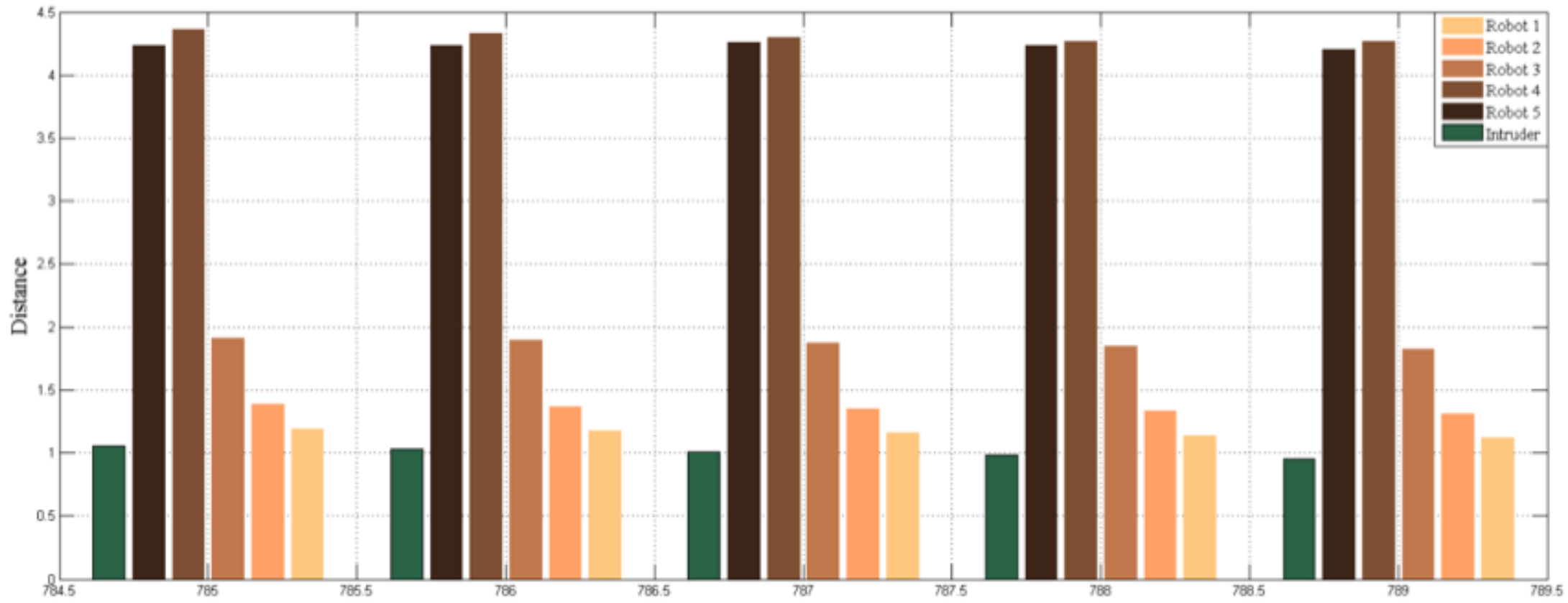

Figure 4.13 (a, b, c): The position of the intruder and the interceptors while the intruder is in its closest position to the curved region boundary for *k=1*.

Fig.4.14, represents the time intervals when the intruder has its minimum distance with the entry of the region in addition to illustrating, the distance between the intruder and each agent in the group. On the other side, Fig.4.15, shows the length of the position of each agent based on P1 and the length of the virtual position of the intruder on the curved path based on P1 either.

Comparing these two figures help us to perceive the proper work of the developed law in a curved path while k=2.

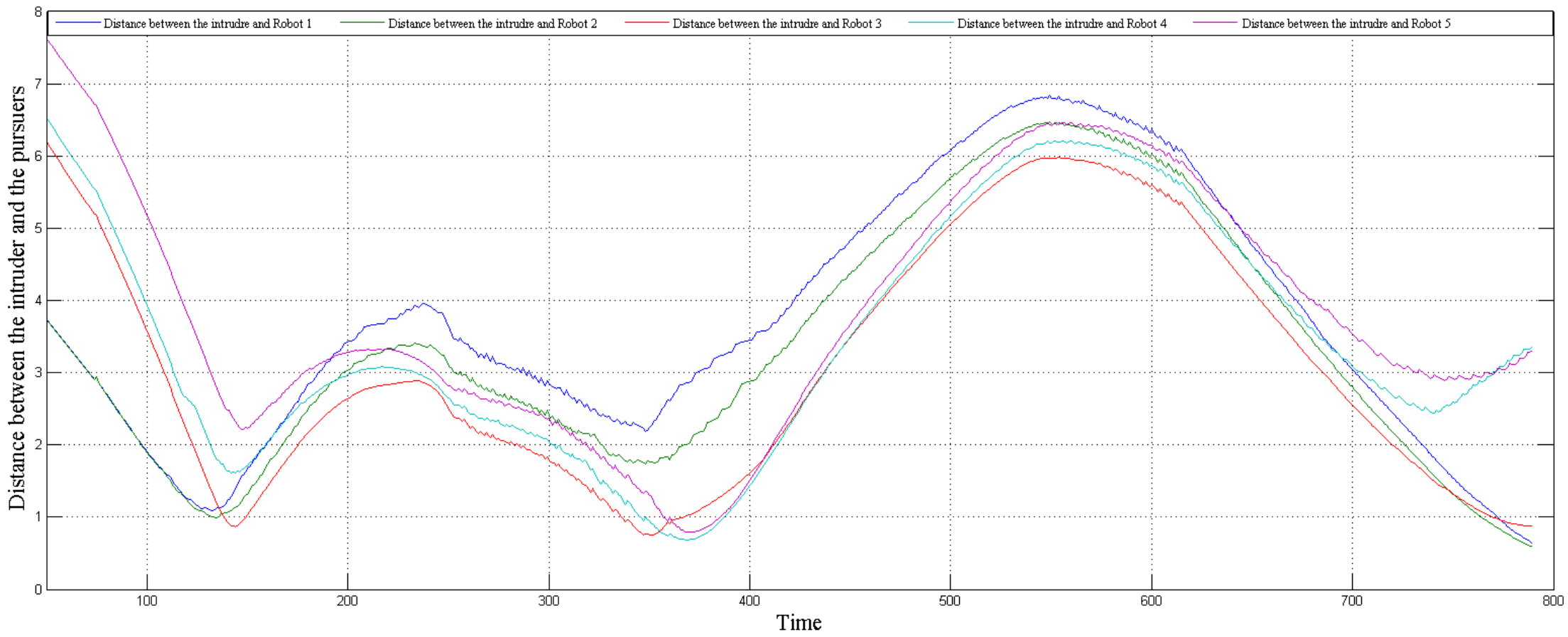

Figure 4.14: Distance between the intruder and each member of the multi-robot team in a curved region boundary when for *k=2*.

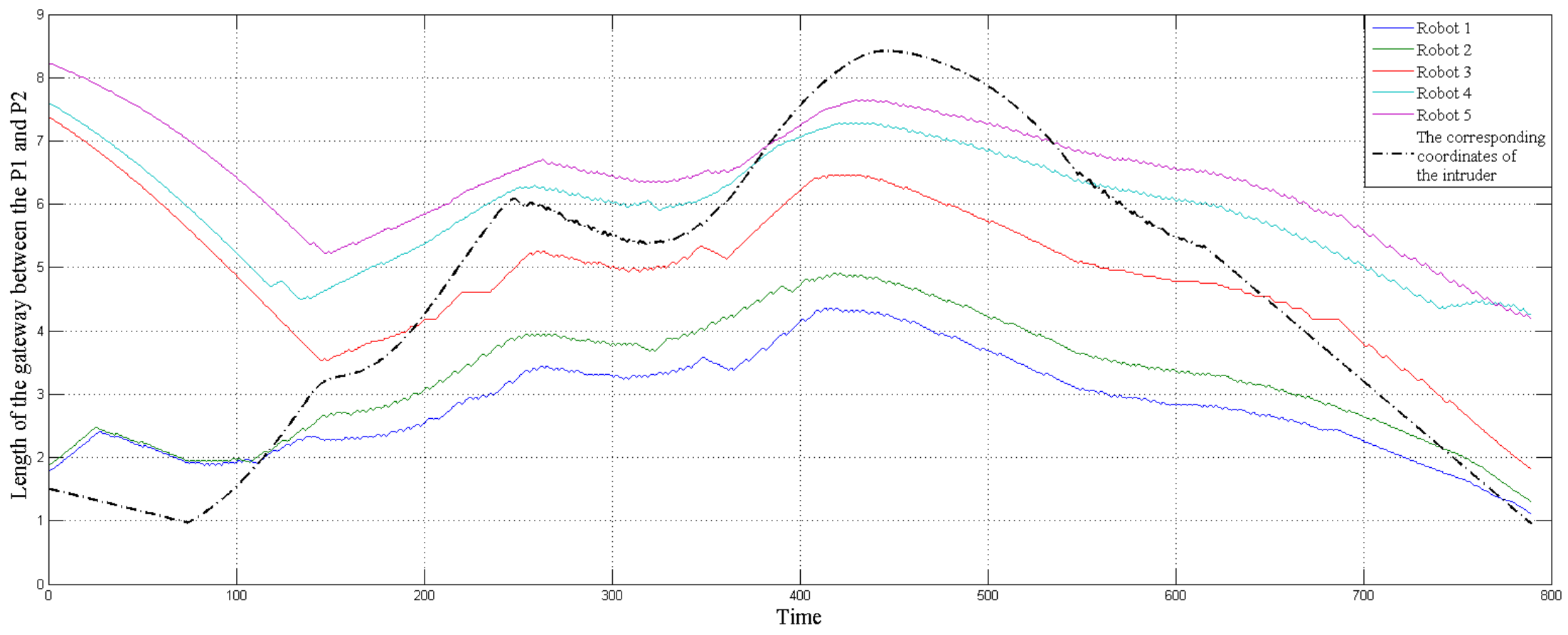

Figure 4.15: *k-intercepting* the intruder in a curved region boundary for *k=2*.

## 4.4 Summary

We proposed a decentralized motion control algorithm for a network of mobile robots to intercept an intruder on the boundary of a protected planar region. A necessary and sufficient condition for the existence of such an algorithm was derived. The proposed algorithm is based on some simple rules that only require information about the closest neighbors of each robot and the intruder. Computer simulations confirmed the efficiency of the developed navigation algorithm.

# Chapter 5

# A decentralized method for dynamic intercepting of an intruder trapped in a siege ring by mobile robots

The story of evasion and pursuit has been a challenging story from eternity. Predators have been pursuing the preys voraciously and preys have been evading tirelessly. In this endless game, each side has advantageous and shortcomings compare to the other side. Chen et al. present a poetic expression in this regard by referring to a Chinese proverb which says: "A lonely tiger in a pasture would be insulted by a group of hyenas"[198]. This is the fact that a tiger is faster, stronger and smarter than a lone hyena, However, the tiger could be placed in a weaker position when it is surrounded by a group of hyenas which means the collaboration for hunting and trapping can compensate for possible shortcomings of the pursuers. There are lots of real life examples in which the slower or smaller predators hunt a faster, bigger and smart prey through an effective cooperation[199]. For example wolf pack hunting behavior [200],[201] or lions on antelope [202]. In the previous examples, we mentioned the hunting application of multi agent target trapping, however, target trapping by a multi agent system is not limited to hunting merely. There are many other interesting applications such as search and rescue, transportation, sensor network deployment, intruder detection and siege and border protection could be addressed [203],[204],[205],[206],[207],[208]. The aim of this research is to present an effective model to protect an area against vandalism by maintaining an intruder in a siege.

Among all the researches in this area, coverage control is the most common approach to protect a region from an unwanted intruder as well as sieging the intruder in a closed area [28]. Barrier coverage, sweep coverage, and IGD could be referred as some examples in this area. In barrier coverage, a group of mobile robots form a static barrier with sensing capability to detect any intruder willing to enter or exit the region [147], [137]. In sweeping coverage problem, a group of mobile robots sweeping along the boundary of the region that needs to be protected to prevent entering or existing any unwanted intruder to or from the region [136],[138]. In IGD, a group of mobile robots protect a region from an unwanted intruder by maximizing the probability of detection of the intruder based on game theory decision making [29].

In this chapter, we consider a team of mobile robots moving along the boundary of a planar region. The robots move in a decentralized fashion, i.e. each robot navigates independently and has information about current coordinates of just several closest other robots. In this scenario, the intruder is visible to the robots during the mission.

Furthermore, we consider the case that, the intruder is detected and surrounded by the multi-robot team and the objective of the multi-robot team is to maintain the intruder in a siege to prevent any sabotaging by the intruder in the environment which means that there should be at least one robot close to the crossing point on the boundary to intercept the intruder.

The proposed problem statement is relevant to various problems of asset guarding in which a team of autonomous unmanned surface vehicles (USVs) patrols and guards an asset in an environment with hostile boats. Such problems require the team of USVs to cooperatively patrol the area around the asset, identify intruders, and actively block them[193], [194], [196].

The necessary and sufficient condition of the proposed decentralized navigation strategy for the multi-robot confirms that the surrounded intruder has been intercepting always.

The remainder of this chapter is organized as follows. Section 5.1 states the problem under investigation. The main results are presented in Section 5.2. Examples illustrating the proposed navigation strategy are given in Section 5.3. Finally, Section 5.4 summarizes the chapter.

## 5.1 Problem formulation

Let R be a closed convex planar region with a piecewise smooth boundary where the robots moving on to siege the intruder. Furthermore, let $S_1$ and $S_2$ be segments of the boundary of the region $R$ between points $P_1 \circlearrowleft P_2$ and $P_2 \circlearrowleft P_1$ respectively where $\circlearrowleft$ denotes the clockwise direction; see Fig.5.1.

It's obvious that the moving hostage $I$ tries to escape from the region $R$ through segments $S_1$or $S_2$ where the robots moving to maintain the intruder $I$ in the region $R$. Let $xI(t)$ and $yI(t)$ be planar coordinates of the intruder. The intruder is moving with an arbitrary time-varying vector velocity $v_I(t) = \begin{pmatrix} \dot{x}_I(t) \\ \dot{y}_I(t) \end{pmatrix}$ satisfying the constraint:

$$\|v_I(t)\| \leq V_I^{max} \ \ \forall t \geq 0 \tag{5.1}$$

where $V_I^{max} > 0$ is a given constant, $\|\cdot\|$ denotes the standard Euclidean vector norm. Moreover, let $n > 1$ be a given positive integer.

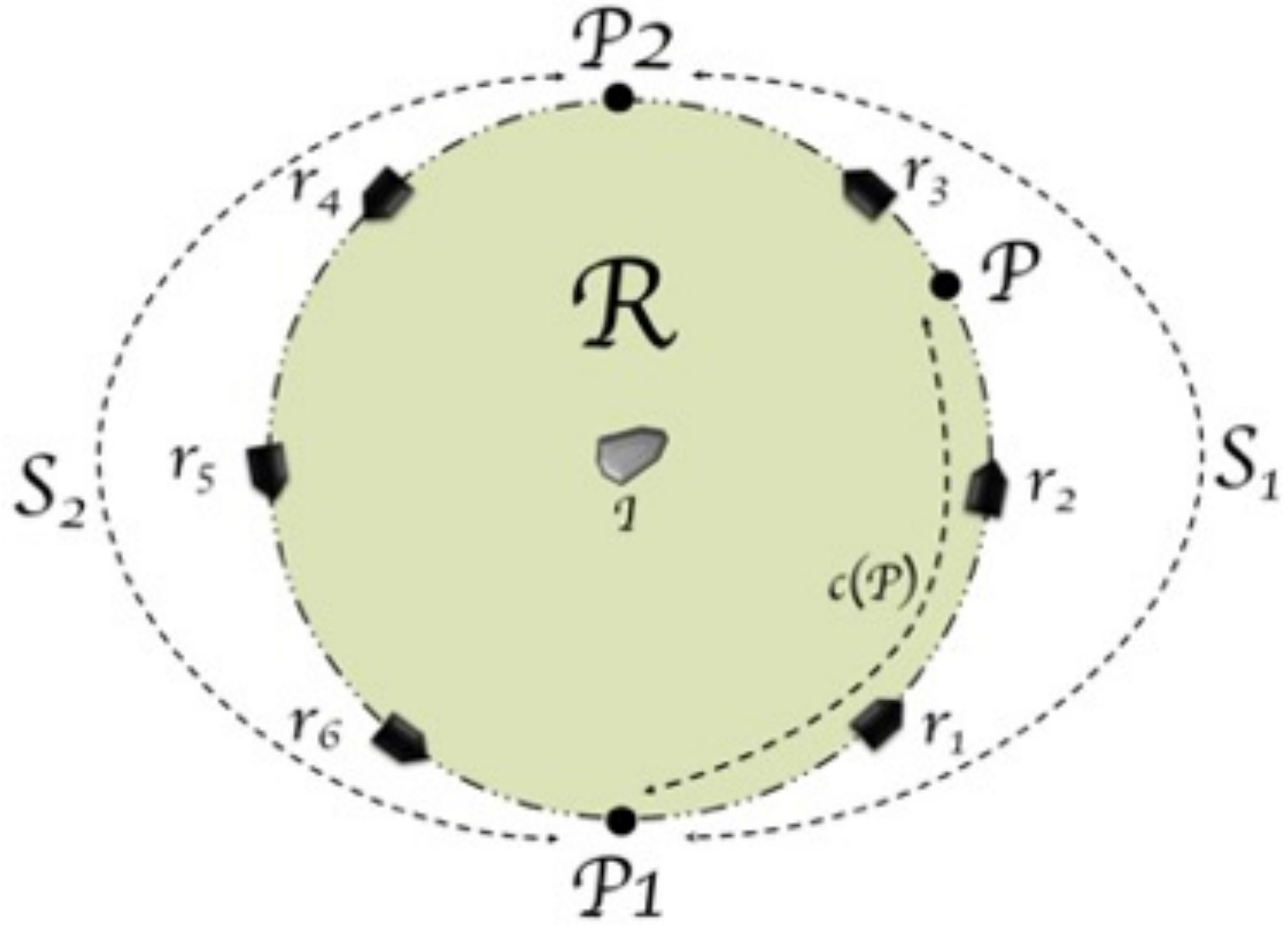

Figure 5.1: Siege ring $R$

We consider n mobile point-wise robots labelled $1,2,\ldots,n$ that prevent the intruder from leaving the region $R$ through the segments $S_1$ and $S_2$. Unlike the intruder that can move in any direction in the plane, the robots can move only along the segments $S_1$ and $S_2$ in the both directions. Furthermore, $P_{r1} = \begin{pmatrix} x_1(t) \\ y_1(t) \end{pmatrix}, P_{r2} = \begin{pmatrix} x_2(t) \\ y_2(t) \end{pmatrix}$, ..., $P_{rn} = \begin{pmatrix} x_n(t) \\ y_n(t) \end{pmatrix}$ denote the planar coordinates of the robots $1, 2, \ldots, n$.

We introduce $c(P)$ for any point $P \in S_1 \vee P \in S_2$ such that $c(P)$ is the length of the curved shape portion of the segment $S$ between the points $P1$ and $P$; see Fig.1.5.

This implies that $c(P_1)$=0 and $c(P_2) = L$, where L is the length of the curves $S_1$ and $S_2$ in either sides where:

$$x = r(\beta)\, cos(\beta)$$
$$y = r(\beta)\, sin(\beta)$$
$$c(P_i) = \int_{P_1}^{P_{ri}} \sqrt{[r(\beta)]^2 + \left[\frac{dr(\beta)}{d\beta}\right]^2}\, d\beta \qquad (5.2)$$

Therefor $c_1(t) := c(P_{r1}), c_2(t) := c(P_{r2}),\ldots, c_n(t) := c(P_{rn})$ denote the length of the curve of each robot $1,2,\ldots,n$ at time $t \geq 0$ to the point $P_1$.

Furthermore, we suppose that the robots labelled according to their coordinates so that

$$0 \leq c_1(t) \leq c_2(t) \leq \ldots \leq c_n(t) \leq L\ \forall t \geq 0 \qquad (5.3)$$

The requirement (2) means that the robots never change their order on the segment $S_1$ or $S_2$.

We assume that the motion of the robots along $S$ is described by the equation

$$\dot{c}_i(t) = u_i(t) \qquad \forall i = 1, \ldots, n \qquad (5.4)$$

where $ui(t)$ is the control input of the robot $i$. We assume that the control inputs $ui(t)$ satisfy the constraint

$$|u_i(t)| \leq V_R^{max} \quad \forall t \geq 0 \qquad (5.5)$$

where $V\ max > 0$ is a given constant.

At any time $t$, each robot $i, 2 \leq i \leq n-1$ knows the coordinates $P_{ri-1}(t), P_{ri+1}(t)$ of the robots $i - 1$ and $i + 1$, respectively. The robots 1 and $n$ know the coordinates $P_{r2}(t), P_{rn-1}(t)$ of the robots 2 and $n-1$, respectively.

Moreover, each robot i knows its own coordinate $P_i(t)$. Furthermore, the intruder becomes visible to the robots at some time $t0 \geq 0$, i.e. all the robots know the planar coordinates $P_I\begin{pmatrix} x_I(t) \\ y_I(t) \end{pmatrix}$ of the intruder for all $t \geq t0$.

## *Definition 5.1:*

Let $\varepsilon > 0$ be a given constant. Suppose that the intruder crosses the segment $S_1$ at time $t^*$i.e. $P_I(t^*) \in S_1$. We say that the multi-robot team barricading the intruder at time $t^*$ if there exists some index $i = 1,2,\dots,n$ such that $|c(\hat{P}_I(t_\star)) - c_i(t_\star)| \leq \varepsilon$.

Where $\hat{P}_I(t_\star) = \begin{pmatrix} \hat{x}_I(t_*) \\ \hat{y}_I(t_*) \end{pmatrix}$ denotes the projection coordinates of the intruder to the siege ring $S$ while it's close to it in either side.

The proposed navigation strategy that is based on the available information is called dynamic-intercepting as each member of the multi-robot team plays the role of a dynamic interceptor when the intruder crosses the siege ring $S$ in either sides.

In other words, dynamic-intercepting means that when the interceptor crosses the segment $S_1$ or $S_2$, there should be at least one robot close enough to the interception point.

## 5.2 Problem statement

In this scenario, we supposed that an intruder has been trapped and surrounded by a team of mobile robots. The mobile robots deployed in a hypothetical circle. $S$ known as the circumference of the hypothetical circle which is divided in two equal segments $S_1$ and $S_2$.

There are no static obstacles in the region that could prevent the intruder from escaping from the region. The objective of the intruder is to escape from the region $R$ through the segments $S_1$ or $S_2$ while avoiding intercepting by any member of the multi-robot team. On the other hand, the objective of the multi-robot team is to intercept the intruder when it crosses the segment $S_1$ or $S_2$ while the intruder tries to cross the segment $S$ in either sides. The problem under consideration in this study is to drive a necessary and sufficient condition under which the intruder would be intercepted for its every motion when it tries to escape the region by crossing the segment $S$. Moreover, based on this condition we design a decentralised navigation strategy for the multi-robot team in which every team member of the multi-robot team act as a dynamic-interceptor to dynamically intercept the intruder when it gets close to the segment $S$ in all the time.

## 5.3 Decentralized *dynamic-interception* navigation method

Let $P \in (S_1 \vee S_2) \in S$ and $x$ be an interior point of the region $R$. Then there is a straight line $\mathcal{L}(x, P)$ which connects points $x$ and $P$. Since $R$ is convex, $\mathcal{L}(x, P)$ is in $R$ and the intersection of $\mathcal{L}(x, P)$ and the boundary of $R$ contains only the point $P$.

Furthermore, let $\alpha(x, P)$ denote the length of $\mathcal{L}(x, P\ )$. On the other hand, let $i$ be an index such that $|ci\ (t)\ -\ c(P\ )|\ \leq\ |cj\ (t)\ -\ c(P\ )|$ for all $j\ =\ 1,,n$. Then, introduce the variable $\xi(t, P\ ) :=\ |ci(t)\ -\ c(P\ )|$. In other words, $i$ is the closest robot to the point $P$ at time $t$ and $\xi(t, P\ )$ is the length of the sub-segment of the

segment $(S_1 \vee S_2) \in S$ between the closest robot's current location $P_{ri}$ and the point $P$.

Then we introduce the function $F\ (s)$ from the interval $[0, \mathcal{L}]$ to the segment $(S_1 \vee S_2) \in S$ such that for any number $s \in [0, \mathcal{L}]$, $F(s)$ is the point $P \in (S_1 \vee S_2) \in S$ where $c(P) = s$.

Furthermore, let $[P_{A_1}, P_{A_2}]$ denote the closed sub-segment of the segment $(S_1 \vee S_2) \in S$ between the points $P_{A_1}$ and $P_{A_2}$. For $i = 1, \ldots, n$, introduce sub-segments $Si(t)-, Si(t)+$ of the segment $(S_1 \vee S_2) \in S$.

Remark: we consider robots $r_1, \ldots, r_m$ are located between points $P_1$ and $P_2$ in segment $S_1$ and robots $r_{m+1}, \ldots, r_n$ are located between points $P_2$ and $P_1$in segment $S_2$, in a counter clockwise fashion where$m < n$.

Therefore, if the robot $r_i \in [r_1, \ldots, r_m]$:

$$S_i(t)^- := \left[F\left(\frac{c_{i-1}(t) + c_i(t)}{2}\right), F\big(c_i(t)\big)\right]$$
$$if\ i = 2, \ldots, m;$$
$$S_i(t)^- := \big[P_1, F\big(c_i(t)\big)\big]$$
$$if\ i = 1;$$
$$S_i(t)^+ := \left[F\big(c_i(t)\big), F\left(\frac{c_i(t) + c_{i+1}(t)}{2}\right)\right]$$
$$if\ i = 1, \ldots, m-1;$$
$$S_i(t)^+ := \big[\,F\big(c_i(t)\big), P_2\big]$$
$$if\ i = m \tag{5.6}$$

However, if the robot $r_i \in [r_{m+1}, \ldots, r_n]$:

$$\acute{S}_i(t)^- := \left[F\left(\frac{c_{i-1}(t)+c_i(t)}{2}\right), F\big(c_i(t)\big)\right]$$

$$if\ i = m+2,\ldots,n;$$

$$\acute{S}_i(t)^- := \left[\,F\big(c_i(t)\big), P_2\right]$$

$$if\ i = m+1.$$

$$\acute{S}_i(t)^+ := \left[F\big(c_i(t)\big), F\left(\frac{c_i(t)+c_{i+1}(t)}{2}\right)\right]$$

$$if\ i = m+1,\ldots,n-1;$$

$$\acute{S}_i(t)^+ := \left[P_1, F\big(c_i(t)\big)\right]$$

$$if\ i = n; \tag{5.7}$$

Moreover, for $i = 1,2,\ldots,n$, introduce the numbers $M_i^-(t)$ and $M_i^+(t)$ as

$$M_i^-(t) := \sup_{P\in(S_i(t)^- \vee \acute{S}_i(t)^-)} \left(\xi(t,P) - \frac{\alpha(P_I(t),P)V_R^{max}}{V_I^{max}}\right);$$

$$M_i^+(t) := \sup_{P\in(S_i(t)^+ \vee \acute{S}_i(t)^+)} \left(\xi(t,P) - \frac{\alpha(P_I(t),P)V_R^{max}}{V_I^{max}}\right) \tag{5.8}$$

Now we can introduce the following decentralized navigation law:

$$\begin{array}{lll} u_i(t) := V_R^{max} & if & M_i^-(t) < M_i^+(t) \\ u_i(t) := -V_R^{max} & if & M_i^-(t) > M_i^+(t) \\ u_i(t) := 0 & if & M_i^-(t) = M_i^+(t) \end{array} \tag{5.9}$$

for all $i = 1,\ldots,n$.

## *Remark 5.1:*

The intuition behind the decentralized navigation law (5.9) can be explained as follows.

The sub-segments $S_i^-(t), S_i^+(t)$ are sets of points of the curve $S_1$ and the subsegments $\acute{S}_i(t)^-, \acute{S}_i(t)^+$ are sets of points of the curve $S_2$ for which the robot $i$ is the closest robot at time $t$. The robot moves with the maximum allowed speed towards the one of these segments that is more" dangerous" at the current time, i.e. it has the biggest possible distance between the intruder and the closest robot at the moment of crossing $S$ by the intruder. This biggest possible distance is described by (5.8).

## *Theorem 5.1*

Consider the multi-robot team (5.5) and the intruder satisfying (5.1). Then there exists a dynamic-intercepting multi-robot team navigation strategy in either sides of the siege ring $(S_1 \vee S_2)$ if and only if

$$\sup_{P \in S_1 \vee S_2} \left( \xi(t_0, P) - \frac{\alpha(P_I(t_0), P) V_R^{max}}{V_I^{max}} \right) \leq \varepsilon \quad (5.10)$$

where $t0 \geq 0$ is the time at which the intruder becomes visible to the robots.

Moreover, if the inequality (5.10) holds, then the navigation law (5.9) is a dynamic-intercepting navigation strategy.

Notice that since the region $R$ is convex, and the segment $S$ is compact, the supremum in (5.10) is achieved for some point $P$.

## *Proof:*

First, we prove that if the inequality (5.10) does not hold, then the intruder can always cross the segment $S$ without intercepting by the multi-robot team. Indeed, if (5.10) does not hold, then there exists a point $P \in S$ such that

$$\begin{cases} \left(\xi(t_0, P) - \frac{|L(P_I(t_0),P)|V_R^{max}}{V_I^{max}}\right) > \varepsilon \quad for \ P \in S_1 \\ \quad or \\ \left(\xi(t_0, \hat{P}) - \frac{|L(P_I(t_0),\hat{P})|V_R^{max}}{V_I^{max}}\right) > \varepsilon \quad for \ \hat{P} \in S_2 \end{cases} \tag{5.11}$$

Now let the intruder move along the straight-line segment $|L(P_I(t_0), P)|$ and $\left|L\left(P_I(t_0), \hat{P}\right)\right|$ connecting the points $P_I(t_0)$ and $P$ or $\hat{P}$ with its maximum speed $V_I^{max}$ respectively. In this case, the intruder reaches the point $P \in S_1$ or $\hat{P} \in S_2$ at the time $t^* = t_0 + \frac{|L(P_I(t_0),P)|}{V_I^{max}}$ or $t^{**} = t_0 + \frac{|L(P_I(t_0),\hat{P})|}{V_I^{max}}$. It obviously follows from (5.11) that, the closest robot to the point $P \in S_1$ or $\hat{P} \in S_2$ cannot be closer than $\varepsilon$ at time $t^*$ or $t^{**}$. Therefore, there is no any dynamic interceptor close enough to the neighborhood of the points $P \in S_1$ or $\hat{P} \in S_2$ at time $t^*$ or $t^{**}$ to intercept the intruder from crossing the segment $S$ in either side.

We now prove that if the inequality (5.10) holds, the intruder is intercepted always by at least one dynamic interceptor of the multi-robot team while it tries to cross the segment $S$.

Indeed, for any trajectory $[PI\,(t), c1(t), \ldots, cn(t)]$ of the intruder-multi-robot introduce the Lyapunov function

$$\begin{cases} \mathcal{W}[P_I(t), c_1(t), \ldots, c_m(t)] := \sup_{P \in S_1} \left(\xi(t_0, P) - \frac{\alpha(P_I(t_0),P)V_R^{max}}{V_I^{max}}\right) for \ S_1 \in [P_1 \circlearrowleft P_2] \\ \vee \\ \mathcal{W}[P_I(t), c_{m+1}(t), \ldots, c_n(t)] := \sup_{P \in S_2} \left(\xi(t_0, P) - \frac{\alpha(P_I(t_0),P)V_R^{max}}{V_I^{max}}\right) for \ S_2 \in [P_2 \circlearrowleft P_1] \end{cases} \tag{5.12}$$

Notice that since the region $R$ is convex, and the segment $S$ is compact, the supremum in (5.12) is achieved for some point $P$. Furthermore, by definition, $\alpha(PI(t), P)$ is the length of the straight segment $L(PI(t), P)$ connecting $PI(t)$ and $P$. Hence, it is obvious that

$$\alpha(P_I(t), P) = \inf_{M(P_I(t),P) \in \mathcal{M}(P(t),P)} |M(P_I(t), P| \tag{5.13}$$

where $\mathcal{M}(P_I(t), P)$ is the set of all smooth paths $M(P(t), P$ inside $R$ connecting $PI(t)$ and $P$, and $|M(P_I(t), P|$ denotes the length of the path $M(P_I(t), P$. In other words, $\mathcal{M}(P_I(t), P)$ is the set of all possible paths of the intruder between $PI\ (t)$ and $P$ . Furthermore, it immediately follows from (5.12), (5.13) and (5.9) that

$$\begin{cases} \mathcal{W}[P_I(t), c_1(t_1), \dots, c_m(t_1)] \le \mathcal{W}[P_I(t), c_1(t_2), \dots, c_m(t_2)] \ \ for\, S_1 \in [P_1 \circlearrowleft P_2] \\ \vee \\ \mathcal{W}[P_I(t), c_{m+1}(t_1), \dots, c_n(t_1)] \le \mathcal{W}[P_I(t), c_{m+1}(t_2), \dots, c_n(t_2)] \ \ for\, S_2 \in [P_2 \circlearrowleft P_1] \end{cases}$$

$$\forall t_2 \ge t_1 \ge t_0 \tag{5.14}$$

Now (5.14) and (5.11) imply that if the intruder reaches a point $P \in S_1$or $\hat{P} \in S_2$ at some time $t^* \ \ge t_0$ or $t^{**} \ge t_0$ respectively, the robot closest to the point $P$ at time $t^*$or $t^{**}$ cannot be further from $P$ than $\varepsilon$. Therefore, the neighboring point of the point $P$ at the segment $S$ contains at least one robot at time $t^*$ or $t^{**}$. This implies that any team member of the multi robot team plays the role of a dynamic-interceptor to maintain the intruder inside the siege ring.

The inequality (5.10) and the navigation law (5.9) can be made computationally simpler under the following assumption.

## *Assumption 5.1:*

The following inequality holds:

$$V_I^{max} \geq V_R^{max} \tag{5.15}$$

For $i = 1,2,\dots,n$, introduce points $D_i(t)^-, D_i(t)^+$ of the segment $S_1$ and $\widehat{D}_i(t)^-, \widehat{D}_i(t)^+$ of the segment $S_2$ for robot $r_i$ as follows:

For $i \in [1,\dots,m] \implies r_i \in S_1$:

$$\begin{aligned}
D_i(t)^- &:= F\left(\frac{c_{i-1}(t)+c_i(t)}{2}\right) && if\ i = 2,\dots,m;\\
D_i(t)^- &:= P_1 && if\ i = 1;\\
D_i(t)^+ &:= F\left(\frac{c_i(t)+c_{i+1}(t)}{2}\right) && if\ i = 1,\dots,m-1;\\
D_i(t)^+ &:= P_2 && if\ i = m
\end{aligned} \tag{5.16}$$

For $i \in [m+1,\dots,n] \implies r_i \in S_2$:

$$\begin{aligned}
\widehat{D}_i(t)^- &:= F\left(\frac{c_{i-1}(t)+c_i(t)}{2}\right) && if\ i = m+2,\dots,n;\\
\widehat{D}_i(t)^- &:= P_2 && if\ i = m+1;\\
\widehat{D}_i(t)^+ &:= F\left(\frac{c_i(t)+c_{i+1}(t)}{2}\right) && if\ i = m+1,\dots,n-1;\\
\widehat{D}_i(t)^+ &:= P_1 && if\ i = n
\end{aligned} \tag{5.17}$$

Moreover, for $i = 1,2,\dots,n$, we introduce a set of numbers $\mathcal{H}$ which includes $H_i(t)^-$ and $H_i(t)^+$ if $r_i \in S_1$ and $\widehat{H}_i(t)^-$ and $\widehat{H}_i(t)^+$ if $r_i \in S_2$ as follows:

$$H_i^-(t) := \frac{c_i(t)-c_{i-1}(t)}{2} - \frac{\alpha(P_I(t),D_i(t)^-)V_R^{max}}{V_I^{max}}$$

$$if\ i = 2,\dots,m;$$

$$H_i^-(t) := c_i(t) - \frac{\alpha(P_I(t),P_1)V_R^{max}}{V_I^{max}}$$

$$if\ i = 1;$$

$$H_i^+(t) := \frac{c_{i+k}(t)-c_i(t)}{2} - \frac{\alpha\left(P_I(t),D_i(t)^+\right)V_R^{max}}{V_I^{max}}$$

$$if\ i = 1,\dots,m-1;$$

$$H_i^+(t) := L - c_i(t) - \frac{\alpha(P_I(t),P_2)V_R^{max}}{V_I^{max}}$$

$$if\ i = m \tag{5.18}$$

$$\widehat{H}_i^-(t) := \frac{c_i(t)-c_{i-1}(t)}{2} - \frac{\alpha(P_I(t),\widehat{D}_i(t)^-)V_R^{max}}{V_I^{max}}$$

$$if\ i = m+2, \dots, n;$$

$$\widehat{H}_i^-(t) := c_i(t) - \frac{\alpha(x_I(t),P_2)V_R^{max}}{V_I^{max}}$$

$$if\ i = m+1;$$

$$\widehat{H}_i^+(t) := \frac{c_{i+k}(t)-c_i(t)}{2} - \frac{\alpha\left(P_I(t),\widehat{D}_i(t)^+\right)V_R^{max}}{V_I^{max}}$$

$$if\ i = m+1, \dots, n-1;$$

$$\widehat{H}_i^+(t) := L - c_i(t) - \frac{\alpha(P_I(t),P_1)V_R^{max}}{V_I^{max}}$$

$$if\ i = n \tag{5.19}$$

For $i = 1, \dots, n$, the simplified navigation law (5.9) becomes:

For $r_i \in S_1$:

$$\begin{aligned} u_i(t) &:= V_R^{max} \quad if \quad H_i^-(t) < H_i^+(t) \\ u_i(t) &:= -V_R^{max} \quad if \quad H_i^-(t) > H_i^+(t) \\ u_i(t) &:= 0 \quad if \quad H_i^-(t) = H_i^+(t) \end{aligned} \tag{5.20}$$

For $r_i \in S_2$:

$$\begin{aligned} u_i(t) &:= V_R^{max} \quad if \quad \widehat{H}_i^-(t) > \widehat{H}_i^+(t) \\ u_i(t) &:= -V_R^{max} \quad if \quad \widehat{H}_i^-(t) < \widehat{H}_i^+(t) \\ u_i(t) &:= 0 \quad if \quad \widehat{H}_i^-(t) = \widehat{H}_i^+(t) \end{aligned} \tag{5.21}$$

Theorem 3.2: For any multi-robot team which satisfies (5.4) and any intruder which is trapped in the siege ring satisfying (5.1), and assumption 5.1., there exist dynamic-intercepting multi-robot team navigation strategy if and only if:

$$max\,\mathcal{H} \leq \varepsilon \tag{5.22}$$

Moreover, if the inequality (5.22) holds, then the navigation law (5.20) and (5.21) is a dynamic-intercepting navigation strategy.

*Proof:*

We prove that if Assumption 5.1 holds, then

$$\begin{cases} \begin{cases} M_i^-(t) = H_i^-(t) \\ M_i^+(t) = H_i^+(t) \end{cases} & if\ r_i \in S_1 \\ \quad and & \\ \begin{cases} M_i^-(t) = \widehat{H}_i^-(t) \\ M_i^+(t) = \widehat{H}_i^+(t) \end{cases} & if\ r_i \in S_2 \end{cases} \tag{5.23}$$

where $M_i^-(t), H_i^-(t), \widehat{H}_i^-(t)$, $M_i^+(t), H_i^+(t), \widehat{H}_i^+(t)$, are defined by (5.8), (5.18) and (5.19). Indeed, let $P3, P4 \in S_i(t)^-$ and $c(P3) < c(P4)$ where $S_i(t)^-$ is defined by (5.6). Then, for any $x$, we obviously have that $\alpha(x, P3) \le \alpha(x, P4) + c(P4) - c(P3)$. This and Assumption 5.1 imply that

$$\xi(t, P_3) - \frac{\alpha(x,P_3)V_R^{max}}{V_I^{max}} \ge \xi(t, P_4) - \frac{\alpha(x,P_4)V_R^{max}}{V_I^{max}} \tag{5.24}$$

For any $x$. This implies that

$$\sup_{P \in S_i(t)^-} \left( \xi(t,P) - \frac{\alpha(x(t),P)V_R^{max}}{V_I^{max}} \right) \tag{5.25}$$

Is achieved at the interval $S_i(t)^-$.

Similarly, let $P5, P6 \in \acute{S}_i(t)^-$ and $c(P5) > c(P6)$ where $\acute{S}_i(t)^-$ is defined by (5.7). Then, for any $x$, $\alpha(x, P6) \le \alpha(x, P5) + c(P5) - c(P6)$ while assumption 5.1 holds:

$$\xi(t,P_6) - \frac{\alpha(x,P_6)V_R^{max}}{V_I^{max}} \geq \xi(t,P_5) - \frac{\alpha(x,P_5)V_R^{max}}{V_I^{max}} \tag{5.26}$$

Which implies that:

$$\sup_{\hat{P}\in\acute{S}_i(t)^-} \left(\xi\left(t,\hat{P}\right) - \frac{\alpha(x(t),\hat{P})V_R^{max}}{V_I^{max}}\right) \tag{5.27}$$

is achieved at the interval $\acute{S}_i(t)^-$.

Therefore:

$$\begin{cases} M_i^-(t) = H_i^-(t) \mid \forall P \in S_i \\ M_i^-(t) = \widehat{H}_i^-(t) \mid \forall \hat{P} \in \acute{S}_i \\ M_i^+(t) = H_i^+(t) \mid \forall P \in S_i \\ M_i^+(t) = \widehat{H}_i^+(t) \mid \forall \hat{P} \in \acute{S}_i \end{cases} \tag{5.28}$$

Hence, (5.28) holds and the statement of Theorem 5.2 follows from Theorem 5.1.

## 5.4 Simulations and the results

In this section, we consider a team of ten mobile robots known as hunters surrounded one intruder as a prey that aims to escape the region $R$ with maximum speed of $V_I^{max} = 4.5$. the mobile robots divided in two groups including 5 members of each which robots 1 to 5 moving on segment $S_1$ where lies between $P_1 \circlearrowleft P_2$ and robots 6 to 10 moving on segment $S_2$ where lies between $P_2 \circlearrowleft P_1$ respectively. The maximum velocity of the robots is $V_r^{max} = 3.0$. Our illustrative examples include the problems of dynamic−intercepting the intruder by the multi-robot team in which, there is at least one robot exist with the minimum allowed distance $\varepsilon$ to any point $P$ where the intruder has the minimum distance to the point $P$ of segment $S$ in both sides. In our examples, $V_I^{max} = 4.2$ and $V_R^{max} = 3.0$, hence, Assumption 5.2 holds. Therefore, we apply Theorem 5.2 and the navigation law (5.20) and (5.21). Fig. 5.2, illustrates the reaction of the robots to the intruder's motion when the intruder tends to exit the region $R$ at the points $a, b, c,$ and $d$. The robots are indexed in anti-clockwise direction from point $P1$ in a circle, where five robots protect the segment $S1$ which is located between points $P1$ and $P2$, and the other five team members protect the segment $S1$ which is located between points $P2$ and $P1$ on the boundary. Fig.5.3.a, shows the evolution of the $y - coordinates$ of the intruder and the robots during the trajectory shown in Fig.5.2, when the intruder gets close to the segment $S1$. Analogously, Fig.5.3.b, shows the evolution of the $y - coordinates$ of the intruder and the robots during the trajectory shown in Fig.5.2 when the intruder gets close to the segment $S2$.

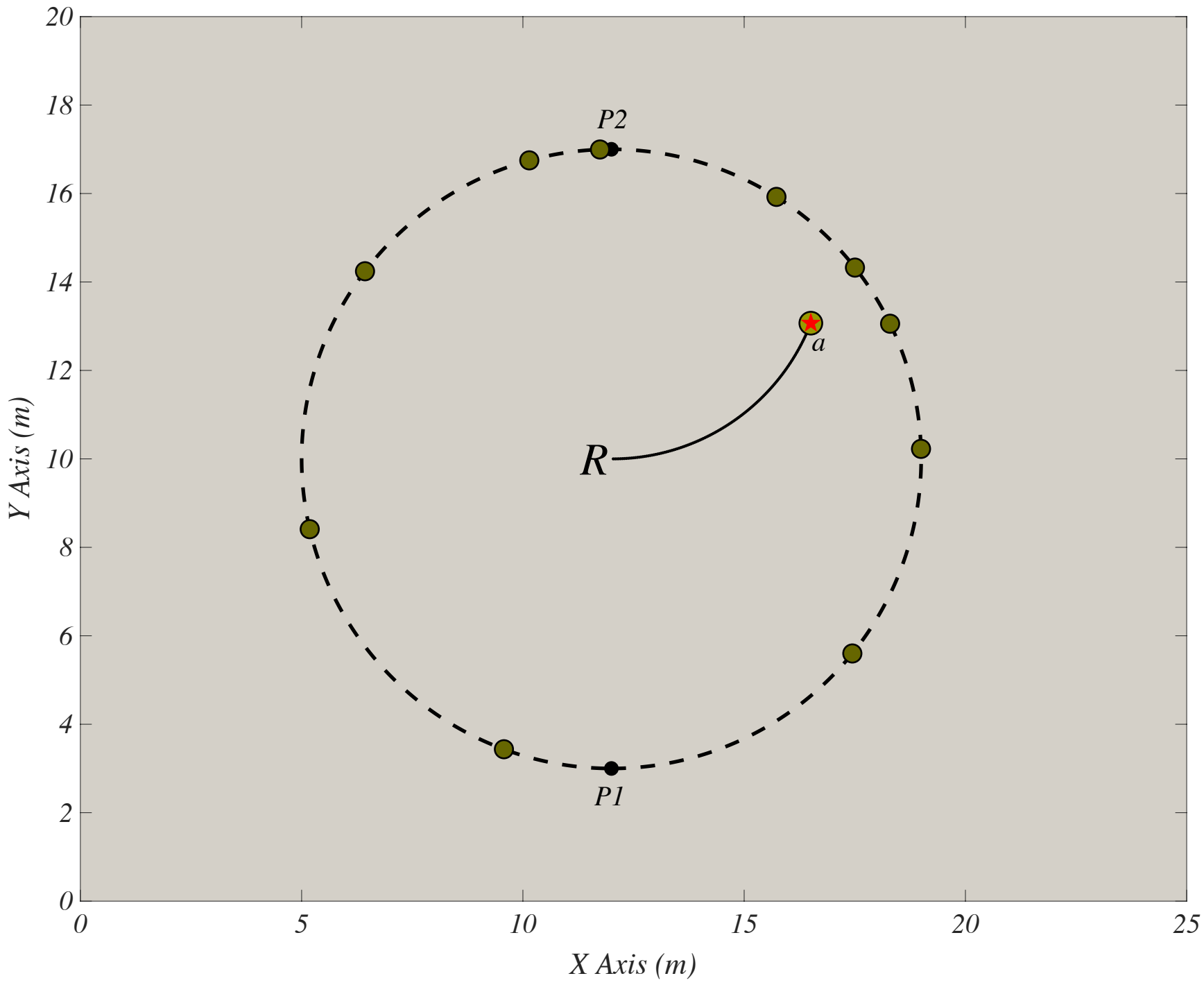

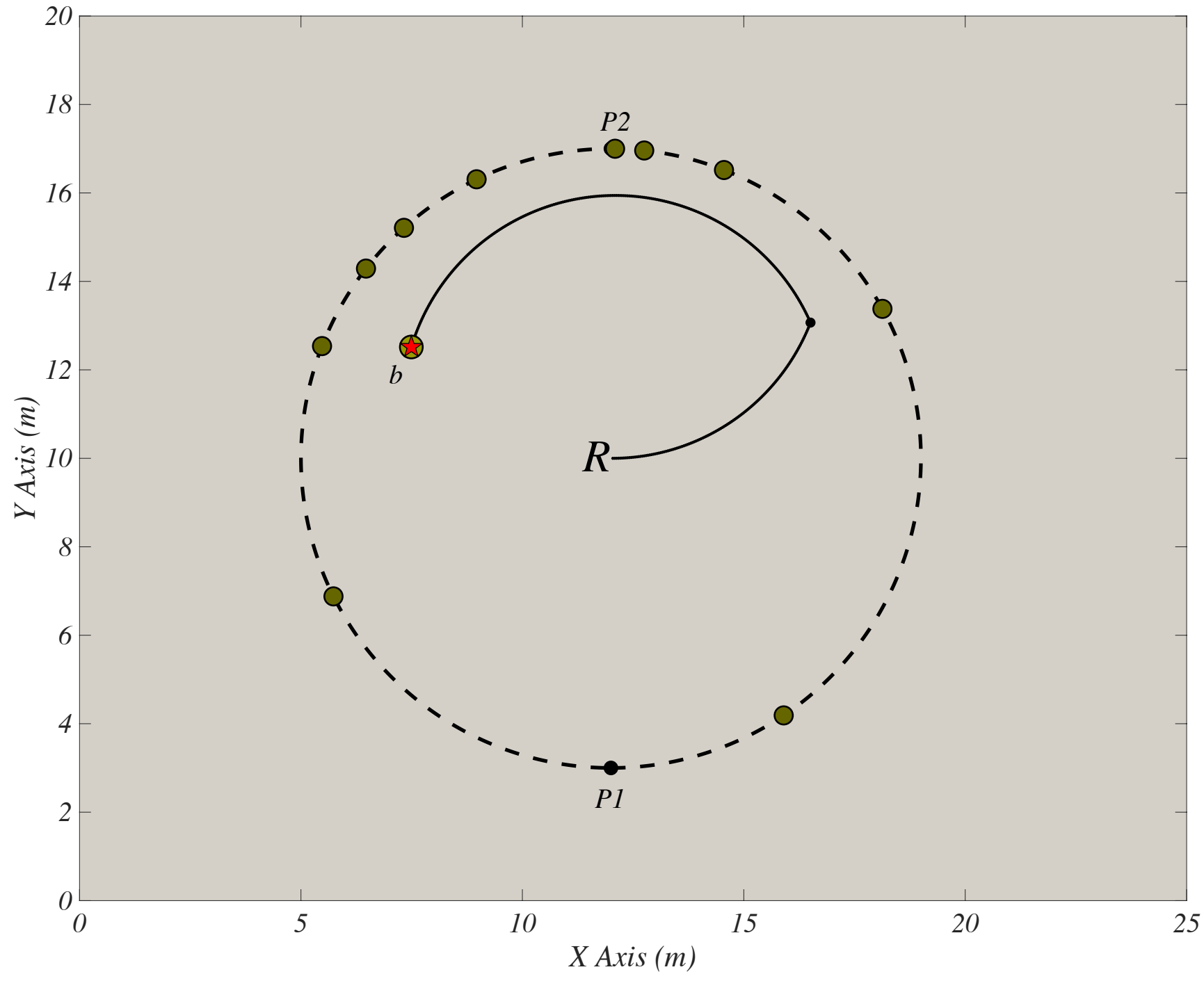

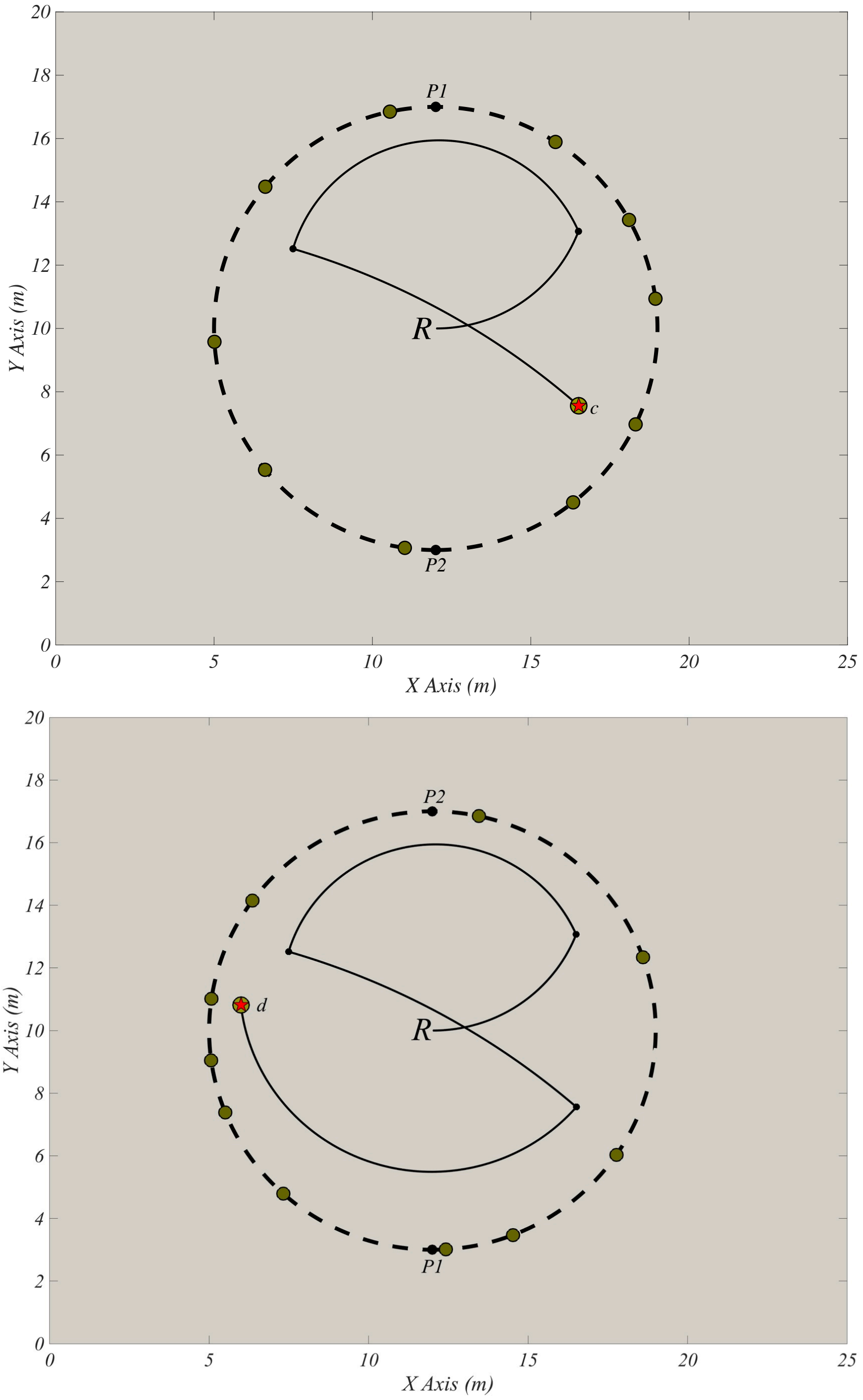

Figure 5.2: Trajectory of the intruder and the robots in the siege ring *R*

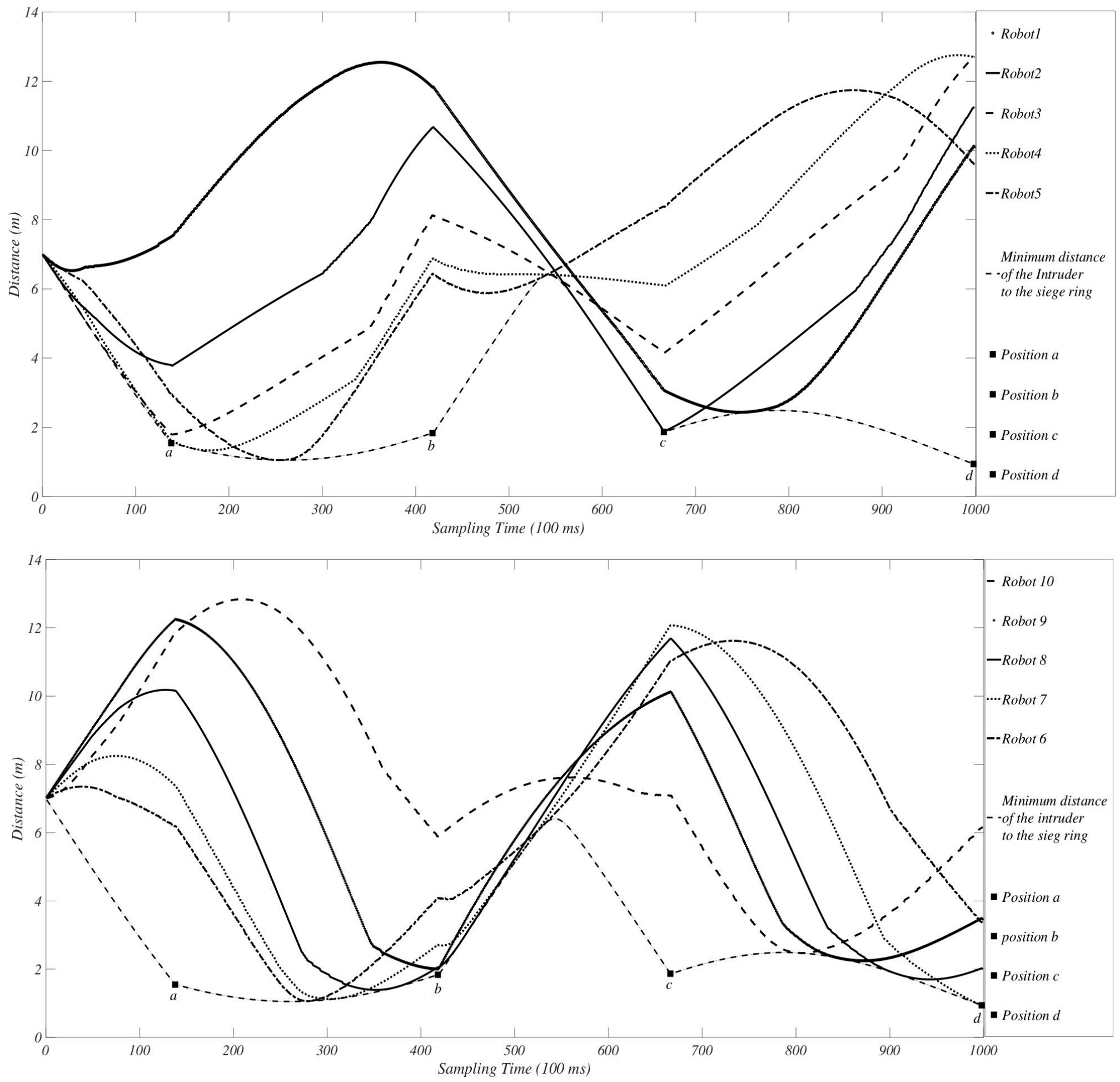

Figure 5.3. a,b: y-coordinates of the robots and the intruder

## 5.5 Summary

We proposed a decentralized motion control algorithm for a network of mobile robots to intercept an intruder on the boundary of a planar region. A necessary and sufficient condition for the existence of such an algorithm was derived. The proposed algorithm is based on some simple rules that only require information about the intruder and the closest neighbors of each robot. Computer simulations confirmed the efficiency of the developed navigation algorithm.

# Chapter 6

# Modified decentralized navigation method for dynamic-intercepting multi intruders trapped in a siege ring by a multi robot team

In this section we study the problem of intercepting multi intruders which are trapped in a siege ring by a multi robot team. In this case, the guardian mobile robots face with multi intruders that any of them tries to exit the region $R$ from some points $P', P'', P''', \dots$ at any time interval $t', t'', t''', \dots$ from segments $S_1$or $S_2$. The intruders are visible to the guardian robots all the times. The team of mobile robots move along the boundary of a planar region $R$ between two hypothetical points $P_1$ and $P_2$ in a decentralized fashion, i.e. each robot navigates independently and has information about current coordinates of just several closest other robots. Similar to the problem in previous chapter, the objective of the multi-robot team is to maintain the intruders in a siege to prevent any sabotaging by the intruder in the environment which means that there should be at least one robot close to the crossing point on the boundary to intercept the intruders. In continue, we explain the problem of the proposed modified model for maintaining the intruders inside the siege ring by a multi robot team in section 6.1. section 6.2, presents the algorithm of the modified model for intruders' interception. The performance of the system is confirmed and analyzed in chapter 6.3, and the chapter is summarized in section 6.4.

## 6.1 Problem statement

In this scenario, we supposed that $m$ intruders have been trapped and surrounded by a team of $n$ mobile robots. The mobile robots deployed in a hypothetical circle. $S$ known as the circumference of the hypothetical circle which is divided in two equal segments $S_1$ and $S_2$. There are no static or dynamic obstacles in the region that could prevent the intruders from escaping from the region. The objective of the intruders is to escape from the region $R$ through the segments $S_1$ or $S_2$ while avoiding intercepting by any member of the multi-robot team. On the other hand, the objective of the multi-robot team is to intercept each individual intruder when it crosses the segment $S_1$ or $S_2$ while any of the intruder tries to cross the segment $S$ in either side. The problem under consideration in this study is to modify the necessary and sufficient condition result from the previous section under which the intruders would be intercepted for their every motion when they try to escape the region by crossing the segment $S$. Moreover, based on this condition we design a decentralized navigation strategy for the multi-robot team in which every team member of the multi-robot team act as a dynamic-interceptor to dynamically intercept the intruders when they get close to the segment $S$ in all the time. Let R be a closed convex planar region with a piecewise smooth boundary where the robots moving on to siege the intruder. Furthermore, let $S_1$ and $S_2$ be segments of the boundary of the region $R$ between points $P_1 \circlearrowleft P_2$ and $P_2 \circlearrowleft P_1$ respectively where $\circlearrowleft$ denotes the clockwise direction; see Fig.1.6.

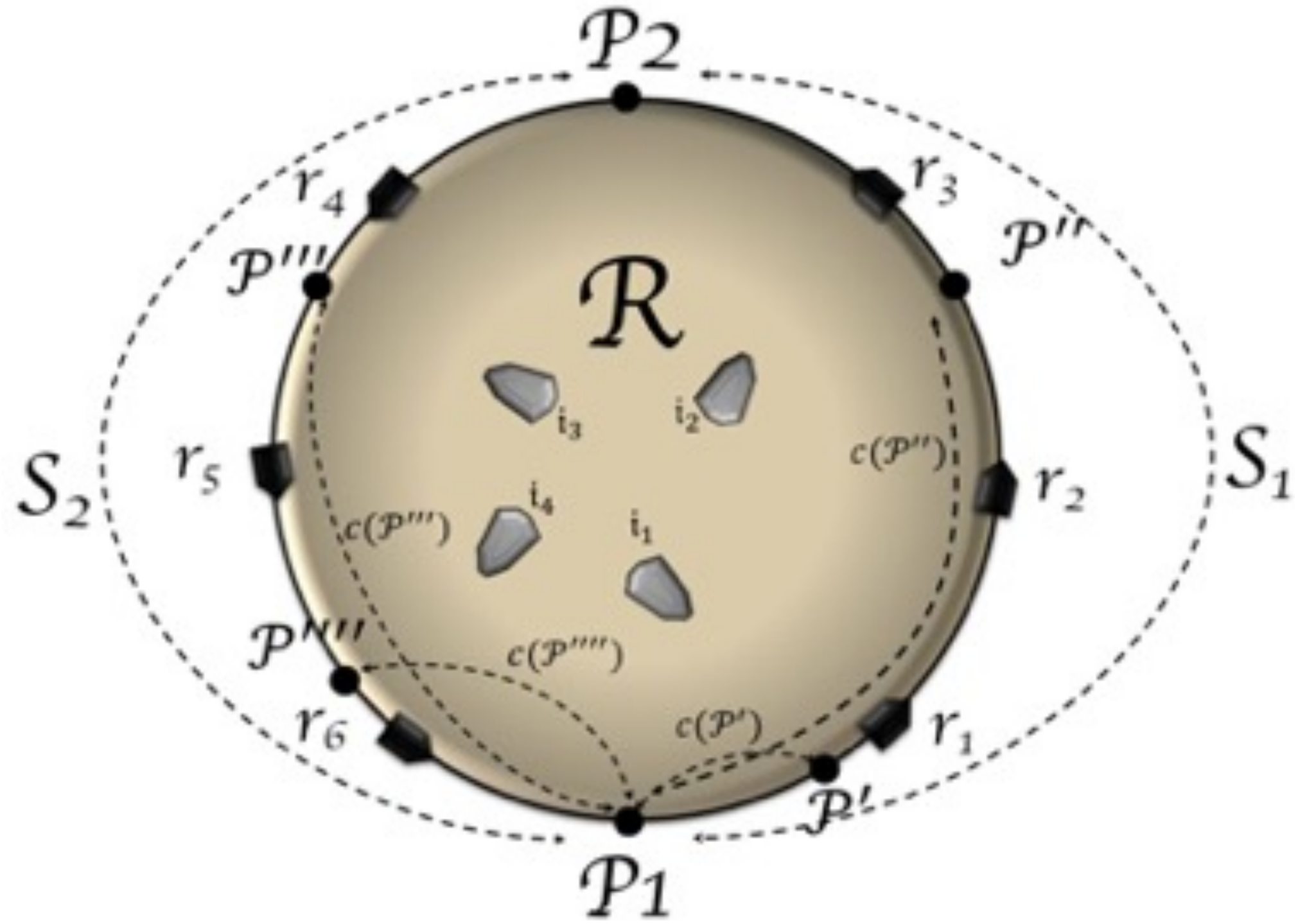

Figure 6.1: Siege ring $R$ surrounded by the robots and trapped intruders

It's obvious that the intruders $\mathrm{i}_1, \mathrm{i}_2, \mathrm{i}_3, \mathrm{i}_4$ try to escape from the region $R$ through segments $S_1$ or $S_2$ where the robots moving to maintain the intruders in the region $R$. Let

$$\begin{cases} X_i^T = [x_{i1}(t), x_{i2}(t), \dots, x_{im}(t)\,] \\ \qquad\qquad and \\ Y_i^T = [y_{i1}(t), y_{i2}(t), \dots, y_{im}(t)\,] \end{cases} \tag{6.1}$$

denote planar coordinates of the intruder.

The intruders are moving with an arbitrary time-varying vector velocity

$V_i^T = [v_{i1}(t), v_{i2}(t), \dots, v_{im}(t)]$, where $v_{ik}(t) = \begin{pmatrix} \dot{x}_{ik}(t) \\ \dot{y}_{ik}(t) \end{pmatrix}$ , $k \in 1,2, \dots, m$, satisfying the constraint

$$\|v_{ik}(t)\| \le V_I^{max} \ \ \forall t \ge 0 \tag{6.2}$$

where $V_I^{max} > 0$ is a given constant, $\|\cdot\|$ denotes the standard Euclidean vector norm. Moreover, let $n > 1$ be a given positive integer.

We consider n mobile point-wise robots labelled $1,2,\dots,n$ that try to maintain the intruders $1,2,\dots,m$ inside the region $R$ while the intruders try to escape the region through the segments $S_1$ and $S_2$. Unlike the intruders that can move in any direction in the plane, the robots can move only along the segments $S_1$ and $S_2$ in the both directions.

Furthermore, $P_{r1} = \begin{pmatrix} x_1(t) \\ y_1(t) \end{pmatrix}, P_{r2} = \begin{pmatrix} x_2(t) \\ y_2(t) \end{pmatrix}$, ..., $P_{rn} = \begin{pmatrix} x_n(t) \\ y_n(t) \end{pmatrix}$ denote the planar coordinates of the robots $1, 2, \dots, n$.

We introduce a set of $C(P)$ for a set of $m$ points $P = (P^1, P^2, \dots, P^m) \in S_1 \lor P = (P^1, P^2, \dots, P^m) \in S_2$ such that $C(P) = (c(P^1), c(P^2), \dots, c(P^m))$ is the length of the curved shape portion of the segment $S$ between the points $P1$ and any point of the set $P$; see Fig.(1). This implies that $c(P_1)$=0 and $c(P_2) = L$, where L is the length of the curves $S_1$ and $S_2$ in either side.

Furthermore, for each guardian robot we define $c(P_i)$ such that:

$$x = r(\beta)\cos(\beta)$$
$$y = r(\beta)\sin(\beta)$$

$$c(P_i) = \int_{P_1}^{P_{ri}} \sqrt{[r(\beta)]^2 + \left[\frac{dr(\beta)}{d\beta}\right]^2}\, d\beta \qquad (6.3)$$

Therefor $c_1(t) := c(P_{r1}), c_2(t) := c(P_{r2}), \ldots, c_n(t) := c(P_{rn})$ denote the length of the curve of each robot $1,2,\ldots,n$ at time $t \geq 0$ to the point $P_1$.

Furthermore, we suppose that the robots labelled according to their coordinates so that

$$0 \leq c_1(t) \leq c_2(t) \leq \ldots \leq c_n(t) \leq L \ \forall t \geq 0 \qquad (6.4)$$

The requirement (2) means that the robots never change their order on the segment $S_1$ or $S_2$.

We assume that the motion of the robots along $S$ is described by the equation

$$\dot{c}_i(t) = u_i(t) \qquad \forall i = 1, \ldots, n \qquad (6.5)$$

where $ui(t)$ is the control input of the robot $i$. We assume that the control inputs $ui(t)$ satisfy the constraint

$$|u_i(t)| \leq V_R^{max} \quad \forall t \geq 0 \qquad (6.6)$$

where $V\ max > 0$ is a given constant.

At any time $t$, each robot $i, 2 \leq i \leq n-1$ knows the coordinates $P_{ri-1}(t), P_{ri+1}(t)$ of the robots $i-1$ and $i+1$, respectively. The robots 1 and $n$ know the coordinates $P_{r2}(t), P_{rn-1}(t)$ of the robots 2 and $n-1$, respectively. Moreover, each robot i knows its own coordinate $P_i(t)$. Furthermore, the intruders become visible to the robots at some time $t_0 \geq 0$, i.e. all the robots know the planar coordinates $P_I^T = (P_{i1}, P_{i1}, \ldots, P_{im})$ such that $P_{ik}\begin{pmatrix} xik\ (t) \\ y_{ik}(t) \end{pmatrix}$of the intruder $k \in 1,2, \ldots, m,$ for all $t \geq t_0$.

*Definition 6.1:*

Let $\varepsilon > 0$ be a given constant. Suppose that the intruder crosses the segment $S_1$ at time $t^*$i.e. $P_I(t^*) \in S_1$. We say that the multi-robot team barricading the intruder at time $t^*$ if there exists some index $i = 1,2, \ldots, n$ such that $|c(\hat{P}_I(t \star)) - ci(t \star)| \leq \varepsilon$.

Where $\hat{P}_I(t \star) = \begin{pmatrix} \hat{x}_I(t_*) \\ \hat{y}_I(t_*) \end{pmatrix}$ denotes the projection coordinates of the intruder to the siege ring $S$ while it's close to it in either side.

The proposed navigation strategy that is based on the available information is called dynamic-intercepting as each member of the multi-robot team plays the role of a dynamic interceptor when the intruder crosses the siege ring $S$ in either side.

In other words, dynamic-intercepting means that when the interceptor crosses the segment $S_1$ or $S_2$, there should be at least one robot close enough to the interception point.

## 6.2 Modified dynamic-intercepting navigation method

Let $P^k \in P \in S$ $(k = 1,2, \ldots, m)$, and $P_{ik}\begin{pmatrix} xik\ (t) \\ y_{ik}(t) \end{pmatrix}$ be an interior point of the region $R$. Then there is a straight line $\mathcal{L}(P_{ik}, P^k)$ which connects points $P_{ik}$ and $P^k$. Since $R$ is convex, $\mathcal{L}(P_{ik}, P^k)$ is in $R$ and the intersection of $\mathcal{L}(P_{ik}, P^k)$ and the boundary of $R$ contains only the point $P^k$.

Furthermore, let $\alpha_k(P_{ik}, P^k)$ denote the length of $\mathcal{L}(P_{ik}, P^k)$. On the other hand, let $i$ be an index such that $|ci\ (t)\ -\ c(P^k)|\ \leq\ |cj\ (t)\ -\ c(P^k\ )|$ for all $j\ =\ 1,,n$. Then, introduce the variable $\xi_k(t, P^k\ )\ :=\ |ci(t)\ -\ c(P^k\ )|$. In other words, $i$ is the closest robot to the point $P^k$ at time $t$ and $\xi_k(t, P^k)$ is the length of the sub-segment of the segment $(S_1 \vee S_2) \in S$ between the closest robot's current location $P_{ik}$ and the point $P^k$.

According to the problem statement, we have multi intruders in the region $R$, therefore, we introduce a set of weighting factors for each intruder as $\Psi = (\psi_1, \psi_2, \ldots, \psi_m)$. To calculate the weighting factors for each intruder, we need to find out which one is the most dangerous and which on is the least one to the given point $P^k \in P$. Its obvious that, the intruder with the maximum distance to the point $P^k \in P$ is less likely to be able to escape the region $R$ from point $P^k$ compare to the other intruders. Furthermore, as the region is considered as a siege ring with a given diameter, we consider that any given intruder with the distance of the diameter of the siege ring to the point $P^k$ is the less dangerous intruder to the point. Then we define a probability function with the probability 1 for the intruder which is located at the point with the distance equal to the diameter of the siege ring to the point $P^k$ as in:

$$\wp_k = \frac{\alpha_k(P_{ik}, P^k)}{\mathfrak{D}_R} \tag{6.7}$$

Where $\mathfrak{D}_R$, denotes the approximate diameter of the siege ring. The probability (6.7) states that for any intruders $k$ and $\tilde{k} \in (1,2, \ldots, m)$ if $\alpha_k(P_{ik}, P^k) > \alpha_{\tilde{k}}(P_{i\tilde{k}}, P^k)$ then $\wp_k > \wp_{\tilde{k}}$ which means that, the intruder $k$ is less likely to be able to cross the point $P^k$ at time $t$ compare to the intruder $\tilde{k}$.

Then we find the weighting factor $\Psi$ as follows:

$$\Psi = 1 - \wp \quad (6.8)$$

In continue, define the new distance which is the inference of mean value of the minimum length of each intruder to any point $P^k$ and its weight as in:

$$\eta_{|k\in(1,2,\ldots,m)}(P_{i=1}^{m}, P^k) = \frac{\sum_{i=1}^{m} \psi_i \alpha_i (P_{ii}, P^k)}{m} \quad (6.9)$$

where $m$ denotes the number of intruders.

Then we introduce the function $F\ (s)$ from the interval $[0, \mathcal{L}]$ to the segment $(S_1 \vee S_2) \in S$ such that for any number $s \in [0, \mathcal{L}]$, $F(s)$ is the point $P \in (S_1 \vee S_2) \in S$ where $c(P^k) = s$.

Furthermore, let $[P_{A_1}, P_{A_2}]$ denote the closed sub-segment of the segment $(S_1 \vee S_2) \in S$ between the points $P_{A_1}$ and $P_{A_2}$. For $i = 1, \ldots, n$, introduce sub-segments $Si(t)-, Si(t)+$ of the segment $(S_1 \vee S_2) \in S$.

## *Remark 6.1:*

we consider robots $r_1, \ldots, r_m$ are located between points $P_1$ and $P_2$ in segment $S_1$ and robots $r_{m+1}, \ldots, r_n$ are located between points $P_2$ and $P_1$in segment $S_2$, in a counter clockwise fashion where$m < n$.

Therefore, if the robot $r_i \in [r_1, \ldots, r_m]$:

$$S_i(t)^- := \left[F\left(\frac{c_{i-1}(t)+c_i(t)}{2}\right), F(c_i(t))\right]$$
$$if\ i = 2, \dots, m;$$
$$S_i(t)^- := [P_1, F(c_i(t))]$$
$$if\ i = 1;$$
$$S_i(t)^+ := \left[F(c_i(t)), F\left(\frac{c_i(t)+c_{i+1}(t)}{2}\right)\right]$$
$$if\ i = 1, \dots, m-1;$$
$$S_i(t)^+ := [\,F(c_i(t)), P_2]$$
$$if\ i = m. \tag{6.10}$$

However, if the robot $r_i \in [r_{m+1}, \dots, r_n]$:

$$\acute{S}_i(t)^- := \left[F\left(\frac{c_{i-1}(t)+c_i(t)}{2}\right), F(c_i(t))\right]$$
$$if\ i = m+2, \dots, n;$$
$$\acute{S}_i(t)^- := [\,F(c_i(t)), P_2]$$
$$if\ i = m+1.$$
$$\acute{S}_i(t)^+ := \left[F(c_i(t)), F\left(\frac{c_i(t)+c_{i+1}(t)}{2}\right)\right]$$
$$if\ i = m+1, \dots, n-1;$$
$$\acute{S}_i(t)^+ := [P_1, F(c_i(t))]$$
$$if\ i = n; \tag{6.11}$$

Moreover, for $i = 1, 2, \dots, n$, introduce the numbers $M_i^-(t)$ and $M_i^+(t)$ as

$$M_i^-(t) := \sup_{P^k \in P \in (S_i(t)^- \vee \acute{S}_i(t)^-)} \left(\xi_k(t, P^k) - \frac{\eta_{|k\in(1,2,\dots,m)}(P_{i=1}^m, P^k) V_R^{max}}{V_I^{max}}\right);$$

$$M_i^+(t) := \sup_{P^k \in P \in (S_i(t)^+ \vee \acute{S}_i(t)^+)} \left(\xi_k(t, P^k) - \frac{\eta_{|k\in(1,2,\dots,m)}(P_{i=1}^m, P^k) V_R^{max}}{V_I^{max}}\right) \tag{6.12}$$

Now we can introduce the following decentralized navigation law:

$$u_i(t) := V_R^{max} \quad if \quad M_i^-(t) < M_i^+(t)$$
$$u_i(t) := -V_R^{max} \quad if \quad M_i^-(t) > M_i^+(t)$$

$$u_i(t) := 0 \qquad if \qquad M_i^-(t) = M_i^+(t) \tag{6.13}$$

for all $i = 1, \ldots, n$.

## *Remark 6.2:*

The intuition behind the decentralized navigation law (6.13) can be explained as follows.

The sub-segments $S_i^-(t), S_i^+(t)$ are sets of points of the curve $S_1$ and the subsegments $\acute{S}_i(t)^-, \acute{S}_i(t)^+$ are sets of points of the curve $S_2$ for which the robot $i$ is the closest robot at time $t$. The robot moves with the maximum allowed speed towards the one of these segments that is more" *dangerous*" at the current time, i.e. it has the biggest possible distance between the intruder and the closest robot at the moment of crossing $S$ by the intruder. This biggest possible distance is described by (6.12).

## *Theorem 6.1:*

Consider the multi-robot team satisfying (6.6) and each of the intruders satisfying (6.2). Then there exists a *dynamic-intercepting* multi-robot team navigation strategy in either sides of the siege ring $(S_1 \vee S_2)$ if and only if

$$\sup_{P \in S_1 \vee S_2} \left( \xi(t_0, P^k) - \frac{\eta_{|k \in (1,2,\ldots,m)}(P_{i=1}^m, P^k) V_R^{max}}{V_I^{max}} \right) \leq \varepsilon \tag{6.14}$$

where $t_0 \geq 0$ is the time at which the intruders become visible to the robots.

Moreover, if the inequality (6.14) holds, then the navigation law (6.13) is a dynamic-intercepting navigation strategy.

## *Remark 6.3:*

Notice that since the region $R$ is convex, and the segment $S$ is compact, the supremum in (6.14) is achieved for some point $P$.

## *Proof:*

First, we prove that if the inequality (6.14) does not hold, then the intruder can always cross the segment $S$ without intercepting by the multi-robot team. Indeed, if (6.14) does not hold, then there exists a point $P \in S$ such that

$$\begin{cases} \left(\xi(t_0, P^k) - \frac{\eta_{|k\in(1,2,\dots,m)}(P_{i=1}^m, P^k) V_R^{max}}{V_I^{max}}\right) > \varepsilon \quad for\ P \in S_1 \\ or \\ \left(\xi\left(t_0, P^{\tilde{k}}\right) - \frac{\eta_{|\tilde{k}\in(1,2,\dots,m)}(P_{i=1}^m, P^{\tilde{k}}) V_R^{max}}{V_I^{max}}\right) > \varepsilon \quad for\ \hat{P} \in S_2 \end{cases} \tag{6.15}$$

Now let the intruders move along the straight-line segments $|L(\mathrm{P}_{i=1}^m(t_0),\ P^k)|$ and $\left|L\left(\mathrm{P}_{i=1}^m(t_0), P^{\tilde{k}}\right)\right|$ connecting the points $\mathrm{P}_{i=1}^m(t_0)$ and $P^k$ or $P^{\tilde{k}}$ with its maximum speed $V_I^{max}$ respectively. In this case, each intruder reaches the point $P^k \in S_1$ or $P^{\tilde{k}} \in S_2$ at the time:

$$\begin{cases} t^* = t_0 + \frac{|L(P_{k\in(1,2,\dots,m)}(t_0),\ P^k)|}{V_I^{max}} \\ or \\ t^{**} = t_0 + \frac{\left|L\left(P_{\tilde{k}\in(1,2,\dots,m)}(t_0), P^{\tilde{k}}\right)\right|}{V_I^{max}} \end{cases} \tag{6.16}$$

where $P_{k\in(1,2,\dots,m)}$ and $P_{\tilde{k}\in(1,2,\dots,m)}$ denote the coordinates of the closest robot to either point $P^k$ or $P^{\tilde{k}}$ at time $t_0$. It obviously follows from (6.15) that, the closest robot to the point $P^k \in S_1$ or $P^{\tilde{k}} \in S_2$ cannot be closer than $\varepsilon$ at time $t^*$ or $t^{**}$ when the closest intruder reaches the points in either side. Therefore, there is no any dynamic interceptor close enough to the neighborhood of the points $P^k \in S_1$or $P^{\tilde{k}} \in S_2$ at time $t^*$ or $t^{**}$ to intercept the closest intruder from crossing the segment $S$ in either side.

We now prove that if the inequality (6.14) holds, the intruders are intercepted always by at least one dynamic interceptor of the multi-robot team while they try to cross the segment $S$. First, we prove the following claim.

Indeed, for any trajectory $[P_{i=1}^m(t), c_1(t), \ldots, c_n(t)]$ of the intruder-multi-robot introduce the Lyapunov function

$$\begin{cases} \mathcal{W}[P_{i=1}^m(t), c_1(t), \ldots, c_l(t)] := \\ \sup_{P^k \in S_1} \left( \xi(t_0, P^k) - \frac{\eta_{|k\in(1,2,\ldots,m)}(P_{i=1}^m, P^k) V_R^{max}}{V_I^{max}} \right) for\ S_1 \in [P_1 \circlearrowleft P_2] \\ \vee \\ \mathcal{W}[P_{i=1}^m(t), c_{l+1}(t), \ldots, c_n(t)] := \\ \sup_{P \in S_2} \left( \xi(t_0, P^{\tilde{k}}) - \frac{\eta_{|\tilde{k}\in(1,2,\ldots,m)}(P_{i=1}^m, P^{\tilde{k}}) V_R^{max}}{V_I^{max}} \right) for\ S_2 \in [P_2 \circlearrowleft P_1] \end{cases} \tag{6.17}$$

Notice that since the region $R$ is convex, and the segment $S$ is compact, the supremum in (6.17) is achieved for some point $P^k$ or $P^{\tilde{k}}$. Furthermore, by definition, $\eta_{|k\in(1,2,\ldots,m)}(P_{i=1}^m, P^k)$ is the length of the straight segment $L(P_{k\in(1,2,\ldots,m)}(t_0),\ P^k)$ connecting any intruder $P_{k\in(1,2,\ldots,m)}(t)$ and $P^k$ at the segment $S_1$ or analogously $\eta_{|\tilde{k}\in(1,2,\ldots,m)}(P_{i=1}^m, P^{\tilde{k}})$ is the length of the straight lines $L(P_{\tilde{k}\in(1,2,\ldots,m)}(t_0), P^{\tilde{k}})$ connecting any intruder$P_{k\in(1,2,\ldots,m)}(t)$ and $P^{\tilde{k}}$ at the segment $S_2$ .

Hence, it is obvious that

$$\begin{cases} \eta_{|k\in(1,2,\ldots,m)}(P_{i=1}^{m}, P^{k}) = \inf_{M(P_{i=1}^{m}, P^{k})\in\mathcal{M}(P(t),P)} |M(P_{i=1}^{m}, P^{k})| \\ \vee \\ \eta_{|\tilde{k}\in(1,2,\ldots,m)}(P_{i=1}^{m}, P^{\tilde{k}}) = \inf_{M(P_{i=1}^{m}, P^{\tilde{k}})\in\mathcal{M}(P(t),P)} |M(P_{i=1}^{m}, P^{\tilde{k}})| \end{cases} \quad (6.18)$$

Where, $\mathcal{M}(P(t), P)$ is the set of all possible paths of the intruders between $P_{i=1}^{m}(t)$ and $P$ . Furthermore, it immediately follows from (6.17), (6.18) and (6.13) that

$$\begin{cases} \mathcal{W}[P_{i=1}^{m}(t), c_1(t_1), \ldots, c_l(t_1)] \leq \mathcal{W}[P_{i=1}^{m}(t), c_1(t_2), \ldots, c_l(t_2)] \; for\, S_1 \in [P_1 \circlearrowleft P_2] \\ \vee \\ \mathcal{W}[P_{i=1}^{m}(t), c_{l+1}(t_1), \ldots, c_n(t_1)] \leq \mathcal{W}[P_{i=1}^{m}(t), c_{l+1}(t_2), \ldots, c_n(t_2)] \; for\, S_2 \in [P_2 \circlearrowleft P_1] \end{cases}$$

$$\forall t_2 \geq t_1 \geq t_0 \quad (6.19)$$

Now (6.19) and (6.15) imply that if the closest intruder reaches a point $P^k \in S_1$or $P^{\tilde{k}} \in S_2$ at some time $t^* \geq t_0$ or $t^{**} \geq t_0$ respectively, the robot closest to the point $P^k$ or $P^{\tilde{k}}$ at time $t^*$or $t^{**}$ cannot be further from$P^k$ or $P^{\tilde{k}}$ than $\varepsilon$. This implies that any team member of the multi robot team plays the role of a dynamic-interceptor to maintain the intruders inside the siege ring.

The inequality (6.15) and the navigation law (6.13) can be made computationally simpler under the following assumption.

## *Assumption 6.1:*

The following inequality holds:

$$V_I^{max} \geq V_R^{max} \quad (6.20)$$

For $i = 1,2,\ldots,n$, introduce points $D_i(t)^-, D_i(t)^+$of the segment $S_1$ and $\widehat{D}_i(t)^-, \widehat{D}_i(t)^+$of the segment $S_2$ for robot $r_i$ as follows:

For $i \in [1, \dots, m] \implies r_i \in S_1$:

$$D_i(t)^- := F\left(\frac{c_{i-1}(t)+c_i(t)}{2}\right) \qquad if\ i = 2, \dots, m;$$
$$D_i(t)^- := P_1 \qquad if\ i = 1;$$
$$D_i(t)^+ := F\left(\frac{c_i(t)+c_{i+1}(t)}{2}\right) \qquad if\ i = 1, \dots, m-1;$$
$$D_i(t)^+ := P_2 \qquad if\ i = m. \tag{6.21}$$

For $i \in [m+1, \dots, n] \implies r_i \in S_2$:

$$\widehat{D}_i(t)^- := F\left(\frac{c_{i-1}(t)+c_i(t)}{2}\right) \qquad if\ i = m+2, \dots, n;$$
$$\widehat{D}_i(t)^- := P_2 \qquad if\ i = m+1;$$
$$\widehat{D}_i(t)^+ := F\left(\frac{c_i(t)+c_{i+1}(t)}{2}\right) \qquad if\ i = m+1, \dots, n-1;$$
$$\widehat{D}_i(t)^+ := P_1 \qquad if\ i = n \tag{6.22}$$

Moreover, for $i = 1, 2, \dots, n$, we introduce a set of numbers $\mathcal{H}$ which includes $H_i(t)^-$ and $H_i(t)^+$ if $r_i \in S_1$ and $\widehat{H}_i(t)^-$ and $\widehat{H}_i(t)^+$ if $r_i \in S_2$ as follows:

$$H_i^-(t) := \frac{c_i(t)-c_{i-1}(t)}{2} - \frac{\eta_{|k\in(1,2,\dots,m)}(P_{i=1}^m, D_i(t)^-)V_R^{max}}{V_I^{max}}$$
$$if\ i = 2, \dots, l;$$
$$H_i^-(t) := c_i(t) - \frac{\eta_{|k\in(1,2,\dots,m)}(P_{i=1}^m, P_1)V_R^{max}}{V_I^{max}}$$
$$if\ i = 1;$$
$$H_i^+(t) := \frac{c_{i+k}(t)-c_i(t)}{2} - \frac{\eta_{|k\in(1,2,\dots,m)}(P_{i=1}^m, D_i(t)^+)V_R^{max}}{V_I^{max}}$$
$$if\ i = 1, \dots, l-1;$$
$$H_i^+(t) := L - c_i(t) - \frac{\eta_{|k\in(1,2,\dots,m)}(P_{i=1}^m, P_2)V_R^{max}}{V_I^{max}}$$
$$if\ i = l. \tag{6.23}$$

$$\widehat{H}_i^-(t) := \frac{c_i(t)-c_{i-1}(t)}{2} - \frac{\eta_{|k\in(1,2,\dots,m)}(P_{i=1}^m, \widehat{D}_i(t)^-)V_R^{max}}{V_I^{max}}$$
$$if\ i = l+2, \dots, n;$$

$$\widehat{H}_i^-(t) \coloneqq c_i(t) - \frac{\eta_{|k\in(1,2,\dots,m)}(P_{i=1}^m, P_2)V_R^{max}}{V_I^{max}}$$

$$if\ i = l + 1;$$

$$\widehat{H}_i^+(t) \coloneqq \frac{c_{i+k}(t) - c_i(t)}{2} - \frac{\eta_{|k\in(1,2,\dots,m)}(P_{i=1}^m, D_i(t)^+)V_R^{max}}{V_I^{max}}$$

$$if\ i = l + 1, \dots, n - 1;$$

$$\widehat{H}_i^+(t) \coloneqq L - c_i(t) - \frac{\eta_{|k\in(1,2,\dots,m)}(P_{i=1}^m, P_1)V_R^{max}}{V_I^{max}}$$

$$if\ i = n. \tag{6.24}$$

For $i = 1, \dots, n$, the simplified navigation law (6.13) becomes:

For $r_i \in S_1$:

$$u_i(t) \coloneqq V_R^{max} \quad if \quad H_i^-(t) < H_i^+(t)$$

$$u_i(t) \coloneqq -V_R^{max} \quad if \quad H_i^-(t) > H_i^+(t)$$

$$u_i(t) \coloneqq 0 \quad if \quad H_i^-(t) = H_i^+(t) \tag{6.26}$$

For $r_i \in S_2$:

$$u_i(t) \coloneqq V_R^{max} \quad if \quad \widehat{H}_i^-(t) > \widehat{H}_i^+(t)$$

$$u_i(t) \coloneqq -V_R^{max} \quad if \quad \widehat{H}_i^-(t) < \widehat{H}_i^+(t)$$

$$u_i(t) \coloneqq 0 \quad if \quad \widehat{H}_i^-(t) = \widehat{H}_i^+(t) \tag{6.27}$$

## 6.3 Simulations and discussion

In this section, we consider a team of ten mobile robots known as hunters surrounded three intruders as a prey that each of them tries to escape the region $R$ with maximum speed of $V_I^{max} = 4.5$. the mobile robots are divided in two groups including 5 mobile robots where robots 1 to 5 move on segment $S_1$ , from $P_1$ to $P_2$ and robots 6 to 10 move on segment $S_2$ which is a hypothetical curved line, connects $P_2$ to $P_1$ respectively. The maximum velocity of each robot is $V_r^{max} = 3.0$ and they are supposed to move in left and right on segments $S_1$ and $S_2$. The robots are not allowed to overtake each other which means that they shouldn't disarrange their formation during their mission.

Fig. 6.2(a,b,c,d) illustrate the reaction of the robots to the intruders' motion when the intruders try to exit the region $R$. The robots are indexed in anti-clockwise direction from point $P1$ in a circle, where five robots protect the segment $S1$ which is located between points $P1$ and $P2$, and the other five team members protect the segment $S1$ which is located between points $P2$ and $P1$ on the boundary. The intruders arbitrary move in the region $R$ to find the best point for escaping the area which is not protected by the robot, however as it's obviously cleared, there is at least one robot at the dangerous point close to the intruders which prevents the intruders escape the region.

To analyze the reaction of the robots to the intruders when they apply the proposed navigation law we refer to the Fig.6.3 and Fig.6.4. In both Fig.6.3 and Fig.6.4 the evolution of the $y - coordinates$ of the intruders and the robots during the mission are shown in detail. As its obvious the intruder 1 has been intercepted by the robot 4 when it has minimum distance with the segment $S_1$ in the first stage. Analogously, intruder 2 has been intercepted by robot 3 and intruder 3 has been intercepted by robots 2 and 3 which they both is heading towards the high-risk point where the intruder 3 tends to escape from. In the second stage, the intruder 1 has been intercepted by the robots 8 and 9, while the intruder 2 has been intercepted by the robot 7 at segment $S_2$. As it shows in the graphs, the intruder 3 has been intercepted by the robot 3 when it's in minimum distance to the segment $S_1$. All intruders have their minimum distance to the boundary of segment $S_1$ in stage 3. In this stage, robot 2 has intercepted the intruder 1, the intruder 2 has been intercepted by the robot 1 and the intruder 3 has been intercepted by the robot 5. And finally, in the last stage, the intrude 2 has been covered by robot 4 in segment $S_1$. In segment $S_2$, the robots 6, 7 and 8 are going to cover the high-risk point where the intruder 3 is likely to escape from and the robots 9 and 10 are going to cover the point that intruder 1 is likely to cross in the boundary.

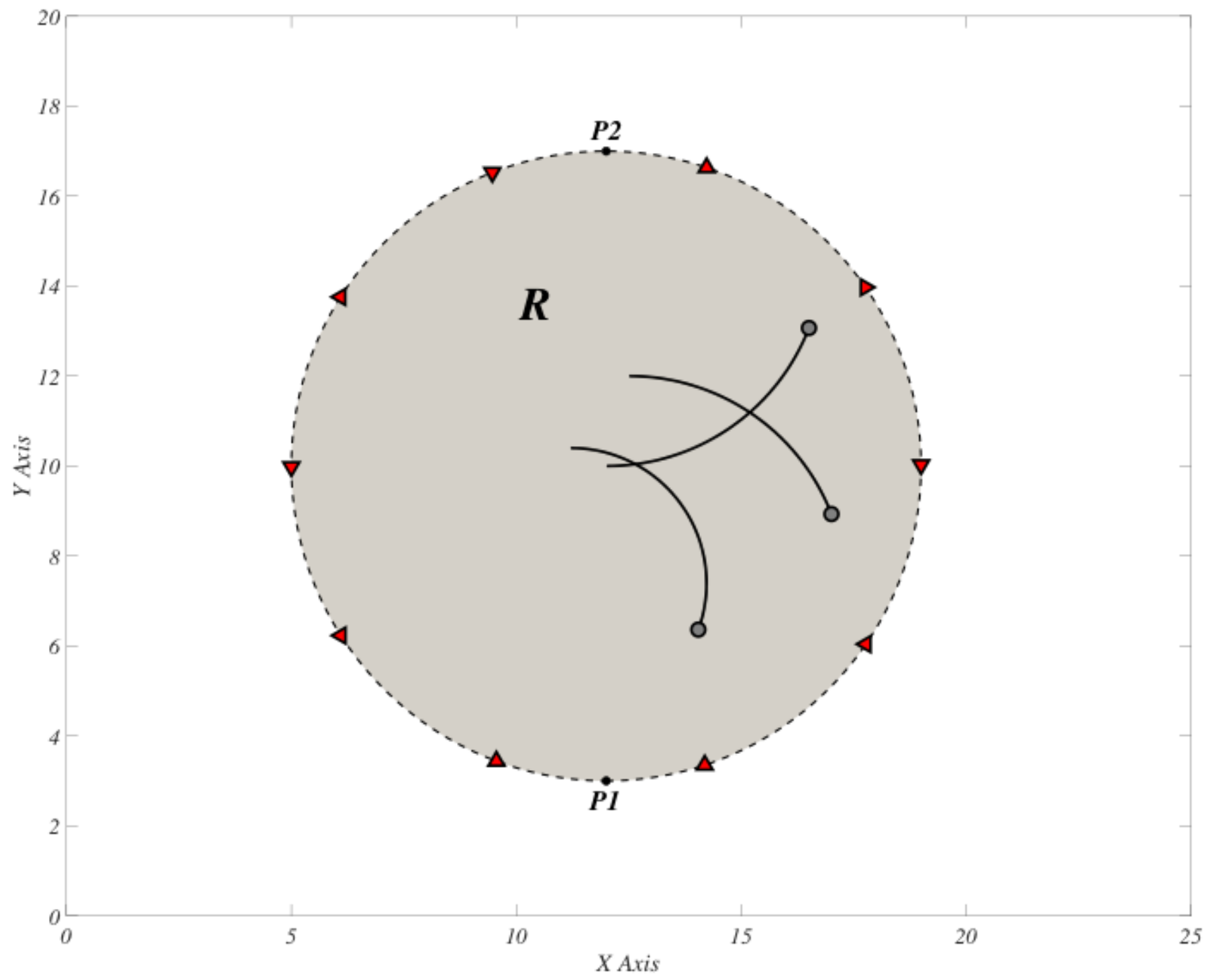

P2
R
P1
Y Axis
X Axis

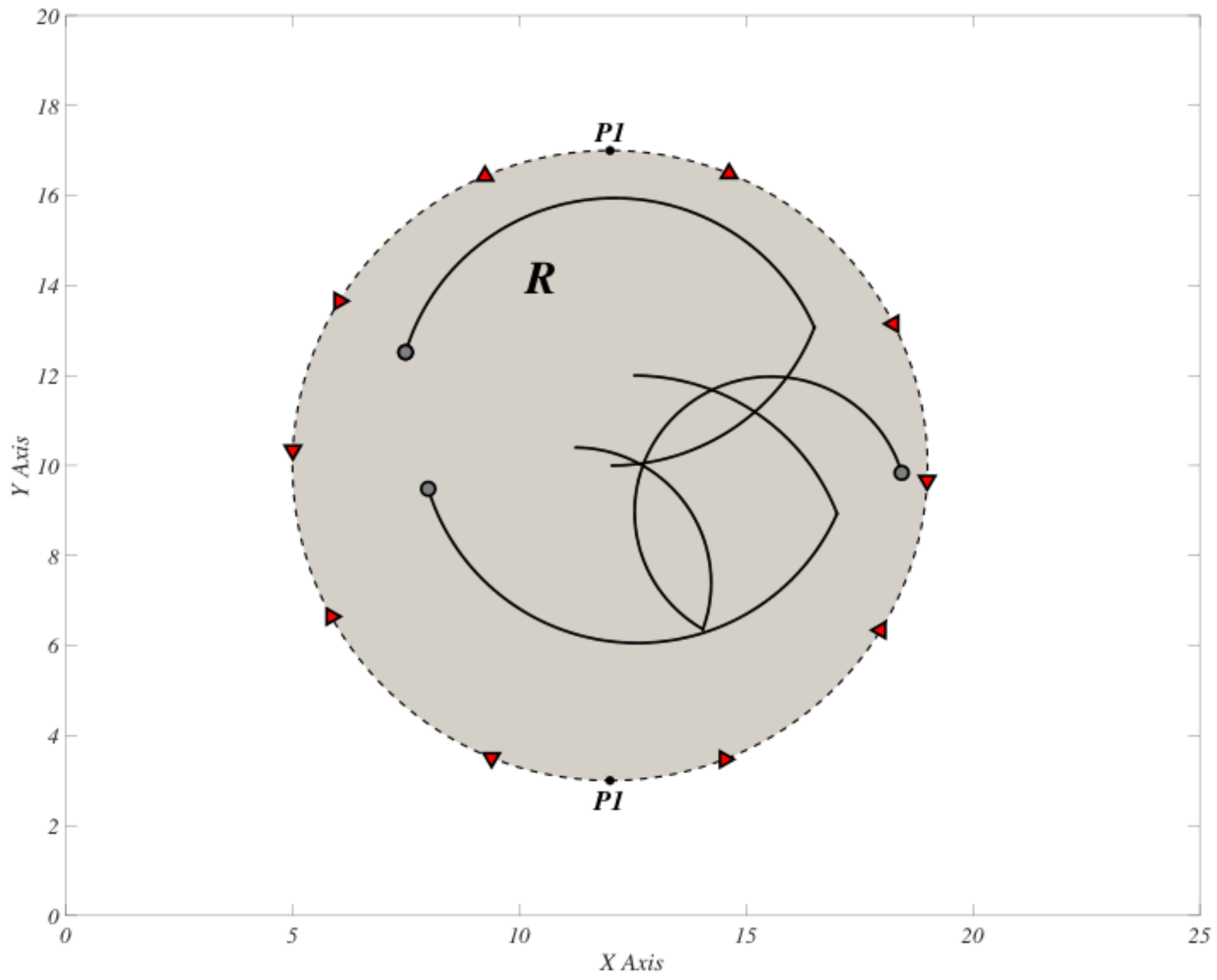

P1
R
P1
Y Axis
X Axis

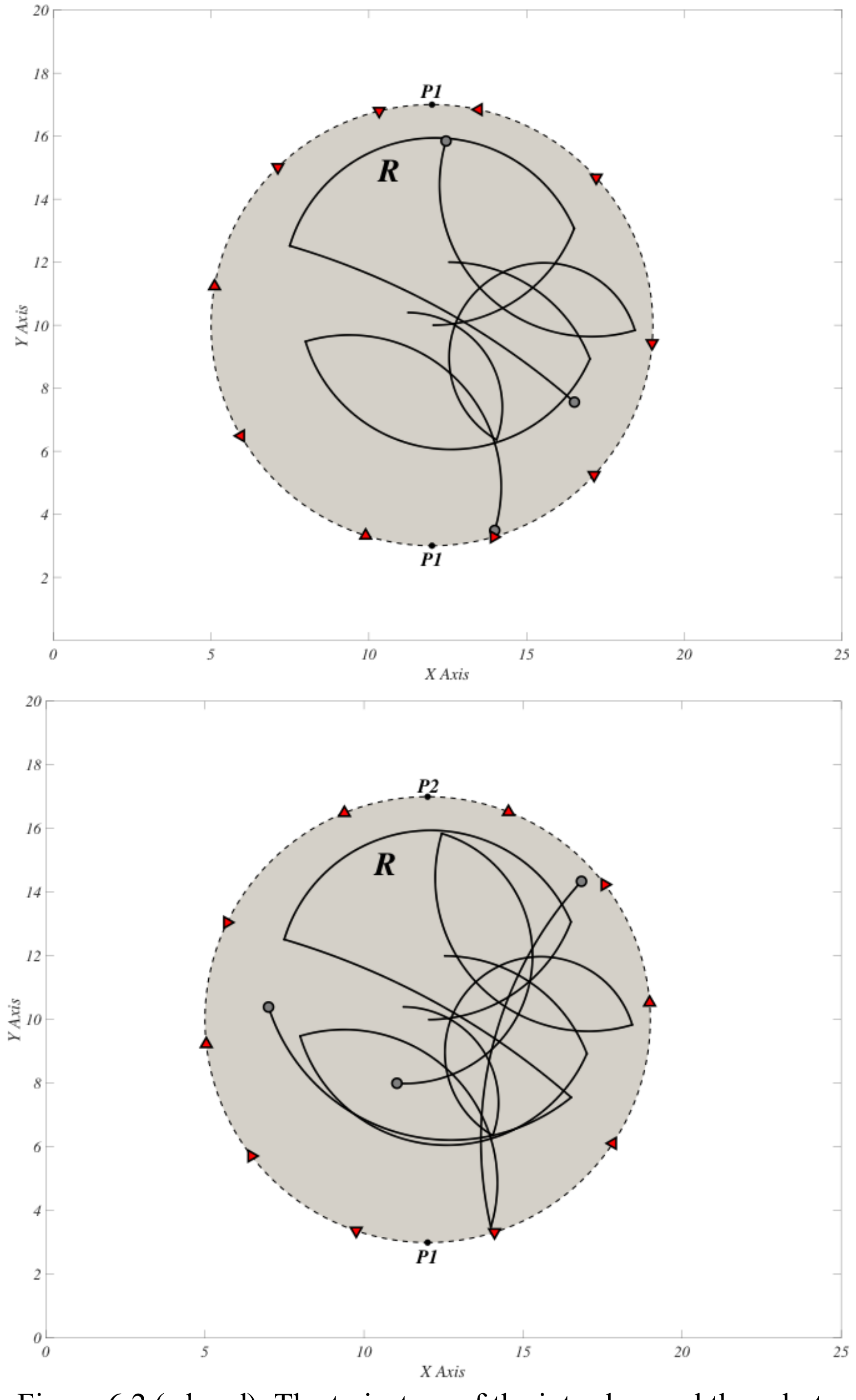

Figure 6.2 (a,b,c,d): The trajectory of the intruders and the robots

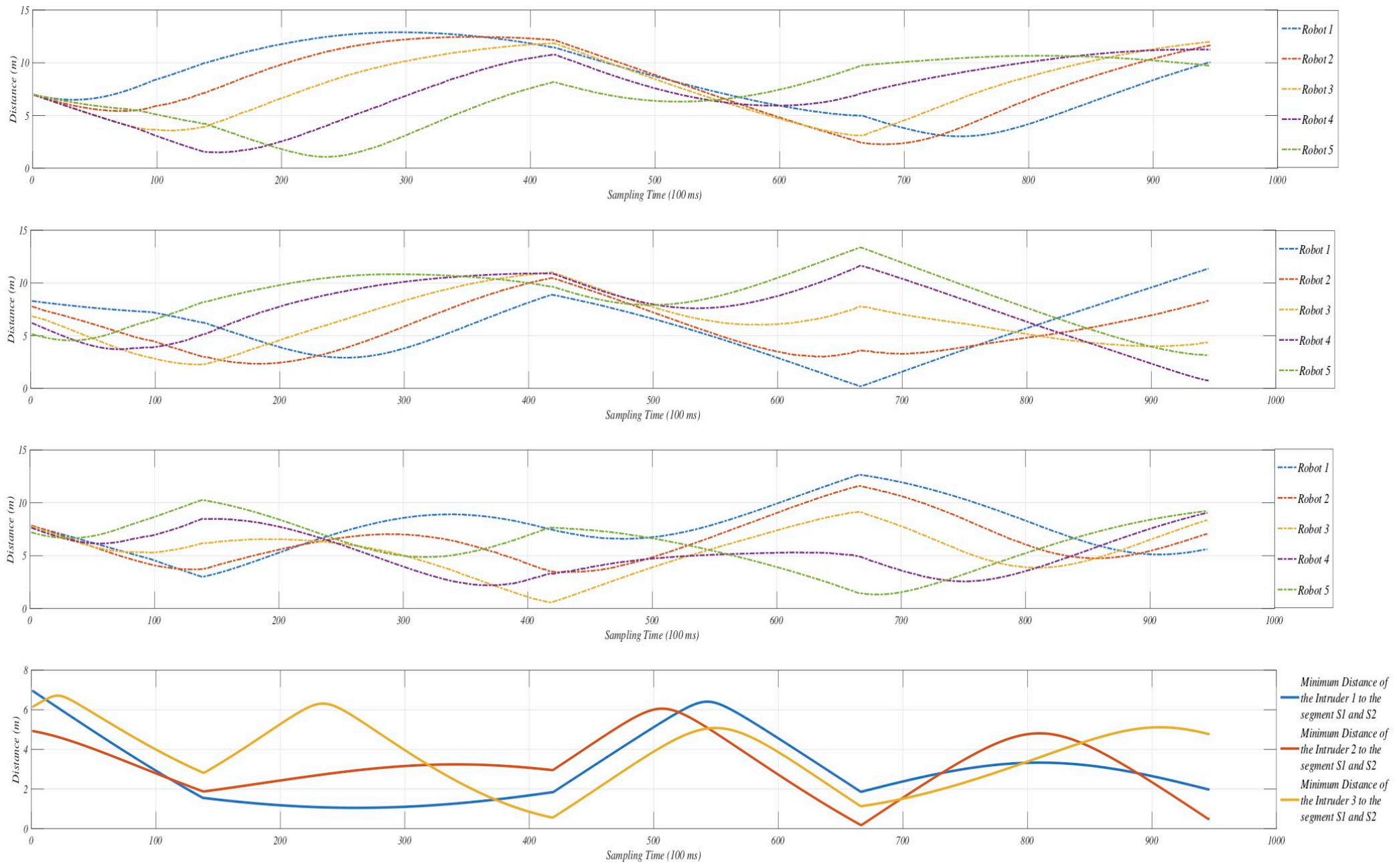

Figure 6.3: Minimum distances of the intruders to the robots 1 to 5 in segment $S_1$

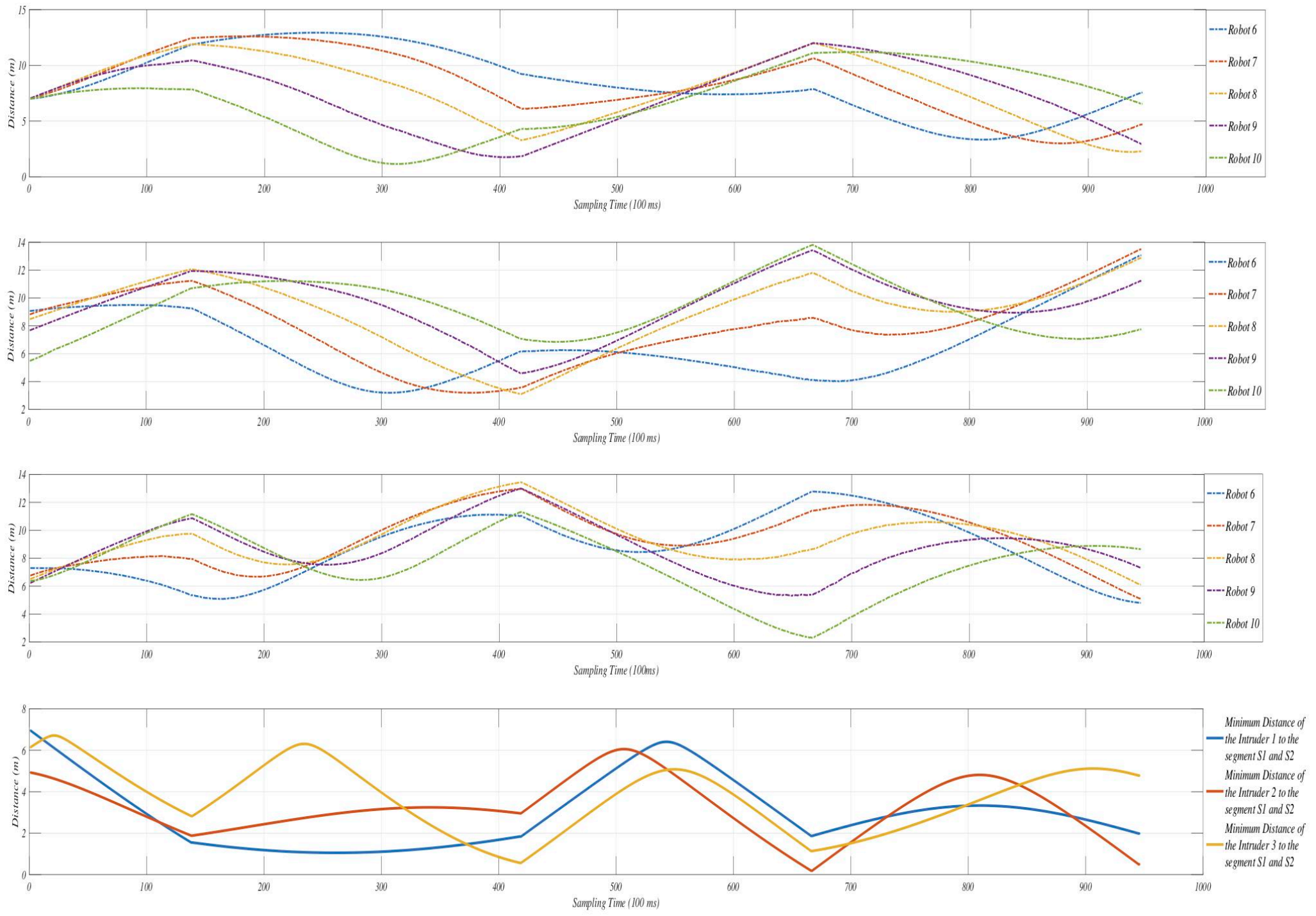

Figure 6.4: Minimum distances of the intruders to the robots 6 to 10 in segment $S_2$

## 6.4 Summary

We proposed a decentralized motion control algorithm for a network of mobile robots to intercept some intruders which move arbitrary to escape the region by crossing the on the boundary of a siege ring without being trapped by the robots. On the other hand, the robots try to intercept the intruders in all the time and keep them inside the region. A necessary and sufficient condition for the existence of such an algorithm was derived. The proposed algorithm is based on some simple rules that only require information about the intruders and the closest neighbors of each robot. Computer simulations confirmed the efficiency of the developed navigation algorithm.

# Chapter 7

# Intelligent game-based navigation and decision-making strategy (IGD) for intruder interception

In this chapter, the problem of intruder detection and trapping is considered. In real-time applications, the noisy communication channels with a limited bandwidth result in disturbance and faulty connections between the team members of a multi robot system [17],[193]. In some cases, e.g. jamming attacks in the battle field which led to maximum damage to communication network systems, the team members possibly lose their communication with each other entirely[209]–[211]. Therefore, we consider the case that the robots have minimum communication or even no communication in the environment. In the other word, based on this method, every agent is able to make a decision based on the local information received from its on-board sensors autonomously. The coordinates of the region are the only priori information that robots have. Indeed, the main focus of this chapter is to introduce a game-based decision-making strategy to make sure every single point of the region has been scanned by at least by one robot from the team at each monitoring period $\mathcal{T}_k$ for $k = 0,1,2,\ldots,$ to detect any possible intruder at the minimum applicable time with the maximum probability of detection. To achieve the goal, an optimal formation method is proposed for the sensors result in minimizing the detection time in the environment rather than standard sweeping method.

To decide for the path planning in each step, the robots have to play both cooperative and non-cooperative game. They play co-op game while they are in the sensing range of each other and communication is applicable. Otherwise, if the communication failed or they are out of the sensing range of each other, then each robot starts playing a non-cooperative game to make the best decision autonomously based on its local information. The rest of the chapter is organized as follows; Section 7.1 presents the problem formulation. Section 7.2 briefly explains a multi-player game theoretic decision-making strategy. Section 7.3 presents the problem modeling and the solution. Section 7.4 compares the simulation results based on game theoretic approach with the sweeping coverage strategy, and finally Section 7.5 summarizes the chapter.

## 7.1 Problem formulation

In this section, we consider the problem of maximizing the probability of intruder detection in a bounded region. In this case, the dynamic of the mobile robots, the specification of the environment and the navigation control strategy is explained.

### 7.1.1 Kinematic of mobile robot and the region's specifications

Let $x_i(t), y_i(t)$ and $\theta_i(t)$ denote the Cartesian coordinate and the heading of the mobile robot $i$ respectively in the plane, where $\theta_i(t) \in (0,2\pi)$ is measured based on the $x$-axis in the counter-clockwise direction. Furthermore, $v_i(t)$ and $\omega_i(t)$ denote linear velocity and angular velocity of the robot $i$ respectively such that satisfy the following condition:

$$v_{min} < v_i(t) < v_{max} \tag{7.1}$$

For all $i = 1, 2 \dots, n, \qquad 0 < v_{min} < v_{max}$.

Then the kinematic equation of the robot $i$ is defined as follows:

$$\dot{x}_i(t) = v_i(t)\, cos\big(\theta_i(t)\big)$$

$$\dot{y}_i(t) = v_i(t)\, sin\big(\theta_i(t)\big)$$

$$\dot{\theta}_i(t) = \omega_i(t) \tag{7.2}$$

We assume each robot can detect any object in the environment which is located within its disk shape sensing range with the radius $R$. It should be noted that, result from the environmental factors and hardware problems the sensing range of a mobile robot is not a perfect disk shape, however, this model could be used as a sufficient approximation of the real sensing range of a mobile robot [212].

In this scenario, an intruder knows as detected if it is located within the sensing range of any of the team member. Furthermore, the intruder has enough information about the points of the boundary with the minimum risk of detection. Moreover, we consider the case that the robots are able to communicate in the region merely, when they are in the sensing range of each.

## *Assumption 7.1:*

The corridor $S \in \mathbb{R}^2$, is a square shape segment consists of nine equal square sub-segments such that, the side of each segment equals $\acute{R} = \sqrt{2}R$, where, $\acute{R}$ denotes the length of the edge of each sub-segment and $R$, denotes the sensing range of each robot which is homogeneous for all of them. On the other word, each sub-segment is inscribed a mobile robot with the sensing range $R$ (see Fig 7.1).

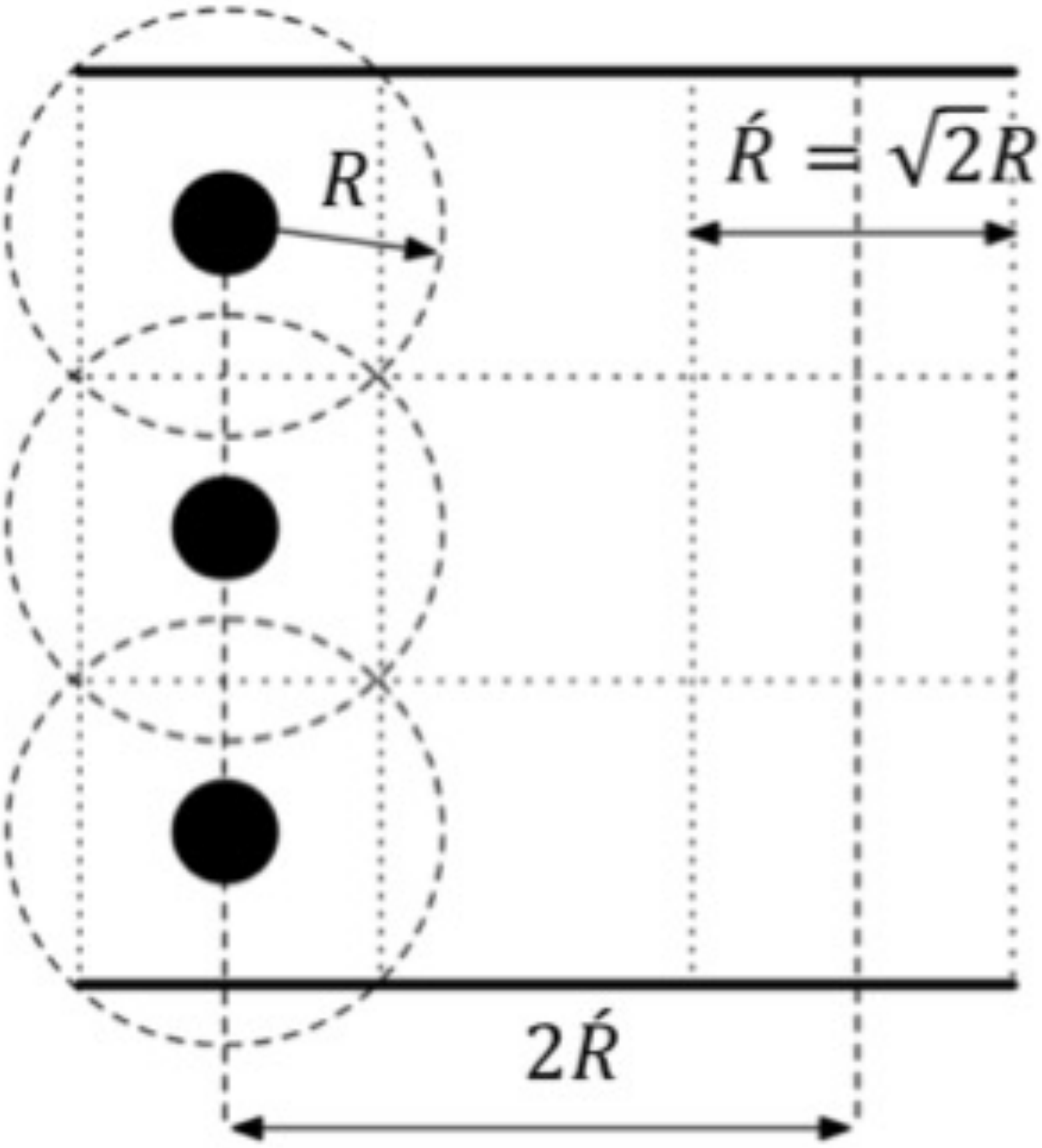

Figure 7.1: The main area of interest

Then we define the area of interest $\mathcal{A}$ based on assumption 2.1 as follow:

$$\begin{cases} A = S \ for \ k = 0 \\ A = 2kS \ for \ k = \{1,2,\dots,n\} \end{cases} \tag{7.3}$$

Based on the all definition if we consider $n$ sensors in be the members of the multi-robot team then for the area of interest $\mathcal{A}$, we define $\mathcal{N}$ sets of sensors that could be calculated as follow:

$$\begin{cases} \mathcal{N} = n \ for \ k = 0 \\ \mathcal{N} = 2kn \ for \ k = \{1,2,\dots,n\} \end{cases} \tag{7.4}$$

Assume the area is a toxic or hazardous area, then the robots should be remotely distributed in the area (i.e., air dropped or launched via artillery). Therefore, the initial deployment of the sensors could be considered as two-dimensional poison distribution model [213]. $\lambda$, denotes the poison point process density. After initialization the robots start moving arbitrary in the area until all of them find each other and be connected. Then they form in a line orthogonal to the closed edges of the corridor $S$.

### 7.1.2 Sensor measurements and control

Let the linear velocity of the robots and the intruder be almost constant as in:

$$v_{min} = v_i - \varepsilon < v_i < v_{max} = v_i + \varepsilon \quad \varepsilon \simeq 0 \tag{7.5}$$

We define a control function based on a distance between $S_i$ & $S_j$ those represent sensors $i$ $and$ $j$ and $T_t$ that represents target and heading of sensors and the intruder as follow:

We introduce a control $U_i$ which is a function of headings the distances between any robot $i$ $and$ $j$ and the target $t$ as follows:

$$U_i = [D_{i,j,t}, \Theta_{i,j,t}] \tag{7.6}$$

Where $D_{i,j,t}$ denotes a set of all measured Euclidean distance of the robot to the intruder and $\Theta_{i,j,t}$ denotes a set of headings of each mobile robot and the intruder in the region $S$ such that:

$$D_{i,j,t} = \{d(S_i, S_j), d(S_i, T_t), d(S_j, T_t)\}$$

$$\Theta_{i,j,t} = \{\theta_i, \theta_j, \theta_t\} \tag{7.7}$$

## 7.2 Game-based decision-making strategy

We describe the rule for a $N$-player game, as in:

$$J_i = \prod_{i=1}^{N} \mathfrak{D}_i \to \mathbb{R}^1 \; for\; i = \{1,2,\ldots,n\} \tag{7.8}$$

In (7.8), $J_i$ represents the cost function and $\mathfrak{D}_i$denotes a set of available decisions for player $i$.

According to the strategy of the game, each player intends to minimize its cost function, which means the following condition should be achieved by each player:

$$J_i(d^*) \leq J_i(d) \quad for\ \forall\ d \in \prod_{i=1}^{N} \mathfrak{D}_i \tag{7.9}$$

where $d^*$ represents the desirable N-tuple decision.

## *Definition 7.1:*

A cooperative game is optimal if and only if, there is no any new joint decision which can decrease the cost function of one without increasing the cost function of the others. On the other word, a N-tuple decision $d^* \epsilon \prod_{i=1}^{N} \mathfrak{D}_i$ known as Pareto-Optimal if and only if:

$$\begin{cases} J_i(d^*) = J_i(d)\ for\ \forall i \epsilon \{1,2,\dots,n\} \\ or \\ J_i(d^*) < J_i(d)\ for\ at\ least\ one\ i \epsilon \{1,2,\dots,n\} \end{cases} \tag{7.10}$$

## *Definition 7.2:*

A non-cooperative game each player makes its own decision without cooperating with the other players in any case. Therefore, each player intends to minimize its cost function regardless of the consequences of the decision has been made to the other players. However, as the rational opponents make their decision in the same way, hence each player aims to minimize its cost function but not hurting the other players [214].

Therefore, a N-tuple decision $d^* \epsilon \prod_{i=1}^{N} \mathfrak{D}_i$is a Nash-Equilibrium if and only if:

$$J_i(d^*) \leq J_i(\acute{d}_i)\ \ \forall\ \acute{d}_i \epsilon \mathfrak{D}_i\,,\ \ \acute{d}_i = J_i\left(d^*{}_1, \dots, d^*{}_{i-1}, \acute{d}_i, d^*{}_{i+1}, \ \dots, d^*{}_N\right) \tag{7.11}$$

Back to the scenario, sensors $i$ and $j$, start playing a cooperative game when they detect each other. In this negotiation, they exchange the coordinates of the sub-segment they scanned before meeting each other in the region. This information prevents re-scanning any sub-segment during each scanning period that results in cost reduction for the team and choosing the best heading for the next step scanning by each robot. On the other hand, when the robots are not connected, each agent plays a non-cooperative game based on its local sensory information and by considering the information it has received by the other team mate from the most recent meeting.

At this stage, each robot gives the priority for scanning to the sub-segment with the high probability of intrusion.

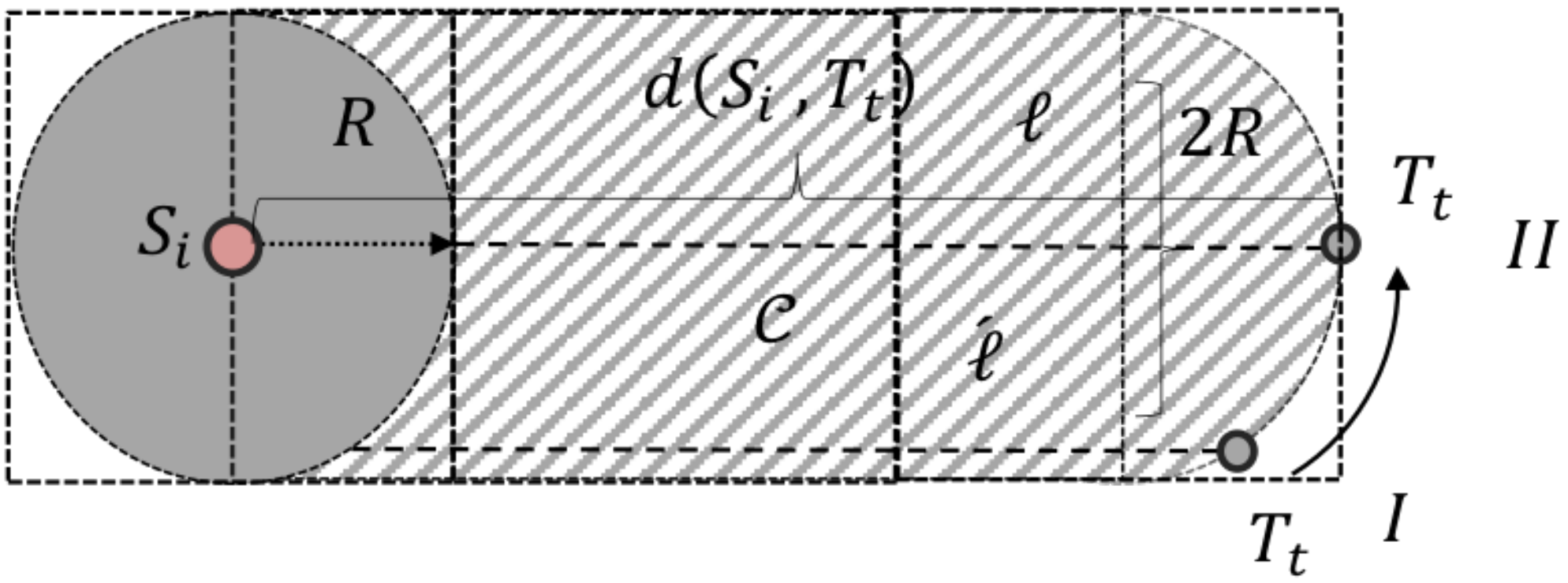

Figure 7.2: Probability of the first detection

Fig.7.2, shows a bounded region $\mathcal{C}$. Let target $T_t$ be static and it is located at point $I$ in the region $\mathcal{C}$. It is obvious that, $\acute{\ell}$ represents the shortest path the sensor $S_i$ must travel in a straight line to reach the target for the first time. Suppose $T_t$ moves through a curve with the radiuses $R$ from point $I$ to $II$. As it shows in the Fig.7.2, $\ell = \acute{\ell}$. Now, introduce line $\ell$ and the area $\mathcal{S}_{\mathcal{C}}$ as follows:

$$\ell = d(S_i, T_t) - R$$

$$\ell = \sqrt{(x_i - x_t)^2 + (y_i - y_t)^2} - R. \tag{7.12}$$

Hence, we can find the probability of the first intrusion as follow:

$$\mathcal{S}_{\mathcal{C}} = \ell \times 2R. \tag{7.13}$$

Then, the Poisson probability distribution of the intrusion for the area $\mathcal{C}$, could be find as in:

$$P(X) = \frac{(\lambda x)^n}{n!} e^{-\lambda x} \tag{7.14}$$

where $x = \mathcal{S}_{\mathcal{C}}$

Now we find the probability of no intrusion:

$$(X = 0) = e^{-\lambda S_C} \tag{7.15}$$

Therefore, if we take probability of no intrusion out of one, the reminder is the probability of intrusion as in:

$$P(X = 1) = 1 - P(X = 0)$$

$$P(X = 1) = 1 - e^{-\lambda S_C}$$

$$P(X = 1) = 1 - e^{-\lambda(\sqrt{(x_i - x_t)^2 + (y_i - y_t)^2} - R) \times 2R} \tag{7.16}$$

The intruder chooses the least dangerous point of the region boundary to enter the region $S$ based on its available information about the robots' current position with the minimum loss. On the other hand, the robots predict the most vulnerable points of the region boundary as they can calculate the probability of intrusion of each single point in the area using (7.16) which leads them to minimize their cost while maximizing the probability of detection either they play cooperative or non-cooperative game.

Fig.7.3, compares the probability of intrusion and the probability of detection in a sub-segment out of the segment $S$, based on (7.16).

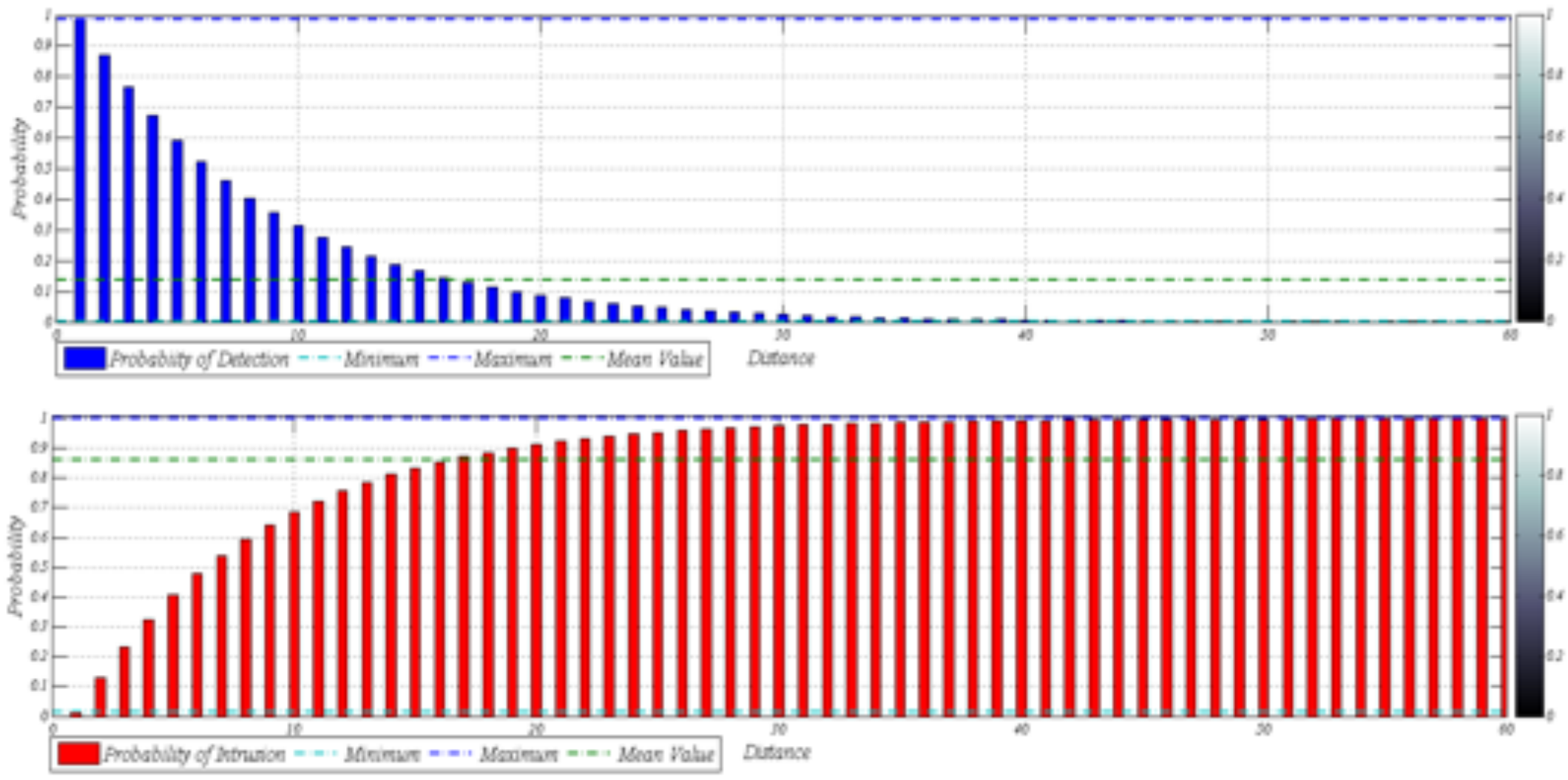

Figure 7.3: Probability of Intrusion and detection in a sub-segment

## 7.3 Game-based navigation model

We propose a new initial formation model for the multi-robot team and we prove that the proposed formation model guarantees the maximum payoff for the robots.

### 7.3.1 Perpendicular and diagonal formation model of the multi-robot team

After initial deployment, the robots wandering in the region to find each other. While all the team members are connected they form in a line perpendicular to the boundaries $\partial\mathcal{L}_1, \partial\mathcal{L}_2$ of the corridor $\mathcal{A}$ (see Fig.7.4).

If $\mathcal{S}(R_j) \cap \mathcal{S}(R_i) \neq \emptyset$, then sensors update their heading and position based on the following updating rules:

$$\theta_j = tg^{-1}\left(\frac{y_j - y_i}{x_j - x_i}\right) \mid d(s_i, \mathcal{L}_1) \leq R$$

$$\hat{x}_i = x_i + v_i t \, cos(\theta_i + \omega_i t)$$

$$\hat{y}_i = y_i + v_i t \, cos(\theta_i + \omega_i t)$$

$$\hat{x}_j = x_j + v_j t \, cos(\theta_j + \omega_j t)$$

$$\hat{y}_j = y_j + v_j t \, cos(\theta_j + \omega_j t)$$

$$\{\hat{x}_i \,\&\, \hat{y}_i \mid \mathcal{S}(R_i) \cap \mathcal{L}_1 = \emptyset\} \text{ and } \; x_j - x_i \geq \varepsilon \qquad (7.17)$$

The next step is to design a formation plan that minimizes detection time and maximizes the probability of detection in every single point of the corridor. Fig.7.4, shows two formation models of the mobile robots in the region. Point $P$, is the point with the minimum risk of detection, where the intruder tends to enter the region.

Therefore, robots have to form in a way that minimizes their distance with any vulnerable point in the region to maximize the probability of detection.

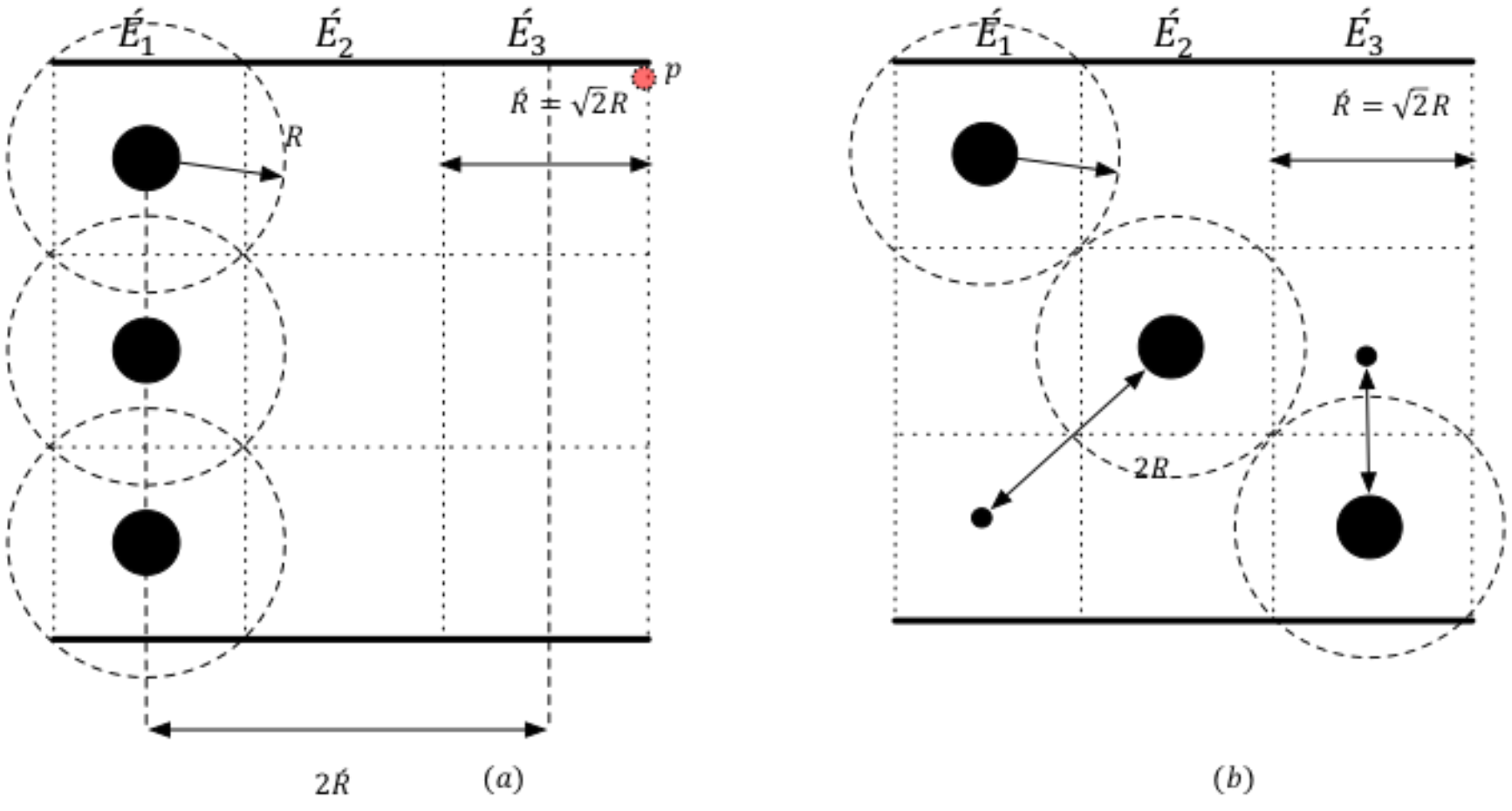

Figure 7.4: Initial formation for the maximum payoff

*Proposition 7.1:*

The optimal coverage in the corridor $\mathcal{A}$ is obtained when all sensors $\mathcal{S}_{i,j}$ are deployed in a diagonal form between $\partial\mathcal{L}_1 and \partial\mathcal{L}_1$.

*Proof:*

According to (7.13), we know that the probability of first detection varies with the distance between the robots and the intruder. Furthermore, an intruder known as detected while it lies within the sensing range of a mobile robot. And finally, the intruder has been received enough information about the current location of the sensors in the corridor $\mathcal{A}$. Hence, a point with the maximum probability of detection is the point with the minimum probability of intrusion such that:

$$P(\acute{X} = 1) \equiv P(X = 0) \tag{7.18}$$

where $P(\acute{X} = 1)$ shows the probability of first detection and could be calculated as follows:

$$P(\acute{X} = 1) = e^{-\lambda(\sqrt{X^2+Y^2}-R)\times 2R} \quad | \, d(S_i, T_t) \geq R \tag{7.19}$$

Then we calculate the probability of first detection when the robots deployed in a perpendicular line to $\partial\mathcal{L}_1 and\ \partial\mathcal{L}_1$ based on (7.19) as follows:

$$P(\acute{X} = 1)_{ort} = e^{-(4\sqrt{2}\lambda R^2)} \tag{7.20}$$

Now we calculate the probability of first detection when the sensors deployed in a diagonal form between $\partial\mathcal{L}_1 and\ \partial\mathcal{L}_1$ as in:

$$P(\acute{X} = 1)_{dig} = e^{-(4\lambda R^2)} \tag{7.21}$$

Hence, comparing (7.20) and (7.21) shows that the probability of first detection when the robots are deployed in diagonal form is 5 times more than when they are deployed in orthogonal form:

$$P(\acute{X} = 1)_{dig} \approx 5P(\acute{X} = 1)_{ort} \tag{7.22}$$

Furthermore, there is an overlap of the sensing area of the neighbors such that $S(R_j) \cap S(R_i) = R^2(\pi\sqrt{2} - 4)$. However, in diagonal formation, the area of intersection of sensors $i\ and\ j$ is minuscule ($\mathcal{S}(R_j) \cap \mathcal{S}(R_i) = \varepsilon$).

### 7.3.2 Game based decision-making algorithm

In an n-player game, each agent $i$ has a strategy set $S_i$ with the elements $(s_1, s_2, \ldots, s_i, \ldots s_n)$ and a payoff function $u_i: S_1 \times S_2 \times \ldots \times S_n \to \mathbb{R}$ [215].

Furthermore, each agent, has its own action profile $a_i$ based on the decision it makes for the next step motion. In this scenario, each agent $i$ has to choose the best action profile out of all available options which leads to maximising the probability of detection. As it shows in the Fig.7.5., each agent has to pick a Nash equilibrium strategy from the strategy set $\mathfrak{D}_i$. For the sake of simplicity, we suppose each robot $S_i$ moves with heading $\theta_i = \frac{k\pi}{4}$ for $k = 0,1,2, \ldots$. If $\acute{E}^i_{nm}(k) < \mathcal{S}_i$ at time $k + 1$.

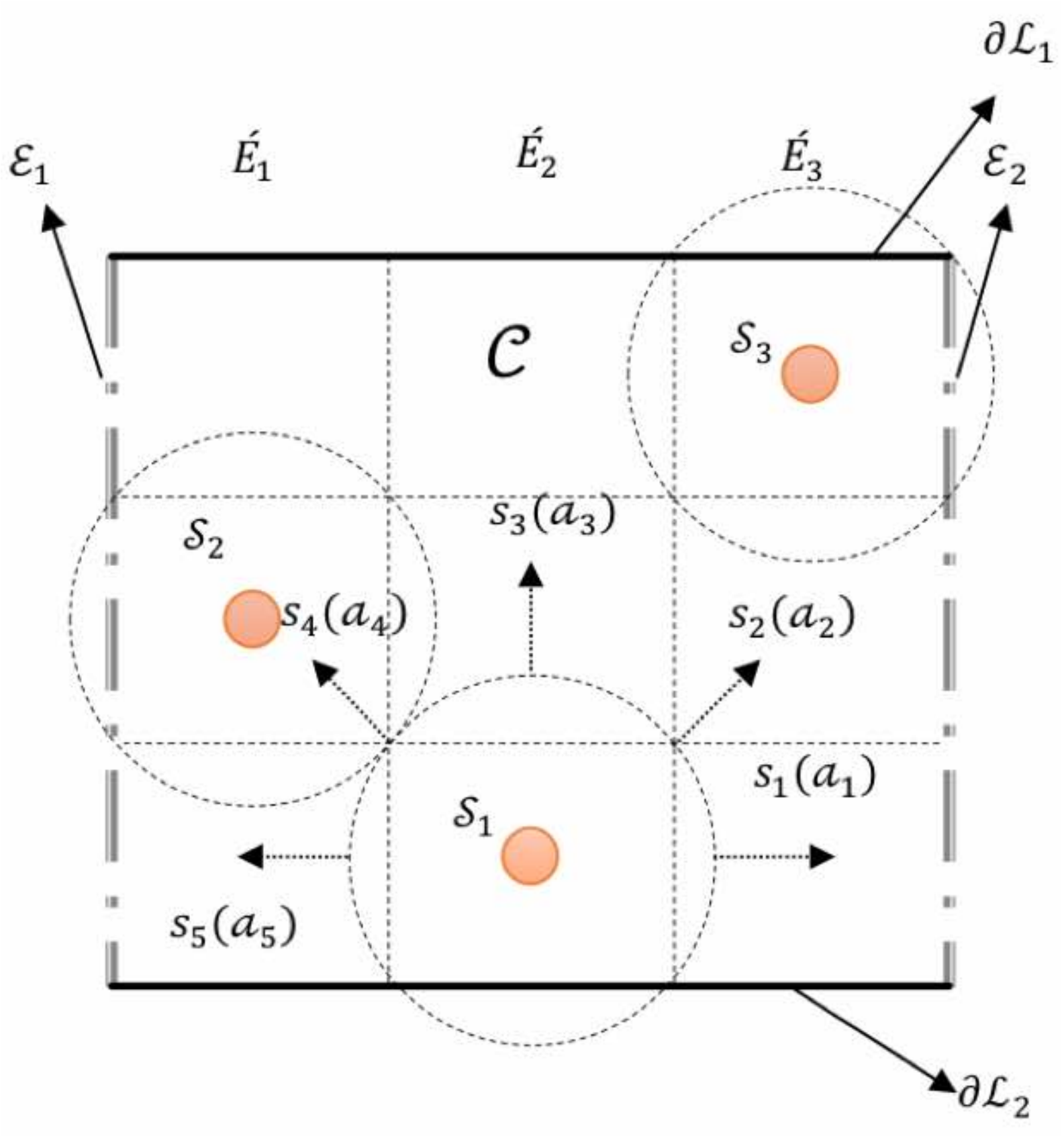

Figure 7.5: The strategy and the action profile of the robots

Where $\acute{E}^{i}_{nm}(k)$ denotes the next cell in the neighborhood of $\mathcal{S}_i$ at time $k+1$ for $n = 1,2,3$ *and* $m = 1,2,3$. Then each robot $\mathcal{S}_i$ checks the condition set $\mathcal{M}_i = (\mu_1, \mu_2, \mu_3, \mu_4)$ for every strategy profile $s_i$ belongs to each action profile $a_i$ for its next step moving. We introduce a set of weights $\mathcal{W}_\text{i} = (w_1, w_2, w_3, w_4)$ which is used in the algorithm for prioritizing the action of each robot, where, $w_1 = (0,1)$ and $w_2 = w_3 \ll w_4$.

## *Algorithm 7.1:*

1. *If* $\acute{E}^{i}_{nm}(k+1) < \mathcal{S}_i \wedge \acute{E}^{j}_{nm}(k+1) < \mathcal{S}_j\ , i \neq j$

   *where $k$ denotes the sojourn time,*

   $$whether\ \acute{E}^{i}_{nm}(k+1) = \acute{E}^{j}_{nm}(k+1) \vee \acute{E}^{i}_{nm}(k+1) \neq \acute{E}^{j}_{nm}(k+1)$$

2. *Each sensor needs to check whether:*

$$É^{i}_{nm}(k) \cap \mathcal{S}_{i,j} = \phi \vee É^{i}_{nm}(k) \cap \mathcal{S}_{i,j} \neq \phi for \ \forall x \in É^{i}_{nm}(k), i \neq j$$

3. *For any action profile* $a_i$ *whether:*

$$P_{\mathcal{E}}(X = 1) > P_{\mathcal{E}}(X = 0) \vee P_{\mathcal{E}}(X = 1) < P_{\mathcal{E}}(X = 0)$$

4. *For sensors* $\mathcal{S}_i$ , $\mathcal{S}_j$ *and* $\mathcal{S}_k$, $(i \neq j \neq k)$,

*if* $(R_i \cap R_k) \wedge (R_j \cap R_k) = \phi$ *and* $(R_i \cap R_j) \neq \phi$

*Both* $\mathcal{S}_i$ , $\mathcal{S}_j$*should predict the best motion strategy* $s_k{}^*$ *of* $\mathcal{S}_k$ *based on their last step information and conditions a, b, and c.*

Then we calculate the *Payoff* for each strategy profiles as follows:

$$\mathfrak{D}_i = (s^i_1, s^i_2, \dots, s^i_n)$$

$$u_i(s_i) = \mu_1 w_1 \sum_{j=2}^{4} \mu_j w_j$$

$$U_i = (\mathrm{u_{i1}}, \mathrm{u_{i2}}, \dots, \mathrm{u_{in}}) \quad (7.23)$$

Finally, the best strategy results from $max(U_i)$ meets the following condition:

$$s^*_i \mid \mathrm{u_i}(s^*_1, \dots, s^*_{i-1}, s^*_i, s^*_{i+1}, \dots, s^*_n) \geq \mathrm{u_i}(s^*_1, \dots, s^*_{i-1}, s, s^*_{i+1}, \dots, s^*_n) \quad (4.24)$$

For all $s \in \mathfrak{D}_i$.

Based on the proposed strategy, the game continues until there is at least one pursuer remains in the region. For example, if sensor $\mathcal{S}_j$will not back to the initial deployment location, the remained sensor(s) $\mathcal{S}_i$ make the new decision based on the new condition which is the absence of one or some of the sensors in the region based on the following algorithm.

*Algorithm 7.2:*

1. *While* $\mathcal{S}(R_j) \cap \mathcal{S}(R_i) = \emptyset$
2. *move towards the centre of the area*
3. *If* $\mathcal{S}(R_j) \cap \mathcal{S}(R_i) = \varepsilon$
4. *check condition set* $\mathcal{M}_i$
5. *creating* $\mathfrak{D}_i$
6. *Finding* $s_i^*$
7. *else if* $\mathcal{S}(R_j) \cap \mathcal{S}(R_i) = \emptyset$ *and* $\acute{E}_{22} \lessdot \mathcal{S}(R_i)$
8. *do (4), (5), (6)*
9. *end*
10. *end*

## 7.4 Simulations

Fig.7.6, shows the initial deployment of the sensors after they have distributed randomly in the region. They are wandering the area until all the entire team would be connected. Then they update their heading and the position based on (7.17). Fig.7.7 (a , b), shows a comparison between the standard swiping method with the proposed *IGD* algorithm

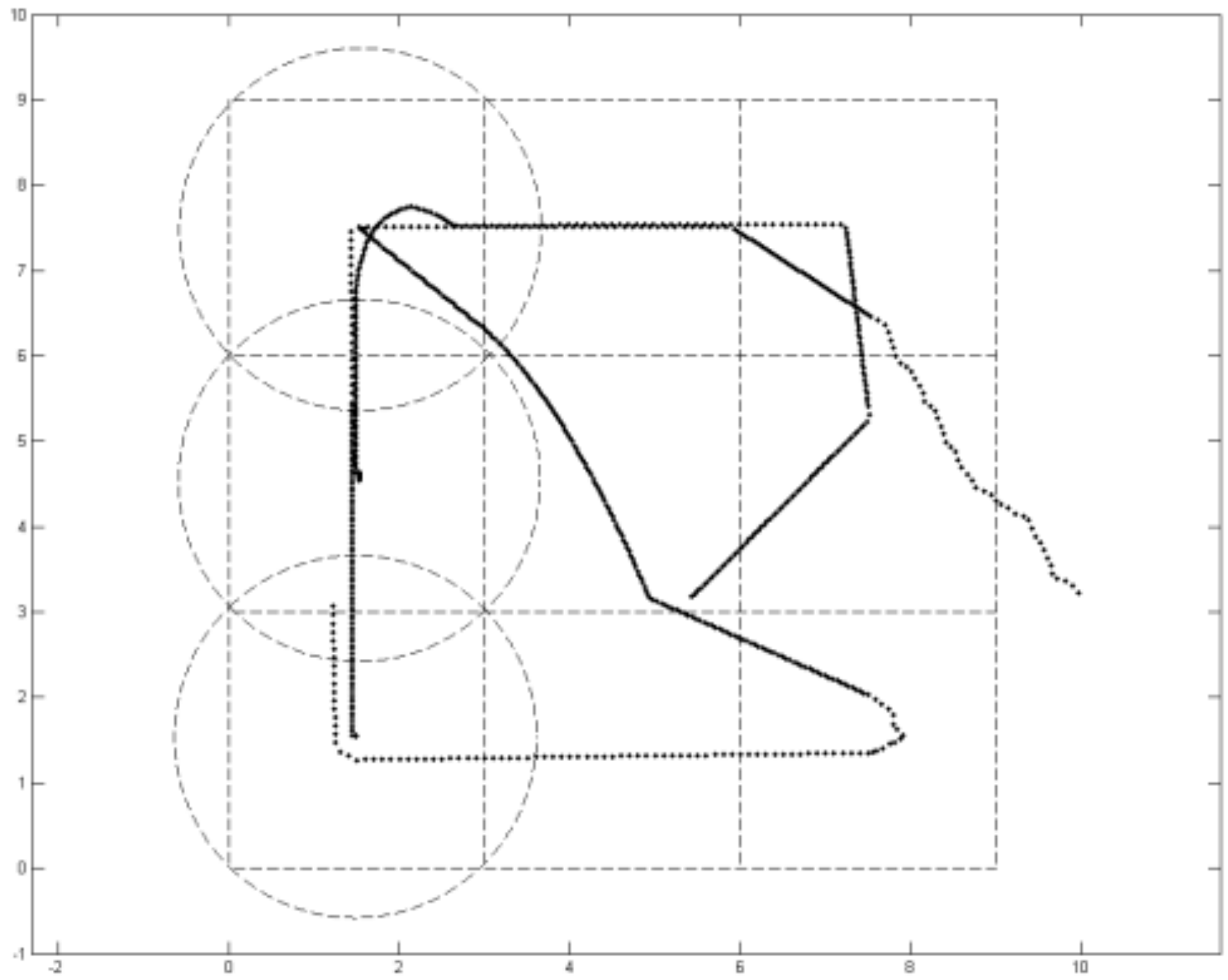

Figure 7.6: Initial deployment of robots in a straight line

It takes $4\mathcal{T}_{\text{steps}}$ the sensors scan all the sub segments of the region and back to their initial position when the multi-robot team uses the standard sweeping coverage method. However, using (IGD) result in one-step reduction of the searching process, that means $3\mathcal{T}_{\text{steps}}$ is required that all the sub-segments to be scanned and the robots back to their initial position. In this simulation $\mathcal{T}_{\text{steps}}$ indicates the period between each motion step. Therefore, energy and time saving could be considered as the first advantages of the proposed method.

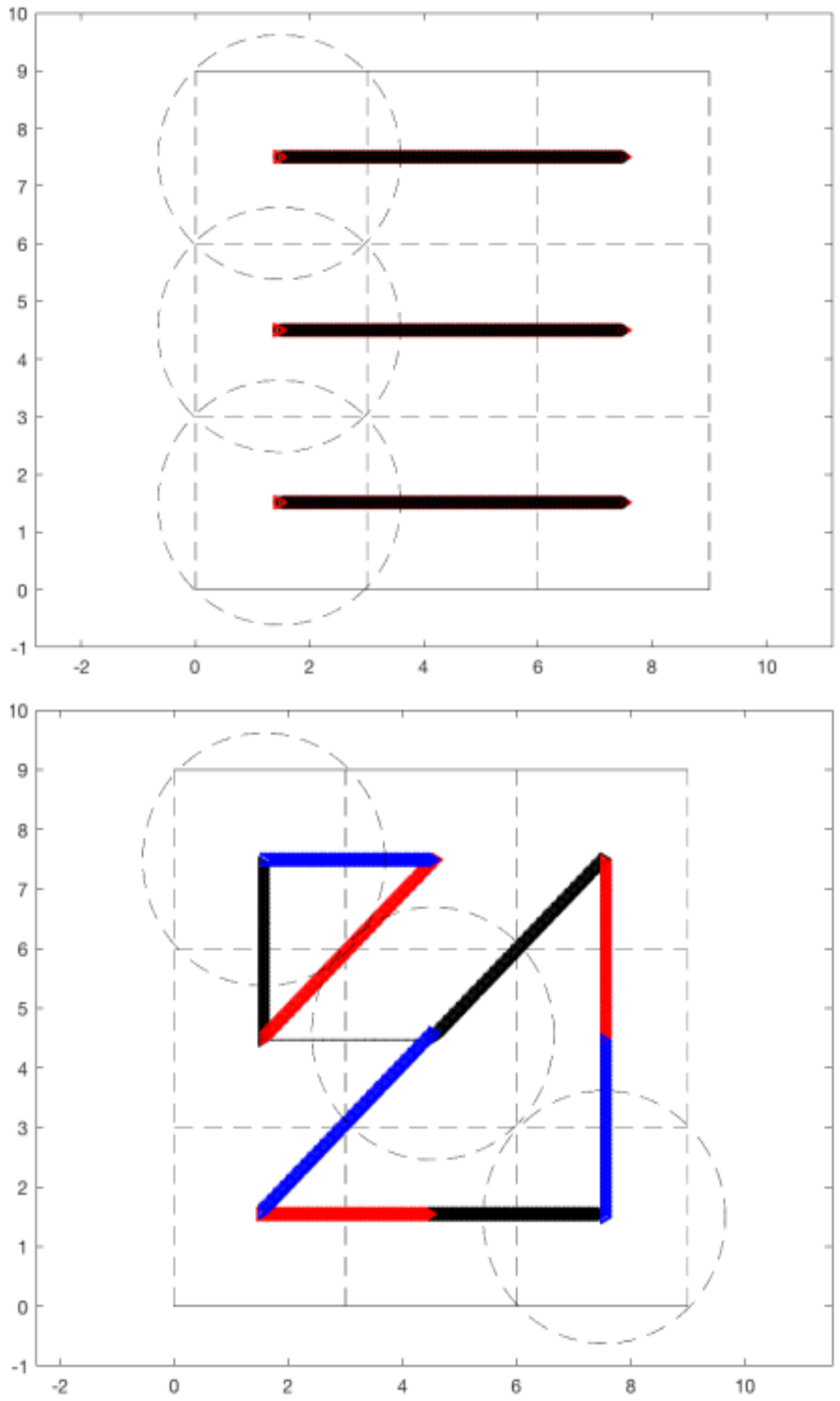

Fig. 7.7 (a,b): $\mathcal{T}_{steps}$ for at least one time scanning each sub-segment in the area of interest

Furthermore, based on the proposed method, in the case of absence of any of the robot $i\ or\ j(s)$, the remained robots adapt with the new circumstances by changing their strategy. This approach, guaranties that, the area always being monitor even if one robot remain in the region. Furthermore, it prevents robots to be confused in such critical circumstances. Fig.7.8 (a,b,c) gives an illustrative example of the way the robots act in a situation of the absence of one of the team member and Fig.7.9, shows the action of one robot in case of the absence of the other team members. In both case scenarios, the remained robot(s) change their policy based on the Algorithm 7.2, which guaranties the area is under surveillance always.

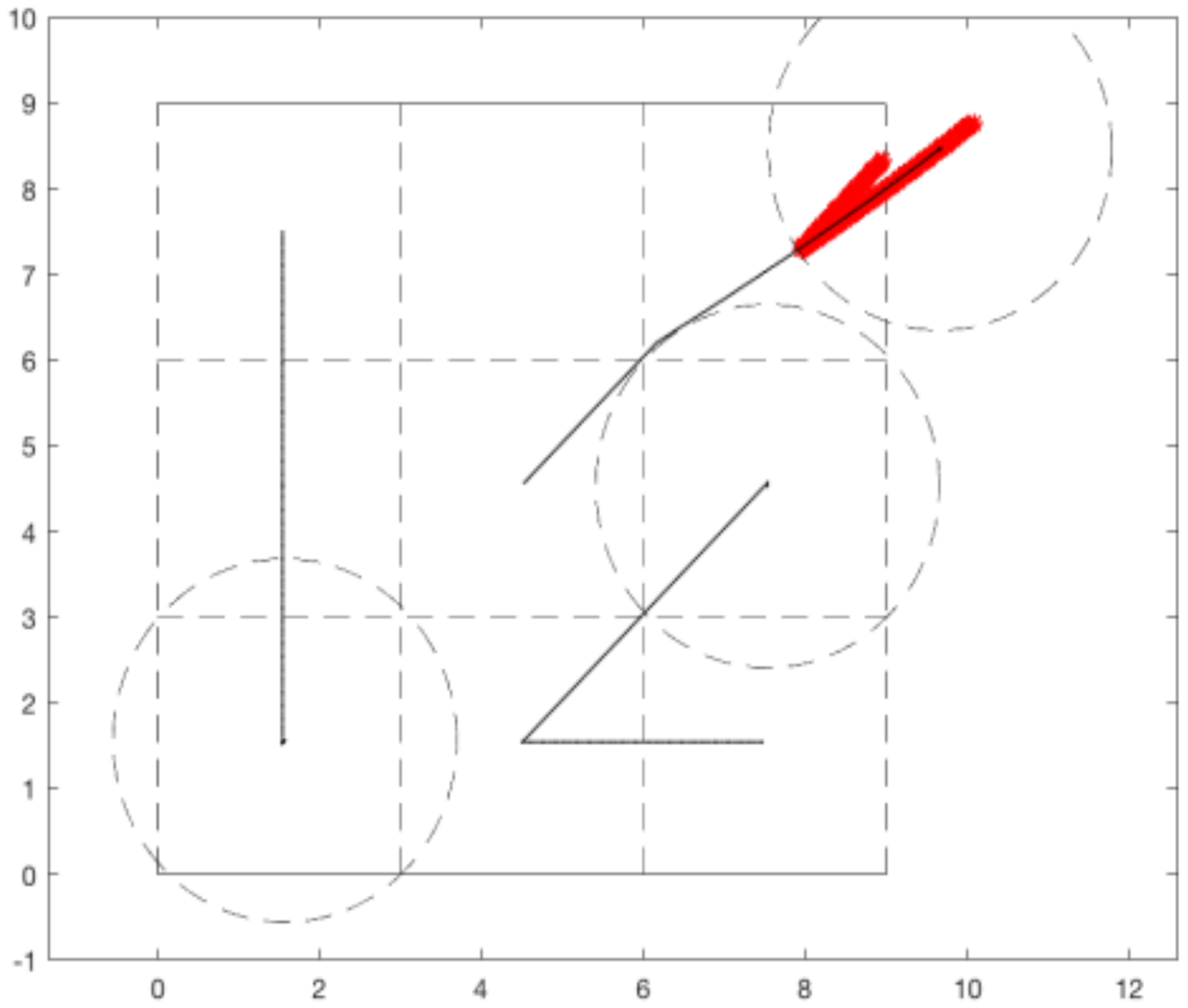

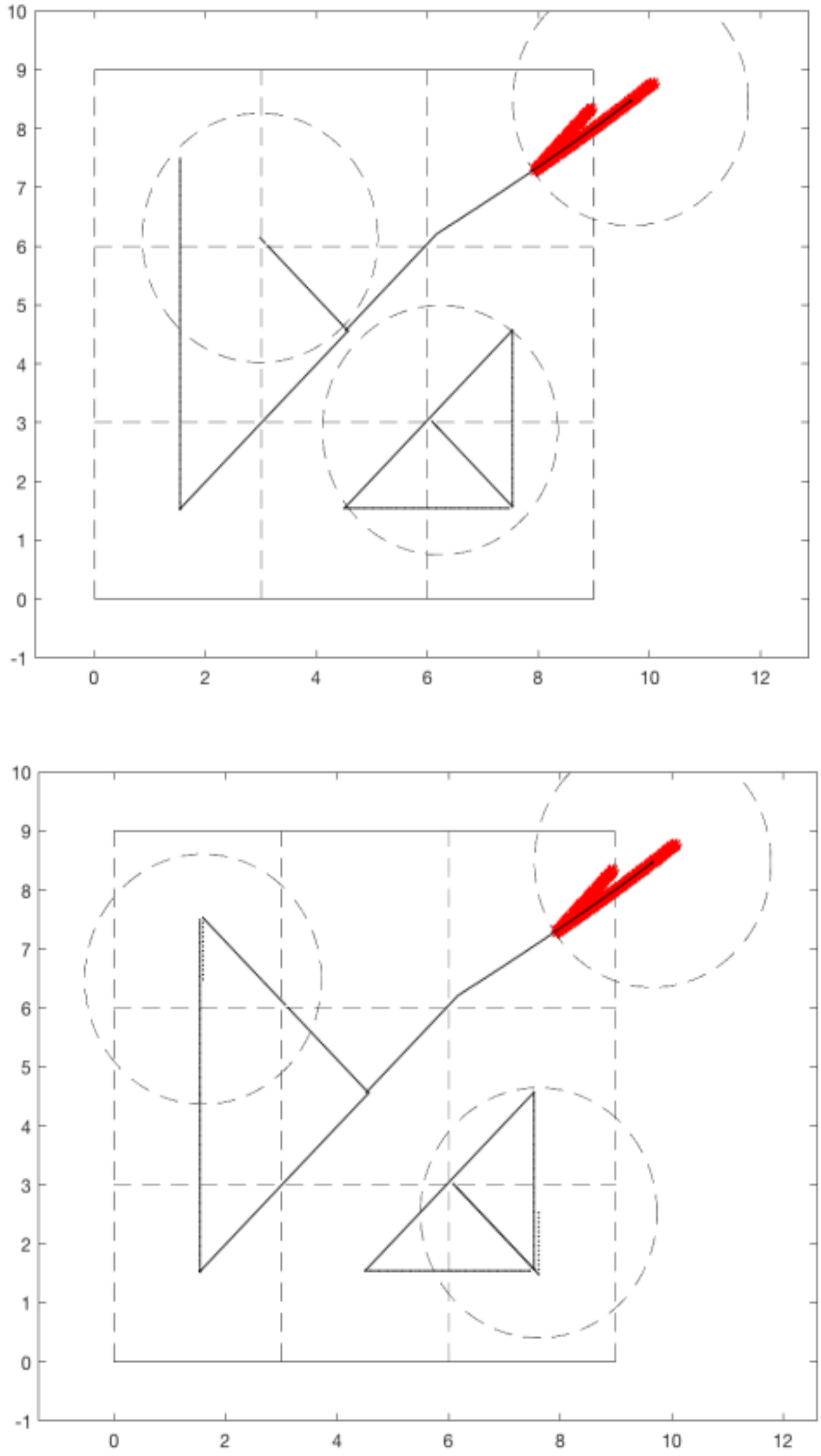

Figure 7.8 (a, b, c): New decision making based on two sensors revert to the initial position

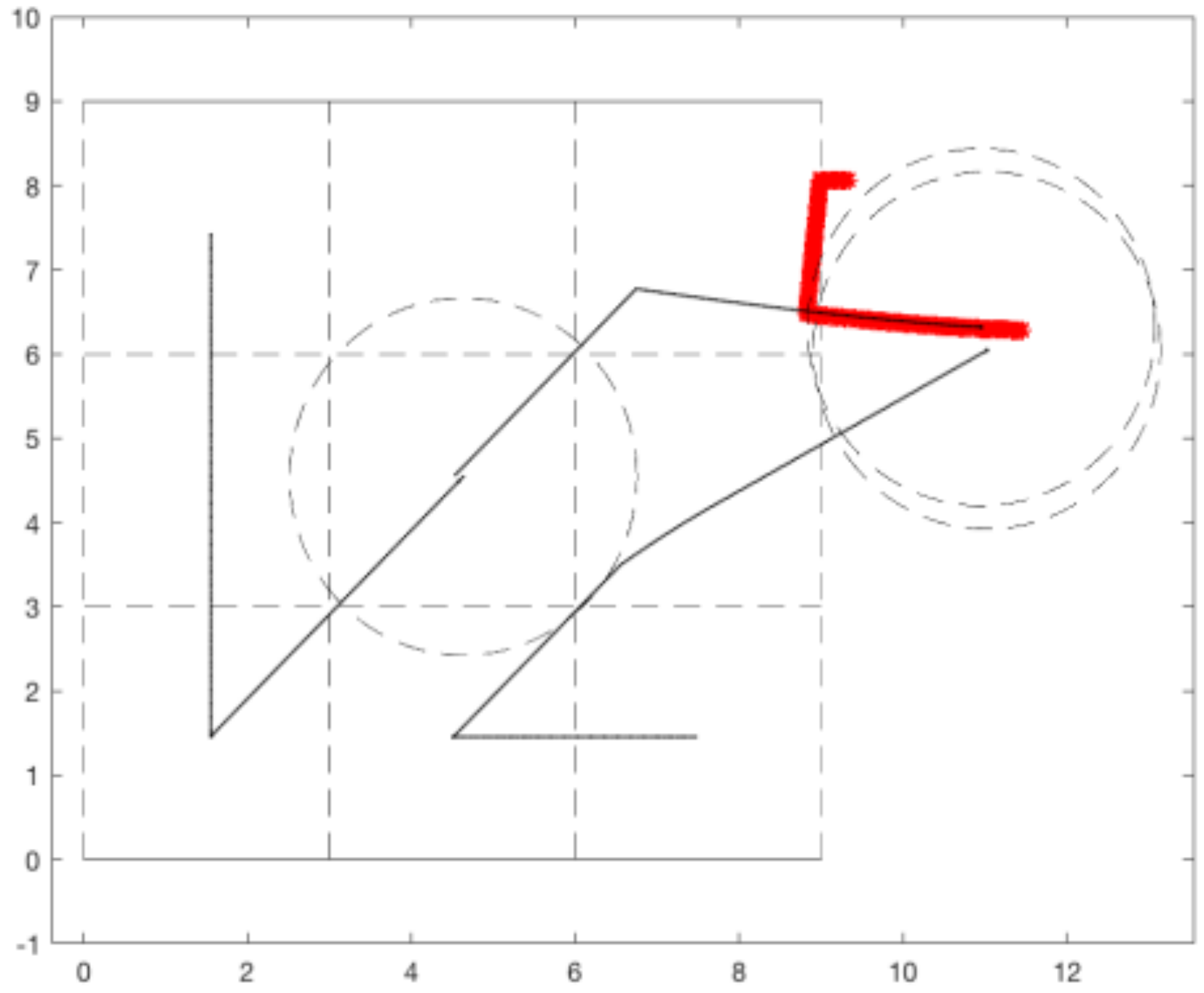

Figure 7.9: New decision based on one sensor revert to the initial position

Table 7.1 shows the achievement of the multi-robot team while using IGD We compared the achievement of IGD comparing to the standard swiping method.

Table 7.1: Comparison between IGD and standard swiping method

| Monitoring Method | IGD | Swiping |
|---|---|---|
| No of Repeats | 100 | 100 |
| Detection Success | 96% | 17% |
| Detection Failure | 4% | 83% |

According to the results, the intruder successfully escaped, 83 times out of 100 repeats, while using the standard sweeping method. However, the robots detected the intruder 96 times out of 100 repeats successfully while using IGD method. So, with the same scenario, the probability of detection while using sweeping coverage is just 17% however, using proposed IGD improved the performance of the robots by 79%.

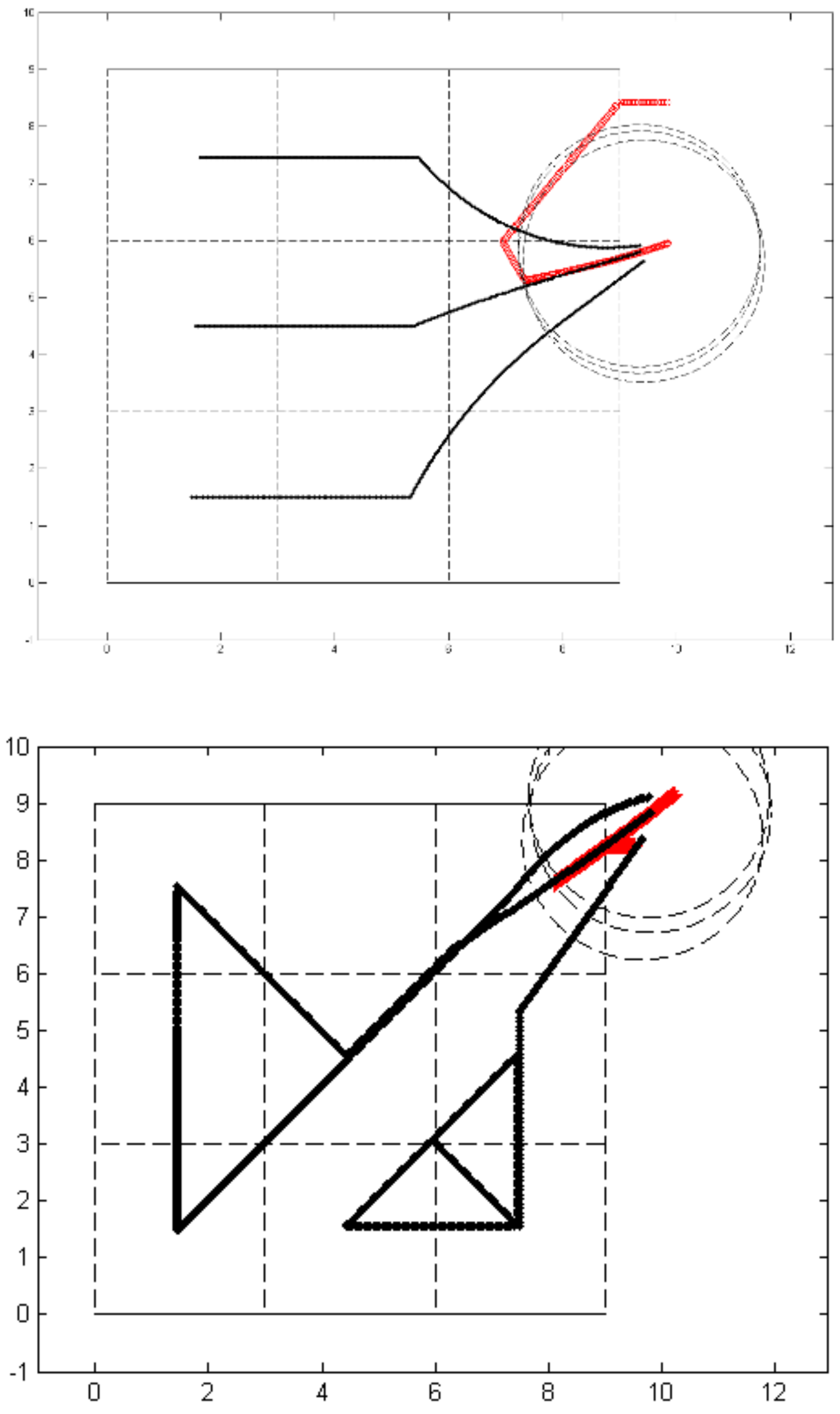

Figure 7.10: Target Trapping and Tracking Comparison between Diagonal and Orthogonal Formation

Moreover, in our proposed method, in the case of the absence of any sensor with any reason, whether damage or intruder tracking, the other sensors are smart enough to adapt themselves to the new circumstances and make the best decision for maximizing the coverage.

## 7.5 Summary

In this chapter, we proposed decentralized intelligent Game-based navigation and decision-making strategy called IGD for the propose of maximizing the probability of intruder detection in a bounded area. Furthermore, the new initial formation method improved the coverage problem of the area of interest. We proved that, using this method by a multi robot team guarantees the non-stop surveillance of the region even if there is just one mobile robot remained in the area. Furthermore, as the robots are able to make a decision autonomously, based on the local sensory information and the most recent information they have been received by the other team mates, therefore, any communication failure result from sabotaging by a given hostile, such as jamming attacks don't affect on the performance of the mobile robots during the mission. The mathematically rigorous proof of the proposed method, in addition to the simulations, confirm the validity of the model.

# Chapter 8
# Semi-decentralised switching navigation method in an obstacle-ridden environment

This chapter presents a semi-decentralized navigation strategy named *Position Estimation Switching Algorithm (PESA)*, for a fleet of mobile robot based on the leader-follower concept. In this method the robots don't have any priori global information about the environment, which means the area is unknown to the robots. The leader is the only one, that collects the local information of the region and plans the safest path in each switching step. Then shares the new heading and coordinates with its nearest neighbor. The other team members hand the updating information down to the nearest neighbors in a communication chain.

In the proposed method, the path planning by the leader repeated in just a few steps. Therefore, continues monitoring, measurement and computation of the safest path is not required. The rest of the chapter is organized as follows; In section 8.1, we present the problem statement. The mathematical analysis and model are described in section 8.2 and in section 8.3, we present the simulations and the results.

## 8.1 Problem statement

We assume each sensor can detect the obstacle within its sensing range with radius $R$. The kinematic model of the mobile sensors is the same as the model that has described in (8.2). In this scenario, the team members should follow the leader. It doesn't mean that every individual should be connected to the leader. However, a solid connection between each neighbor is required to create a chain-link communication channel. The main objective of this study is to develop a navigation algorithm for a safe maneuvering of the multi-robot team without collision with a static convex obstacle.

## 8.2 Position estimation switching method

We consider robot $x_{r1}$as the leader of the team which is initially located at a given point $p_r$with a distance of $\varepsilon > 0$ perpendicular to the obstacle in the region $R$. Furthermore, let the $\varepsilon > 0$ be the minimum allowed distance between the robots to the obstacle. Moreover, the following condition holds for any neighboring points $p_{rj}$ to the point $p_{ri}$ as in:

$$\parallel p_{ri} - p_{oi} \parallel \leq \parallel p_{rj} - p_{oj} \parallel \tag{8.1}$$

$\forall\, i, j \quad where \quad i \,\epsilon\, (1, .., n) \,\&\, j \,\epsilon\, (i - n, i - n + 1, \dots, i - 1, i, i + 1, \dots, i + n - 1, i + n)$

where $i$ and $j$ represent permutations of the point $p_r$.

Then we introduce $S_i(x, y)$ $(for \ i = 1, \dots, n)$, as a sub-segment of the entire perimeter $S \epsilon O$ which is scanning by the robot $x_{r1}$ in each switching period $\mathcal{T}$.

Moreover, introduce, $h_{1i}$ and $h_{2i}$ which represent the tangent lines between the sensing range of the robot and the left-hand side and the right-hand side of the obstacle $O$ respectively.

Furthermore, points $p_1$and $p_2$ are supposed to be the maximum visible point of the obstacle at time $\acute{t} \epsilon \mathcal{T}$ by the robot $x_{r1}$such that:

$$\begin{cases} \ell_{1i} = sup(\| p_{ri} - p_{oj} \|) \ \forall\, j \in (i, i+1, \dots, i+n-1, i+n) \\ \qquad\qquad and \\ \ell_{2i} = sup(\| p_{ri} - p_{oj} \|) \ \forall\, j \in (i-n, i-n+1, \dots, i-1, i) \end{cases} \tag{8.2}$$

then we define $\mathcal{L}_{1i}$ as follows:

$$\mathcal{L}_{1i} = inf\ (\| p_1 - p_2 \|) \tag{8.3}$$

We define a set of vertical lines $(\mathcal{L}_{2i}, \mathcal{L}_{3i})$ such that:

$$\begin{cases} \overline{\mathcal{L}_{2i}} = \varepsilon \ \ and \ \mathcal{L}_{2i} \perp \mathcal{L}_{1i}\ at\ \ p_3 \epsilon \mathcal{L}_{1i} \\ d(p_1, p_3) = d(p_2, p_3) = \mathcal{L}_{1i}/2 \\ \mathcal{L}_{3i} \perp \mathcal{L}_{1i}\ at\ p_2 \epsilon \mathcal{L}_{1i} \end{cases} \tag{8.4}$$

Finally, we introduce another line $\mathcal{L}_{4i}$ which represents the intersecting lines $\mathcal{L}_{1i}, \mathcal{L}_{2i}, and\ \mathcal{L}_{3i}$ at the points $p_1, p_4, and\ p_5$ respectively.

Now we consider the right-angle triangle $\Delta p_1 p_2 p_5$ to find the value of the angle $\theta_{di}$ as in:

$$\theta_{di} = tan^{-1} \frac{\mathcal{L}_{3i}}{\mathcal{L}_{1i}} \tag{8.5}$$

Fig.8.1, shows all required measurement including all the distances and angles that robot the leader measures to find the right angle for the next collision free movement while maintaining the minimum allowable distance with the obstacle.

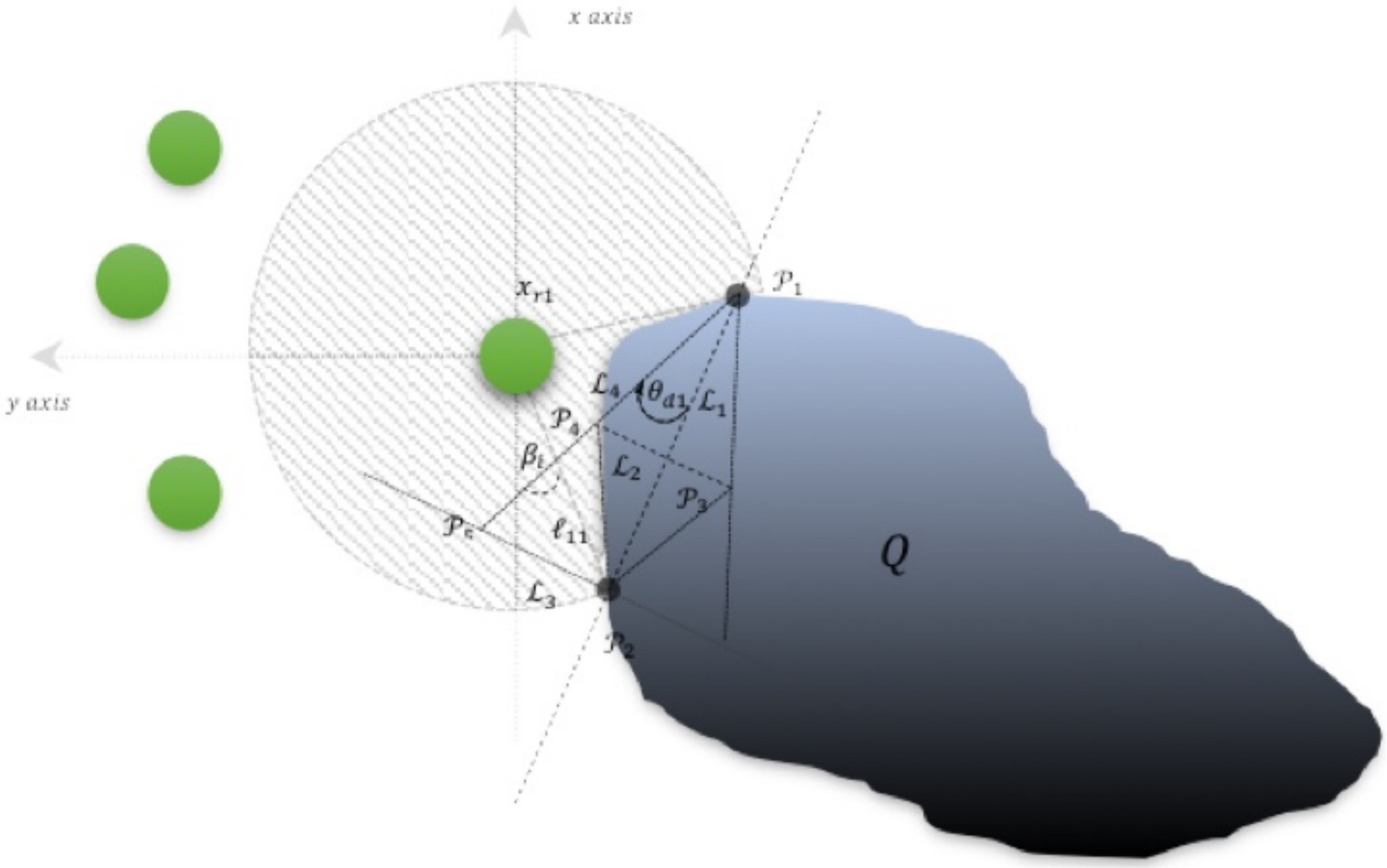

Figure 8.1: The method of measuring required angles and distances by robot the leader

## *Proposition 8.1:*

Supposed that the leader $x_{r1}$ moves a distance from position $\mathcal{O}_1$ to position $\mathcal{O}_2$ . We prove that if the leader moves as long as $\mathcal{L}_{1i}/_2$ in the direction of $\beta_i = \theta_{di}$ based on the line $\ell_{1i}$ $(i = 1,2,3 \ldots)$ which indicates a straight line, lies between the coordinates of the current position of the robot $x_{r1}$ and its intersection point with the obstacle, then the following condition is being satisfied:

$$\varepsilon - \mu_0 \leq inf(\ell_{ri}) \leq \varepsilon + \mu_0 \qquad (8.6)$$

In (8.6), $\ell_{\mathrm{ri}}$ represents the distance between robot and obstacle $\forall o_i \in [\mathcal{O}_i, \mathcal{O}_{i+1}]$ and $\mu_0$ denotes the tolerance of the maximum allowed distance from the robots to the obstacle which results from the measurement error.

## *Proof:*

We find the length of the line $\mathrm{d_j}$ as in (see Fig.8.2):

$$d_j = \sqrt{\left(\mathcal{L}_{1j}/_2\right)^2 - \left(\ell_{rj}\right)^2} \quad (8.7)$$

First, we suppose that the following inequality hold:

$$d_j \leq \ell_{1i} \quad (8.8)$$

Therefore, based on both (8.1) and (8.8), it's obvious that the robot will be settled at $\mathcal{O}_2$ while it satisfies (8.6).

But in the case that (8.8) does not hold, then:

$$d_j > \ell_{1i} \quad (8.9)$$

For sure there exist some points $j' < j$, where:

$$inf(\ell_{ri}) < \varepsilon - \mu_0 \;\; for\ i = j' \quad (8.10)$$

Now, we have to prove that inequality (8.9) is not true.

Fig.8.2, shows the worst-case scenario in which robot faces with an obstacle that its outer face is flat e.g. a flat wall, then obviously, $d_j$ is maximum ($d_j = \ell_{1i}$).

This results that:

$$max(d_j) = max(\ell_{1i}) = R \xrightarrow{yields} inf(\ell_{ri})\varepsilon - \mu_0 \qquad for\ i = j' \quad (8.11)$$

In consequence of (8.11), the length of $d_j$ never exceeds $\ell_{1i}$, therefore, inequality (8.9) is false.

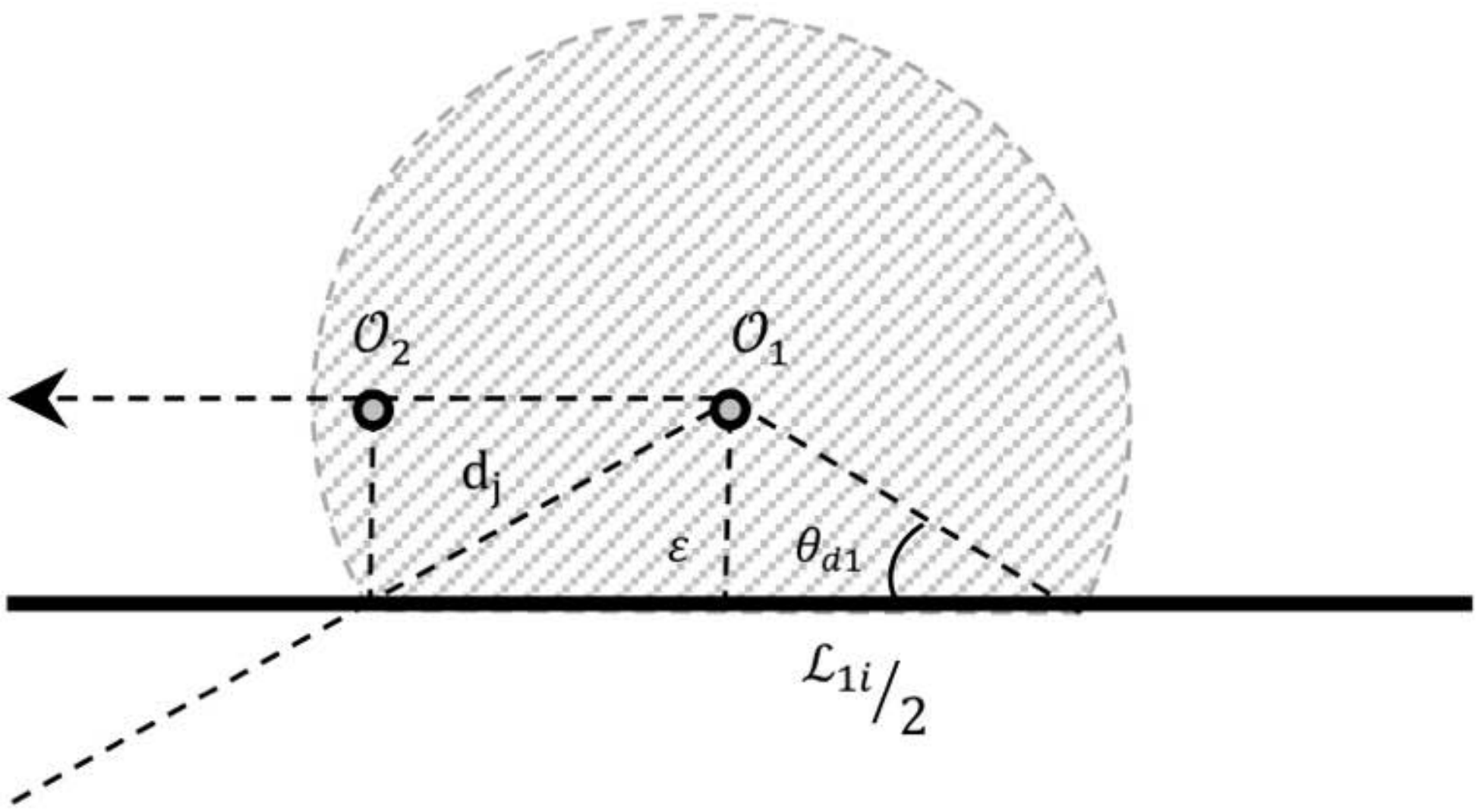

Figure 8.2: The situation where $d_j$ is maximum

Therefore, according to the Proposition 8.1 the path which is planned by the leader $x_{r1}$based on the navigation law (8.4), (8.5), is the safest path that maintain minimum allowed distance of the robots to the obstacle. Now, it's the turn of every individual teammate to update its own heading and distance with the robot ahead to avoid any neither collision to the neighbours nor to the obstacle.

The heading updating rules is defined as follows:

For $k = 1,2,3 \dots, n$ and $i = 1,2,3, \dots$

$$\begin{cases} \alpha_{(k,i)} = tan^{-1}\left(\frac{y_{(k-1)i} - y_{ki}}{x_{(k-1)i} - x_{ki}}\right) for\ k > 1 \\ \alpha_{(k,i)} = \beta_i\ for\ k = 1 \end{cases} \tag{8.12}$$

Furthermore, to avoid any collision with the other team members, each robot is not allowed to travel more than a maximum specified distance in the region as follows:

$$\begin{cases} d_{ki}\left(x_{kr}, x_{(k+1)r}\right) = \| \mathcal{O}_{(k-1)i} - \mathcal{O}_{ki} \|\ for\ k = 2, \dots, n \\ d_{ki}(x_{kr}) = {}^{\mathcal{L}_{1i}}/_{2}\ for\ k = 1 \end{cases} \tag{8.13}$$

Where $k = 1$ indicates leader of the team, and index $i$ denotes the permutation of each robot at the end of each switching step.

Fig.8.3, shows in detail, the way the leader estimates the next position and plans the paths towards the next position in each switching step.

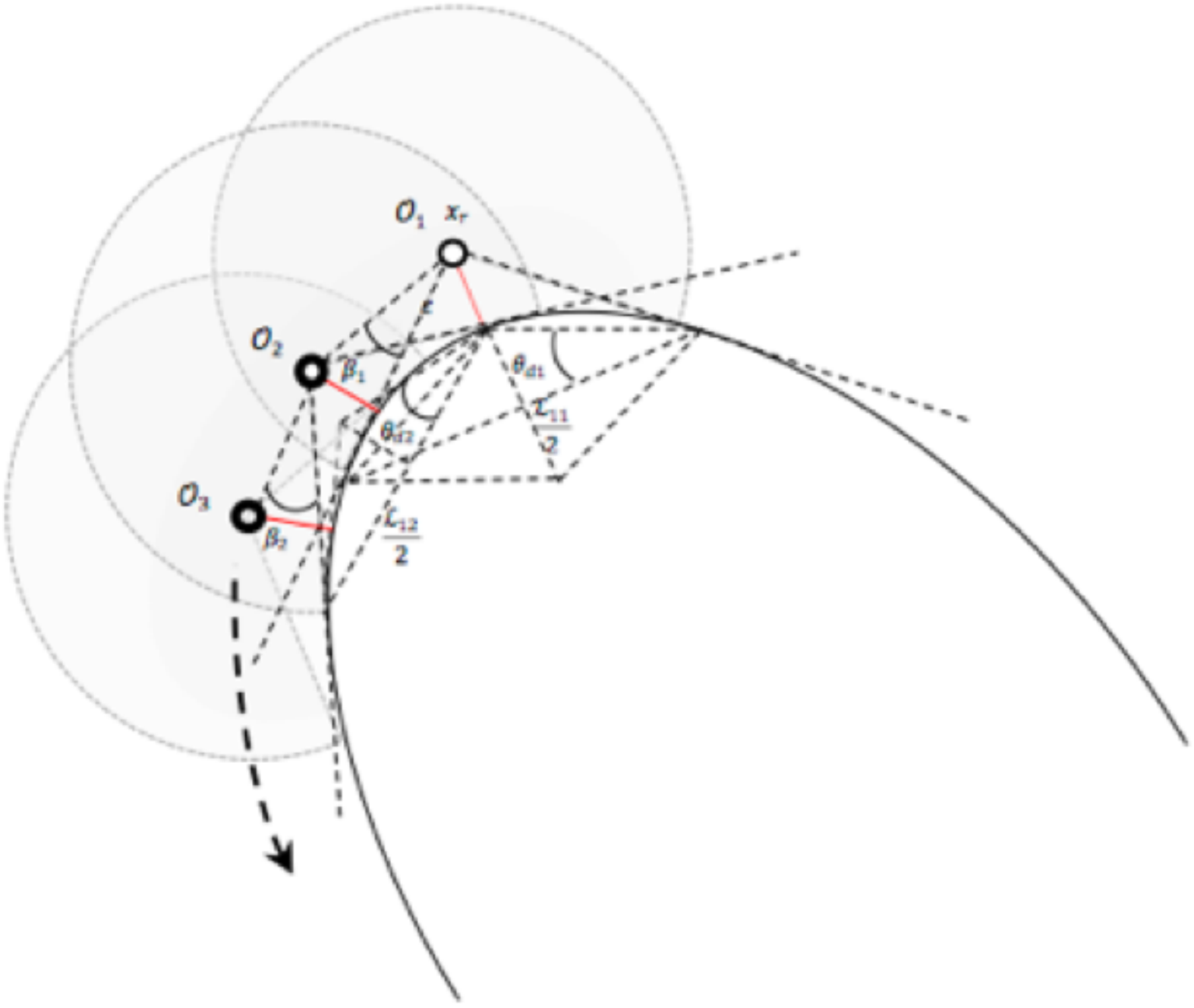

Figure 8.3: Next position estimation in each switching step

## 8.3 Simulations

In this section, we present the result obtained from the simulations of PESA method which is applied by 5 pointwise robots to avoid a convex obstacle in an unknown region. Fig.8.4 (a, b, c) shows the environment, the obstacle and a group of mobile robots. Furthermore, the path the team moves through to avoid the obstacle is illustrated. Moreover, Fig.8.5, shows the measuring and the switching steps of decision-making to estimate each position by the robot the leader.

Both mean value and standard deviation confirms that the leader moves along a path with maintain the distance of $\varepsilon \pm \mu_0$ to the obstacle.

Also, as shown in table 8.1, the mean value shows a reasonable error that is equal to 4.21%. Furthermore, the maximum deviation is 14.9% and the minimum is 1%. It seems the maximum deviation is a bit high but as we can see in both Fig.8.5 and Table 8.1, the minimum deviation which represent the closest distance of the robot the leader with the obstacle is just 1% and which occurred one time merely. If we suppose the $\mu_0 = 5\%$, the results confirm the PESA is highly reliable and robust algorithm for the navigating a multi robot team in a region while avoiding a convex obstacle.

Table 8.1: The statistics data of PESA performance

| Measurement and position estimation steps | 1 | 2 | 3 | 4 | 5 | 6 | 7 |
|---|---|---|---|---|---|---|---|
| Minimum Allowable Distance '$\varepsilon$' (m) | 0.9965 | 0.9965 | 0.9965 | 0.9965 | 0.9965 | 0.9965 | 0.9965 |
| Minimum distance of the robot the leader from the obstacle in each step (m) | 0.9866 | 1.02 | 1.076 | 1.134 | 1.145 | 1.072 | 1.034 |
| Mean Value | 1.044 | 1.044 | 1.044 | 1.044 | 1.044 | 1.044 | 1.044 |
| Standard Deviation | 0.9608 | 0.9608 | 0.9608 | 0.9608 | 0.9608 | 0.9608 | 0.9608 |
| | 1.128 | 1.128 | 1.128 | 1.128 | 1.128 | 1.128 | 1.128 |
| Error | -0.0099 | 0.0235 | 0.0795 | 0.1375 | 0.1485 | 0.0755 | 0.0375 |

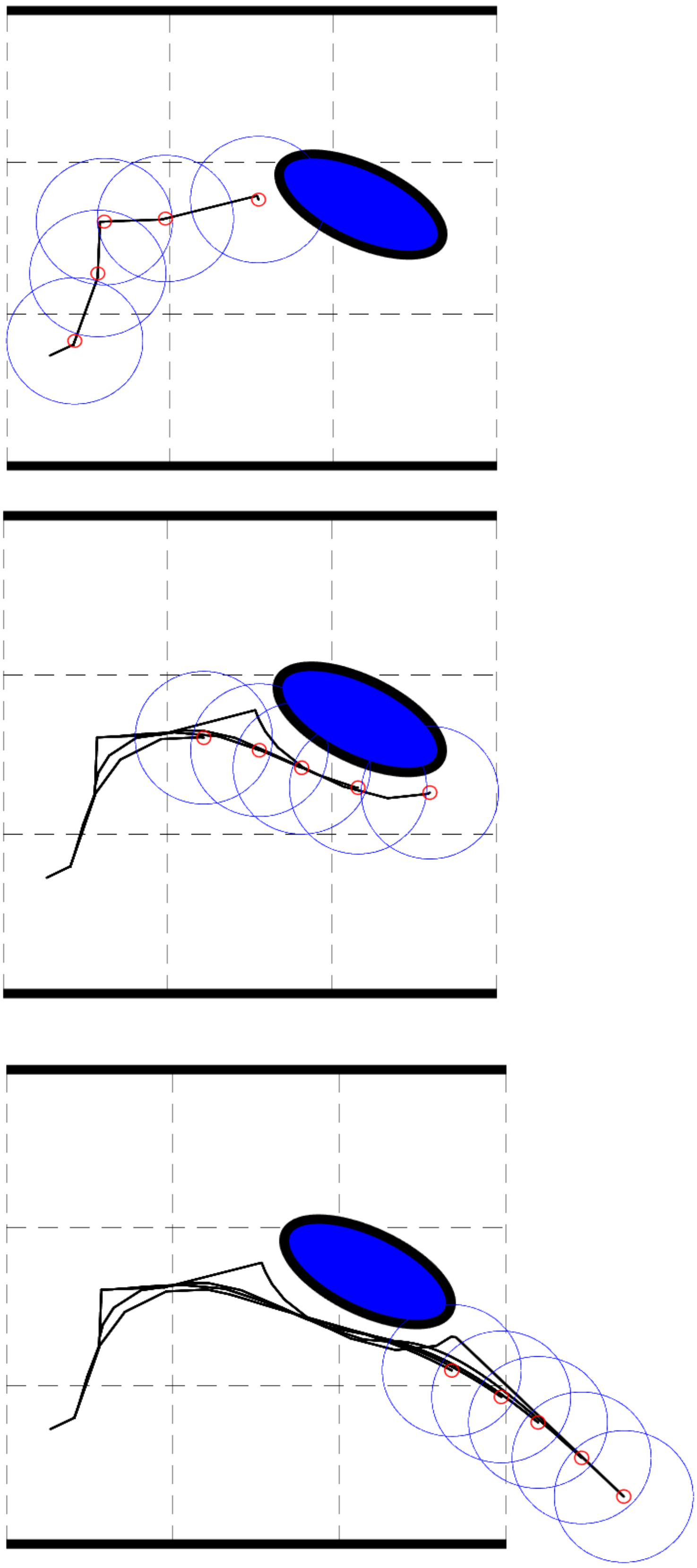

Figure 8.4(a, b, c): Obstacle avoidance by the team of mobile robots

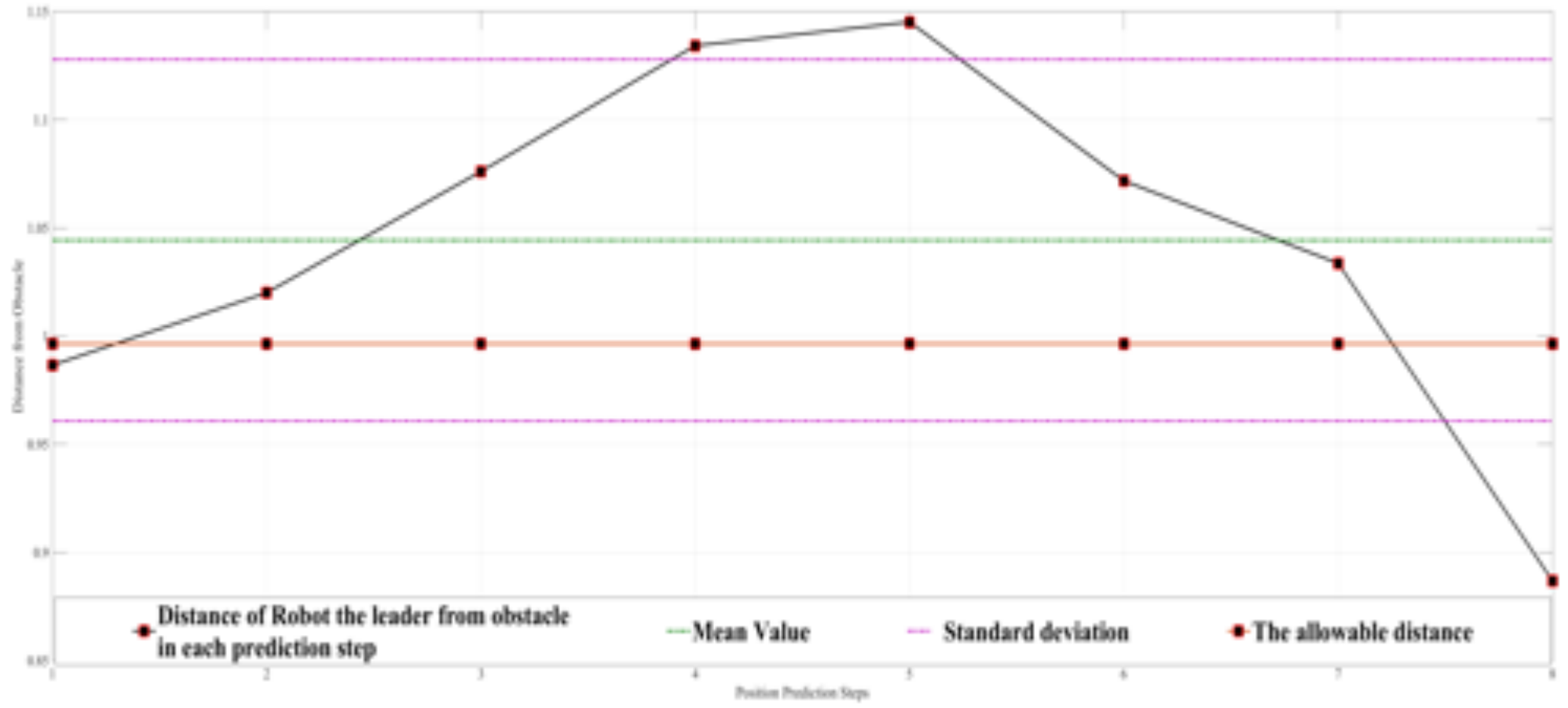

Figure 8.5: Nearest distance of the robots and the obstacle

## 8.4 Summary

In this chapter, we supposed a group of pointwise mobile robots in an unknown smooth environment, occupied by a convex and static obstacle. The team of mobile robots supposed to find a safe path to avoid the obstacle while satisfying the minimum allowed distance with the obstacle. In this case, we proposed a navigation method called Position *Estimation Switching Algorithm* (PESA) which is based on leader-follower pattern. The leader is assigned to estimate the next position and the safest path and communicate the updated heading and coordinates with the nearest neighbour. The members transfer the updating information in chain-link communication channel through the entire of the team. Furthermore, the measurement and the calculation are implemented in just a few steps. Therefore, the team, traverses a distance within each switching steps without measurement and computation.

# Chapter 9

# Modified semi-decentralised switching navigation method in an obstacle-ridden environment

In this work, we present a modified version of the Position Estimation Switching Algorithm (PSEA) that was presented in chapter 5 for navigating a multi-robot team in an environment occupied by multiple obstacles. The pattern of the method is based on leader-follower based formation control algorithm. In this method, the leader has been received the local information by its on-board sensors from the region and then transfers data to the closest neighbor. It becomes the continuous duty of each team mate to share received data with their closest neighbor in a single strand chain communication network.

The distance between the obstacle and the leader has measured by the leader of the multi-robot team in a finite number of direction [30]. Then, leader updates its heading towards the safest path, which satisfies the minimum-allowed distance with the obstacle. Each robot updates its heading, based on the updated data that has been received by its neighbor through the strand chain communication network in each switching period. Consequently, the continuous measurement and computation of the distance, velocity, or repulsive forces are not required.

The remainder of the chapter is organized as follows; Section 9.1 presents the problem statements; Section 9.2 proposes the navigation model and the mathematical analysis; In Section 9.3 the simulations and the results are presented; and finally, Section 9.4 presents the summary of the chapter.

## 9.1 Problem statement

We consider a robot $i$ with the heading $\theta_i(t)$, and the Cartesian coordinates $x_i(t), y_i(t)$ in the plane, where, the heading $\theta_i(t) \in (0,2\pi]$ is measured based on the attached x-axis of a given to the robot $i$, in the counter-clockwise direction. Furthermore, we suppose the linear velocity $v_i(t)$ and the angular velocity $\omega_i(t)$ of the robot $i$ satisfy the following conditions:

$$v_{min} < v_i(t) < v_{max} \tag{9.1}$$

$$\omega_{min} < \omega_i(t) < \omega_{max} \tag{9.2}$$

Then, we introduce a controlled $U_i$ that is a function of robots' polar angles and the distances to the obstacles at time $t$ as follows:

$$U_i = \left[D^{O_j}_{r,\ t}, \Theta^{O_j}_{r,\ t}\right] \tag{9.3}$$

Where $D^{O_j}_{r,\ t}$ indicates a set of all measured Euclidean distances from the robots to the obstacles and $\Theta^{O_j}_{r,t}$ represents a set of headings of the team members at any time $t$.

The main objective of this chapter is to modify the PESA method result in a safe maneuvering of a multi-robot team in a region occupied by multiple convex obstacles with no risk of collision between each teammate.

## 9.2 Modified PESA

Supposed the robots randomly distributed in the environment. At the first stage, the robots line up themselves in a queue regarding the location of the leader which supposed to be the closest one to the obstacle with a given distance $\eta_i$ to the closest neighbor. The initial motion process of the robots has the Markov property as it depends only on the current time.

Furthermore, we assume a sensing range with a radius $R$ for each robot to detect any object in the region. For the sake of simplicity, we consider $R$ is a radius of a perfect disk, however, in reality it can't be a perfect disk result from the hardware and the environmental factors [185].

### *Assumption 9.1:*

In this scenario, we consider a multi-robot team $x_{ri}$ for $i = 1, 2, \dots, n$, that randomly distributed in the environment. The index 1 specifies the robot the leader $x_{r1}$ that is located in point $p_{r1}$. We introduce a point $p_{k^*}^{j^*}$ in the outer face of the obstacle$j^*$ for $j^* \in j$ such that:

$$dist\left(p_{r1}, p_{k^*}^{j^*}\right) = \inf\left(dist\left(p_{ri|_{i\neq 1}}, p_k^j\right)\right), k^* \in k \tag{9.4}$$

where, $k$ indicates a set of all finite number of points of the outer face of the obstacles which are visible to the team leader at any time $t$.

Moreover, introduce $\Gamma = (\gamma_1, \gamma_2, \dots, \gamma_m)$ which represents a set of angles that each of them subtended by the maximum visible curve by the leader and lies between points $\wp_1^j, \wp_2^j$ of each obstacle. Then, we introduce two lines $\ell_1^j$ and $\ell_2^j$ from the Cartesian coordinate of the team leader, $p_{r1}$ to the points $\wp_1^j, \wp_2^j$ of each obstacle as follows:

$$\begin{cases} \ell_1^j = inf(\| p_{r1} - \wp_1^j \|) \\ \ell_2^j = inf(\| p_{r1} - \wp_2^j \|)^{\forall\, j = 1,2,\dots,m} \end{cases} \tag{9.5}$$

and

$$\sphericalangle \ell_2^{j'} \ell_1^{j''} = \gamma_{m'} \; for \; \gamma_{m'} \in \Gamma \, , m' = 1,2,\dots,m \tag{9.6}$$

Introduce a line $\mathcal{L}_{1j}$ which connects the points $\wp_1^j$ to $\wp_2^j$ and satisfies the following condition:

$$\mathcal{L}_{1j} = inf\,(\| \wp_1^j - \wp_2^j \|) \tag{9.7}$$

Set of lines $(\mathcal{L}_{2j}, \mathcal{L}_{3j})$ are supposed to be a set of perpendicular lines to $\mathcal{L}_{1j}$ as in:

$$\begin{cases} \overline{\mathcal{L}_{2j}} = \varepsilon \; and \; \overline{\mathcal{L}_{3j}} = 2\varepsilon \\ \mathcal{L}_{2j} \perp \mathcal{L}_{1j} \; at \; \wp_3^j \epsilon \mathcal{L}_{1j} \\ \mathcal{L}_{3j} \perp \mathcal{L}_{1j} \; at \; \wp_2^j \epsilon \mathcal{L}_{1j} \\ dist(\wp_1^j, \wp_3^j) = dist(\wp_2^j, \wp_3^j) = \mathcal{L}_{1j}/_2 \end{cases} \tag{9.8}$$

The set of lines $\mathcal{L}_{1j}$, $\mathcal{L}_{3j}$ and $\mathcal{L}_{4j}$ indicate the catheti and the hypotenuse of the right-angle triangle $\varDelta \wp_1^j \wp_2^j \wp_5^j$ (see Fig.6.1), such that:

$$\alpha_j = tan^{-1} \frac{\mathcal{L}_{3j}}{\mathcal{L}_{1j}} \tag{9.9}$$

## *Assumption 9.1:*

Supposed that $\mathfrak{D} = (d^{12}, d^{23}, \dots, d^{(m-1)\,m})$ represents a set of all paths that connect point $\wp_1^{j'}$ of obstacle $j'$ to point $\wp_3^{j''}$ of obstacles $j''$ in counter-clockwise direction. The multi-robot team can move between any two obstacles with a given distance $d^{(k-1)\,k} \in \mathfrak{D}_e \subseteq \mathfrak{D}$, if and only if:

$$\frac{d^{(k-1)k}}{2r} \gg 1 + \varepsilon \tag{9.10}$$

Furthermore, to avoid any confusion for the leader when it faces with more than two obstacles in the region, we introduce distance $d^{(\tilde{k}-1)\,\tilde{k}}$ which satisfies the following condition:

$$d^{(\tilde{k}-1)\,\tilde{k}} = sup\left(d^{(k-1)k} \in \mathfrak{D}_e\right), k = 1,2,\ldots,m \qquad (9.11)$$

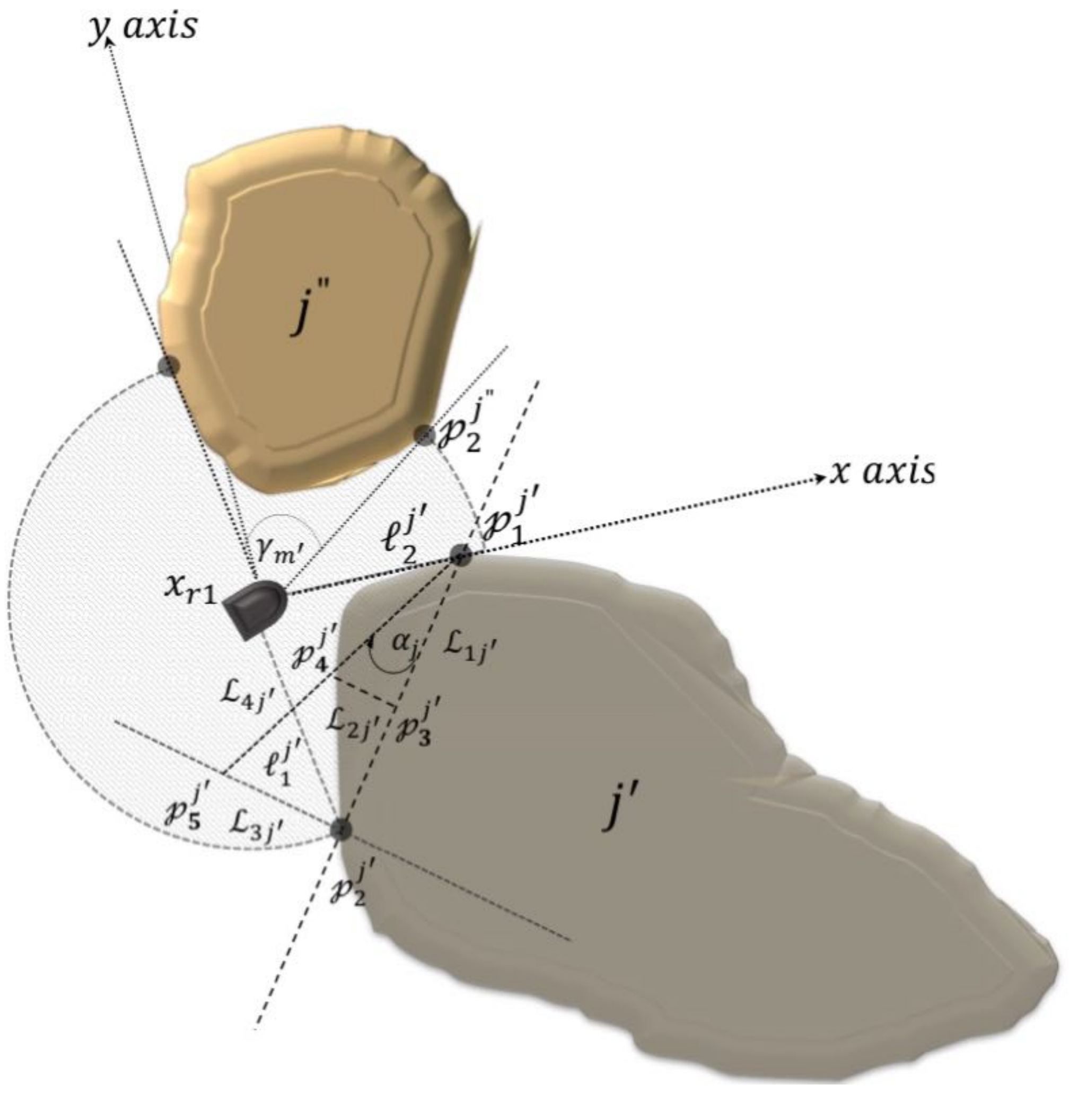

Figure 9.1: Measuring distances by the leader.

## *Proposition 9.1:*

If the robot $x_{r1}$ moves from the location $O_1$ to the location $O_2$ among obstacles $j'$ and $j''$ where $dist\left(p_{r1}, p_{k^*}^{j'}\right)|_{O_1} < dist\left(p_{r1}, p_{k^{**}}^{j''}\right)|_{O_1}$ in the same direction of the vector $\overrightarrow{\tilde{\mathcal{L}}_i}$ as in:

$$\tilde{\mathcal{L}}_i = \frac{1}{2}\sqrt{{\mathcal{L}_{1j'}}^2 + {\mathcal{L}_{1j''}}^2 + 2\mathcal{L}_{1j'}\mathcal{L}_{1j''}cos(\alpha_1 + \alpha_2)} \qquad (9.12)$$

and the length of:

$$\tilde{\ell}_i = inf\ (\frac{\mathcal{L}_{1j'}}{2}, \frac{\mathcal{L}_{1j''}}{2}) \qquad (9.13)$$

Based on the x-axis of the attached reference frame to it which is shown by $\ell_1^{j'}$, then:

$$inf\left(dist\left(p_{r1}, p_{k^*}^{j'}\right)|_{O_2}, dist\left(p_{r1}, p_{k^{**}}^{j''}\right)|_{O_2}\right) \geq \varepsilon - \mu_0 \qquad (9.14)$$

In which, $\mu_0$ denotes the sensor measurement error.

## *Proof:*

First, we consider the case that, the leader moves from location $O_1$ to the location $O_3$ in the direction of $\alpha_1$ based on $\ell_1^{j'}$ and $\frac{\mathcal{L}_{1j'}}{2}$ in length. It's obvious that at the location $O_3$, where the robot is settled, the distance between the robot and the obstacle $j'$ is more than $\varepsilon$. Similarly, if the robot moves from position $O_1$ to the position $O_4$ in the direction of $\delta_{j'j''} - \alpha_2$ based on $\ell_1^{j'}$ and $\frac{\mathcal{L}_{1j''}}{2}$ in length, the distance of the robot and the obstacle $j''$ is more than $\varepsilon$ as well.

Now, introduce vector $\overrightarrow{\tilde{\mathcal{L}}_i}$ that represents the resultant vector of $\vec{l}_i^{j'}$ and $\vec{l}_i^{j''}$ (see Fig. 9.2). Considering (9.11) holds, we find the angle $\beta_{1i}$ as follows:

$$\beta_{1i} = tan^{-1}\left(\frac{\mathcal{L}_{1j'} \sin \alpha_1 + \mathcal{L}_{1j''} \sin \alpha_2}{\mathcal{L}_{1j'} \cos \alpha_1 + \mathcal{L}_{1j''} \cos \alpha_2}\right) \quad (9.15)$$

then, it's obvious that:

$$\alpha_1 \leq \beta_{1i} \leq \delta_{j'j''} - \alpha_2 \quad (9.16)$$

Therefore, satisfying the navigation law (9.13) guaranties that the leader would settle at the location $\mathcal{O}_2$ with a distance more than $\varepsilon$ to both obstacles $j'$ and $j''$. This is the complete proof of the Proposition 9.1.

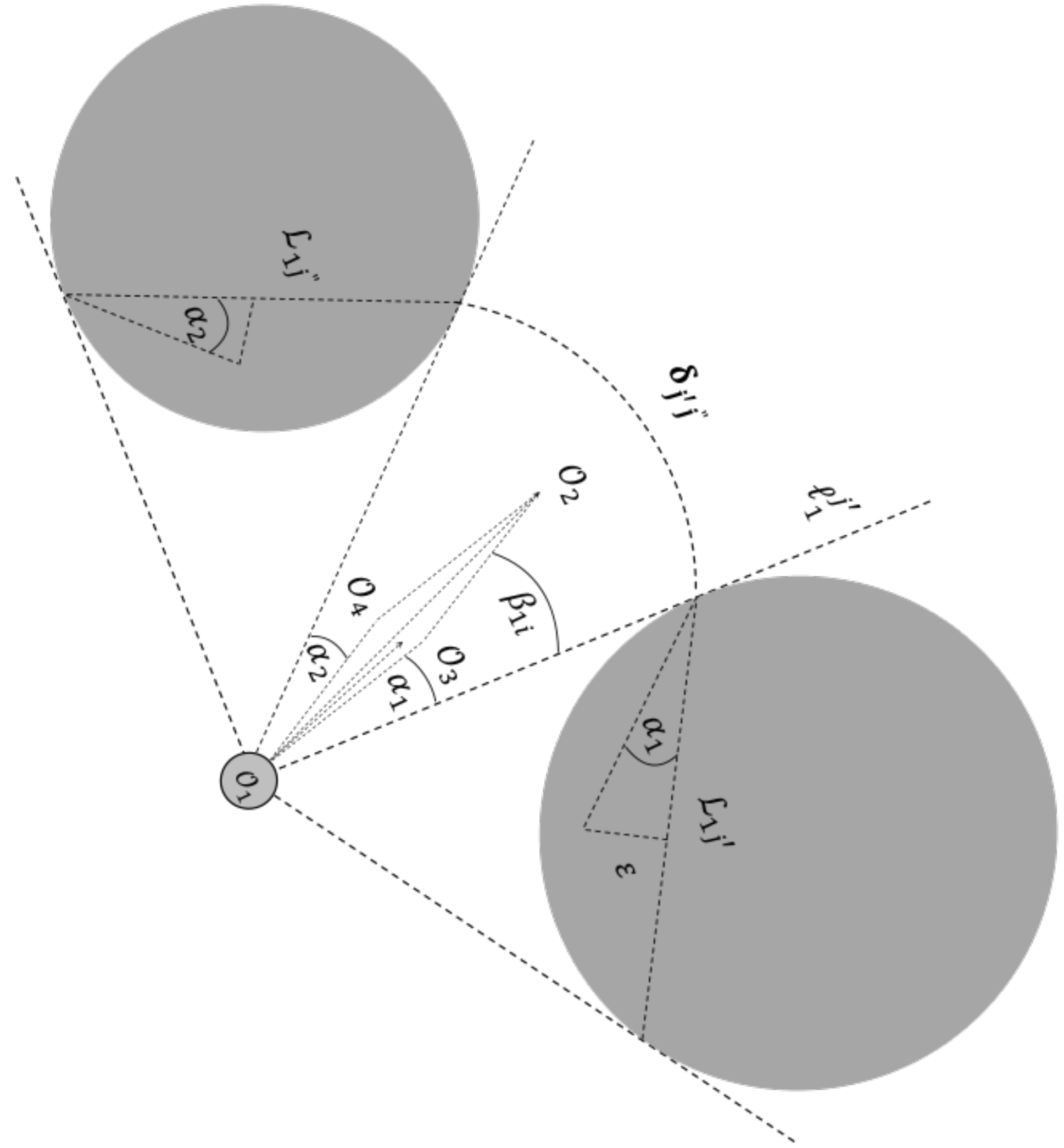

Figure 9.2: Computation of the switching position.

Since the safest path is planned by the leader, this is the responsibility of each follower to update its heading and distance, according to the data received from the closest neighbor to avoid collision consequently. Therefore, each follower updates its heading based on the following rule:

$$\beta_{ki} = tan^{-1}\left(\frac{y_{(k-1)i}-y_{ki}}{x_{(k-1)i}-x_{ki}}\right) \tag{9.17}$$

Furthermore, each robot estimates the next position as in:

$$dist_{ki}\left(x_{kr}, x_{(k+1)r}\right) = \| \mathcal{O}_{(k-1)i} - \mathcal{O}_{ki} \| \tag{9.18}$$

where index $k$ in updating rules (9.17) and (9.18) denotes the number of the members of the team including the leader and the followers and the index $i$ denotes the permutation of the leader and the followers at each switching step.

## 9.3 Simulations

Fig.9.3 shows, the environment where occupied by 3 convex and static obstacles and the multi-robot team includes 4 followers and a leader travel in the region whilst avoiding the obstacles. The measurement error is considered as $\mu_0 = 0.1\varepsilon$. It is obvious that the leader can choose the safest path in the area when it applies the navigation law (9.12), and (9.13), as well as when the other teammates follow the leader safely in the region. Furthermore, applying updating rules (9.17) and (9.18) guarantee a collision free navigation between the team members. The distances between the leader and each obstacle are shown in Fig.9.4 by solid lines. As Fig.9.4 confirms, the distance between the leader and the obstacles satisfies the minimum-allowed distance $\varepsilon$ in each switching step. The only exception happened in the switching step 11, however, the difference (0.08 Decimeter) is acceptable considering the value of $\mu_0$. Table 1 shows the multi-robot team successfully avoids all the obstacles while updating their headings and distances in just 11 switching steps and not continuously, which results in a fast motion in the region. Furthermore, the distance between the leader and the obstacle 2 in switching step 11 which is 0.82 decimeter doesn't contradict with (9.14).

Table 9.1: Distances from the obstacles in each switching steps

| Switching steps | Distance from Obstacle 1 | Distance from Obstacle 2 | Distance from Obstacle 3 |
|---|---|---|---|
| 1 | 11.68 | 4.33 | 2.19 |
| 2 | 11.25 | 3.84 | 2.1 |
| 3 | 10.71 | 3.34 | 2.01 |
| 4 | 9.94 | 2.84 | 1.85 |
| 5 | 8.74 | 2.39 | 1.58 |
| 6 | 7.09 | 2.08 | 1.32 |
| 7 | 5.19 | 1.76 | 1.41 |
| 8 | 3.41 | 1.77 | 1.56 |
| 9 | 1.77 | 1.99 | 1.83 |
| 10 | 1.37 | 1.13 | 3.18 |
| 11 | 1.32 | 0.82 | 4.48 |

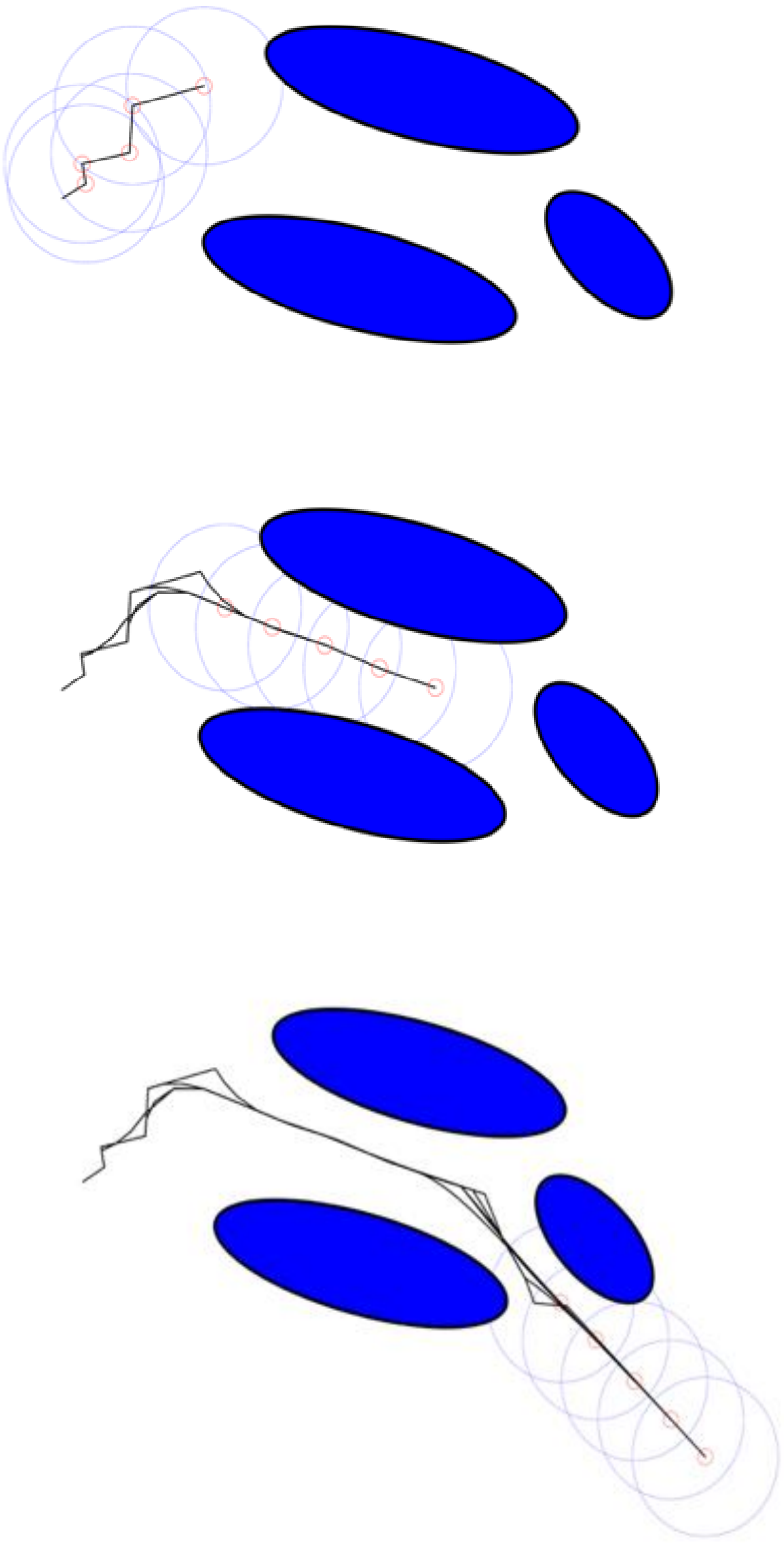

Figure 9.3: Obstacle avoidance by the team of mobile robots.

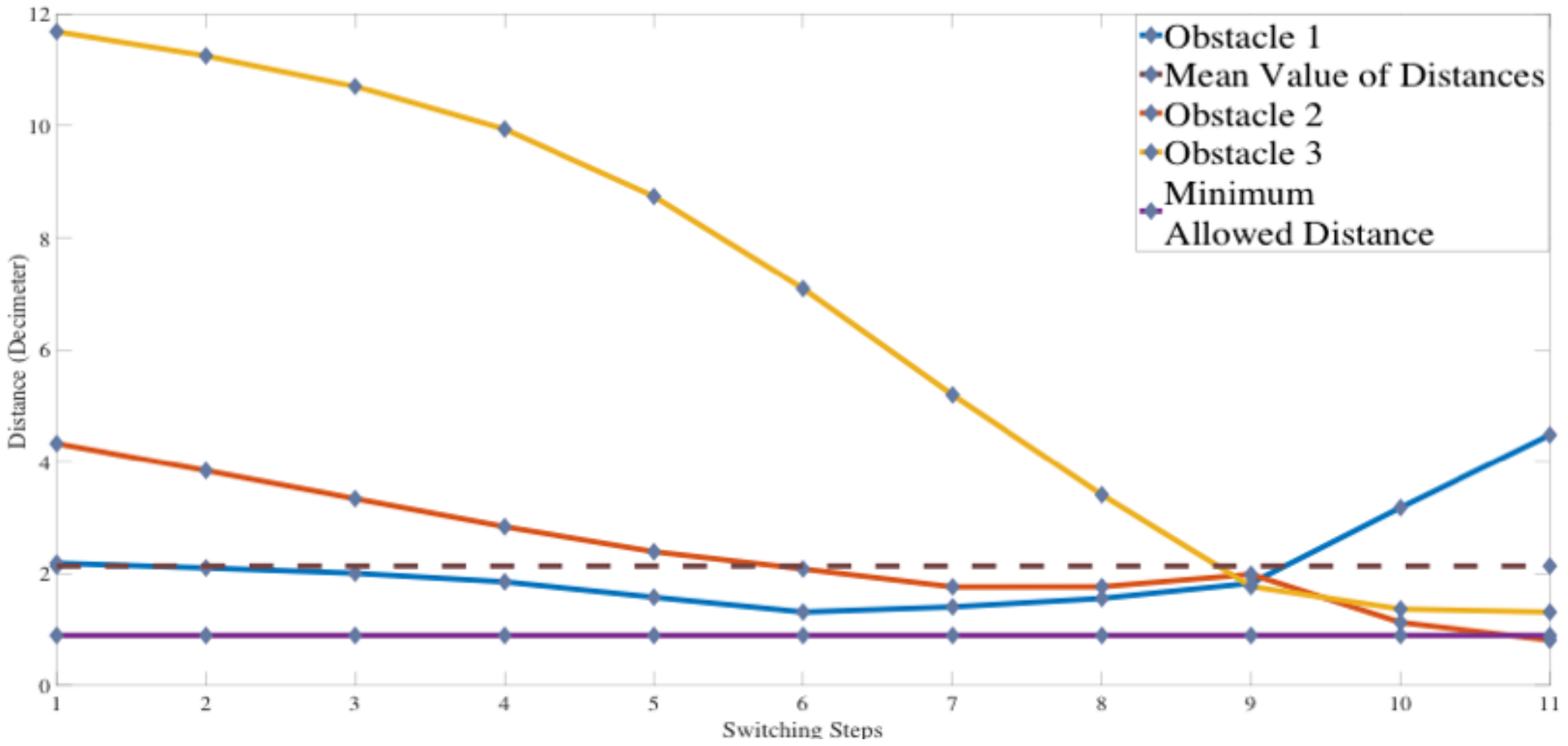

Figure 9.4: Distance of the leader and the obstacles.

## 9.4 Summary

In this study, we considered a fleet of pointwise mobile robots in an unknown environment occupied by smooth convex obstacles. In case of the mission of the multi-robot team which was a safe traverse in the cluttered environment, we proposed a semi-decentralized navigation method based on the leader-follower concept, that is a modified version of the PESA, presented in chapter 5. the advantage of the proposed model is reducing the measurement and computation time in addition to guarantees a safe maneuvering of a multi-robot team. Both mathematical analysis and the simulations results confirm the robustness and validation of the proposed algorithm.

# Chapter 10

# Decentralised switching navigation method in an obstacle-ridden environment

This chapter presents a decentralized navigation algorithm for a team of mobile robots to traverse an unknown obstacle-ridden environment to detect and trap a target located in the region. The proposed navigational strategy guarantees that the robots maintain the minimum distance allowed to the obstacles while avoiding them to trap the target. The area was occupied by many obstacles with multiple shapes that were randomly distributed in the region; therefore, each robot had to find the safest path between the obstacles based on a decision-making algorithm when there was more than one path to choose from. Unlike the conventional method of collecting information by mobile robots based on sampling in short and pre-set periods, in the proposed method robots collected information at indeterminate intervals leading to reductions in the sensing period, computation and consequent energy consumption. The mathematical proof and the computer simulations confirmed the reliability and robustness of the proposed method. The remainder of the chapter is organized as follows: in Section 10.1, we present the problem formulation and mathematical analysis; Section 10.2 presents the simulations and discussion; and finally, Section 10.3 presents a brief summary of the chapter.

## 10.1 Problem formulation

In this section we present a navigation algorithm for a fleet of mobile robots to move safely in an unknown obstacle-ridden area. The obstacles were supposed to be static with multiples shape and were randomly distributed in the region. The local Cartesian coordinates and heading of the robots were represented by the sets:

$$X = (x_1(t), x_2(t), \dots, x_n(t)) ,$$

$$Y = (y_1(t), y_2(t), \dots, y_n(t)) ,$$

$$\Theta = (\theta_1(t), \theta_2(t), \dots, \theta_n(t)).$$

where $n$ denotes the number of robots in a team. The polar angle $\theta_i(t) \in (0,2\pi]$, for $i = 1,2, \dots, n$, was measured based on the x-axis of a reference frame, which was attached to each robot in an anti-clockwise direction.

Furthermore, the sets

$$V = \left(v_1(t), v_2(t), \dots, v_n(t)\right),$$

$$\mathcal{W} = \left(\omega_1(t), \omega_2(t), \dots, \omega_n(t)\right).$$

stand for the linear velocity and angular velocity of each robot, respectively. In the present study, we considered the constant linear and angular velocities for the robots in any time that satisfied the following constraints:

$$v_{min} < v_1(t) = v_2(t) = \dots = v_n(t) < v_{max} \quad (10.1)$$

$$\omega_{min} < \omega_1(t) = \omega_2(t) = \dots = \omega_n(t) < \omega_{max} \quad (10.2)$$

Then, we defined a controlled $U_i$ that was a function of the robots' polar angles and the distances of the robots with the obstacles at time $t$:

$$U_i = \left[D_{r,\ t}^{O_j}, \Theta_{r,\ t}^{O_j}\right] \quad (10.3)$$

where, $D_{r,\ t}^{O_j}$ denotes a set of all measured Euclidean distances between all agents and the obstacles, and $\Theta_{r,t}^{O_j}$ denotes a set of headings concerning the obstacles that were measured by each member of the network based on the x-axis of the attached Cartesian coordinate system to each individual robot.

## *Assumption 10.1:*

We supposed a team of multi robot $x_{ri}$ for $i = 1, 2, \dots, n$, were randomly deployed in the area of interest. Therefore, the robot $x_{r1}$was located at point $p_{r1}$. We considered a point $p_{k^*}^{j'}$ in the outer face of the obstacle $j' \in j$ where:

$$dist\left(p_{r1}, p_{k^*}^{j'}\right) = \inf\left(dist\left(p_{ri|_{i\neq 1}}, p_k^j\right)\right), k^* \in k \qquad (10.4)$$

In (10.4), $k$ denotes a set of finite numbers of all visible points of the outer face of the obstacles to the sensors at time $t$.

We defined a set of angles $\Gamma = (\gamma_1, \gamma_2, \dots, \gamma_m)$, which were each subtended by the maximum visible curve lying between points $\wp_1^{\mathrm{j}}, \wp_2^{\mathrm{j}}$ of each obstacle to the robot that detected them. Then, we defined two lines $\ell_1^{\mathrm{j}}$ and $\ell_2^{\mathrm{j}}$ from the point $p_{r1}$ of the leader to the points $\wp_1^{\mathrm{j}}, \wp_2^{\mathrm{j}}$ of each obstacle as in:

$$\begin{cases} \ell_1^j = inf(\parallel p_{r1} - \wp_1^j \parallel) \\ \ell_2^j = inf(\parallel p_{r1} - \wp_2^j \parallel)^{\forall\, j = 1,2,\dots,m} \end{cases} \qquad (10.5)$$

and

$$\measuredangle \ell_2^{j'} \ell_1^{j''} = \gamma_{m'}\ for\, \gamma_{m'} \in \Gamma\, , m' = 1,2,\dots,m \qquad (10.6)$$

The points $\wp_1^{\mathrm{j}}, \wp_2^{\mathrm{j}}$ were assumed to be connected by a line $\mathcal{L}_{1\mathrm{j}}$, which held the following condition:

$$\mathcal{L}_{1j} = inf\ (\parallel \wp_1^j - \wp_2^j \parallel) \qquad (10.7)$$

and a set of lines $(\mathcal{L}_{2j}, \mathcal{L}_{3j})$ that were perpendicular to $\mathcal{L}_{1j}$ as follows:

$$\begin{cases} \overline{\mathcal{L}_{2j'}} = 2\varepsilon \\ \mathcal{L}_{2j'} \perp \mathcal{L}_{1j'}\ at\ \wp_3^{j'} \epsilon \mathcal{L}_{1j'} \\ \mathcal{L}_{3j'} \perp \mathcal{L}_{1j'}\ at\ \wp_2^{j'} \epsilon \mathcal{L}_{1j'} \\ dist(\wp_1^{j'}, \wp_3^{j'}) = dist(\wp_2^{j'}, \wp_3^{j'}) = {}^{\mathcal{L}_{1j'}}/_{2} \end{cases} \quad (10.8)$$

As shown in Fig.10.1, $\mathcal{L}_{1j'}$, $\mathcal{L}_{3j'}$ and $\mathcal{L}_{4j}$ represent the catheti and the hypotenuse of the right-angle triangle $\Delta\wp_1^{j'}\wp_2^{j'}\wp_5^{j'}$, where:

$$\alpha_{j'} = tan^{-1}\frac{\mathcal{L}_{3j'}}{\mathcal{L}_{1j'}} \quad (10.9)$$

## *Definition 10.1:*

We considered $\mathfrak{D} = (d^{12}, d^{23}, \dots, d^{(m-1)\,m})$ as a set of all minimum distances between the $\wp_1^{j'}$ and $\wp_3^{j''}$ of any two obstacles $j'$and $j''$in an anti-clockwise direction. The robots could traverse between any two obstacles with the distance $d^{(k-1)\,k} \in \mathfrak{D}_e \subseteq \mathfrak{D}$, if and only if:

$$\frac{d^{(k-1)k}}{2r} \gg 1 + \varepsilon \quad (10.10)$$

According to (10.11), the robots had different options to choose from if $m > 2$ caused more computation, more energy consumption and bewilderment, except if the leader chose $d^{(\tilde{k}-1)\,\tilde{k}}$, which satisfied the following condition:

$$d^{(\tilde{k}-1)\,\tilde{k}} = sup\big(d^{(k-1)k} \in \mathfrak{D}_e\big), k = 1,2, \dots, m \quad (10.11)$$

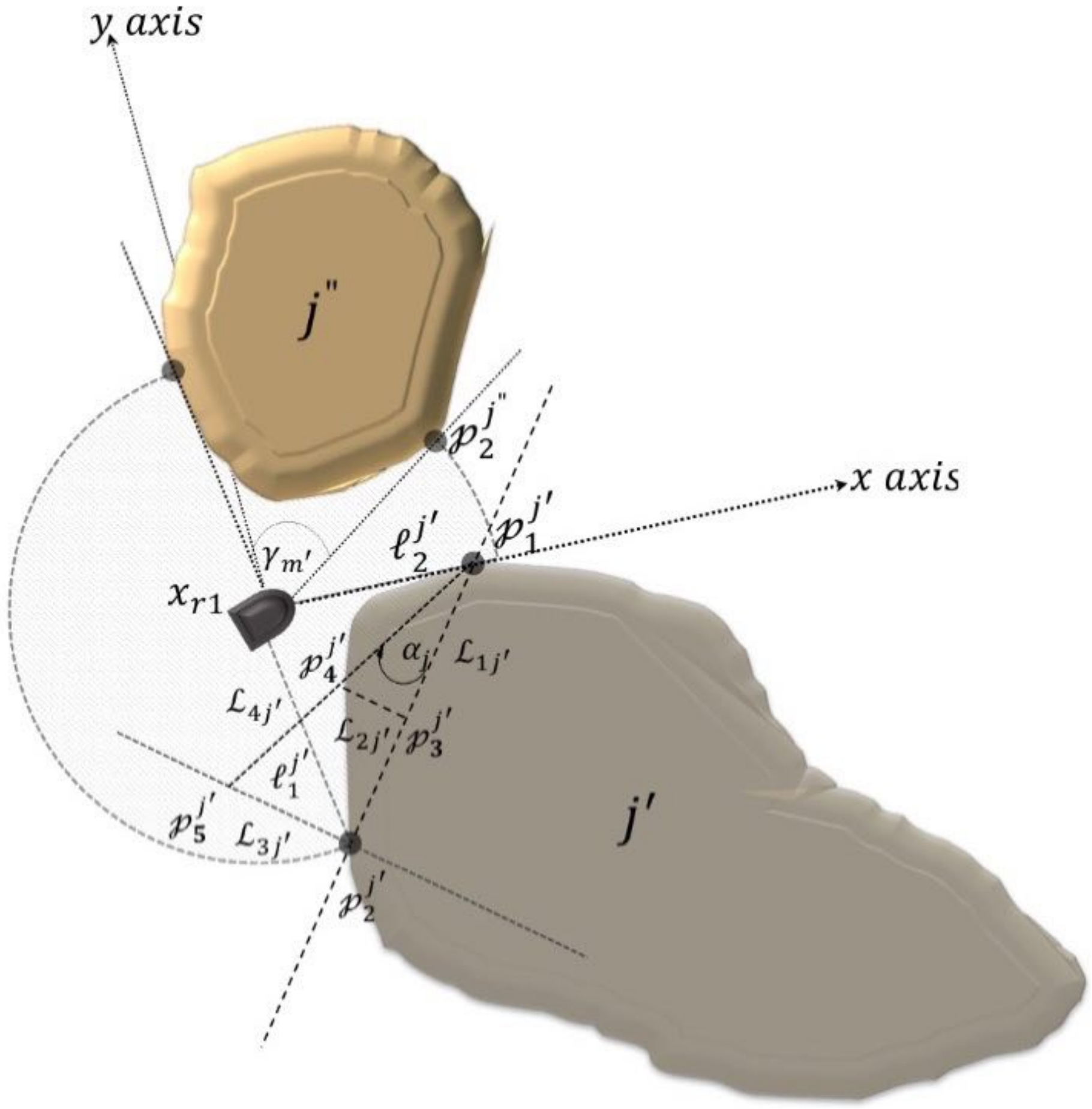

Figure 10.1: Measuring distances by each robot

## *Proposition 10.1:*

If the robot $x_{r1}$ started moving from location $\mathcal{O}_1$ towards location $\mathcal{O}_2$ between two obstacles $j'$ and $j''$ where $dist\left(p_{ri}, p_1^{j'}\right)|_{\mathcal{O}_1} < dist\left(p_{ri}, p_2^{j''}\right)|_{\mathcal{O}_1}$ with the length of the vector $\overrightarrow{\tilde{\mathcal{L}}_i}$ as in:

$$\tilde{\mathcal{L}}_i = \sqrt{(\frac{\mathcal{L}_{1j'}}{4} + \ell_{1j''}\, cos(\alpha_3))^2 + (\varepsilon + \ell_{1j''}\, sin(\alpha_3))^2} \tag{10.12}$$

and the angle of:

$$\beta_{1i} = \tan^{-1}\left(\frac{\varepsilon + \ell_{1j''}\, sin(\alpha_3)}{\frac{\mathcal{L}_{1j'}}{4} + \ell_{1j''}\, cos(\alpha_3)}\right) \tag{10.13}$$

Based on $\ell_1^{j'}$ , which is considered the x axis of the attached frame to the leader robot, then:

$$inf\left(dist\left(p_{ri}, p_1^{j'}\right)|_{O_2}, dist\left(p_{ri}, p_2^{j''}\right)|_{O_2}\right) \geq \varepsilon - \mu_0 \qquad (10.14)$$

In (10.14), $\mu_0$ represents the tolerance of the allowed distance result from the measurement error.

*Proof:*

According to Fig.10.2, if the robot moved from location $O_1$ to location $O_3$, with the heading $\alpha_1$ based on $\ell_1^{j'}$ and the length of $\mathrm{l}_{1j'}$ then the distance between the robot and the obstacle $j'$ would be greater than $\varepsilon$. In the same way, if the leader moved to location $O_4$ with the heading

$$\alpha_3 = \delta_{j'j''} - \alpha_2 \qquad (10.15)$$

where

$$\delta_{j'j''} = \measuredangle\, p_1^{j'} p_2^{j''} \qquad (10.16)$$

based on $\ell_1^{j'}$ and the length $\mathrm{l}_{1j''}$, the minimum allowed distance between the robot and the obstacle $j''$ would be satisfied.

The vector $\overrightarrow{\tilde{\mathcal{L}}_\iota}$ denotes the resultant vector of $\vec{\mathrm{l}}_{1j'}$ and $\vec{\mathrm{l}}_{1j''}$ (see Fig. 10.2).

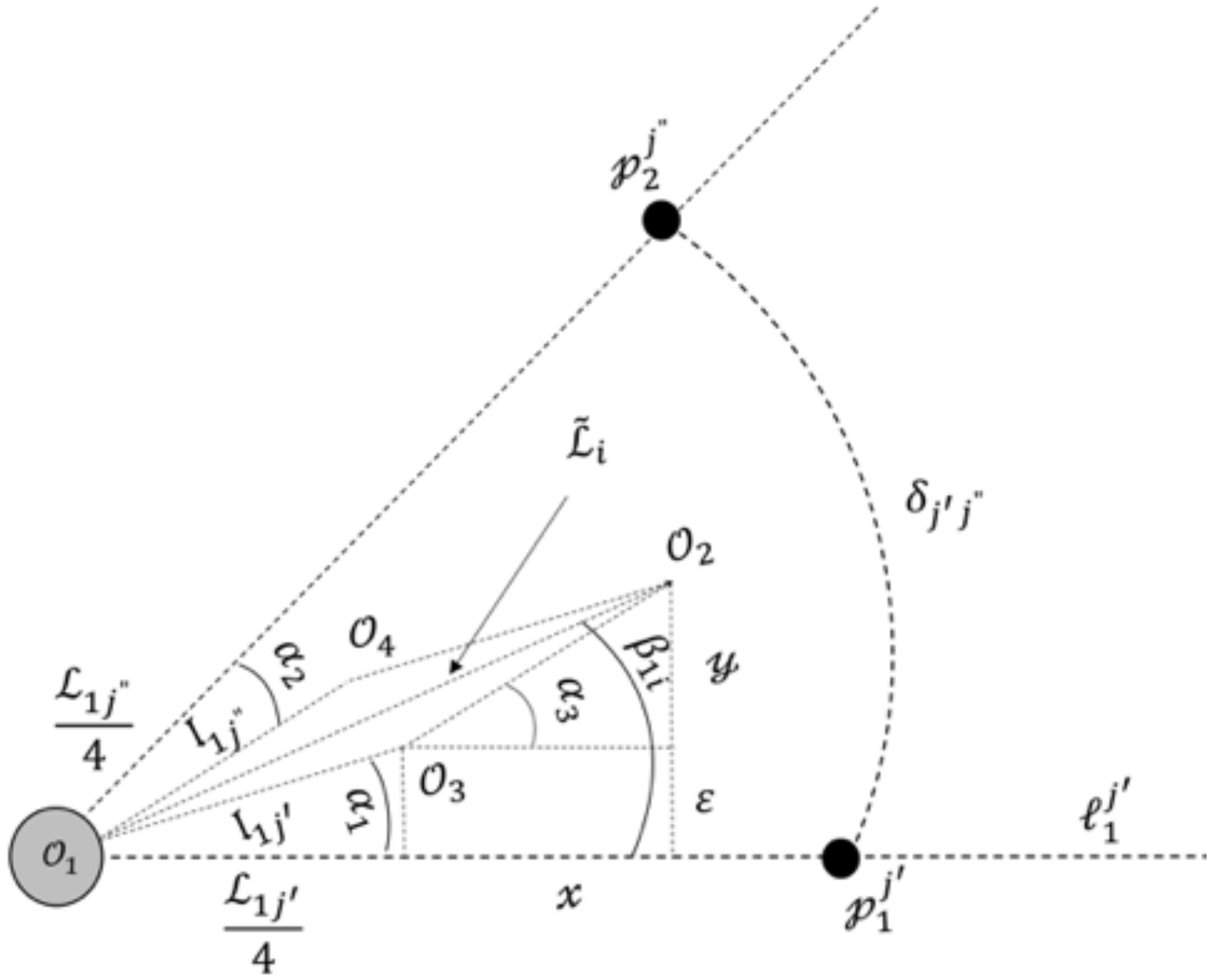

Figure 10.2: Finding the safest path by the robots

Then, we found the lengths $x$ and $y$ as follows:

$$\begin{cases} x = \text{I}_{1j''}\ cos(\alpha_3) \\ y = \text{I}_{1j''}\ sin(\alpha_3) \end{cases} \tag{10.17}$$

and consequently, the angle $\beta_{1i}$ while (10.10), holds as:

$$\beta_{1i} = tan^{-1}\left(\frac{\varepsilon + y}{\frac{\mathcal{L}_{1j'}}{4} + x}\right) \tag{10.18}$$

Therefore,

$$\alpha_1 \leq \beta_{1i} \leq \delta_{j'j''} - \alpha_2 \tag{10.19}$$

and (10.12) and (10.13) guarantee that the robot $i$ would settle at location $\mathcal{O}_2$ while satisfying (10.14).

Furthermore, navigation law (10.12) guarantees that the point $\mathcal{O}_2$ never locates out of the sensing range of the robot, which results in no collision with any possible undetected obstacles until the next switching time.

### 10.1.1 Decision-making rule

If there are more than two obstacles in the sensing range of the robots, they need to choose the safest path among them. We supposed $\mathcal{J} = (j', j'', j''', \dots)$ was a set of obstacles detected by the robot $x_{ri}$ and $\mathcal{D}_j = (d_{j'j''}, d_{j''j'''}, \dots)$ represented a set of minimum distances between any two closest neighbor obstacles. Then, the robot $x_{ri}$ could choose two obstacles that satisfied the following condition:

$$d_{j^* j^{**}} \equiv sup\ (\mathcal{D}_j) \qquad (10.20)$$

### 10.1.2 Energy consumption

Based on the proposed algorithm, since the safest path was planned by each robot during each switching time, they did not require the sample data from the region until they reached the next planned point.

According to the model proposed in [216], the power consumed by the sensors is a function of the sampling period, as in:

$$p_s(f_s) = c_{s0} + c_{s1} f_s \qquad (10.21)$$

where, $p_s$ denotes the sensing power that varied for different types of sensors with different frequencies and $c_{s0}$ and $c_{s1}$were constant coefficients dependent on the sensors. Thus, by decreasing the sampling period the energy consumed by the sensors was decreased, leading to increased battery life. Therefore, the proposed navigational model resulted in decreased energy consumption and increased battery life because the robots used the sensors less frequently to find the safest path in the region.

## 10.2 Simulations

Fig.10.3 shows how the multi-robot team moved in the region while avoiding obstacles. We considered $n = 3$, which represented the number of pointwise robots in the area that were occupied with four static obstacles. The mission of the team was to trap the target while avoiding the obstacles in the region. The robots did not have any a priori information of the region. The only information they had was the position of the target, which was supposed to be static. Each robot attempted to find the safest path based on navigation laws (10.12) and (10.13) in a decentralized fashion, which meant they were working autonomously. We considered $\varepsilon = 2$ decimetres and the tolerance was $\mu_0 \simeq 0.2\varepsilon$.

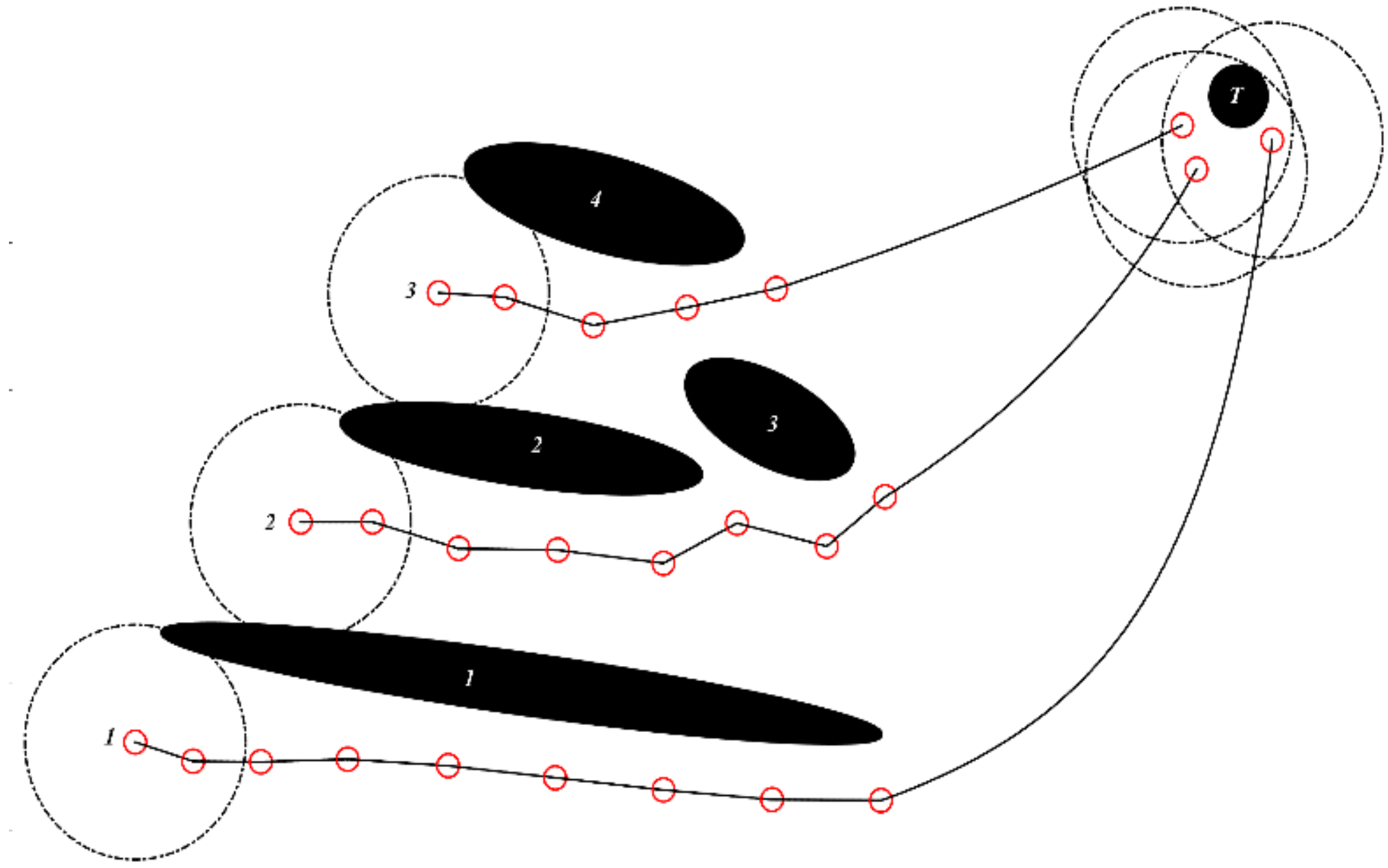

Figure 10.3: Obstacle avoidance by the mobile robot team

Furthermore, there was no possibility of collision between the robots as we considered no more than one robot traversed in a certain path between the obstacles. In the obstacle free area, they started communicating if they were in the sensing range of another robot. In the case of failure of a teammate, if it was located within the sensing range, the others considered it a static obstacle.

In Fig.10.4, the solid line represents the distance of robot 1 to obstacle 1. The dashed lines show the distance of robot 2 to obstacles 1, 2 and 3, and the dotted lines represent the distance of robot 3 to obstacles 2, 3 and 4 during their mission to trap the target. As Fig10.4 shows, each robot moved between two switching points with no computation, which means that each robot moved blindly between any two switching points. Furthermore, the switching number of switching points varied based on the number of obstacles located in the path of each robot.

Table 1 confirms that the minimum distance between the obstacles and the robots were satisfied.

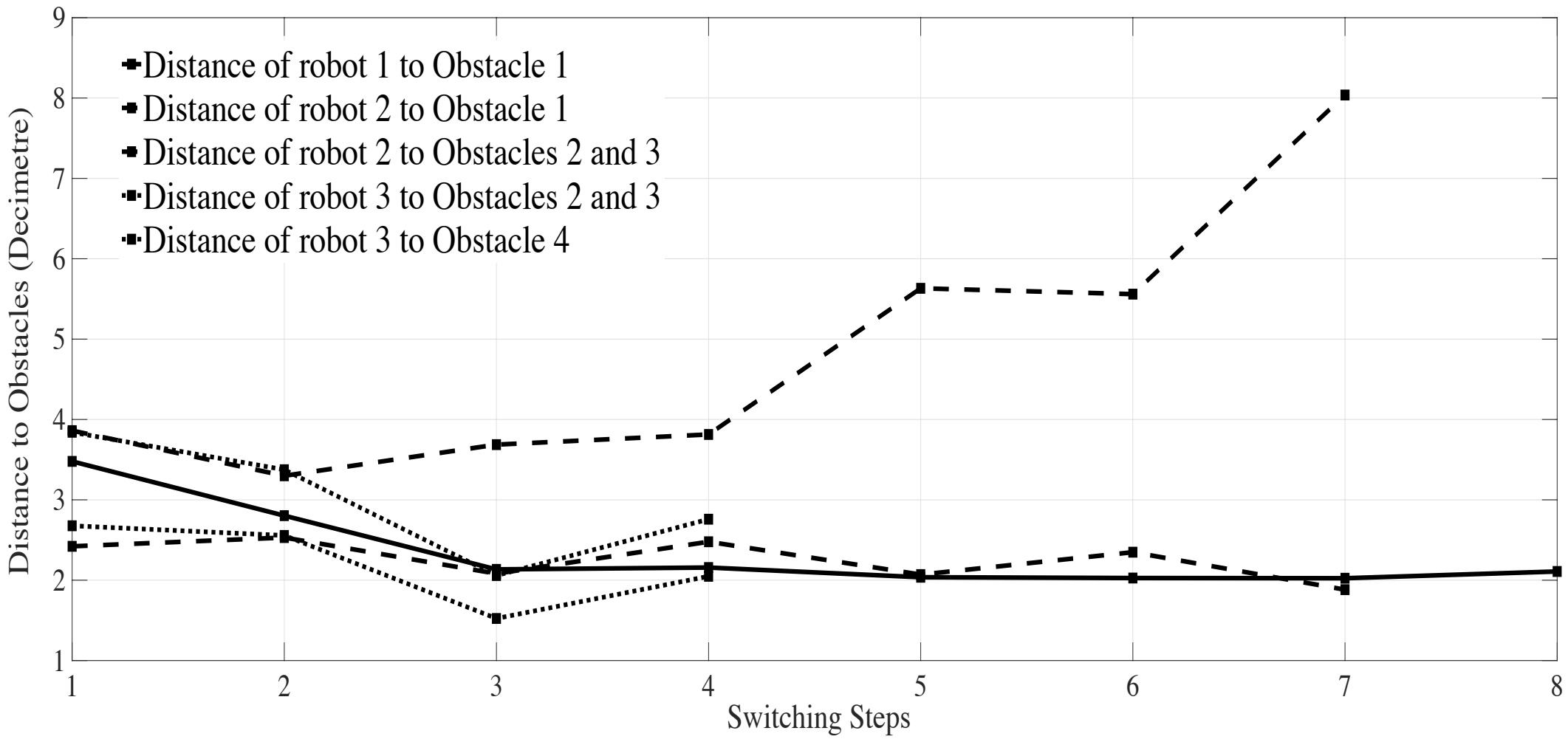

Fig 10.4: Distance of the robots to the obstacles

According to Table 10.1, robot 3 had the minimum distance with obstacle 4 at its third switching time, which equalled 1.52 decimetres. Referring to the $\mu_0$, the distance was acceptable. As we can see from Table 10.1, the robots never exceeded the minimum allowed distance with the obstacles at any time.

Reducing energy consumption was another achievement of the proposed model. Considering the portion of 1.9%–5.1% energy consumption of the sonars and 14.8%–28.8% of the microcontroller in the mobile robot pioneer 3d-x as an example [216] then the sonars should be on and off only for a few seconds periodically, which has a significant improvement on this portion. Furthermore, based on the data captured from the simulations, the shortest distance between the first switching point and the last switching point of robot 1 was 28 decimetres. Conversely, standard service information packets were sent to the mobile robot pioneer 3d-x every 100 milliseconds [217]. In the present study, we considered that v = 500 mm/s. Thus, for robot 1, the number of times the microcontroller could process the data received by the sensors was reduced from 256 times to 8 times, which is an impressive reduction in computation resulting in great reduction in energy consumption and consequently increased battery life.

Table 10.1: Minimum distance of robots and obstacles in each switching period in decimetres

| | SW1 | SW2 | SW3 | SW4 | SW5 | SW6 | SW7 | SW8 |
|---|---|---|---|---|---|---|---|---|
| Robot 1 to Obstacle 1 | 3.48 | 2.80 | 2.13 | 2.16 | 2.04 | 2.03 | 2.02 | 2.11 |
| Robot 2 to Obstacle 1 | 3.86 | 3.30 | 3.69 | 3.81 | 5.63 | 5.56 | 8.04 | |
| Robot 2 to Obstacle 2 & 3 | 2.42 | 2.53 | 2.08 | 2.48 | 2.07 | 2.35 | 1.88 | |
| Robot 3 to Obstacle 2 & 3 | 3.84 | 3.37 | 2.06 | 2.76 | | | | |
| Robot 3 to Obstacle 4 | 2.68 | 2.56 | 1.52 | 2.05 | | | | |

## 10.3 Summary

In the present chapter, we considered a group of pointwise robots in a region occupied by static obstacles. The mission of the team was to traverse the area while avoiding the obstacles to trap a static target. We developed a novel decentralised navigation algorithm where the robots did not have any a priori information about the area except for information about the position of the target. In the proposed method, the robots found the safest path to the target autonomously based on the real time information they received by the on-board sensors in just a few switching steps.

Both mathematical analysis and simulations results confirmed the robustness and validity of the proposed algorithm. As explained in Section 4, the navigation laws (10.12) and (10.13) guaranteed that each individual in the fleet of multi-robots moved through the planned path based on the proposed navigation laws to reach the target while avoiding static obstacles, with minimum computation in the region. Furthermore, we proved that the power and sampling period could be reduced significantly using the proposed method.

# Chapter 11
# Virtual source/sink force field navigating method in an environment occupied by dynamic obstacles

In this chapter, we present a navigation algorithm for a mobile robot to avoid any type of obstacles in an unknown region. The proposed method is based on virtual force field, where, any object, whether static or dynamic supposed to be a source of repulsive forces and any spaces between every two obstacles in the region supposed to be a virtual sink of attractive forces. Then, we mathematically prove that, the polar angle of the source/sink vectors and the amplitude of the resultant force vector, if set in a certain interval, implies the orientation and the maximum distance that a robot is allowed to move in a cluttered area with no risk of collision to any type of obstacles.

The reminder of the chapter is organised as follows; Section 11.1, describes the problem statement; Section 11.2, presents the Source/Sink force field navigation strategy; The simulations in presented in Section 11.3; and finally, Section 11.4, gives a summary of the chapter.

## 11.1 Problem statement

A nonlinear model of a unicycle mobile robot could be considered as follows:

$$\dot{x}_i(t) = v_i(t)\,cos\big(\theta_i(t)\big)$$

$$\dot{y}_i(t) = v_i(t)\,sin\big(\theta_i(t)\big)$$

$$\dot{\theta}_i(t) = u_i(t)$$

$$\dot{\omega}_i(t) = \theta_i(t) \tag{11.1}$$

Various type of autonomous vehicle such as ground-base, aerial vehicles, missiles, etc., used the non-holonomic model (11.1),see e.g.[140] [104] [107], [218]–[221], and any references thein. In (11.1), $x_i(t)$ , $y_i(t)$ and $\theta_i(t)$ represent Cartesian coordinates and the polar angle of the robot in a given bounded region $\partial D$ ($D \subset \mathbb{R}^2$), respectively.

Furthermore, the linear and the angular velocity of the non-holonomic mobile robot $i$ are denoted by $v_i(t)$ and $\omega_i(t)$, where, the linear velocity vector $\vec{v}_i(t)$ varies between $(0, V^{max})$ and the polar angle $\theta_i$ takes value in the range of $(0,2\pi)$.

The region $D$ is a smooth area, has been occupied by impenetrable dynamic objects, which are able to rotate in any direction, deforming and merging with each other.

In this case, $\vec{v}_{o_j}$denotes the linear velocity of each obstacle and satisfies the following constraints:

$$\vec{v}_{o_j}(t) \leq \vec{v}_i(t) for\ j = 1,2, \dots, m \tag{11.2}$$

Moreover, we assume the outer face of each obstacle has been bounded by a smooth Jordan curve. Therefore, the effect of cups and cavity would be eliminated to the sensor measurements. A reference frame has been assumed to be attached to the centre of mass of the robot $i$ and the heading of the robot is considered towards the $x$-axis of the reference frame. Therefore, the robot measures a set of polar angles $\alpha_i^j$ and a set of distances $d_i^j(\alpha_i^j, t)$ in each time where $i = 1, 2, \dots, m$ denotes the number of visible points on the obstacle $j$ at time $t$.

### *Remark 11.1:*

Set $O = \{o_1, o_2, \dots, o_m\}$ includes a set of nonhomogeneous obstacles able to merge with each other. Therefore, any merged obstacles known as a single obstacle to the robot.

Furthermore, the obstacles move freely in the region, so the next direction of the obstacles is not predictable by the robot in each measuring step.

## 11.2 Source/sink force field navigation strategy

In this scenario, the path and the orientation of each obstacle has not been pre-defined. Therefore, the obstacles move arbitrary in the region $\partial D$ where $O_j|_t \neq O_{j'}|_{t'}\ \forall\, t \neq t'$.

Let $\mathcal{L}_j = \{l_j^i\}$ represents a set of lines of the rays from a sensor to a set of visible points $\wp_j = \{\mathscr{p}_j^i\}$ of each obstacle where $i = 1, \dots, n$ denotes the number of visible points and $j = 1, \dots, m$ denotes the number of obstacles in some time $t$. Furthermore, let the lines $l_j^1$ and $l_j^n$ be the tangent lines to the Jordan curve of each obstacle corresponding to the points $\mathscr{p}_j^1$ and $\mathscr{p}_j^n$ of the obstacle $j$.

### *Assumption 11.1:*

We say any neighbour obstacles $j^* < j^{**} \epsilon j$ are disjoint if and only if the following condition holds:

$$\left\|\mathscr{p}_{j^*}^n - \mathscr{p}_{j^{**}}^1\right\| \geq 2\sqrt{3}r \tag{11.3}$$

Where $j^*$ and $j^{**}$ denotes the permutation of the neighbour obstacles and $r$ represents the circumradius of the hypothetical circumscribed circle of the robot $i$.

### 11.2.1 Virtual force formation control

Each obstacle supposed to be a source of repulsive forces, therefore, it applies a repulsive force vector $\vec{f}_{ri}^{j}$ from the point $\wp_i^j$ to the robot at some time $t$. On the other hand, any obstacle-free areas between the obstacles that satisfies the inequality (11.3), is considered as source of attractive forces, therefore, it applies an attractive force vector $\vec{f}_{ai}^{j}$ to the robot at some time $t$ as well. As each point $\wp_i^j \epsilon \wp_j$ of the obstacles and the robot as well (see.Fig.11.1), considered to be a particle then the virtual repulsive force is defined as follows:

$$\mathcal{F}_r = \sum_{j=1}^{m} \sum_{i=1}^{n} k_r \frac{q_i^j q_{ro}}{(d_i^j)^2} \tag{11.4}$$

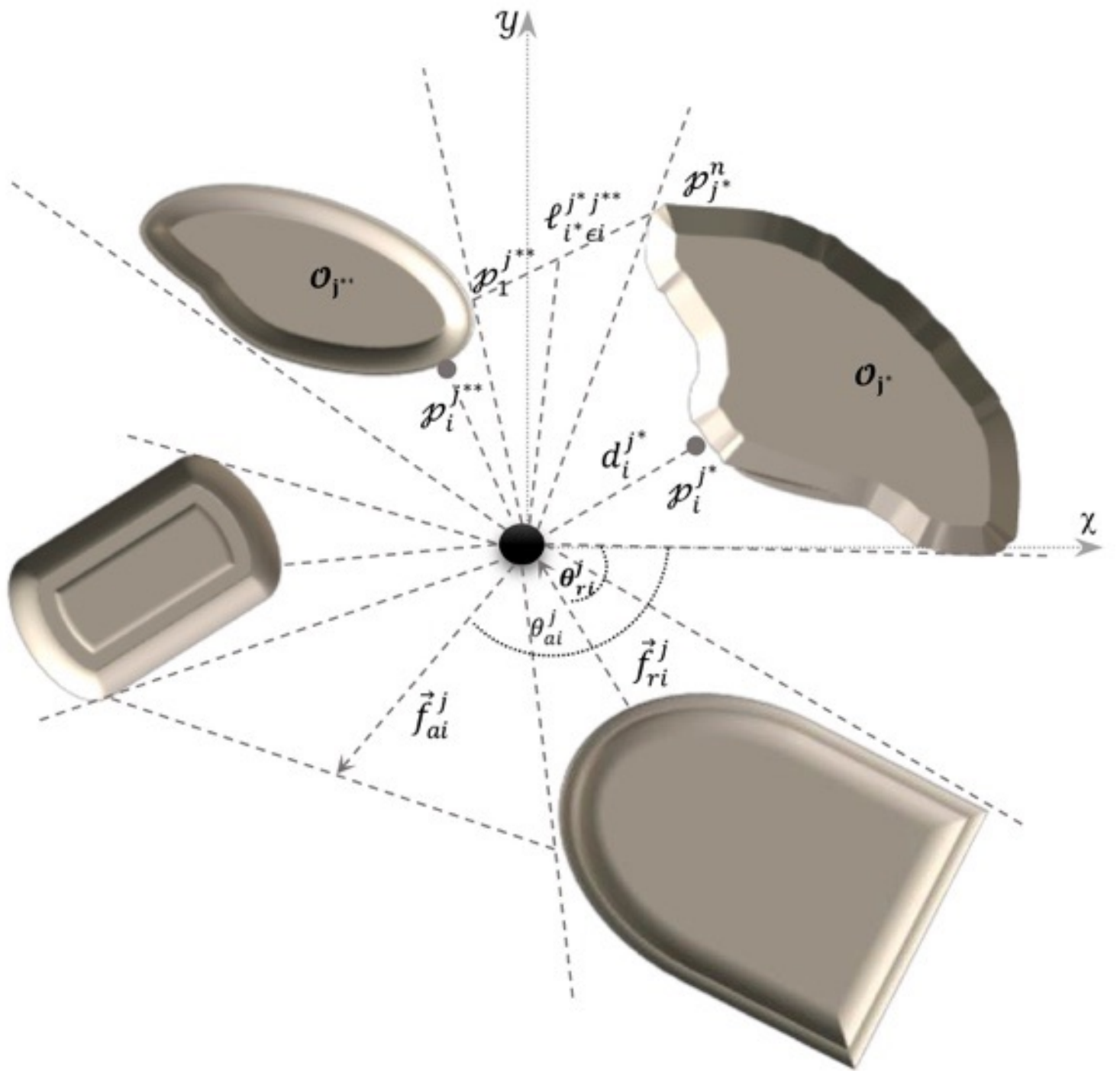

Figure 11.1: Robot’s sensing and measurements

Where $\mathcal{F}_r$ represents the resultant force of the all virtual repulsive forces exerted from each point $p_i^j \epsilon \wp_j$ of the obstacles to the robot. We considered $q_i^j$ and $q_{ro}$ as the charges of the particles on the obstacles and the pointwise robot respectively.

Now we need to find the virtual attractive forces, exerted from the obstacles to the robot. As shown in Fig.11.1, let $L_i^{j^* j^{**}}$ denotes a set of lines that connect the point $p_1^{j^{**}}$ of any obstacle $j^{**} \epsilon j$ to point $p_n^{j^*}$ of any obstacle $j^* \epsilon j$. Furthermore, let line $\ell_{i^* \epsilon i}^{j^* j^{**}} = \inf(L_i^{j^* j^{**}})$, be the infimum distance points $p_n^{j^*}$ and $p_1^{j^{**}}$. Then, we define $n$ virtual particles to exert attractive forces to the robot as $p_n^{j^*} < p_k^{j^* j^{**}} < p_1^{j^{**}}$, $k = 1,2,\dots,n-1$ with the coordinates:

$$\begin{pmatrix} \mathcal{X}_{p_k^{j^* j^{**}}} \\ \mathcal{Y}_{p_k^{j^* j^{**}}} \end{pmatrix} = \begin{pmatrix} \mathcal{X}_{p_{k-1}^{j^* j^{**}}} + \frac{\ell_{i^* \epsilon i}^{j^* j^{**}}}{n} \cos\gamma \\ \mathcal{Y}_{p_{k-1}^{j^* j^{**}}} + \frac{\ell_{i^* \epsilon i}^{j^* j^{**}}}{n} \sin\gamma \end{pmatrix} \tag{11.5}$$

and the attractive force equation as in:

$$\mathcal{F}_a = \sum_{j=1}^{m} \sum_{i=1}^{n} k_a \frac{q_i^j q_{ro}}{(\check{d}_i^j)^2} \tag{11.6}$$

Where, $\mathcal{F}_a$, denotes the resultant force of the all virtual force vectors attract the robot from the obstacle-free regions.

We decompose both repulsive and attractive forces in (11.4) and (11.6) as in:

$$\vec{\mathcal{F}}_{rxy} = \begin{cases} \mathcal{F}_{rx} = \sum_{j=1}^{m} \sum_{i=1}^{n} k_r \frac{q_i^j q_{ro}}{(d_i^j)^2} \cos(\theta_{ri}^j) \\ \mathcal{F}_{ry} = \sum_{j=1}^{m} \sum_{i=1}^{n} k_r \frac{q_i^j q_{ro}}{(d_i^j)^2} \sin(\theta_{ri}^j) \end{cases} \tag{11.7}$$

$$\vec{\mathcal{F}}_{axy} = \begin{cases} \mathcal{F}_{ax} = \sum_{j=1}^{m} \sum_{i=1}^{n} k_a \frac{q_i^j q_{ro}}{(\check{d}_i^j)^2} \cos(\theta_{ai}^j) \\ \mathcal{F}_{ay} = \sum_{j=1}^{m} \sum_{i=1}^{n} k_a \frac{q_i^j q_{ro}}{(\check{d}_i^j)^2} \sin(\theta_{ai}^j) \end{cases} \tag{11.8}$$

where $\theta_{ri}^{j} = tan^{-1}\left\|\frac{y_{p_j^i}-y_{ro}}{x_{p_j^i}-x_{ro}}\right\|$ and $\theta_{ai}^{j} = tan^{-1}\left\|\frac{y_{p_k^{j^*j^{**}}}-y_{ro}}{x_{p_k^{j^*j^{**}}}-x_{ro}}\right\|$.

Furthermore $q_i^j$ and $q_{ro}$, represent the electrical charges of each virtual point and the robot respectively. We supposed the robot as a pointwise vehicle, therefore, we presume $q_i^j = q_{ro}$. Moreover, $k_r$ and $k_a$ are given positive constants and considered to be equal.

Thereafter, we define a new constant $\psi$ as in:

$$\psi = k_r q_i^j q_{ro} = k_a q_i^j q_{ro} \tag{11.9}$$

The resultant force vector would be calculated as follows:

$$\vec{F} = \psi(\vec{\mathcal{F}}_{rxy} + \vec{\mathcal{F}}_{axy}) \tag{11.10}$$

The robot updates its heading and the linear velocity based on the amplitude $|F|_t$and the polar angle $\hat{\theta}$ of the resultant force vector at time $t$ as in:

$$\begin{cases} \dot{\mathcal{X}}_{ro|\hat{t}} = |F|_t \cos(\hat{\theta}t) \\ \dot{\mathcal{Y}}_{ro|\hat{t}} = |F|_t \sin(\hat{\theta}t) \end{cases}, \hat{t} > t \tag{11.11}$$

# *Theorem 11.1:*

Considering region $\partial D$ is occupied with a set of dynamic obstacles and a pointwise mobile robot $i$,where, constraint (11.2) and Assumption 11.1 holds. Furthermore, the robot updates its heading and the linear velocity based on updating rule (11.11) at any time $t$. The collision between the robot $i$, and the dynamic obstacles would be avoided in the region $\partial D$ if:

$$\psi \leq \inf\left(\frac{d_i^j}{2} - \varepsilon\right) \tag{11.12}$$

*Proof:*

As shown in Fig.8.2, Let $d_{i^*}^{j^*}$ bet the minimum distance of the robot to the point $\wp_{i^*} \epsilon \wp_{j^*}$ of the obstacle. Moreover, let $l_1^{j^*}$ and $l_n^{j^*}$ representing the tangent lines to the maximum visible points $\wp_1^{j^*}$ and $\wp_n^{j^*}$ of the obstacle $O_{j^*}$ to the robot with the angle $\beta$. We consider the case that $|F| = \left|d_{i^*}^{j^*}\right|$ is the maximum desired distance that the robot moves from its initial position $\wp_{\mathrm{ro}|_\mathrm{t}}$ in the region with the polar angle $\hat{\theta}$, where, $0 < \hat{\theta} < \beta$ to the next position $\wp_{\mathrm{ro}|_{\hat{\mathrm{t}}}}$. in this case, $t$ and $\hat{t}$ represent the first and the second sojourn time that is the time required for measurement and data gathering. On the other hand, as the obstacle moves arbitrary in the region, therefore, any single point on the outer face of the obstacle would be on the circumference of a circle with the radius $(\hat{r} + d_{i^*}^{j^*})$ at $\hat{t}$. Thus, if the obstacle moves toward the point $\wp_1^{j^*}$ with its maximum velocity $\vec{\mathcal{V}}^{max}{}_{o_{j^*}} = \vec{\mathcal{V}}^{max}{}_i$, (see Fig.11.2), then:

$$\Delta t = (\hat{t} - t) = \frac{(2\hat{r}\left|d_{i^*}^{j^*}\right| + \pi\hat{r}^2)}{\mathcal{V}^{max}{}_{o_{j^*}}} \qquad (11.13)$$

Hence, there is at least one point $\wp_{\hat{\imath}}^{j^*} \epsilon \wp_{j^*}$ at some time $\acute{t} < \hat{t}$ where:

$$\wp_{\mathrm{ro}|_\mathrm{t}} \cap \wp_{\acute{t}}^{j^*} \neq \emptyset \qquad (11.14)$$

unless (11.12) holds. Therefore, (11.12) guarantees a collision free motion of the robot in the interval of each sojourn time, regardless of the motion orientation of the obstacles. This is the complete proof of the Theorem 11.1.

In the worst-case scenario, if the robot and the closest obstacle moving towards each other, then there would be a minimum allowed distance between the robot and the point $\wp_{\hat{t}}^{j^*} \epsilon \wp_{j^*}$ as in:

$$\left\| \wp_{ro|_t} - \wp_{\hat{t}}^{j^*} \right\| \geq \varepsilon \tag{11.15}$$

According to Theorem 11.1, if $|F|_t \geq \frac{|d_i^j|}{2}$ , then there is possibility of collision between the robot and the obstacle therefore, we propose the following modification to the constant $\psi$ as follows:

$$\psi = \begin{cases} 1 \quad if \qquad |F|_t < \frac{|d_i^j|}{2} \\ inf\left(\frac{d_i^j}{2} - \varepsilon\right) \ if \ \ |F|_t \geq \frac{|d_i^j|}{2} \end{cases} \tag{11.16}$$

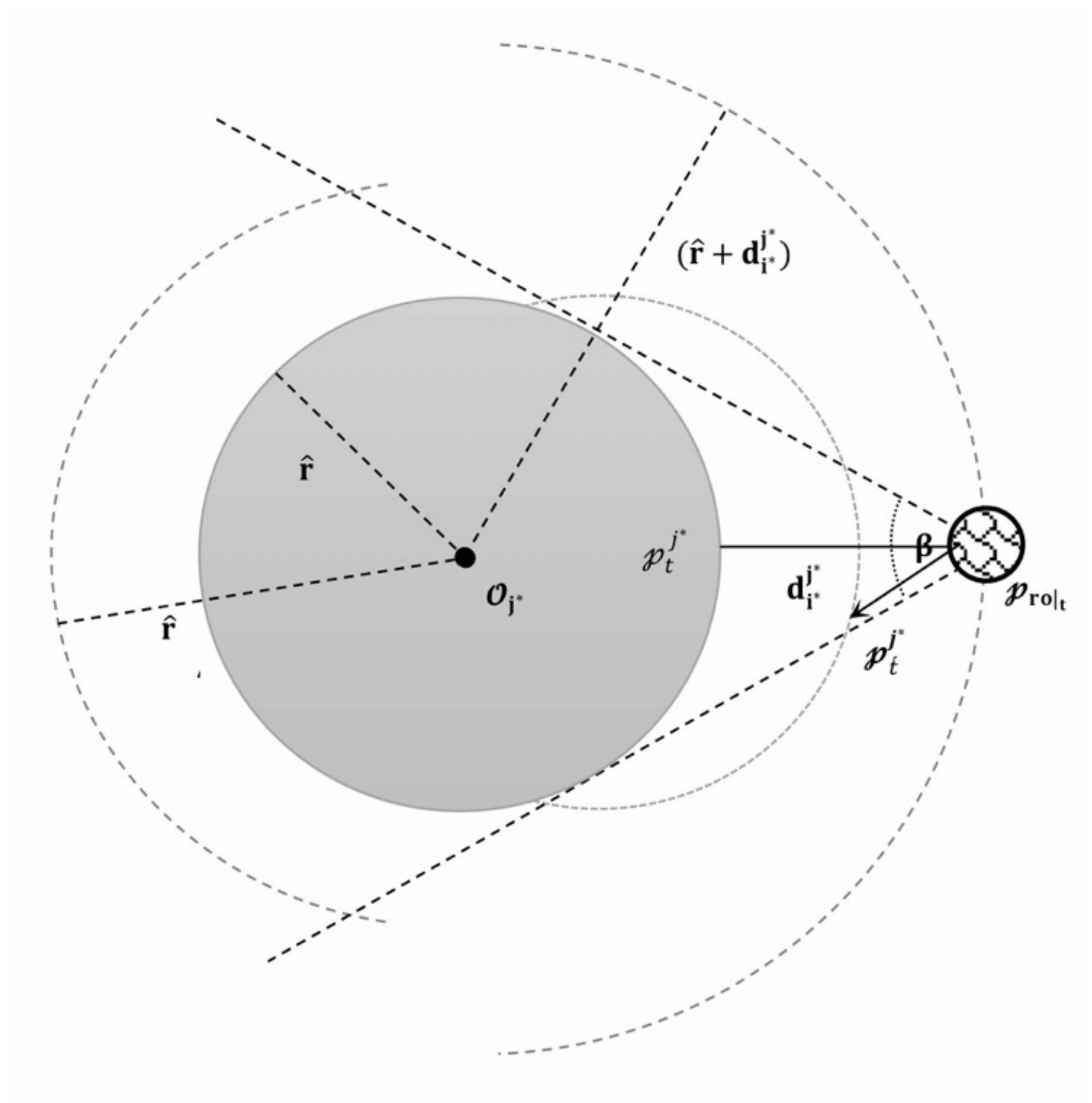

Figure 11.2. Circle shape obstacle moving towards the initial position of the robot

## 11.3 Simulations

In this section, MATLAB simulations confirms the validation of the proposed algorithm. Fig.11.3 (a, b, c, d), shows the path which planned by the robot to avoid collision with the dynamic obstacles in the region. The robot surrounded by 4 dynamic obstacles (3 ellipsoids and a circle shape) which arbitrary move in the area with the capability of merging and rotation in any direction while the constraint (11.2) is applied. In this scenario, neither the pointwise robot nor the obstacles have priori information of the region and the robot collect the local information based on the real-time sensor measurement in each sojourn time. The sojourn time ($t_{son}$) equals 0.01s in each measurement step and the $\varepsilon = 5\,Decimiter$. Fig.8.4, shows the minimum distances of the robot with the obstacles in each sojourn time. As shown in the Fig.11.4, the infimum of the minimum distances between the robot and the obstacles is $7.17\,Decimeter$ which is far enough to avoid any collision with the closest obstacle.

It is obvious that, the robot chooses the best direction with a proper velocity to avoid the collision regardless the motion direction of the obstacles, while using the proposed navigation algorithm.

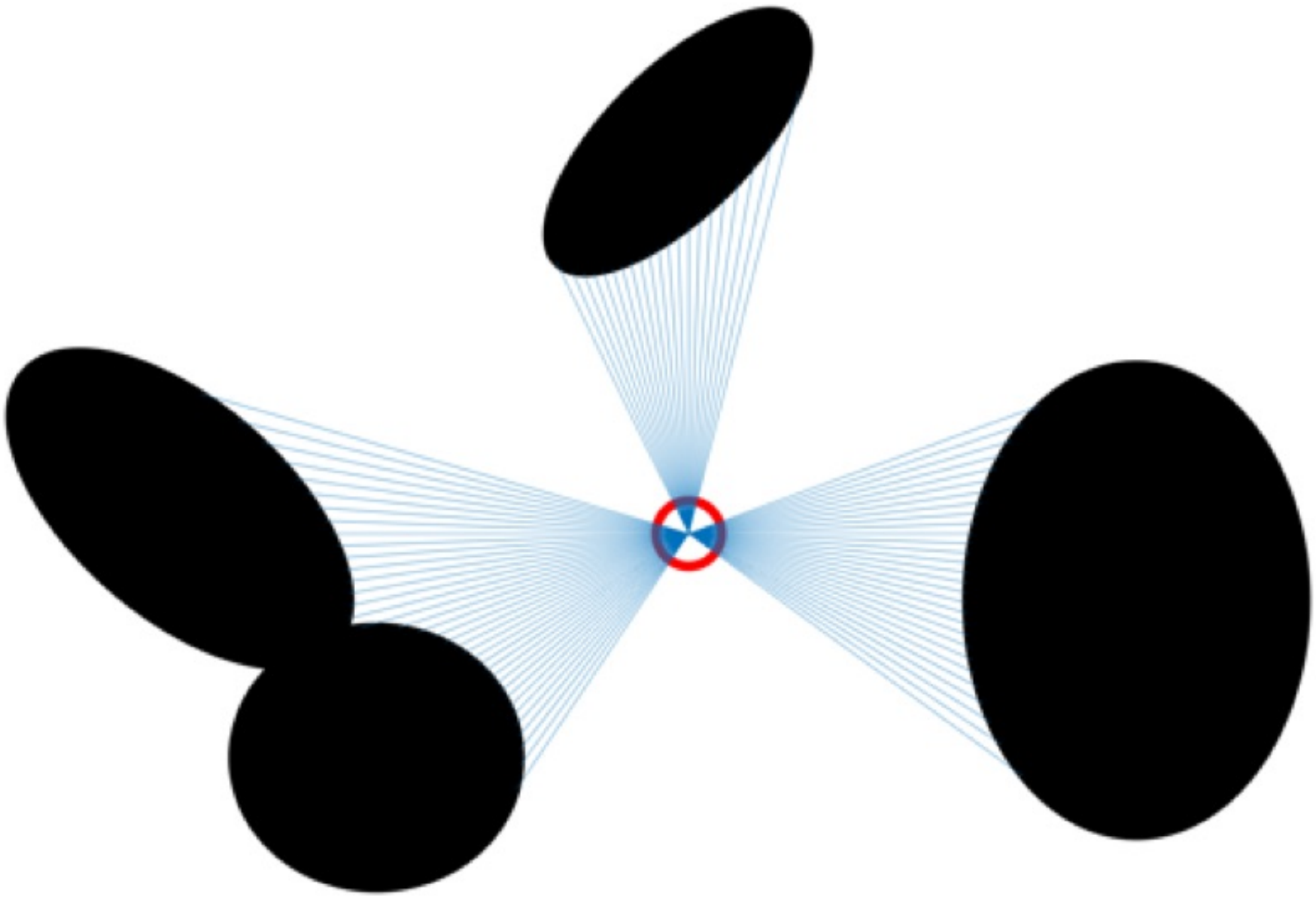

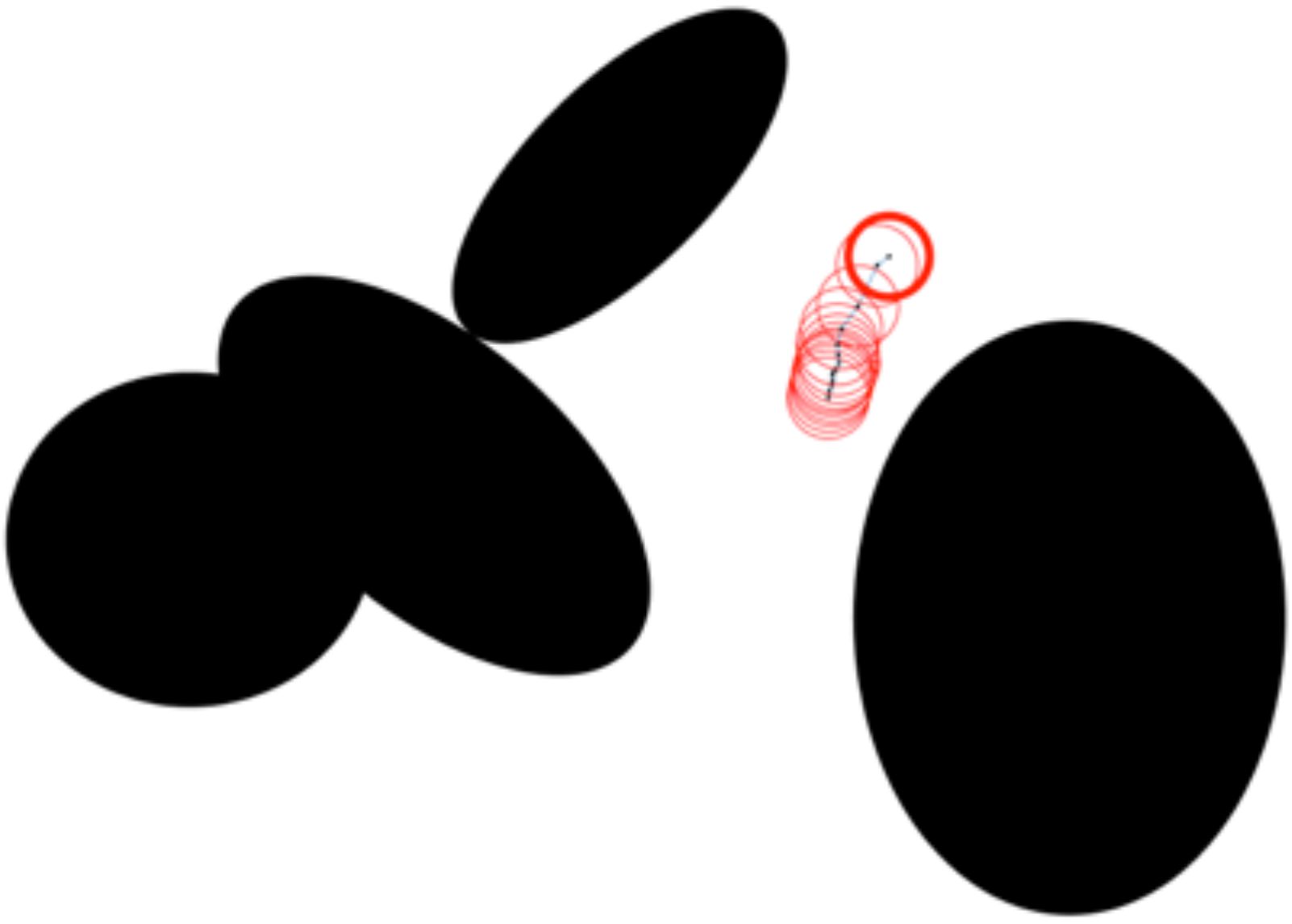

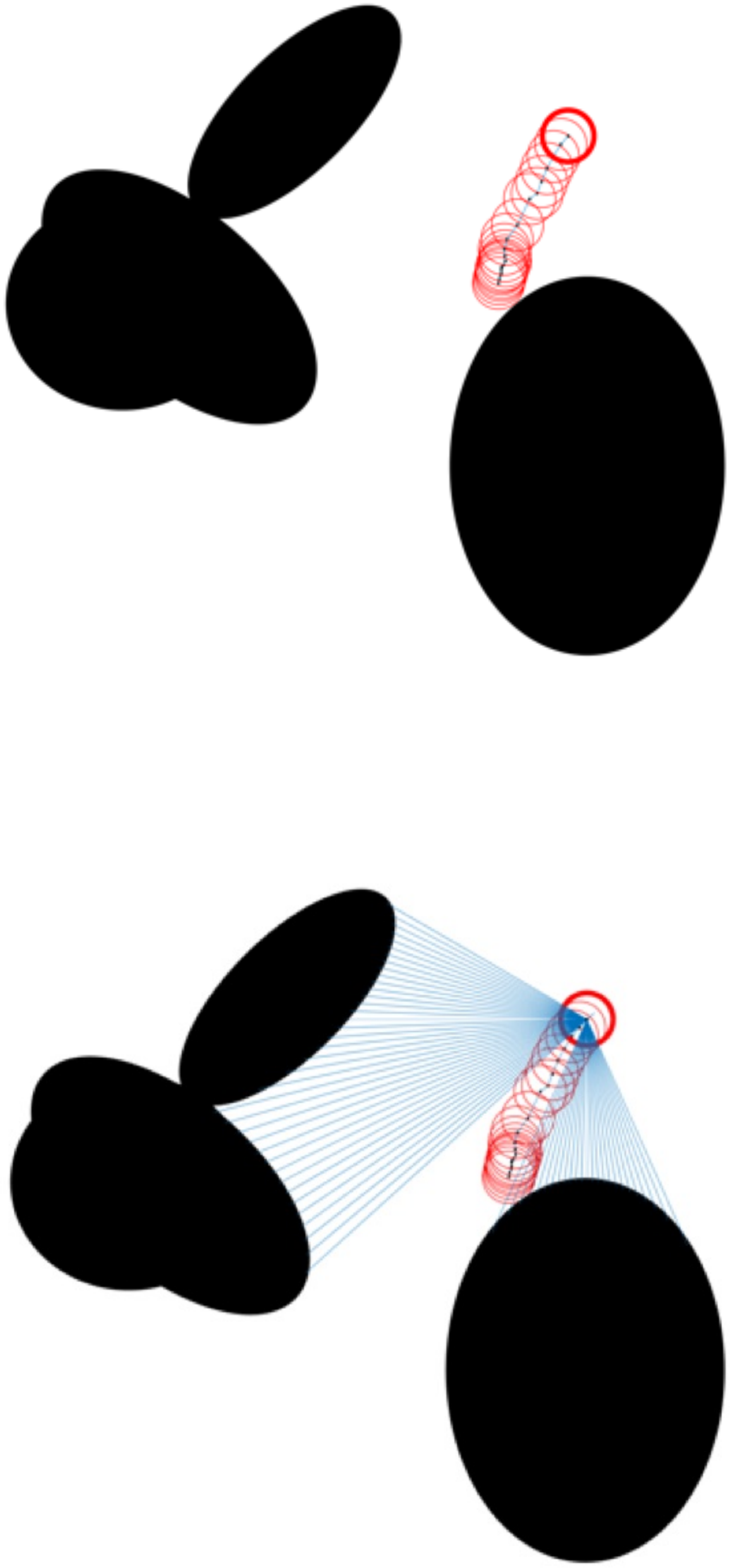

Figure 11.3 (a, b, c, d): Moving obstacles collision avoidance

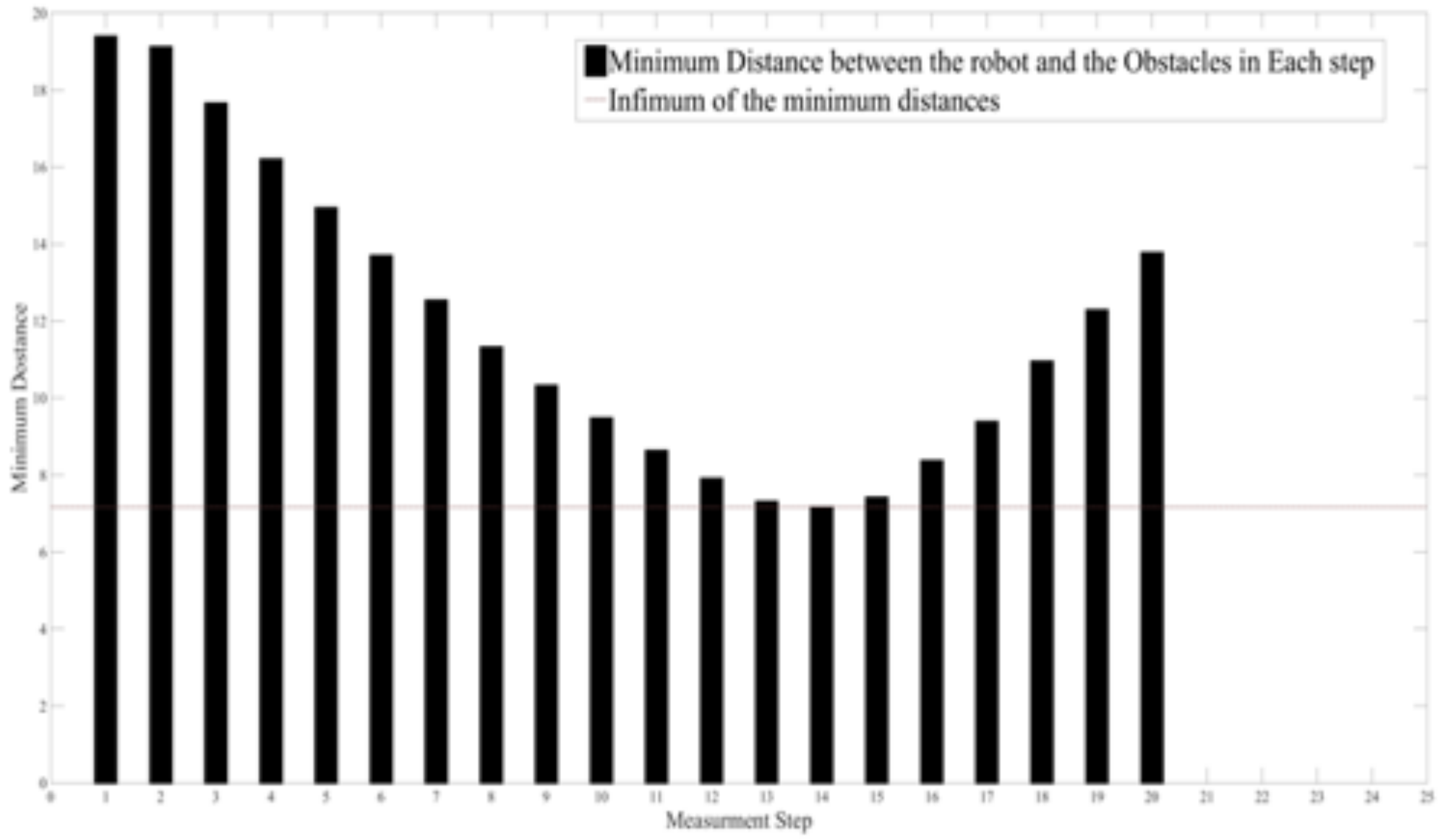

## 11.4 Summary

In this chapter, we presented a navigation strategy based on virtual force field method to navigate a unicycle robot in an unknown bounded region which is occupied by the dynamic obstacles. The obstacles supposed to move arbitrary in any direction with the capability of deformation, rotation, and merging. There is no any special limitation except the linear velocity of the obstacles that should not exceed the maximum velocity of the robot. In the proposed method, each obstacle was assumed to be a virtual repulsive force field and any obstacle-free area between the nearest neighbour obstacles considered as a sink of attractive force field. We mathematically proved the robot always plans the safest path to move between the obstacles while uses the proposed navigation algorithm. Furthermore, the simulations, confirmed the reliability and the robustness of the proposed navigation law.

# Chapter 12
# Conclusion

The main focus of this desertion is mainly concern with the problem of navigation of autonomous vehicles which are responsible for intrusion detection and target tracking in an unknown obstacle-ridden environment.

The first contribution of this work is the problem of *e-intercepting* proposed in Chapter 3. In this case, a bounded region $R$ is considered to be protected by a multi robot team and an intruder is trapped inside the region which tends to escape the region by crossing the boundary of the region from any given point $P$. The mission of the robots is to intercept the intruder when it's in closest distance to the boundary in a way that the maximum distance of at least one robot and the intruder in the neighbouring point of the given point $P$ on the boundary does not exceed $\varepsilon$ in all time. The simulations result of the algorithm confirms the validation and performance of the proposed method.

A decentralised *k-intercepting* strategy with all the necessary and sufficient conditions to a multi-robot system for protecting a boundary region against any unwanted intrusion. The proposed model guarantees that at there is always, at least $k$ robots are intercepting the intruder while the intruder is getting close to the region boundary (see Chapter 4). A rigorous mathematical proof is provided for the model in addition to the simulations and illustrative examples verifying the robustness of the novel proposed method in intrusion detection and boundary protection missions. The *k-intercepting* navigation law is an original intercepting strategy with a non-demanding computational quiddity which makes it easy to implement in real time boundary protection applications.

In continue, Chapter 5 presented a decentralized navigation method to investigate the problem of hunting a prey which is trapped in a hypothetical siege ring which is created by a team of guardian robots. The objective of the prey is to escape the siege ring however the multi-robot team willing to maintain the prey inside the ring using the proposed navigation strategy. The strategy is a developed interpretation of the e-intercepting method which is proposed in Chapter 4. In this algorithm, the siege ring is divided in two equal sections. Furthermore, the multi-robot team is also divided in two groups, each includes the same number of robots. The mobile robots in each group are responsible for intercepting the intruder in just one section of the siege ring. The proposed navigation method guarantees that the intruder interception of every individual point on the siege ring when the intruder tries to cross the boundary to escape the region. The simulations results confirm the validity, robustness and the reliability of the proposed algorithm.

Chapter 6 proposed modified version of the intruder's interception in the scenario of hunting and escaping. The model presented in this chapter investigate the case that, there are multiple preys trapped in the siege ring which is created by multiple guardian robots which are moving on a curve to maintain the intruders inside the region. The region is unknown and smooth and there is neither static nor dynamic obstacles in the region. Unlike the robots that move on a curved path just in left and right side, the intruders can move arbitrary inside the region to find the best point to cross the boundary and escaping the region. however, rigorous mathematical proof of the model in addition to the simulations results, confirm that it's impossible for the intruders to escape the region from any point on the boundary without being intercepted by at least one robot in all time.

In Chapter 7. An intelligent game-based strategy (IGD) is developed for the purpose of intrusion detection by a multi-robot team in a bounded region. In this case scenario, the robots are considered to have a limited communication or even no communication with each other result from jamming attacks by a given hostile. Furthermore, unlike *k-intercepting navigation* method in which the intruder considered to be visible by the robot in the entire duration of the mission, in this navigation strategy the intruder is invisible, unless it is within the sensing range of any of the team members. In the proposed navigation strategy, mobile robots considered as players of the strategy of the game which play cooperative or non-cooperative game regarding the communication situation, which means, if the members connected, they share the most recent information they collected from the area of interest and play a cooperative game to maximise the payoff for the entire team based on pareto-optimality decision strategy. On the other ways, if they are not connected, each member needs to make a Nash equilibrium decision to minimise the cost of the game in the interest of entire team as well. And consequently, maximizing the probability of detection of the intruder, not even on the boundary, but also inside the region. As a result, the novel navigation method allows each individual member of the team to operate fully autonomously in a security operation. The initial formation proposed in this method increased the probability of intrusion detection significantly. Comparing the proposed game-based strategy with the swiping coverage strategy has indicated a dramatic improvement in the performance of the multi-robot team with 79% success in intrusion detection.

In Chapter 8, a semi-decentralised navigation problem of a multi-robot team in an area occupied by a convex static obstacle is discussed. The proposed navigation strategy is developed based on the concept of leader-follower strategy. In this method the area is unknown, and the robots does not have any priori information about the region. Furthermore, they will not receive any global information about the environment from an external source. In the proposed navigation method, the leader measures and calculate the desired heading and estimates the next position which is called the switching position. furthermore, the time elapsed by the leader for measuring and calculation in each switching position, called the sojourn time.

the updating information transfers from each robot to its nearest neighbour in a strand chain communication network from the leader to the last member in the chain. This position estimation switching algorithm (PSEA), guarantees a collision free navigation in the region. In Chapter 9., a modified version of the PSEA navigation strategy is presented. The modified PSEA, is designed for a group of mobile robots to navigate in an obstacle-ridden environment which is occupied by multiple static obstacles. Furthermore, the proposed method allows the robots move in the region with no risk of collision, regardless the shape of the obstacles. Additionally, a complementary decision-making rule has been proposed, to let the robots choose the best path when, confronting to more than one pathway.

A fully decentralised navigation strategy for the purpose of target trapping in an obstacle-ridden environment is presented in Chapter 10. In the proposed method each mobile robot plans the best path autonomously, regardless the position and the situation of the other team members to trap the target. The obstacles are considered to be static, therefore, if any of the robots stop working in the region, the other treat it as a static obstacle. Furthermore, the proposed navigation method results in energy saving and increasing the battery life of the mobile robots.

Finally, a novel navigation strategy is proposed in Chapter 11, regarding the problem of ground-based autonomous vehicle path planning in an area which is occupied by dynamic obstacles. The method which is called Virtual Source/Sink Force Field Navigating strategy. The proposed method navigation law which is applied to a single autonomous vehicle. The dynamic obstacles could be in any shape with the capability of merging, rotation and moving in any direction in the area. The area of interest is unknown to the robot and the prediction of the next motion orientation of the obstacles is impossible by the robot. The only constraint of this method is the velocity of the obstacles that should not exceed the maximum velocity of the robot. The proposed method guaranties that the robot can move in the area with no risk of collision to the obstacles regardless the motion direction and the orientation of the obstacles. Furthermore, any merged obstacles are treated as a single obstacle by the robot.

The entire proposed algorithms have been proved mathematically and the robustness of them have been validated by the simulations at the end of each chapter.

# Future work

In this report, we proposed two methods for the problem of intrusion detection in a bounded environment. In case of *k-intercepting* problem *e-intercepting* problem and the problem of hunters and preys which are presented in chapters 3,4,5 and 6 respectively, the region supposed to be static 2D environment with no obstacles. Therefore, the proposed models could be extended to a cluttered dynamic 3D environment such as see or air as a future work. Furthermore, the problem of multi-intruder attack didn't consider in the *k-intercepting* and *e-intercepting* which are proposed in Chapters 3 and 4. Moreover, this method is designed for ground-based mobile robots that could be modified for marine or aerial vehicles as well. Similarly, the problem of multi attack and path planning in a cluttered area are not considered in the proposed IGD method. Furthermore, intruder(s) could be considered as rational players in the game as well. In this case a more complex payoff matrix is required to decrease cost function for the benefit of the pursuers. On the other hand, this method could be applied to the blanket coverage problem. In the blanket coverage method, the mobile robots are moving to form in an optimal deployment to fully monitor the area which means every single point of the region is sensed by at least one robot and detect any unwanted intruder consequently [222]–[226]. The proposed model could be applied to the blanket coverage problem to minimise the number of sensor nodes as well as make a dynamic pattern to optimise the coverage based on the different situations. The model proposed in Chapter 9, was modified in Chapter 9 and Chapter 10, however, the case of multiple dynamic obstacles and multiple dynamic and static targets can be considered as important subjects which required more research. And finally, in Chapter 11, a model proposed for problem of dynamic obstacle avoidance merely, for one robot. Modifying the model to adapt it to a multi-robot team navigation control in a region occupied by static and dynamic obstacles could be addressed as a potential future work in this matter. Furthermore, the proposed models in this report could be extended to work on non-linear non-holonomic models which describe motion of many mobile robots, missile, underwater and marine vehicles [227]–[232].